\documentclass{sapthesis}
\usepackage[utf8]{inputenc}
\usepackage{amsmath}
\usepackage{amssymb}
\usepackage{dutchcal}
\usepackage{graphicx}
\usepackage{color}
\usepackage{algpseudocode}
\usepackage[thinc]{esdiff}
\usepackage{hyperref}
\usepackage[
backend=biber,
style=alphabetic,
backref,
hyperref
]{biblatex}
\usepackage{color}
\usepackage{textcomp}
\usepackage{braket}
\usepackage{comment}
\usepackage{listings}
\usepackage{bm}
\usepackage[chapter]{algorithm}

\lstdefinestyle{mystyle}{
    backgroundcolor=\color{backcolour},   
    commentstyle=\color{codegreen},
    keywordstyle=\color{magenta},
    numberstyle=\tiny\color{codegray},
    stringstyle=\color{codepurple},
    basicstyle=\ttfamily\footnotesize,
    breakatwhitespace=false,         
    breaklines=true,                 
    captionpos=b,                    
    keepspaces=true,                 
    numbers=left,                    
    numbersep=5pt,                  
    showspaces=false,                
    showstringspaces=false,
    showtabs=false,                  
    tabsize=2
}

\title{Low energy excitations of vector spin glasses}
\author{Flavio Nicoletti}
\IDnumber{1597815}
\authoremail{flavio.nicoletti@uniroma1.it}
\copyyear{2023}
\submitdate{2022/2023}
\advisor{Prof. Federico Ricci-Tersenghi (Università di Roma 'La Sapienza')}
\advisor{Prof. Silvio Franz (LPTMS, Université Paris-Saclay)}
\date{May 2023}
\course{PhD in Physics: Joint International Cotutelle between Università di Roma 'La Sapienza' and Université Paris-Saclay}
\courseorganizer{Scuola di Dottorato 'Vito Volterra' and EDPIF (École Doctorale Physique en Île de France)}
\copyrightstatement{\copyright 2023 Flavio Nicoletti. All rights reserved
\newline

The research presented in this manuscript has benefited of financial support from Simons Foundation, grants No. 454941, S. Franz and No. 454949, G. Parisi.}
\cycle{XXXV}
\versiondate{May 3, 2023}
\thesistype{PhD thesis}
\reviewerlabel{Reviewers}
\reviewer{Prof. Pierluigi Contucci}
\reviewer{Prof. Allan Peter Young}
\examdate{May 9 2023}

\examiner{Prof. Irene Giardina}
\examiner{Prof. Ada Altieri}
\examiner{Prof. Pierluigi Contucci}
\examiner{Prof. Alberto Rosso}
\examiner{Prof. Giacomo Gradenigo}
\examiner{Prof. Andrea Gabrielli}

\begin{document}

\newcommand{\Tr}{\operatorname{Tr}}
\newcommand{\Supp}{\operatorname{Supp}}
\newcommand{\vc}[1]{\boldsymbol{#1}}
\newcommand{\cH}{\mathcal{H}}
\newcommand{\cHp}{\bm{\mathcal{H}}^\prime}
\newcommand{\bS}{\bm{S}}
\newcommand{\bb}{\bm{b}}
\newcommand{\bh}{\bm{h}}
\newcommand{\bH}{\bm{H}}
\newcommand{\bu}{\bm{u}}
\newcommand{\bv}{\bm{v}}
\renewcommand{\Im}{\operatorname{Im}}
\newcommand{\tP}{\widetilde{P}}
\newcommand{\IPR}{\text{IPR}}
\newcommand{\Tonset}{T_\text{onset}}
\renewcommand{\Re}{\text{Re}\,}
\renewcommand{\Im}{\text{Im}\,}

\maketitle

\tableofcontents

\chapter*{Abstract}

\section*{Abstract in English}

The work of this thesis concerns the problem of linear low energy excitations of vector spin glass models. An analytical and numerical study is carried out, considering a fully connected random-field Heisenberg model at zero temperature, a fully-connected vector p-spin glass model and a sparse random-field Heisenberg model. We test these models against the low temperature behavior of finite dimensional glassy systems, in particular we show that they posses phases where the density of states is gapless with quasi-localised modes. In the case of the sparse model, we show that the density of states follows a quartic law at low frequency, consistently with several recent measures of this quantity that can be found in the literature of computer glasses. In all the three models, the spin glass transition is characterised in terms of the behavior of the softest excitations. We found that in the fully connected models the zero temperature spin glass transition in a field is a delocalisation transition of the softest modes. In the sparse case, a weaker form of delocalisation appears at the transition. These results broaden our understanding of the zero temperature critical point, by showing how spin glass ordering affects the way the system responds to small magnetic perturbations.

\section*{Abstract in Italiano}

Il lavoro contenuto in questa tesi riguarda il problema dell'eccitazioni di bassa energia dei modelli di vetri di spin vettoriali. Viene proposto uno studio analitico e numerico di tre modelli: il primo consiste in un vetro di spin di Heisenberg con campo magnetico esterno random con grafo di interazioni denso, il secondo in un modello p-spin di Heisenberg con grafo di interazioni denso, il terzo infine in un modello di Heisenberg con campo magnetico esterno random e grafo di interazioni diluito.
Questi modelli sono valutati rispetto al comportamento dell'eccitazioni dei sistemi vetrosi a basse temperature: in particolare, nella tesi si mostra che questi modelli posseggono delle fasi in cui la densità degli stati è senza gap e i modi sono quasi-localizzati. Nel modello sparso la densità degli stati ha una dipendenza quartica dalla frequenza, in accordo con molteplici misure di questa quantità reperibili dalla letteratura sui modelli vetrosi computazionali. In tutti e tre i casi di studio, la transizione nella fase del vetro di spin è caratterizzata rispetto al comportamento dei modi soffici. Troviamo che nei modelli densi la transizione del vetro di spin è una transizione di delocalizzazione dei modi soffici. Nel caso sparso, la delocalizzazione alla transizione si manifesta in forma più debole. Questi risultati ampliano la nostra comprensione del punto critico di temperatura nulla, mostrando come l'emergenza di un ordinamento da vetro di spin modifichi la risposta del sistema a piccole perturbazioni magnetiche.

\section*{Abstract en Français}

Le travail de cette thèse concerne le problème des excitations linéaires à basse énergie des modèles de verre de spin vectoriel. Une étude analytique et numérique est menée, considérant un modèle de Heisenberg à champ aléatoire entièrement connecté à température nulle, un modèle de verre de spin p vectoriel entièrement connecté et un modèle de Heisenberg dilué à champ aléatoire. Nous testons ces modèles par rapport au comportement à basse température des systèmes vitreux de dimension finie, en particulier nous montrons qu'ils possèdent des phases où la densité d'états est sans lacunes avec des modes quasi-localisés. Dans le cas du modèle dilué, nous montrons que la densité d'états suit une loi quartique à basse fréquence, en accord avec plusieurs mesures récentes de cette quantité que l'on peut trouver dans la littérature des modèles computationnels de verres. Dans les trois modèles, la transition de verre de spin est caractérisée en termes de comportement des excitations les plus douces. Nous avons constaté que dans les modèles entièrement connectés, la transition de verre de spin à température zéro dans un champ est une transition de délocalisation des modes les plus douces. Dans le cas dilué, une forme plus faible de délocalisation apparaît à la transition. Ces résultats élargissent notre compréhension du point critique à température zéro, en montrant comment l'ordre du verre de spin affecte la façon dont le système répond à de petites perturbations magnétiques.


\dedication{A Mamma, Papà e Saverio, per ogni cosa: in particolare, per il supporto costante ricevuto in questi mesi cruciali.\newline\newline Ai Bros, amici di sempre e famiglia acquisita, per esserci sempre stati.\newline\newline Ai ragazzi del LPTMS, per aver reso il mio soggiorno in Francia indimenticabile.}

\chapter*{Acknowledgements}

This manuscript condensates three-year efforts of my PhD project. These three years have been rich of intellectual and social stimuli, thanks to several people encountered in this period. It has been the most interesting and horizons-broadening period of my life. In the following, I wish to make some special mentions. 
\newline
\newline
First of all, My supervisors, Prof. Federico Ricci-Tersenghi from "Sapienza" and Prof. Silvio Franz from "Paris-Saclay". They had a crucial role in my scientific and intellectual growth over the last three years. It has been a privilege and an honour to have the chance to learn from and work with them.
\newline
\newline
I would like to thank the two reviewers of this thesis, Prof. Pierluigi Contucci and Prof. Allan Peter Young, for their kind reports. They gave me interesting suggestions which allowed me to improve the manuscript on some passages. They demonstrated a sincere appreciation of my work.
\newline
\newline
I acknowledge interesting scientific discussions with Raoul Santachiara, Marco Tarzia, Gianmarco Perrupato, Marco Benedetti, Jacopo Niedda, Enrico Ventura, Vincenzo Schimmenti, Fabian Aguirre Lopez, Mauro Pastore and many others encountered in Summer Schools and Conferences.
\newline
\newline
I acknowledge Simons Foundation for involving me, through my supervisor S. Franz, in the activities of their collaboration 'Cracking the glass problem' through financial support.

\chapter*{Introduction}
\addcontentsline{toc}{chapter}{Introduction}

\pagenumbering{roman}

\section*{Introduction in English}

Spin glass models are an example of the power of simple but complex ideas. From their very first introduction in the seventies, by the hands of P. W. Anderson \cite{anderson1972anomalous} and subsequent models \cite{edwards1975theory, sherrington1975solvable}, they have attracted the interest of many scientist, first in the condensed matter community and secondly from other communities, also external to physics. In particular, the breakthroughs of Giorgio Parisi \cite{parisi1979toward, parisi1979infinite, parisi1980order, parisi1980sequence, parisi1980magnetic} have opened the Pandora box of complex systems. Beside the natural connection between spin glasses, structural glasses and granular systems, applications to neuroscience \cite{hopfield1982neural, amit1989modeling}, optimization problems \cite{mezard2002analytic}, protein folding \cite{wolynes2005energy} and many others confer to the original simple problem the status of archetype of disorder and complexity.

While the first efforts were mainly concentrated on discrete models, spin glass models with continuous degrees around the nineties acquired great importance thanks to the introduction of p-spin spherical models. These models are mean field toy models of the glass transition: they have a phenomenology that features many dynamical behaviors found in experiments \cite{castellani2005spin}. Beside spherical models, vector spin glasses have been studied since the beginning \cite{de1978infinite, gabay1981coexistence, gabay1982symmetry, cragg1982instabilities, moore1982critical}. Vector models are more closer to real systems than discrete ones: in particular, they allow for the study of magnetic response at low temperatures, featuring arbitrarily small excitations: thus, vector models are ideal proxies to understand the spectrum of harmonic excitations in disordered systems. 
In recent years, there has been a renown interest in vector models \cite{coolen2005finitely, sharma2010almeida, baity2015inherent, baity2015soft, sharma2016metastable, yeo2004complexity, lupo2017approximating, lupo2018comparison, lupo2019random}. While dense models of vector spin glasses have been studied with greater detail, very few results are available on sparse vector spin glasses. Many properties of these models remained so far undisclosed: in particular, the relation between the spin glass transition and the statistical properties of low energy excitations. 

In this thesis work, we consider two related problems: the behavior of excitation spectra at zero temperature, i.e. how the system responds to small perturbations around a stable configuration, and the properties of the zero temperature critical point.

In part I we provide a background, covering spin glasses, structural glasses and random matrices. The introduction of these topics is functional to the discussion made in the following of the manuscript.

In part II we discuss a fully connected vector spin glass model, subjected to the action of a random field: we show that the zero temperature spin glass critical point features a delocalisation transition for the lowest modes of the density of states, from a phase with localised states at the lower edge of the spectrum to a delocalised spin glass phase. We show that these results, when applied to a vector p-spin model, prove the existence of a stable band of the energy landscape with localised excitations.

In part III, we consider a sparse spin glass model under a random field, defined on a random regular graph. For sufficiently weak external field, localised soft excitations, featuring a quartic DoS at low frequency, appear.
These excitations are ubiquitous in finitely-connected disordered systems. Finally, we show that at the spin glass transition a delocalisation transition can occur, even though in a weaker form than that observed in the dense case. 

\section*{Introduzione in Italiano}

I modelli di vetro di spin sono un esempio di come da dell'idee semplici possa scaturire un'inaspettata complessità. Fin dalla loro prima introduzione negli anni Settanta, per mano di P. W. Anderson \cite{anderson1972anomalous} e dei modelli successivi \cite{edwards1975theory, sherrington1975solvable}, questi modelli hanno attirato l'interesse di molti scienziati, in primo luogo nella comunità della materia condensata e in secondo luogo in altre comunità, anche esterne alla fisica. In particolare, le scoperte di Giorgio Parisi \cite{parisi1979toward, parisi1979infinite, parisi1980order, parisi1980sequence, parisi1980magnetic} hanno aperto il vaso di Pandora dei sistemi complessi. Oltre alla naturale connessione tra vetri di spin, vetri strutturali e sistemi granulari, le applicazioni alle neuroscienze \cite{hopfield1982neural, amit1989modeling}, ai problemi di ottimizzazione \cite{mezard2002analytic}, al ripiegamento delle proteine \cite{wolynes2005energy} conferiscono al semplice problema originale lo status di archetipo del disordine e della complessità.

Mentre i primi sforzi si sono concentrati principalmente su modelli discreti, i modelli di spin glass con gradi continui hanno acquisito grande importanza intorno agli anni Novanta grazie all'introduzione dei modelli sferici p-spin. Questi sono modelli di campo medio della transizione vetrosa: hanno una fenomenologia che presenta molti comportamenti dinamici riscontrati negli esperimenti \cite{castellani2005spin}. Oltre ai modelli sferici, i vetri di spin vettoriali sono stati studiati fin dall'inizio \cite{de1978infinite, gabay1981coexistence, gabay1982symmetry, cragg1982instabilities, moore1982critical}. I modelli vettoriali sono più vicini ai sistemi reali rispetto a quelli discreti; in particolare, permettono di studiare la risposta magnetica a basse temperature, con eccitazioni arbitrariamente piccole: sono quindi ideali per comprendere lo spettro delle eccitazioni armoniche nei sistemi disordinati. 
Negli ultimi anni si è assistito a un rinnovato interesse per i modelli vettoriali \cite{coolen2005finitely, sharma2010almeida, baity2015inherent, baity2015soft, sharma2016metastable, yeo2004complexity, lupo2017approximating, lupo2018comparison, lupo2019random}. Mentre i modelli densi sono stati studiati con maggiore dettaglio, sono disponibili pochissimi risultati sui vetri di spin vettoriali sparsi. Molte proprietà di questi modelli sono rimaste finora sconosciute: in particolare, la relazione tra la transizione di spin glass e le proprietà statistiche delle eccitazioni a bassa energia. 

In questo lavoro di tesi, consideriamo due problemi correlati: il comportamento degli spettri di eccitazione a temperatura zero, cioè come il sistema risponde a piccole perturbazioni intorno a una configurazione stabile, e le proprietà del punto critico a temperatura zero.

Nella prima parte forniamo un inquadramento di base, che riguarda i vetri di spin, i vetri strutturali e le matrici casuali. L'introduzione di questi argomenti è funzionale alla trattazione del seguito del manoscritto.

Nella seconda parte discutiamo un modello di vetro di spin vettoriale con rete d'interazioni completamente connessa, soggetto all'azione di un campo aleatorio: dimostriamo che il punto critico del vetro di spin a temperatura zero presenta una transizione di delocalizzazione per i modi soffici della densità di stati, da una fase con stati localizzati al margine inferiore dello spettro a una fase delocalizzata del vetro di spin. Mostriamo che questi risultati, applicati a un modello vettoriale p-spin, dimostrano l'esistenza di una banda stabile del paesaggio energetico, con eccitazioni localizzate.

Nella terza parte, consideriamo un modello di spin glass sparse sotto un campo stocastico, definito su un grafo aleatorio regolare. Per un campo esterno sufficientemente debole, appaiono eccitazioni localizzate soffici, caratterizzate da una densità degli stati con dipendenza quartica a bassa frequenza.
Queste eccitazioni sono onnipresenti nei sistemi disordinati con connettività sparsa. Infine, mostriamo che alla transizione spin glass si può manifestare una transizione di delocalizzazione, in una forma più debole rispetto a quanto osservato nel caso diluito.

\section*{Introduction en Français}

Les modèles de verre de spin sont un exemple de la complexité inattendue qui peut naître d'idées simples. Depuis leur première introduction dans les années 1970, par P. W. Anderson \cite{anderson1972anomalous} et les modèles ultérieurs \cite{edwards1975theory, sherrington1975solvable}, ces modèles ont suscité l'intérêt de nombreux scientifiques, d'abord dans la communauté de la matière condensée et ensuite dans d'autres communautés, y compris celles qui ne relèvent pas de la physique. En particulier, les découvertes de Giorgio Parisi \cite{parisi1979toward, parisi1979infinite, parisi1980order, parisi1980sequence, parisi1980magnetic} ont ouvert la boîte de Pandore des systèmes complexes. Outre le lien naturel entre les verres de spin, les verres structuraux et les systèmes granulaires, les applications aux neurosciences \cite{hopfield1982neural, amit1989modeling}, aux problèmes d'optimisation \cite{mezard2002analytic}, au repliement des protéines \cite{wolynes2005energy} confèrent au simple problème initial le statut d'archétype du désordre et de la complexité.

Alors que les premiers efforts portaient principalement sur des modèles discrets, les modèles de verre de spin à degrés continus ont pris de l'importance vers les années 1990 avec l'introduction des modèles de spin p sphériques. Il s'agit de modèles à champ moyen de la transition vitreuse : leur phénoménologie présente de nombreux comportements dynamiques observés dans les expériences {castellani2005spin}. Outre les modèles sphériques, les verres de spin vectoriels ont été étudiés dès le début \cite{de1978infinite, gabay1981coexistence, gabay1982symmetry, cragg1982instabilities, moore1982critical}. Les modèles vectoriels sont plus proches des systèmes réels que les modèles discrets ; en particulier, ils nous permettent d'étudier la réponse magnétique à basse température, avec des excitations arbitrairement petites : ils sont donc idéaux pour comprendre le spectre des excitations harmoniques dans les systèmes désordonnés. 
Ces dernières années ont vu un regain d'intérêt pour les modèles vectoriels \cite{coolen2005finitely, sharma2010almeida, baity2015inherent, baity2015soft, sharma2016metastable, yeo2004complexity, lupo2017approximating, lupo2018comparison, lupo2019random}. Alors que les modèles denses ont été étudiés plus en détail, très peu de résultats sont disponibles sur les verres à vecteurs de spin épars. De nombreuses propriétés de ces modèles sont restées inconnues jusqu'à présent : en particulier, la relation entre la transition du verre de spin et les propriétés statistiques des excitations de basse énergie.

Dans ce travail de thèse, nous considérons deux problèmes liés : le comportement des spectres d'excitation à température zéro, c'est-à-dire comment le système répond à de petites perturbations autour d'une configuration stable, et les propriétés du point critique à température zéro.

Dans la première partie, nous fournissons un cadre de base, couvrant les verres de spin, les verres structurels et les matrices aléatoires. L'introduction de ces sujets est fonctionnelle pour la suite du manuscrit.

Dans la deuxième partie, nous discutons d'un modèle de verre de spin vectoriel avec un réseau d'interaction entièrement connecté, soumis à l'action d'un champ aléatoire : nous montrons que le point critique du verre de spin à température nulle présente une transition de délocalisation pour les modes doux de la densité d'états, d'une phase avec des états localisés au bord inférieur du spectre à une phase délocalisée du verre de spin. Nous montrons que ces résultats, appliqués à un modèle vectoriel de spin p, démontrent l'existence d'une bande stable dans le paysage énergétique avec des excitations localisées.

Dans la troisième partie, nous considérons un modèle de verre de spin dispersé sous un champ stochastique, défini sur un graphe aléatoire régulier. Pour un champ externe suffisamment faible, des excitations localisées douces apparaissent, caractérisées par une densité d'états avec une dépendance quartique à basse fréquence.
Ces excitations sont omniprésentes dans les systèmes désordonnés à faible connectivité. Enfin, nous montrons qu'une transition de délocalisation peut se produire à la transition de verre de spin, sous une forme plus faible que celle observée dans le cas dilué.

\pagenumbering{arabic}

\part{Background}

\chapter{Spin Glasses}

Spin glasses are archetipal models of complex systems \cite{mez1987, altieri2023introduction}. Introduced in the 70s to idealise the behavior of magnetic alloys, it was soon clear that their thermodynamic and dynamic behavior had a more general nature. Starting from their natural connection to structural glasses \cite{castellani2005spin, cavagna2009supercooled}, spin glass models nowadays embrace a plethora of different fields of knowledge: from optimization problems \cite{mez1987} to neural networks \cite{hopfield1982neural, amit1989modeling}, from applications to finance \cite{potters2020first}, to ecology \cite{altieriunderstanding}. 

In this chapter we give a survey of the earliest results in spin glass theory, discussing mean field models on fully connected graphs both in the discrete and in the vector case. A section at the end of the chapter is devoted to spin glass models on random sparse graphs, a more recent field of research.

\section{Disordered Magnetic Alloys}

\noindent Starting from the 50s there was a growing interest within the condensed matter community in the properties of magnetic alloys \cite{Zen51a,Zen51b,Zen51c,Rud54,Kas56,Yos57}, obtained by doping\footnote{Impurities diffuse inside the paramagnetic metal at fusion temperature. After the right quantity of magnetic impurities has been let diffuse, the sample is brought to room temperatures and solidifies.} a paramagnetic matrix of a noble metal with magnetic impurities of a transition metal: some notable examples are given by alloys of Manganese (Mn) and Copper (Cu) and Iron (Fe) and Gold (Au).
When the concentration of magnetic impurities is very high (far greater than $10\%$), the resulting solid is a ferromagnet at low temperatures. Conversely, if the concentration is sufficiently low (less than $10\;\%$), there are compelling evidences of a new kind of magnetism.  Even though the sample does not develop spontaneous magnetisation, no matter how low the temperature, several experimental results suggested the presence of a freezing phase transition occurring at low temperatures. In \cite{DENOB59, ZIM60} the specific heat of CuMn alloys was measured: while at high enough temperature the system follows Curie-Weiss prediction, it is found a peak at low temperatures, which broadens under the application of an external magnetic field \cite{DENOB59}, and a linear scaling with temperature, with no dependence on Mn concentration \cite{ZIM60}; the measured specific heat is anomalously large with respect to that of pure Cu. In \cite{Can71} workers show that the magnetic susceptibility has a cusp peak at a temperature in an interval $1-100\;K$ for concentrations of impurities ranging in $0.01\;\%\;-\;10\%$. The magnetic susceptibility follows the Curie Law $\chi\propto 1/T$ of paramagnets at high temperatures, but strongly deviates from it under the freezing temperature, being roughly constant. The susceptibility peaks at a lower temperature than the specific heat and the cusp is smoothed when an external magnetic field is present.
All these anomalous behaviors were also related to an exotic form of antiferromagnetism \cite{adk74}, but later experiments excluded this hypothesis \cite{Mur78}.


\noindent After this brief account of the most relevant experiments, let us discuss the first microscopic theoretical models of these systems. The puzzling low temperature behaviors of CuMn, AuFe and similars was related to s-d interactions \cite{Mar60}, coupling electrons of unfilled inner shells with conduction electrons.
The electrons in the conduction band of the alloy are polarised by the presence of the impurities: as a consequence, they mediate interactions between the magnetic moments of impurities \cite{Rud54, Kas56, Yos57}, namely RKKY (Ruderman-Kittel-Kasuya-Yoshida) interactions
\begin{equation}
\label{eq:RKKY_0}
    \mathcal{J}(\vec{r}_i,\vec{r}_j)\overset{r_{ij}/a\rightarrow\infty}{\sim}\;\frac{2\cos(k_F\;r_{ij})}{r_{ij}^3}
\end{equation}
where $r_{ij}\,=\,|\vec{r}_i-\vec{r}_j|$ is the distance between the centers of mass of Mn atoms and $a$ is the reticular step. Besides the fast $r^{-3}$ decay, the cosine term in eq. \eqref{eq:RKKY_0} tells us that both ferromagnetic and antiferromagnetic couplings are possible. Moreover, since Fermi wavevector is $k_F=O(1/a)$, spatial oscillations of RKKY interactions are very quick, so if the concentration of magnetic ions is low enough, ferromagnetic and antiferromagnetic couplings are equally probable; on the contrary, if there is a high density of magnetic impurities, many ions interact within a reticular step, and thus ferromagnetism is favoured.

Magnetic impurities are let diffuse freely during the preparation of the sample, so their final positions are random: thus, both the absolute value and the sign of \eqref{eq:RKKY_0} are random. The equal occurrence of ferromagnetic and antiferromagnetic interactions and their randomness are the fundamental features that make these materials different from ordinary magnets. The first phenomenon is called \emph{frustration} and represents the impossibility for local degrees of freedom of simultaneously satisfying contradictory constraints \cite{toulouse1987theory, vannimenus1977theory}. In magnetic systems, frustration occurs in presence of ferromagnetic and antiferromagnetic couplings or in particular lattices: for instance, the two-dimensional antiferromagnet in the triangular lattice is frustrated\footnote{However, in this case it is possible to map the antiferromagnet to an unfrustrated ferromagnet.}. The second crucial ingredient is \emph{quenched disorder}: the magnetic ions are in random positions that are frozen with respect to the time-scale of the dynamics of the system, so that the related interactions can be regarded as random external parameters. Frustration and disorder are not necessarily related: there are frustrated systems with no disorder, like the two-dimensional antiferromagnet in the triangular lattice, and systems with disorder but no frustration, like in one-dimensional spin glasses.

\section{The Edwards-Anderson model}

In 1970 P.W. Anderson introduces the term \emph{Spin Glasses} for the disordered magnets discussed above, to make an analogy with structural glasses, since in both systems there seems to be an unidentified low temperature phase \cite{And70}.

Anderson claims that the low temperature behavior of spin glasses is due to the interactions between Mn and Cu ions, rather than to conduction electrons. He proposes an idealised model of RRKY interactions \eqref{eq:RKKY_0}, proposing the following Hamiltonian for a system of $N$ Heisenberg spins 

\begin{equation}
    \label{eq:And_1}
    \mathcal{H}_J[{\bf S}]\,=\,-\frac{1}{2}\sum_{\Vec{x},\Vec{y}\;\in\;\mathcal{R}_3}\;\mathcal{J}_{xy}\Vec{S}_{x}\cdot\Vec{S}_{y}
\end{equation}

\noindent where $\mathcal{R}_3$ is the three-dimensional regular lattice of side $L=N^{1/3}$ and the couplings $\mathcal{J}_{{\bf x}{\bf y}}$ are taken as quenched random variables, with an unspecified distribution that should reproduce the salient features of RKKY interactions \eqref{eq:RKKY_0}. The distributions from which couplings are drawn should have moments that decay with distance not faster than RKKY interactions \eqref{eq:RKKY_0} and admits both positive and negative values. The simplest idealisation consists in cut-offing interactions to nearest neighbours and to draw couplings from a gaussian distribution.
In order to model the emergence of ferromagnetism at high concentrations of magnetic ions, the gaussian should have a non-zero mean value, representing a ferromagnetic bias.

\noindent In \cite{And70}, Anderson proposes a mean field approach without averaging over the disorder. In the spirit of his pioneering work on localisation \cite{And58}, the system is represented as a set of independent clusters, each with its own critical temperature. This interpretation had a discrete success on condensed matter community: the most accepted idea was that each of these subsystems is either a ferromagnet or an anti-ferromagnet, and as temperature is lowered enough long range interactions between different clusters appear \cite{beck1971, Can72, smith1974}. This approach had a discrete success, being able to reproduce the magnetic susceptibility cusp. Nevertheless, it did not give a true understanding of the low temperature phase, since no order parameter was yet defined.

\noindent The first proposal of an order parameter for Spin Glasses came later in 1975-76 \cite{edwards1975theory, edwards1976theory}. Edwards and Anderson propose a model based on the idea that at low temperatures spins freeze along random directions, dependent on the specific realisation of the disorder. Even though these microscopic amorphous magnetisation profiles are unknown and unrelated to any global spin symmetry\footnote{In Ising ferromagnets, for instance, the low temperature phase emerges as a result of the spontaneous breaking of the global inversion symmetry.}, long-range order can be detected by measuring how the orientation of a spin changes with time. They assume that the average autocorrelation of a spin does not depend on the sample in the thermodynamic limit: they introduce the overlap

\begin{equation}
\label{eq:qEA_1}
    q\,=\,\lim_{t \rightarrow \infty}\frac{1}{N}\sum_{\Vec{ x}}\;\int_0^t dt'\;\Vec{S}_{\bf x}(t')\cdot\Vec{S}_{\bf x}(0).
\end{equation}

\noindent If the system is ergodic, \eqref{eq:qEA_1} can be rewritten in terms of an ensemble average on Gibbs measure

\begin{equation}
\label{eq:qEA_2}
    q\,=\,\frac{1}{N}\sum_{\Vec{x}}\;|\langle\Vec{S}_{{\bf x}}\rangle|^2\equiv \frac{1}{N}\sum_{\Vec{x}}\;|\Vec{m}_{x}|^2.
\end{equation}

\noindent At high temperature, thermal fluctuations are too strong and no ordering can occur, thus $|\Vec{m}_{x}|=0$ and $q=0$. At a critical temperature $T_c$ there is a second order phase transition such that, for any $T<T_c$
\begin{equation}
\Vec{m}_x\neq 0 \qquad \Vec{m}\equiv \frac{1}{N}\underset{\Vec{x}}{\sum}\;\Vec{m}_{x}=0 \qquad q>0
\end{equation}

\noindent The spin glass phase can be defined with a dynamical point of view as the low temperature region with persistent correlations in spin orientations. The simplest parameter capturing spins freezing is $q$ in \eqref{eq:qEA_1}, \eqref{eq:qEA_2}, which is thus named \emph{Edwards-Anderson Order Parameter}. 

In their work of 1975, they study the thermodynamics of a spin glass model with Hamiltonian \eqref{eq:And_1} and gaussian couplings with zero mean and variance $J^2\rho$, being $\rho_0$ the mean bond occupation number. In the following, we sum up their results in \cite{edwards1975theory}.
Since disorder is quenched, the free-energy of the system must be computed as an average over the free-energies of many different samples
\begin{equation}
\label{eq:quenched_free_energy}
    f(T)\,=\,-\;\frac{T}{N}\;\lim_{N\rightarrow \infty}\int\;\mathcal{D}\boldsymbol{J}\;P(\boldsymbol{J})\log\;Z_N(T;\boldsymbol{J})\equiv \int\;\mathcal{D}\boldsymbol{J}\;P(\boldsymbol{J})\;f(T;\boldsymbol{J})
\end{equation}
where $Z_N(T;\boldsymbol{J})$ is the canonical partition function of a sample with disorder realisation of bonds $\boldsymbol{J}$. In order for the thermodynamic limit to exist\footnote{Imagine to divide the system in macroscopic subsystems: any extensive observable averaged over these subsystem concentrates around the expected value, thanks to the law of large numbers. This must be true with or without disorder in the subsystems}, sample-to-sample fluctuations must not affect the values assumed by extensive quantities: this property is called \emph{self-averaging}. In formulae
\begin{equation}
\label{eq:self-averaging}
    \lim_{N\rightarrow\infty}|f_{J}^{(N)}-\overline{f_J}|=0\qquad\text{or}\qquad \overline{f_J^2}=\overline{f_J}^2
\end{equation}
where we have introduced the notation $\overline{(\cdot)}$ for the average over disorder.

The disorder average in \eqref{eq:quenched_free_energy} cannot be done directly: it is needed to average the logarithm of a function of the couplings. In general, one cannot assume that the partition function is self-averaging\footnote{This is the case for spin glasses with pairwise interactions. Conversely, in spin glasses with p-body interactions ($p>2$) the partition function is actually self-averaging.}, $\overline{\log Z}\neq \log \overline{Z}$. In order to circumvent this difficulty, Edwards and Anderson propose the so called \emph{Replica Trick}:
\begin{equation}
    \label{eq:Replica_Trick}
    \overline{\log Z_N(T;\boldsymbol{J})}\,=\,\lim_{n\rightarrow 0}\;\frac{\overline{Z_N(T;\boldsymbol{J})^n}-1}{n}
\end{equation}
which is based on the identity
\begin{equation*}
    x^n\,=\,1+n\log x+O(n^2).
\end{equation*}
If one assumes gaussian couplings, the disorder average can be performed\footnote{One can make use of the standard identity $\langle e^x \rangle\equiv e^{\frac{1}{2}\operatorname{Var}x}$.} yielding
\begin{equation}
\label{eq:dis_avg}
    \overline{Z^n}\,=\,\underset{\{\boldsymbol{S}_a\}}{\Tr}\;\overline{e^{-\beta\sum_{a=1}^n\;\mathcal{H}_J[\boldsymbol{S}_a]}}\,=\,\underset{\{\boldsymbol{S}_a\}}{\Tr}e^{\frac{\beta^2 J^2}{2}\sum_{a}\sum_{(ij)}^{'}S_i^aS_j^aS_i^bS_j^b}
\end{equation}
where $\Tr$ stands for a sum over spin configurations and $\sum_{(ij)}^{'}$ is a sum over nearest neighbour links. If the disorder average is performed before the thermal average, the system is mapped into an effective theory where replica are coupled through quartic interactions.
Edwards and Anderson make a mean field approximation\footnote{They make use of a standard variational principle and replace the quartic form in \eqref{eq:dis_avg} with a best quadratic.} and complete the computation. In particular, they find that their order parameter satisfies the self-consistent equation
\begin{equation}
q\,=\,\int \frac{d^3 r}{(2\pi)^{3/2}}\left[\coth\left(\frac{\sqrt{2}}{3}\beta J \sqrt{q}\right)-\frac{3}{\sqrt{2}\beta J \sqrt{q}}\right]^2    
\end{equation}
which admits a non-zero solution under $T_c=\sqrt{2}J/3$. The magnetic susceptibility, which is related to $q$ via
\begin{equation}
    \chi\,=\,\beta\,(1-q)
\end{equation}
is found to have a cusp at the freezing temperature, with a Curie-Weiss behavior for $T>T_c$ and a quadratic scaling in $T_c-T$ slightly below $T_c$. The specific heat as well is peaked with a cusp at the critical temperature. Edwards and Anderson theory, based on the order parameter \eqref{eq:qEA_2}, already at the mean field level reproduces key features of Spin Glasses, observed experimentally.

\section{SK model}

Given the success of EA mean field theory, defining and solving models where mean field approximation is exact was the natural next step.
In 1975 Sherrington and Kirkpatrick propose an infinite-range model for spin glasses with Ising spins $S_i=\pm 1$ \cite{sherrington1975solvable}:
\begin{equation}
    \label{eq:SK_1}
    \mathcal{H}_{SK}[{\bf S}]\,=\,-\frac{1}{2}\sum_{ij}^{1,N}\;J_{ij}S_iS_{j}-H\sum_{i=1}^N\;S_i
\end{equation}
where $J_{ij}$ are gaussian variables with
\begin{equation*}
    \overline{J_{ij}}=\frac{J_0}{\sqrt{N}}\qquad \overline{J_{ij}^2}=\frac{J^2}{N}
\end{equation*}
and $H$ is an external magnetic field. In this section, we report their results in the article for the $J_0=0$ case, corresponding to magnetic alloys in the zero impurities concentration limit.
They make use of Replica Trick \eqref{eq:Replica_Trick} to compute the free energy of their model: the sample average of the replicated partition function \footnote{The computation exploits the relation $\langle e^{J\;z}\rangle_J=e^{\frac{1}{2}\;Var(J)z^2}$, where the variable $J$ is a gaussian variable with zero mean.} reads

\begin{equation*}
    Z_J^n\,=\,\sum_{\{\boldsymbol{S}\}_{a=1}^{n}}\;\exp\left[\frac{\beta^2Nn}{4}+\frac{\beta^2N}{2}\sum_{(ab)}^{1,n}\;\left(\sum_{i=1}^N\;\frac{S_i^aS_i^b}{N}\right)^2+\beta H\sum_{i=1}^N\sum_{a=1}^nS_i^a\right].
\end{equation*}

The presence of disorder has mapped the original Hamiltonian \eqref{eq:SK_1} into an effective Hamiltonian that couples replica. To complete the computation, sites should be decoupled: this can be achieved through an Hubbard-Stratonovich rappresentation\footnote{$e^{\frac{\lambda^2}{2}}\,=\,\int_{.\infty}^{+\infty}\frac{dx}{\sqrt{2\pi}}\;e^{-\frac{x^2}{2}+\lambda x}$}, returning

\begin{eqnarray}
    \label{eq:SK_Q}
    & (Z_J)^n\,=\,\int\;\prod_{(ab)}^{1,n}\;dQ_{ab}\left(\frac{N\beta^2}{2\pi}\right)^{1/2}\exp\Bigr[-N A(\mathbb{Q})\Bigl] \\
    & \nonumber \\
    \label{eq:Action_SK}
    & A[\mathbb{Q}]\,=\,-\frac{n\beta^2}{4}+\frac{\beta^2}{4}\;Tr\;\mathbf{\mathbb{Q}}^2-\log \underset{\{S\}_R}{\sum}\;e^{-\beta \mathcal{H}_{R}[\mathbf{Q},\;\{\boldsymbol{S}\}_R]} \\
    & \nonumber \\
    \label{eq:Eff_Ham_SK}
    & \mathcal{H}_R[\mathbb{Q},\;\{\boldsymbol{S}\}_R]\,=\,-\beta\;\sum_{(ab)}^{1,n}Q_{ab}S_aS_b-H\sum_{a=1}^n\;S_a 
\end{eqnarray}

where $\{S\}_R$ indicates the $2^n$ configurations of single-site replica. The $n\times n$ overlap matrix $\mathbb{Q}$ is the order parameter of the model. 
By exchanging the order of the two limits $N\rightarrow\infty$ and $n\rightarrow 0$, the free energy density of the system can be computed with the saddle point method

\begin{equation}
    \label{eq:fe_0}
    f(T, H)\,=\,\lim_{n\rightarrow 0}\;\frac{T}{n}A[\mathbb{Q}_*]
\end{equation}

where saddle point matrix $\mathbb{Q}_*$ satisfies the self-consistency equation

\begin{equation}
    \label{eq:ordPar_eq_0}
    Q_{ab}^*\,=\,\frac{\underset{\{S\}_R}{Tr} S^aS^b\;e^{-\beta \mathcal{H}_R[\mathbb{Q}_*,\;\{\boldsymbol{S}\}_R]}}{\underset{\{S\}_R}{Tr}e^{-\beta \mathcal{H}_R[\mathbb{Q}_*,\;\{\boldsymbol{S}\}_R]}}\equiv \langle S^aS^b \rangle_{R}.
\end{equation}

The action $A[\mathbb{Q}]$ is invariant under exchange of rows or columns of the overlap matrix: thus the group of permutations of $n$ elements $P_n$ is a symmetry of the problem; this group is often called the \emph{Replica Group} and any function of the overlap matrix left invariant by its action is \emph{Replica Symmetric} (RS).

For positive integer $n>0$, the minimum of $A$ can be found through the following parametrization of the overlap matrix

\begin{eqnarray}
    \label{eq:ordParAnsatz_RS}
    & Q_{ab}^*&=q\text{,}\qquad a<b \nonumber \\
    & Q_{aa}^*&=0
\end{eqnarray}

This is the only form of the matrix left invariant by the action of the replica group. After analytically continuing the solution of $dA/dq=0$ to $n=0$, the self-consistency equation of the order parameter becomes

\begin{equation}
    \label{eq:ordPar_eq_RS}
    q\,=\,\int_{-\infty}^{+\infty}\;\frac{dz}{\sqrt{2\pi}}e^{-\frac{z^2}{2}}\tanh^2(\beta \sqrt{q} z+\beta H)
\end{equation}

while the free energy reads

\begin{equation}
\label{eq:fe_RS}
f(\beta, H)\,=\,-\frac{\beta}{4}(1-q)^2-\int_{-\infty}^{+\infty}\;\frac{dz}{\sqrt{2\pi}}e^{-\frac{z^2}{2}}\log 2 \cosh(\beta \sqrt{q} z+\beta H)
\end{equation}

The solution of this model provided by Sherrington and Kirkpatric captures many qualitative features of the spin glass behavior \cite{sherrington1975solvable}, \cite{kirkpatrick1978infinite}. In particular, it correctly reproduces the cusp of the magnetic susceptibility and of the specific heat at the critical temperature $T_c=J$. The order parameter $q$ is zero for $T>T_c\equiv J/k_B$, returning the Curie-Weiss paramagnetic solution, but as $T<T_c$, $q>0$ and the system is in the spin glass phase.   
Despite these encouraging results, Sherrington and Kirkpatric notice that their solution cannot be correct down to zero temperature. Indeed, they encounter a serious physical inconsistency: the entropy becomes negative at low temperatures. The entropy of a discrete system is non-negative by definition, since $S[P]\equiv -\langle \log P\rangle>0$. In particular, they find the zero temperature result
\begin{equation*}
    S(0)=-\frac{1}{2\pi}\approx -0.17
\end{equation*}
Comparing their Monte-Carlo simulations with theoretical predictions of the RS solution \eqref{eq:fe_RS}, they notice that in the same region where the entropy becomes negative all thermodynamic observables are inconsistent with theoretical predictions. They claim that these strange results are consequence of the exchange of limits  $N=\infty$ and $n=0$ in the saddle point evaluation.

\section{Instability of SK solution: the dAT line}

In their work of 1978 \cite{de1978stability}, de Almeida and Thouless (dAT) show that the RS solution of SK model becomes unstable at low enough temperature for any $H\geq 0$. In order to study the stability of the RS saddle point \eqref{eq:ordParAnsatz_RS}, they consider the Hessian of the replica action \eqref{eq:Action_SK}, that can be easily computed
\begin{equation}
\label{eq:Hessian_SK_Action}
    \frac{\partial A(\mathbb{Q})}{\partial Q_{ab}\partial Q_{cd}}\,\equiv D_{(ab)(cd)}\,=\delta_{(ab)(cd)}-J^2\beta^2(\langle S_a S_b S_c S_d\rangle-\langle S_a S_b\rangle\langle S_c S_d\rangle)
\end{equation}
where averages $\langle \cdot \rangle$ are with respect to the Gibbs measure of the Hamiltonian in \eqref{eq:Eff_Ham_SK}. The Hessian is a matrix of order $n (n-1)/2$. When evaluated at the RS saddle point, as a consequence of RS ansatz \eqref{eq:ordParAnsatz_RS} it acquires a much simpler form
\begin{eqnarray}
    \label{eq:Hessian_RS}
    && D_{(ab)(ab)}\,=\,1-\beta^2 J^2(1-q^2)\equiv P \nonumber \\
    && D_{(ab)(ac)}\,=\,-\beta^2 J^2 q\,(1-q)\equiv tilde{Q}, \qquad c\neq b \nonumber \\
    && D_{(ab)(cd)}\,=\,-\beta^2 J^2 (r-q^2)\equiv R,\qquad (ab)\neq (cd)
\end{eqnarray}
where q satisfies the saddle point eq. \eqref{eq:ordPar_eq_RS} and
\begin{equation}
    \label{eq:r}
    r=\frac{1}{\sqrt{2 \pi}}\int\;dh\;e^{-\frac{h^2}{2}}\tanh^4(\beta J \sqrt{q}+\beta H)+O(n).
\end{equation}
When $H=0$, one has $Q=R=0$, the Hessian is diagonal and hence the stability is given by requiring $P>0$, which means $\beta<J$: the instability of RS solution extends to the whole spin glass phase. In presence of a field $H$, it is necessary to determine the eigenvalues and eigenvectors of \eqref{eq:Hessian_RS}. This can be done by considering the symmetries under replica indices permutations of the elements of \eqref{eq:Hessian_RS}. There are three different invariant families, represented by the three different values $P, Q, R$: the related replica permutations must respect the constraints on the indices in \eqref{eq:Hessian_RS}. Given \eqref{eq:Hessian_RS}, the eigenvalue equation can be written as
\begin{equation}
    \label{eq:eig_eq}
    \sum_{(cd)}D_{(ab)(cd)}\eta_{(cd)}\,=\,R\sum_{\{(cd)\neq (ab)\}}\eta_{cd}+2 \tilde{Q}\;\sum_{d}\eta_{ad}+P\;\eta_{ab}=\lambda\;\eta_{ab}
\end{equation}
For any $n>0$, there are three distinct eigenvalues, related to the three symmetry group under which the elements of the Hessian are invariant. 


Here we do not report the details of the calculation, which can be found in \cite{de1978stability}, but only the final result. In the limit $n\rightarrow 0$, two of the three eigenvalues become equal. The final result is
\begin{eqnarray}
\label{eq:dAT_eigs}
    && \lambda_s=P-4\tilde{Q}+3R \\
    && \lambda_r=P-2\tilde{Q}+R
\end{eqnarray}
dAt found that $\lambda_s>0$ for any temperature and field, whereas at low temperatures and fields $\lambda_r$ crosses zero and becomes negative. This eigenvalue in literature is called \emph{Replicon}. It is connected with the Spin Glass susceptibility through
\begin{eqnarray}
&& \chi_{SG}\,\equiv \frac{1}{N}\sum_{(ij)}\overline{\chi_{ij}^2}\,=\,\,\frac{\beta J\chi_{SG}^{(0)}}{1-\beta J \chi_{SG}^{(0)}} \\
&& \lambda_r\,=\,1-\beta J \chi_{SG}^{(0)}  
\end{eqnarray}
where $\chi_{SG}^{(0)}\equiv \overline{\chi_{00}^2}$ is the single-site spin glass susceptibility. It is called like this because when the replicon becomes zero, $\chi_{SG}$ diverges, as it is expected in second order phase transitions.
By substituting the expressions in \eqref{eq:Hessian_RS},\eqref{eq:r} in the second of \eqref{eq:dAT_eigs} they find the stability condition
\begin{equation}
    \frac{1}{\beta^2 J^2}>\int_{-\infty}^{+\infty} \frac{dh}{\sqrt{2\pi}}\frac{\exp\left(-\frac{h^2}{2}\right)}{\cosh^4(\beta J\sqrt{q}+\beta H)}
\end{equation}
This condition, when the '$>$' is replaced by an equality, defines an instability line for the RS solution of the SK model in the $(h, T)$ plane, called \emph{dAT} line. The asymptotic expressions of the line in the limits $H\rightarrow 0$ and $T\rightarrow 0$ are
\begin{eqnarray}
    \label{eq:dAT_boundaries}
    && H(T)\sim \frac{2}{\sqrt{3}}\;J\;\left(1-\frac{T}{J}\right)^{3/2}\qquad H\rightarrow 0 \\
    && H(T)\sim \sqrt{2}\;J\sqrt{\log\left(\frac{4J}{3\sqrt{2\pi}T}\right)}\qquad T\rightarrow 0
\end{eqnarray}
Close to the zero-field transition temperature the dAT line is non-analytic, whereas close to zero-temperature the instability line slowly diverges: in SK model there is no zero-temperature phase transition in the external field. This last prediction appears to be quite unphysical, since physically a very large field freezes the orientation of spins, no matter how low the temperature. This result holds only at this level of mean field approximations. In improvements of mean field theory, such as the Bethe theory discussed later in section \ref{sec:BP_and_cav}, the dAT line of SK model is finite at zero temperature. For mean field models with infinite-range interactions one should resort to Heisenberg vector spins in order to observe a zero temperature field phase transition: this will be discussed later in section \ref{sec:vector_models_chap1}.

\noindent The result found by dAT has deep consequences: since the RS solution of SK model is unstable also in presence of a field, it seems that there is a phase transition unrelated to the spontaneous breaking of a spin symmetry. The destabilisation of the RS saddle point at low temperatures can only imply that under the dAT line replica permutation symmetry spontaneously breaks.

\section{TAP approach}

Not much after the seminal work of Sherrington and Kirkpatric, a different approach was proposed by Thouless, Anderson, Palmer (TAP) \cite{thouless1977solution}. They suggest to build a mean field theory for the given sample, i.e. before averaging out the disorder. Given that the problems encountered by SK were believed to be due to the exchange of the two limits $n\rightarrow 0$, $N\rightarrow \infty$, it was natural to build a mean field theory following a different strategy.
Their approach is based on a high temperature expansion of the free-energy functional of the system: we briefly sketch its details, making use of an equivalent technique \cite{georges1991expand}, \cite{plefka1982convergence} instead of their original derivation. The free energy functional is 

\begin{equation}
\label{eq:free-energy_functional}
    -\beta F[\boldsymbol{m}|\boldsymbol{b}]\,\equiv A[\boldsymbol{m}|\boldsymbol{b}]=\,\log \Tr\exp\left(-\beta \mathcal{H}_J[\boldsymbol{S}]+\beta \sum_{i=1}^N b_i (S_i-m_i)\right).
\end{equation}
External fields $\{b_i\}$ are Lagrange multipliers that fix the thermal average of each spin $S_i$ to $m_i$ in a self-consistent way: by imposing $\partial F/\partial b_i\,=\,0$, one gets
\begin{equation}
    \label{eq:spins_self_cons}
    \langle S_i \rangle_b \,=\, m_i.
\end{equation}
Physical magnetisations are obtainable by setting $\boldsymbol{b}=0$. The Georges-Yepidia expansion consists in expanding \eqref{eq:free-energy_functional} in powers of $\beta$ around $\beta=0$. For Hamiltonians of fully connected spin models, since couplings are weak in $N$ large, the series truncates at a finite order. In ordered models the expansion stops at $O(\beta)$ because $J\sim 1/N$, while in the disordered case $J \sim 1/\sqrt{N}$ and one must include also the $\beta^2$ term into the expansion. After performing the computation, the final result is \emph{TAP free energy}
\begin{eqnarray}
    \label{eq:TAP_FE_SK}
   && F_{TAP}[\boldsymbol{m}]\,=\,A_0[\boldsymbol{m}]-\left(\frac{1}{2}\sum_{ij}J_{ij} m_i m_j-H\sum_{i=1}^N\;m_i\right) -\frac{N\chi^2}{4\beta} \\
   && A_0[m]\,=\,\frac{1}{2\beta}\sum_{i}\left[(1+m_i)\log\Bigl(\frac{1+m_i}{2}\Bigr)+(1-m_i)\log\Bigl(\frac{1-m_i}{2}\Bigr)\right] \\
   && \chi\equiv \beta\left(1-\frac{1}{N}\sum_{i=1}^N\;m_i^2\right)=\beta (1-q)
\end{eqnarray}
The interpretation of addenda in \eqref{eq:TAP_FE_SK} is straightforward: the first addendum is the the total entropy of a system of binary spins constrained to magnetisation profile $\{m_i\}$, in the limit of infinite temperature; the second is the infinite-temperature internal energy of a system of spins with magnetisations $\{m_i\}$; finally, the third term is \emph{Onsager term}, expressing the energetic contribution of mutual correlations between spins. The first two terms appear also in spins systems without disorder, whereas the term term is peculiar of frustrated disordered systems. 
By computing $\partial F_{TAP}/\partial m_i=0$, one finds the renown \emph{TAP equations}
\begin{eqnarray}
    \label{eq:TAP_eqs}
    && m_i\,=\,\tanh \beta h_i \\
    && \nonumber \\
    && h_i\,=\,\sum_{j:j\neq i}J_{ij}m_j+H-\chi\;m_i
\end{eqnarray}
These equations remind of the mean field equation of Curie-Weiss model:
\begin{eqnarray*}
    && m\,=\,\tanh(\beta h) \\
    && h\,=\,J m + H
\end{eqnarray*}
TAP equations feature an additional term $-\chi m_i$, called \emph{Onsager reaction term}: for a given spin $i$, it represents the mean reaction of all the other spins to the polarising effect of $i$. Thus, $\{h_i\}$ are usually called \emph{cavity fields}: each $h_i$ is the field that would act on site $i$ in a system where spin $S_i$ has been removed.

\noindent In the paramagnetic phase $T>T_c$ eq.\eqref{eq:TAP_FE_SK} coincides with the result from the original diagrammatic expansion in TAP article \cite{thouless1977solution}. When $H=0$, the resummation of the expansion diverges as $T<J\equiv T_c$ \cite{plefka1982convergence}. They extend it to temperature close to $T_c^{-}$ by enforcing TAP equations: the new convergence criterion found by them is $J (1-q)\leq T$. Indeed, as $T<T_c$, non trivial solutions to TAP equations appear and consequently, a non-zero EA overlap emerges: TAP are able to estimate it close to the critical temperature, finding a linear scaling $q\approx 1-\frac{T}{T_c}$ for Edwards-Anderson overlap, in agreement with calculations from SK. In the paramagnetic phase and close to the critical temperature, all predictions coming from TAP and SK approaches are mutually compatible. 
Close to $T=0$, they make precise phenomenological assumptions, based on numerical simulations performed by themselves:
\begin{eqnarray}
\label{eq:TAP_mods}
    && P_h(h)\sim h/H_0^2,\qquad h\rightarrow 0 \\
    && q\sim 1-a\;T^2,\qquad T\rightarrow 0
\end{eqnarray}
where $P_h(h)$ is the $T=0$ cavity fields pdf. Notice that from the second of eqs \eqref{eq:TAP_eqs} one deduces that cavity fields are gaussian variables with zero mean and variance $J^2 q$: since in this case $P_h(h)\sim const$, the first assumption in \eqref{eq:TAP_mods} is equivalent to ask that at low temperatures very small cavity fields are rare. This is the concept of \emph{pseudo-gap}: in a later paper \cite{muller2015marginal}, it was shown that the first eq. \eqref{eq:TAP_mods} is a fundamental property of glassy systems with marginal stability. As to the second in \eqref{eq:TAP_mods}, it is actually a consequence of the first: indeed, by enforcing TAP equations close to zero temperature ($\beta\rightarrow \infty$)
\begin{eqnarray*}
     \overline{m_i^2}\,&&=\,\int\;dh\;P_h(h)\tanh(\beta h_i)^2\,\simeq 1-\int\;dh\;P_h(h)e^{-2\beta h} \\
    &&\simeq 1-\int\;dh\;h\;e^{-2\beta h} =\,1-c\;T^2
\end{eqnarray*}

Basing on eqs \eqref{eq:TAP_mods}, they find that the inconsistencies of SK theory (in particular the entropy becoming negative) disappear. 

In conclusion TAP approach correctly predicts the presence of a freezing phase transition, finding results that agree qualitatively-and in some cases also quantitatively with previous predictions yielded by EA and SK and correcting inconsistencies found in the SK analysis. Nevertheless, TAP solution is still far from being a satisfactory solution of SK mean field model. Indeed, assumptions \eqref{eq:TAP_mods} are phenomenological, and are not predicted by their theoretical framework. Moreover, their theory does not account for the intermediate temperature region $0\ll T \ll T_c$ and cannot explain the instability of SK solution in presence of a field predicted by dAT.

\section{Breaking Replica Symmetry: Parisi Solution}
\subsection{1RSB}
In the years following the works of EA, SK, TAP and dAT, it was clear that a correct mean field theory had to deal with the breaking of replica symmetry.
The correct solution was found by Parisi in 1979: he proposed a RSB scheme in a series of three articles \cite{parisi1979toward, parisi1979infinite, parisi1980sequence, parisi1980magnetic}. Parisi assumes a RSB ansatz of the form ($a\neq b$)

\begin{equation}
    \label{eq:1RSB-ansatz}
    Q_{ab}\,=\,\begin{cases}
       & Q_{ab}=q_0,\qquad I(a/m)=I(b/m) \\
       & Q_{ab}=q_1,\qquad I(a/m)\neq I(b/m)
    \end{cases}
\end{equation}
where $1<m<n$ is such that $n/m$ is an integer. Eq.\eqref{eq:1RSB-ansatz} corresponds to an overlap matrix with $n/m$ blocks of order $m$ centered on the main diagonal, with elements $a\neq b$ all equal to $q_0$; outside these blocks, every entry is equal to $q_1\neq q_0$. The parametrisation in \eqref{eq:1RSB-ansatz} is called \emph{One Step Replica Symmetry Breaking} (1RSB) ansatz.
It is worth to mention a remark made by de Almeida and Thouless regarding the use of Saddle Point method when $n\rightarrow 0$. The analytic continuation of this method to non-integer $n<1$ values implies that, in the $n\rightarrow 0$ limit, the correct saddle point, in the sense of a non-negative spectrum of the Hessian, is obtained by maximising the Action (and thus the free energy) with respect to $\mathbb{Q}$. Consider, for instance, the second addendum in \eqref{eq:Action_SK}: in the RS case, this term for finite integer $n$ becomes
\begin{equation*}
    \frac{1}{n}\Tr \mathbb{Q}^2\,=\,\frac{(n-1)}{2}q^2
\end{equation*}
This curvature term is non-negative for any $n>1$, suggesting that in this case the correct saddle-point is a minimum of the Action. However, as $n<1$, this term becomes negative!
Having said this, the free energy of SK model becomes\footnote{We set $J=1$.}
\begin{eqnarray}
    \label{eq:fe_SK_1RSB}
    && f_{1RSB}(\beta, H)=\underset{q_0,q_1,m}{\max}\Phi(q_0, q_1, m) \\
    && \Phi(q_0, q_1, m)=-\frac{\beta^2}{4}[1+mq_0^2+(1-m)q_1^2-2q_1]-\log 2 \\
    &&-\frac{1}{m}\int d\mathcal{G}(z_0;q_0)\log\int d\mathcal{G}(z_1;q_1-q_0)\cosh^m[\beta(z_0+z_1+H)] \nonumber
\end{eqnarray}
where d$\mathcal{G}(z;\sigma^2)$ is a gaussian measure with zero mean and variance $\sigma^2$. The crucial novelty of Parisi approach is to consider the block dimension $m$ as a variational parameter of the problem: in earlier attempts \cite{blandin1978theories, bray1978replica}, a block construction with a fixed value of $m$ similar to \eqref{eq:1RSB-ansatz} was considered. The smaller overlap between $q_0$ and $q_1$ must be zero in absence of an external magnetic field, for continuity with RS solution.
Let us now consider the overlap pdf
\begin{equation}
\label{eq:overlap_pdf_replica}
    P_n(q)\equiv \frac{2}{n(n-1)}\sum_{(ab)}\delta(q-Q_{ab})
\end{equation}
When \eqref{eq:overlap_pdf_replica} is evaluated in the 1RSB ansatz with $n>1$, it becomes
\begin{equation}
    P_n(q)\,=\,\frac{n-m}{n-1}\delta(q-q_0)+\frac{m-1}{n-1}\delta(q-q_1)
\end{equation}
As $n<1$, this expression makes sense as a probability distribution\footnote{Probabilities are non-negative by definition.} if and only if $n<m<1$. The continuation of $m$ to real values is legitimated by that of $n$. At $n=0$ one has
\begin{equation}
    P(q)=m\delta(q-q_0)+(1-m)\delta(q-q_1)
\end{equation}
Then, $m$ acquires the meaning of probability weight of the overlap $q_0$. One can wonder if the inconsistencies found by SK in the RS ansatz get any better. Parisi shows that within the $1RSB$ scheme, all observables but the entropy agree with measures from computer simulations performed by SK \cite{sherrington1975solvable}. As to the entropy, he finds a lesser negative value
\begin{equation*}
    S_{1RSB}(T=0, h=0)\simeq -0.01
\end{equation*}
to be compared with $S_{RS}(T=0, h=0)\simeq -0.17$ found by SK. This results is a clear improvement with respect to the RS case, suggesting that a generalisation of 1RSB scheme may give a correct saddle point. PAR understood that a hierarchical iteration of 1RSB ansatz \eqref{eq:1RSB-ansatz} was the key for a mean field theory of spin glasses. 
\subsection{Full RSB}
In \cite{parisi1980sequence} he proposes the \emph{k-RSB} scheme, defined as
\begin{equation}
    \label{eq:kRSB-ansatz}
    Q_{ab}\,=\,\begin{cases}
       & Q_{ab}=q_i,\qquad I(a/m_i)=I(b/m_i) \\
       & I(a/m_{i+1})\neq I(b/m_{i+1}),\qquad $i=0,\dots, k$
    \end{cases}
\end{equation}
The overlap matrix is parametrised through a sequence of blocks nested into each other, in a way to reproduce $k$ times the 1RSB scheme. In the $kRSB$ ansatz, the order parameters of the model are the overlaps $q_0, q_1, \dots, q_K$ and the block sizes $n=m_0\geq m_1\geq \dots \geq m_K>1$. The pdf of the overlaps is
\begin{equation}
    \label{eq:P(q)_kRSB}
    P_n(q)\,=\,\sum_{i=0}^K\;\frac{m_i-m_{i+1}}{n-1}\delta(q-q_i)
\end{equation}
for $n>1$. Again, when $n<1$, one must reverse the order of the $\{m_i\}$ in order to have a non-negative probability, so we have now weights $n=m_0\leq m_1 \leq \dots m_K<1$. At $n=0$ we get
\begin{equation}
\label{eq:overlap_pdf_kRSB}
    P(q)\,=\,\sum_{i=0}^K\;(m_{i+1}-m_i)\delta(q-q_i)
\end{equation}
We can define a step-wise function
\begin{equation}
    \label{eq:x(q)}
    q(x)\,=\,q_i\theta(x-m_i)\theta(m_{i+1}-x)
\end{equation}
where $\theta(x)$ is the Heavised Function. With this definition, we can write in the limit $n$ going to zero any thermodynamic quantity, smooth function of the overlap, as an integral in the interval $[0,1]$, since for any positive $l$
\begin{eqnarray}
\label{eq:overlap_moments}
    && \lim_{n\rightarrow 0}\frac{1}{n}\Tr\mathbf{Q}^l=\int_{0}^1\;q(x)^l\;dx\,=\, \int_{0}^1\;q^l\;P(q)dq\equiv \overline{q^l} \\
    && P(q)\,=\,\diff{x(q)}{q}\qquad x(q)\equiv \int_0^1\;P(q')dq'
\end{eqnarray}
We introduced Parisi function $x(q)$, which is the inverse\footnote{In the sense of generalised functions.} of the order parameter $q(x)$. Functions $x(q)$ and $P(q)$ are stepwise with support in $[q_{min},q_{max}]$: by maximising the free energy functional at growing values of $k$, it was deduced that the overlaps satisfy $q_{0}\equiv q_{min}<q_1<\dots<q_{k-1}<q_{k}\equiv q_{max}$ \cite{parisi1980sequence}. In order to be consistent with EA picture, the maximum overlap is identified as the Edwards-Anderson overlap \eqref{eq:qEA_2}.
The free energy of the $k$-RSB system is
\begin{eqnarray}
\label{eq:kRSB_fe}
    && f_{kRSB}\,=\,\underset{q(x)}{\max}\;\Phi_{kRSB}[q(x)] \\
    && \nonumber \\
\label{eq:kRSB_fe_functional}
    && \Phi_{kRSB}[q(x)]\,=\,-\frac{\beta}{4}\left(1-2q(1)+\int_0^1q^2(x)dx\right) \\
    && -\frac{1}{\beta m_1}\int_{-\infty}^{+\infty}d\mathcal{G}_{q_0}(z_0)\log I(q_0, z_0) \nonumber \\
     && \nonumber \\
 \label{eq:kRSB_iterate_I}
    && I(q_{i-1}, z_{i-1})\,=\,\int_{-\infty}^{+\infty}\;d\mathcal{G}_{q_i-q_{i-1}}(z_i-z_{i-1})[I(q_i, z_i)]^{m_{i+1}/m_i} \\
    && \nonumber \\
\label{eq:final_condition_I}
    && I(q_{k}, z_{k})\,=\,\log 2\,\cosh(\beta(z_{k}+h))
\end{eqnarray}
It is observed that for growing $k$, the illness of the entropy tends to disappear. Already for $k=2$, the entropy is equal to $-0.004$, two order of magnitudes smaller than the initial RS result: thus, it seems that to obtain the solution of $SK$ model, the limit $k\rightarrow\infty$ must be performed. All terms involving integrals of functions of $q(x)$ are easy to extend to this limit: the step-wise function $q(x)$ becomes a continuous function of the interval $[0, 1]$. As to the $k$-fold integral in the expression of the free energy \eqref{eq:kRSB_fe_functional}, it is sufficient to notice that 
\begin{eqnarray}
    \label{eq:backward_diffusion}
    && I(q, z_{i-1}) = \exp\left(\frac{q}{2}\frac{d}{dz}\right)[I_{i+1}(z)]^{m_{i+1}/m_i}\Bigl|_{z=z_{i-1}}
\end{eqnarray}
since the Gaussian is the Green function of heat equation. So, the $i$-RSB fold is the result of a backward diffusion in "time" $q$ of the "observable" $[I(q, z_i)]^{m_{i+1}/m_i}$, in the interval $[q_{i-1},q_{i+1}]$. The limit to infinite breakings, called \emph{Full Replica Symmetry Breaking} (fRSB), is the limit to continuity of this process: the resulting free-energy is
\begin{eqnarray}
\label{eq:fe_SK_fRSB}
    && f(\beta, h)\,=\,\underset{q(x)}{\max}\;\Phi[q(x)] \\
    && \Phi[q(x)]\,=\,-\frac{\beta}{4}\left(1-2q_{EA}+\int_0^1q^2(x)dx\right)-\frac{1}{\beta}\int_{-\infty}^{+\infty}d\mathcal{G}_{q_0}(z)\varphi(q_0, z) \nonumber
\end{eqnarray}
where the maximisation is over the space of non-decreasing functions $q(x)$ in the interval $[q_0,q_{EA}]$.
We have introduced the function $\varphi(q, z)=\log I(q, z)^{1/x(q)}$, which in the same interval satisfies Parisi equation
\begin{equation}
\label{eq:Parisi_EQ}
    \frac{\partial \varphi}{\partial q}\,=\,-\frac{1}{2}\left[\frac{\partial^2 \varphi}{\partial z^2}+x(q)\left(\frac{\partial \varphi}{\partial z}\right)^2\right]
\end{equation}
with the boundary condition \eqref{eq:final_condition_I}. Notice that the function $I(q, z)=\exp x(q) \varphi(q, z)$ satisfies a backward diffusion equation.
The overlap pdf can be written as
\begin{equation}
\label{eq:form_of_overlap_distribution_RSB}
    P(q)\,=\,x(q_m^+)\delta(q-q_m)+\tilde{P}(q)+[1-x(q_M^-)]\delta(q-q_M)
\end{equation}
where $\tilde{P}(q)$ is a smooth function in the interval $(q_0, q_1)$. When the dAT line is approached from the SG phase, the two delta peaks gets closer and closer until they collapse in a unique spike: the RS saddle point becomes stable and the $P(q)$ acquires the trivial form observed in unfrustrated magnets. The overlap distribution $P(q)$, or equivalently $q(x)$ or $x(q)$, is the true order parameter of SG phase transition. Parisi solution was found to be marginally stable in the whole Spin Glass phase \cite{de1983eigenvalues} and only less than two decades ago it was rigorously proven that it is indeed the correct solution of SK model \cite{talagrand2003spin}.

\section{Physical Meaning}

\subsection{Pure states}
It is an established result of statistical mechanics that the configuration space of any system can be decomposed in thermodynamics pure states. A pure state is a region of configuration space where Gibbs measure concentrates in the thermodynamic limit. Usually, pure states are defined in terms of different boundary conditions or through the application of specific magnetic fields. The decomposition of Gibbs measure in pure states reads
\begin{equation}
\label{eq:pure_states_decomposition}
    P(\boldsymbol{S})\,=\,\sum_{\alpha}w_{\alpha}P_{\alpha}(\boldsymbol{S})
\end{equation}
where $w_{\alpha}$ are probability weights summing to unity. Usually, at high-temperature there is only a single pure states, called \emph{Gibbs state}. However, in some systems at low enough temperatures the Gibbs measure splits in many different pure states\footnote{When there is more than a pure state, the system ceases to be ergodic.}, following a thermodynamic continuous phase transition. The simplest system exhibiting such a behavior is Ising Ferromagnet: at the onset of the transition, Gibbs measure decomposes in two "up" and "down" pure states, as a consequence of the spontaneous breaking of the global inversion symmetry. Generally, the appearance of many pure states is related to a second order phase transition, in which a global symmetry of the system spontaneously breaks.
Disordered systems are special because it is the addition of frustration and disorder that leads to the appearance of exponentially many\footnote{The term "exponential" refers to the concept of Complexity, which will be discussed in next chapter.} pure states.

\subsection{Equivalence between replica and pure states}
A necessary condition for a pure state $\alpha$ is the so called \emph{clustering property}: connected correlations evaluated with the Gibbs measure restricted to $\alpha$ decay to zero at large distances
\begin{equation}
\label{eq:clustering}
    \frac{1}{N^r}\sum_{i_1\dots i_r}\langle s_{i_1}\dots s_{i_r}\rangle_{\alpha}^{conn}\underset{\delta r\rightarrow \infty}{\rightarrow} 0
\end{equation}
When it comes to mean field models, the absence of a spatial structure makes clustering property \eqref{eq:clustering} reduce to
\begin{equation}
\label{eq:pure_states_factorization_Gibbs}
    \langle S_{i_1}\dots S_{i_r}\rangle_{\alpha}=\prod_{k=1}^r\;\langle S_{i_k}\rangle_{\alpha}=\prod_{k=1}^r\;m_i^{\alpha}
\end{equation}
 In the context of mean field theories, pure states are completely identified by amorphous magnetisation profiles $\boldsymbol{m}^{\alpha}$. The most relevant information regarding two pure states is their similarity: this can be measured through the overlap
\begin{equation}
\label{eq:pure_states_overlaps}
    q_{\alpha\beta}\equiv \frac{1}{N}\boldsymbol{m}^{\alpha}\cdot\boldsymbol{m}^{\beta}=\frac{1}{N}\sum_{i=1}^N m_i^{\alpha}m_i^{\beta}
\end{equation}
which is also a measure of their distance in configuration space, given that $d_{\alpha\beta}^2\equiv \lVert \boldsymbol{m}^{\alpha}-\boldsymbol{m}^{\beta} \rVert^2 =2(q_{EA}-q_{\alpha\beta})$.
Since states are not coupled in the decomposition of Gibbs measure, the self-overlap $q_{\alpha\alpha}$ is equal to the Edwards-Anderson order parameter for any pure state. We can define the probability distribution of the overlaps
\begin{equation}
    \label{eq:pdf_overlaps_givensample}
    P_J(q)\,=\,\sum_{\alpha\beta}\;w_{\alpha}w_{\beta}\;\delta(q-q_{\alpha\beta})
\end{equation}
between the pure states of a given sample, identified by a realisation of the couplings. Thanks to clustering property \eqref{eq:pure_states_factorization_Gibbs}, we can compute any spin correlation as an average over the overlaps:
\begin{equation}
    \label{eq:q_k_moment_given_sample}
    q_J^{(k)}\,=\,\frac{1}{N^k}\sum_{i_1\dots i_k}\langle s_{i_1}\dots s_{i_k}\rangle^k\,=\,\sum_{\alpha\beta}w_{\alpha}w_{\beta}q_{\alpha\beta}^k\,=\,\int dq P_J(q)q^k
\end{equation}
The physical values $q^{(k)}\equiv\overline{q_J^{(k)}}$ of these correlations are obtained by performing a sample average: by comparison with replica
\begin{equation*}
    \overline{\langle s_i^a s_i^b\rangle^k}=(Q_{ab}^*)^k=q^k
\end{equation*}
it can be shown that the sample average of \eqref{eq:pdf_overlaps_givensample} is equal to the distribution of replica overlaps \cite{mez1987}
\begin{equation}
\label{eq:replica_states_equivalence}
\overline{P_J(q)}\,=\,\sum_{\alpha\beta}\overline{w_{\alpha}w_{\beta}\delta(q-q_{\alpha\beta})}\,=\,\lim_{n\rightarrow 0}\frac{2}{n(n-1)}\sum_{(ab)}\delta(q-Q_{ab})
\end{equation}
This equivalence between pure states and replica is discussed in the paper \cite{mezard1984nature} by Parisi, Mézard, Virasoro. We conclude this section by reporting some of the most important physical consequences of Parisi solution.

\subsection{Properties of Parisi solution}

\subsubsection{Ultrametricity}
Consider now three different replicas $a, b, c$ and suppose one wants to measure the joint pdf of their overlaps $Q_{ab}, Q_{bc}, Q_{ac}$. In physical terms, this corresponds to study the statistics of relative distances in phase space between triples of pure states, extracted according to their weights. Within Parisi solution, a short calculation \cite{mezard1984nature} returns the following striking result
\begin{eqnarray}
\label{eq:Ultrametricity}
    && P(q, q', q'')\,=\,\frac{1}{2}P(q)x(q)\delta(q-q')\delta(q'-q'')+\frac{1}{2}[P(q)P(q')\theta(q-q')\delta(q'-q'') \nonumber \\
    && +P(q')P(q'')\theta(q'-q'')\delta(q''-q)+P(q'')P(q)\theta(q''-q)\delta(q-q')]
\end{eqnarray}
This equation states that in phase space the sets of distances of any triple of pure states forms a triangle that is either equilateral or isosceles with a shorter third site. Basically, the triangular inequality
\begin{equation*}
    d_{ab}\leq d_{ac}+d_{cb}
\end{equation*}
holding in Euclidean spaces is replaced by the stronger inequality
\begin{equation}
\label{eq:UltraM_ineq}
    d_{ab}\leq \max(d_{ac}, d_{cb})
\end{equation}
in the phase space of a RSB system. Any set of real numbers satisfying eq. \eqref{eq:UltraM_ineq} is called \emph{Ultrametric} (UM). In phase space, ultrametricity prescribes a precise organisation of pure states into clusters, grouping pure states with the same overlap in clusters. This structure can be represented through a Caley tree of depth $k+1$, if the system is $k$-RSB: at the root, we define the cluster of pure states with overlap $q_0\equiv q_{min}$ (maximally different), at the first level (offsprings of the root) states with overlap $q_0<q_1<q_2$, at the $l$-th level states with overlap $q_l$. Finally, at the $k$-th we have $q_k\equiv q_{EA}$ and at the leaves the configurations belonging to each state. Note that the branching of each node is random and that in the full RSB limit the tree becomes continuous. It can be shown that ultrametricity implies that free energies fluctuations at each level of the tree are distributed exponentially \cite{mez1987}: the average number of clusters at overlap $q$ and with free energies in $f$, $f+df$ reads
\begin{equation}
    d\mathcal{N}(f)\,=\,\exp(x(q)f)\;df
\end{equation}
In particular, at the last level $x(q_{EA})=1$ and
\begin{equation}
    d\mathcal{N}_{\alpha}(\epsilon)\propto \exp(\beta \epsilon-F^{\alpha}))dE^{\alpha}\equiv e^{\beta S_{\alpha}}dE^{\alpha}
\end{equation}
as it should be.
Note that these distributions are universal: all the details of the particular SG model are contained in function $x(q)$.

\subsubsection{Susceptibilities}

Another important consequence of Parisi solution concerns the response of the system to external perturbations. The breaking of erdogicity-i.e. the existence of infinitely many pure states-in the context of Parisi solution translates in the existence of a spectrum of time-scales in the relaxation dynamics of the system, such that relaxation becomes increasingly slow as time passes and new epochs (time-scales) are entered. This phenomenon is called \emph{aging} \cite{svedlindh1987relaxation}. Aging dynamics affects also the response of the system: suppose to measure the out-of-equilibrium magnetic susceptibility $\chi(t)$, considering the two limits $N\rightarrow \infty$ and $t\rightarrow \infty$ in different order. It holds
\begin{eqnarray}
    \label{eq:intra_state_VS_eq_responses}
    && \lim_{N\rightarrow\infty}\lim_{t\rightarrow\infty}\chi_N(t)\equiv \beta (1-\overline{q_J})\,=\,\chi_{eq} \\
    && \lim_{t\rightarrow\infty}\lim_{N\rightarrow\infty}\chi_N(t)\equiv \beta (1-q_{EA})\,=\,\chi_{LR}
\end{eqnarray}
Whenever there is RSB, $\overline{q_J}<q_{EA}$, implying that $\chi_{LR}<\chi_{eq}$: $\chi_{LR}$ is the linear response susceptibility inside a pure state, whereas $\chi_{eq}$ is the equilibrium susceptibility. The interpretation of the two susceptibilities in a large but finite sized system is the following: at small times the system has explored only the configurations of the state where it was initialised and $\chi(t)\approx \chi_{LR}$, but asymptotically $\chi(t)\approx \chi_{eq}$. At intermediate times, families of states belonging to different ultrametric clusters are explored, and $\chi(t)\approx \chi(q)$ in the time scale corresponding to level $q$. As to spin-glass susceptibilities, by exploiting pure states decomposition \eqref{eq:pure_states_decomposition} it can be shown that
\begin{equation}
    \chi_{SG}\,=\,N\operatorname{Var} q_J
\end{equation}
so that it is infinite in the thermodynamic limit whenever the system is RSB.

\subsubsection{$P_J(q)$ is not self-averaging}
Another important result concerns sample fluctuations of $P_{J}(q)$ \cite{mezard1984nature}. Take four distinct replica and consider a pair of overlaps corresponding to two distinct couples among them: since in the usual framework they are non-interacting, the joint pdf of the overlaps factorizes
\begin{equation*}
    P_J(q_{12},q_{34})\,=\,P_J(q_{12})P_J(q_{34})
\end{equation*}
Anyway, after performing the sample average the result is 
\begin{equation}
\label{eq:PJ_not_self_averaging}
    \overline{P_J(q, q')}\,=\,\frac{1}{3}\overline{P_J(q)}\delta(q-q')+\frac{2}{3}\overline{P_J(q)}\overline{P_J(q')}
\end{equation}
implying that \emph{the sample-dependent overlap distribution is not self-averaging}. This results tells us that the organisation of states of each sample is unique: non-trivial sample-to-sample fluctuations are present and one must take them into account when performing, for instance, numerical simulations.

\newpage




\section{Vector Models}
\label{sec:vector_models_chap1}

The history of mean field theories of SG started with the study of SK model, which defines an Ising spin system with frustrated infinite-range interactions. Generalisations of SK model to systems of vector spins were studied in the same years. The most important advantage to deal with vector spins is that in these systems it is possible to study a zero temperature RSB transition in the external field. Beside the theoretical interest, a mean field theory of vector spin glasses is of great interest in order to reproduce features of real systems that are absent in discrete models: the most important one, that is also the object of this thesis, is the study of low energy excitations. In this section we will discuss general features of vector spin glasses, stressing on the effect of an external magnetic field on the RSB transition. We consider the generalisation of SK model to vector spins with arbitrary number of components $m$
\begin{equation}
    \label{eq:Hamiltonian_Vector_Models}
    \mathcal{H}_J[\boldsymbol{S}]\,=\,-\frac{1}{2}\sum_{i, j}\;J_{ij}\Vec{S}_i\cdot\Vec{S}_j-\Vec{H}\cdot\sum_{i=1}^N\;\Vec{S}_i
\end{equation}
where each spin has fixed norm $|\Vec{S}_i|=1$.

\subsection{Isotropic case}

Let us begin with the case $\Vec{H}=0$: in absence of an external field, the system is isotropic, being invariant under the group $O(m)$ of $m$-dimensional global rotations. 
The replica computation for generic $m$ was firstly performed by de Almeida and Thouless \cite{de1978infinite} with a RS ansatz
\begin{equation}
\label{eq:RS_ansatz_mVectors}
    Q_{ab}^{(\alpha\beta)}\,=\,\delta^{(\alpha\beta)}[p\delta_{ab}+q(1-\delta_{ab})]
\end{equation}
following the same steps as SK. After eliminating the diagonal overlaps since saddle point equations trivially return $p=1/m$, the RS Replica Action reads
\begin{eqnarray}
    && A_{RS}(q)\,=\,\frac{\beta^2 J^2 m n (n-1)}{4}q^2-\frac{\beta^2 J^2 n}{4 m}+\frac{\beta^2 J^2 n}{2}q-\log W(q) \\
    && W(q)\,=\,S_m(1)\int_{0}^{\infty}\;\frac{dh}{(2\pi)^{m/2}}\;h^{m-1}\;\exp\Bigl(-\frac{h^2}{2}\Bigr)\;\mathcal{K}_m(\beta J\sqrt{q} h)^n.
\end{eqnarray}
where $S_m(1)$ is the total solid angle of the $m$-dimensional sphere of radius one and $Y_m(x)$ is a decorated modified Bessel function of order $m/2-1$:
\begin{equation*}
    S_m(1)\,=\,\frac{2\pi^{m/2}}{\Gamma(m/2)}
\end{equation*}
\begin{equation}
\label{eq:mVector_K}
    \mathcal{K}_m(x)\,=\,(2\pi)^{m/2}\frac{I_{m/2-1}(x)}{x^{m/2-1}}.
\end{equation}
The RS free energy after the $n\rightarrow 0$ limit reads
\begin{eqnarray}
    && f(\beta)\,=\,\max_{q}\Phi_{\beta}(q) \nonumber \\
    && \Phi_{\beta}(q)=-\frac{J^2\beta m}{4}\Bigl(\frac{1}{m}-q\Bigr)^2-\frac{1}{\beta}\int_0^{\infty}\frac{dh}{Z_m}h^{m-1}e^{-\frac{h^2}{2}}\;\log Y_m(\beta J\sqrt{q} h)
\end{eqnarray}
where $Z_m\,=\,(2\pi)^{m/2}/S_m(1)\,=\,2^{m/2-1}\Gamma(m/2)$. The saddle point equation giving the overlap $q$ is
\begin{equation}
    q\,=\,\frac{1}{m}\int_{0}^{\infty}\frac{dh}{Z_m}h^{m-1}e^{-\frac{h^2}{2}}\left[\frac{I_{m/2}(\beta J\sqrt{q} h)}{I_{m/2-1}(\beta J\sqrt{q} h)}\right]^2
\end{equation}
The saddle point value in the high temperature region, as in the SK model, is $q=0$, yielding
\begin{equation}
\label{eq:fe_para_mVector_noextfield}
    f(\beta)\,=\,-\frac{J^2\beta}{4 m}-\frac{1}{\beta}\log S_m(1)
\end{equation}
At $T=T_c\,=\,J/m$ a $q\neq 0$ saddle point appears: the free energy is non-analytic and the system undergoes a freezing phase transition. However, as in the SK case, for any finite $m$ the $RS$ solution is unstable for $T<T_c$. Generalising the analysis made in \cite{de1978stability} to vector models, close to $T=T_c^{-}$ they find that the Replicon eigenvalue
\begin{equation*}
    \lambda_R\,\approx\,1-\frac{J}{m T}
\end{equation*}
which becomes negative for $T<T_c$. In the $m\rightarrow \infty$ limit, the vector model converges to the $p=2$ spherical model \cite{kosterlitz1976spherical}, which is RS-stable. 

The correct solution satisfies the fRSB ansatz, with eqs. \eqref{eq:fe_SK_fRSB} and \eqref{eq:Parisi_EQ} generalised to the $m$-dimensional case. The boundary condition for Parisi equation is now
\begin{equation*}
    f(q_{EA}, h)\,=\,\log Y_m(\beta h)
\end{equation*}
In the isotropic case, all the properties of the overlap distribution are identical to the ones found in the case $m=1$. When there is a non-zero external field, either uniform or random with non-zero mean, the RSB scenario becomes more complicated, since one has longitudinal and transverse fluctuations with respect to the external field.

\subsection{Anisotropic case}

In presence of an external field, the RS saddle point is
\begin{equation}
\label{eq:RS_ansatz_mvector_extfield}
    Q_{ab}^{(\alpha\beta)}\,=\,\begin{cases}
        & q_{\perp}\;\delta_{\alpha\beta}\qquad a\neq b, \alpha,\beta\neq 1 \\
        & \frac{\delta_{\alpha\beta}}{m}\qquad a=b, \alpha,\beta\neq 1 \\
        & q_{\parallel}\qquad a\neq b, \alpha=\beta=1 \\
        & r\qquad\qquad a=b, \alpha=\beta=1
    \end{cases}
\end{equation}
where the external field is taken as directed along direction $\alpha=1$.
Take any spin and decompose it into its projection along and perpendicularly to the external field. When the field is strong, the system will be strongly magnetised along the field direction, destroying the interaction frustration. Upon lowering the field, the effect of interactions becomes increasingly important, until the system freezes in a spin glass state. Note that in the anisotropic case it is expected for the spin glass phase
\begin{eqnarray}
    & q_{\parallel}\neq 0\qquad m_{\parallel}\neq 0 \\
    & q_{\perp}\neq 0\qquad m_{\perp}=0
\end{eqnarray}
In the plane $(|\Vec{H}|, T)$, there are two instabilities lines:
\begin{itemize}
    \item The Gabay-Tolouse (GT) line \cite{gabay1981coexistence, gabay1982symmetry}: under this line, the transverse overlap distribution undergoes a fRSB transition. For the longitudinal overlaps, one only has a \emph{weak} RSB transition: the function $q(x)$ shows a weak dependence on the Parisi parameter $x$. Close to zero field, the GT line behaves as
    \begin{equation}
    \label{eq:GT_line_closeToZeroField}
        \left(\frac{H}{J}\right)^2\,\approx\,\frac{4(m+2)^2}{m[(m+2)^2-2]}\left(\frac{T_c^{(0)}-T}{T_c^{(0)}}\right)
    \end{equation}
    \item The de Almeida Thouless (dAT) line \cite{cragg1982spin, elderfield1982parisi}: this is the generalization of the instability line found in \cite{de1978stability}. Under the dAT line, the longitudinal overlap distribution acquires the fRSB form \eqref{eq:form_of_overlap_distribution_RSB}. Close to zero field, one has
    the same scaling found in the $m=1$ case
    \begin{equation}
    \label{eq:dAT_line_mVector_closeToZeroField}
        \left(\frac{H}{J}\right)^2\,\approx\,\frac{4}{m(m+2)}\left(\frac{T_c^{(0)}-T}{T_c^{(0)}}\right)^3
    \end{equation}
\end{itemize}
The dAT line thus in the anisotropic case figures as a crossover from a weak to a strong fRSB regime, rather than a true instability line. It has been pointed out in \cite{moore1982critical, gabay1982symmetry} that, as far as criticality is concerned, the RSB at the GT line of a $m$ vector model is equivalent to the RSB of an isotropic $m-1$-vector model: the authors find that the expansion of the Parisi function $q_{\perp}(x)$ of the $m$-vector model is the same as that of $q(x)$ of the isotropic $m-1$ vector model. Thus, the RSB of an anisotropic vector model with $m$-dimensional spins can be understood in terms of a $m-1$ dimensional vector model that enters the SG phase strongly\footnote{According to the distinction between weak and strong RSB that we made a few lines before.} at the GT line and a SK model that freezes at the dAT line.
The two behaviors at small magnetic field, eqs. \eqref{eq:GT_line_closeToZeroField}, \eqref{eq:dAT_line_mVector_closeToZeroField} are observed experimentally in \cite{monod1982equilibrium, Lauer1982, Fogle1983, Cam1983}, showing the goodness of the RSB transition as a theoretical description of the SG phase.

\noindent One aspect peculiar of vector models is that, for $m>2$, both GT and dAT lines converge to finite critical fields at $T=0$: thus, in these cases there is a zero-temperature SG phase transition in the external field strength.

\subsection{Random Field}

Let us consider now the case of quenched and uncorrelated random external fields $\{\Vec{H}_i\}_{i=1}^N$ in eq. \eqref{eq:Hamiltonian_Vector_Models}
\begin{equation}
    \label{eq:Hamiltonian_Vector_Models_Random_Fields}
    \mathcal{H}[\boldsymbol{S}]\,=\,-\frac{1}{2}\sum_{i, j}^{1, N}\;J_{ij}\Vec{S}_i\cdot\Vec{S}_j-\sum_{i=1}^N\;\Vec{H}_i\cdot\Vec{S}_i.
\end{equation}
If we take external fields with non-zero mean, we fall back into the anisotropic case previously discussed. Instead, it is interesting to consider random fields with zero mean: in this situation, even though the $O(m)$ symmetry is broken for any possible sample, the system is \emph{on average isotropic}. Indeed, since external fields have zero mean, the system does not develop a global magnetization, and so all physical quantities are isotropic. In real system, a spatially random external field can be well represented by an external field that oscillates very fast and irregularly in space. 

Before discussing the random field SG, it is worth to briefly discuss the main features of Random Field Ferromagnets. The Random Field Ising Model (RFIM) was introduced by Larkin in 1970 in order to model the pinning of vortices in superconductors \cite{larkin1970effect}. The Random Field Ising Model (RFIM) has been studied extensively in the last fifty years \cite{fytas2018review}: it consists of a system of Ising spins with ferromagnetic interactions and random fields with zero mean and variance $H^2$. 
In the plane $(H, T)$, there is an instability line which separates a high-temperature or/and high-field paramagnetic phase and a low temperature-low field ferromagnetic phase. Random fields introduce frustration, so the ferromagnetic phase displays features similar to those of spin glasses, such as a slow relaxation dynamics \cite{belanger1998experiments}. One can wonder if spin glass long range order can occur in such systems: in mean field theories on fully connected graphs, the RFIM cannot display any RSB, because the overlaps do not figure as order parameters of the Replica Action, but only replica magnetisation. Conversely, in finite dimensional systems and in mean field theories defined on random graphs with finite connectivity this is not true and spin glass ordering is possible. For the RFIM, it was shown recently \cite{krzakala2010elusive} that RSB never occurs: the Spin-Glass susceptibility is always upper bounded by the Ferromagnetic susceptibility. In vector models with random field this is not true: indeed, just a few years ago it was shown \cite{lupo2019random} that the Random Field XY model can develop a spin glass phase intermediately between the paramagnetic and the ferromagnetic phase.

\noindent Let us switch back to the random field spin glass and let us consider a gaussian random field with zero mean and covariance matrix $H^2\mathbb{I}$. A replica computation of the free energy including also the gaussian field average yields
\begin{equation}
    \Phi_{\beta, H}(q)\,=\,-\frac{J^2\beta m}{4}\Bigl(\frac{1}{m}-q\Bigr)^2-\frac{1}{\beta}\int_0^{\infty}\frac{dh}{Z_m}h^{m-1}e^{-\frac{h^2}{2}}\;\log Y_m(\beta J\sqrt{q+H^2} h)
\end{equation}
\begin{equation}
    q\,=\,\frac{1}{m}\int_{0}^{\infty}\frac{dh}{Z_m}h^{m-1}e^{-\frac{h^2}{2}}\left[\frac{Y'_m(\beta\sqrt{J^2q+H^2} h)}{Y_m(\beta\sqrt{J^2q+H^2} h)}\right]^2
\end{equation}
The RS equation in presence of an external field admits a $q>0$ solution. In particular, for low $H$ one finds $q\approx H^{2/3}$, as expected for mean field theories. The most important physical effect related to the random external field is the depletion of small local fields: for sufficiently strong $H$, typical local fields of the sample and fluctuations about them scale as $H$. The depletion of small fields corresponds to a depletion of high local responses, making the system increasingly stable against external perturbations.

In the random field scenario, the GT line disappears and the dAT line reacquires its role of instability line. The stability analysis of the RS saddle point of the Replica Action is done both in \cite{sharma2010almeida}: in the $n\rightarrow 0$ limit, nine distinct eigenvalues of the Hessian are found. As in the $m=1$ case, the vanishing of the smallest eigenvalue, the Replicon, defines the dAT line: the related stability condition reads
\begin{equation}
    \label{eq:dAT_line_mVector}
    \frac{1}{\beta^2 J^2} > \int_0^{\infty}\frac{dh}{Z_m}h^{m-1}e^{-\frac{h^2}{2}}\left[\left(1-\frac{1}{m}\right)\frac{g_m(\beta\sigma h)^2}{(\beta \sigma h)^2}+\frac{1}{m} g'_m(\beta\sigma h)^2\right]
\end{equation}
\begin{equation}
    \sigma=\sqrt{J^2q+H^2}\qquad g_m(x)\,=\,\frac{Y'_m(x)}{Y_m(x)}\equiv \frac{I_{m/2}(x)}{I_{m/2-1}(x)}.
\end{equation}
For $m>2$ one can explicitly compute the $T=0$ critical field
\begin{equation}
\label{eq:zero_T_crit_field}
    \frac{\sigma^2(0)}{J^2}=\int_0^{\infty}\frac{dh}{Z_m}h^{m-3}e^{-\frac{h^2}{2}}\Longrightarrow H_c(0)\,=\,\frac{J}{\sqrt{m(m-2)}}
\end{equation}
The critical field decreases with increasing $m$ and goes to zero in the spherical limit, as it should be, since the system in the $m\rightarrow \infty$ limit after proper rescaling converges to the spherical model which is known to be RS.

\section{Spin Glasses on random graphs}

So far we have discussed spin glass models with infinite-range interactions, such that each spin interacts with all the others spins, with no underlying space structure. An improved mean field theory can be obtained by studying spin glass models such that in the thermodynamic limit each spin interacts with a finite neighborhood, chosen at random. We refer to this models as \emph{Diluted Spin Glass} models.

\subsubsection{Brief digression on Random Graphs}

A graph is specified by the couple $(\mathcal{V}, \mathcal{E})$ consisting of the \emph{vertex} or \emph{node} set $\mathcal{V}$ and the \emph{edge} or \emph{link} set $\mathcal{E}$. The \emph{adjacency} or \emph{connectivity} matrix of a graph is defined by
\begin{equation}
\label{eq:adjacency}
    A_{ij}=\begin{cases}
        & 1\qquad [ij]\in\mathcal{E}\quad i\text{,}j\in\mathcal{V} \\
        & 0\qquad \text{otherwise}
    \end{cases}
\end{equation}
The adjacency matrix and its powers contain all the topological information of the network. For instance, the $l$-th power of the adjacency matrix yields all the paths of length $l$ connecting two sites. A random graph is an instance of the ensemble $\mathcal{G}(N, M)$, where $N$ is the number of vertices, $M$ the number of edges and links are assigned to nodes according to some statistical rule. Consider the degree $d$ of a node, which is the number of other nodes to which it connects: by choosing different probability distributions for it, one can define different random graphs ensembles. The simplest and also most important examples are
\begin{itemize}
    \item The \emph{Random Regular Graph} (RRG): each node has a fixed connectivity $d=c$. Thus, if $N$ is the number of nodes, the number of links is exactly $M=N c/2$.
    \item The \emph{Erdos-Renyi Graph} (ERG): the connectivity of each node is a poissonian variable with mean $c$. The average number of link is $\overline{M}=N c/2$.
\end{itemize}
Any graph with $c=O(1)$ is a \emph{sparse} graph. Note that the fully connected graph is the special case $c=N-1$; in general, if $c=O(N)$ we deal with \emph{dense} graphs. Graphs such RRG and ERG belong to the class of Bethe Random Graphs: these graphs for $N$ going to infinity are equivalent to the infinite Caley tree. A tree is a graph without loops, i.e. there is a unique path connecting any couple of nodes: thus, in RRG and ERG loops should become infinitely long in the thermodynamic limit. Indeed, it can be shown that the loops on such graphs have typical lengths
\begin{eqnarray}
\label{eq:Random_Graphs_Radii}
    && \ell_{RRG}(N)\,=\,\frac{\log N}{\log(c-1)} \\
    && \ell_{ERG}(N)\,=\,\frac{\log N}{\log c}
\end{eqnarray}
An important property of random graphs is the shortest path lengths distribution. In RRG, it can be rigorously shown that for $N$ large the probability that any two nodes are connected by a shortest path of length at most $L-1$ is a discrete Gompertz distribution \cite{tishby2022mean}:
\begin{eqnarray}
\label{eq:path_length_cumulative_distribution}
    & \mathcal{P}(\ell<L)\;=\;\eta\;\exp(-\eta\;(e^{b\ell}-1))\theta(\ell)+\theta(-\ell) \\
    & \eta\,=\,\frac{c}{N(c-2)}\qquad b=\log(c-1) \nonumber
\end{eqnarray}
For the ERG one obtains the same formula if $b=\log(c)$ is used (verificare se è vero). The probability of having a path of exactly length $\ell$ and the expected number of nodes at the same distance are
\begin{eqnarray}
\label{eq:path_length_density}
    & P(\ell)\,=\,\mathcal{P}(\ell-1<L)-\mathcal{P}(\ell<L) \\
    & \mathcal{N}(\ell)\,=\,N\,P(\ell)
\end{eqnarray}
Note that when $\eta e^{b\ell}\ll 1$, corresponding to $\ell \ll \ell_{RRG}$ (see eq. \eqref{eq:Random_Graphs_Radii}), one finds the tree growing law
\begin{equation}
    \mathcal{N}(\ell)\,=\,c(c-1)^{\ell-1}\theta(\ell)+\delta_{\ell,0}+O\Bigl(\frac{1}{N}\Bigr)
\end{equation}
which becomes exact only for $N\rightarrow \infty$.

\subsection{Bethe-Peierls approximation}

Mean field (MF) approximations for finite-dimensional systems are based on the assumption that spatial fluctuations of macroscopic observables are negligible\footnote{It is known that close to critical points mean field approximations often fail, like in the case of Ising model in three dimensions. For any model with a second order phase transition there is an upper critical space dimension over which mean field approximation works at the critical point.}. 
In order to achieve that, typically one assumes that all connected correlations are zero in the thermodynamic limit: this implies (eq. \eqref{eq:pure_states_factorization_Gibbs} in presence of a single pure state)
\begin{equation*}
    P(\boldsymbol{S})\,\underset{N\rightarrow\infty}{\approx}\,\prod_{i=1}^N\;\eta_i(S_i)
\end{equation*}
\begin{equation*}
    \langle f(\boldsymbol{S})\rangle\,\underset{N\rightarrow\infty}{\approx}\,f(\boldsymbol{m})
\end{equation*}
where here we are considering generic spin variables, $\eta_i(S_i)$ is the marginal on site $i$ and $f$ is a generic smooth function of the configurations.
If these approximations hold, each site feels exactly the same field from its neighbors. Indeed, consider a ferromagnetic system with pairwise interactions: the magnetisation satisfies
\begin{equation*}
    m\,=\,\langle f(\beta J\sum_{j}S_j)\rangle
\end{equation*}
In the MF approximation, this simplifies in
\begin{equation*}
   m\,=\,f(\beta J\,c\,m) \equiv f(\beta h)
\end{equation*}
In disordered systems, because of disorder the mean field criterion can be restated by saying that \emph{statistically} each site feels the same field from its neighborhood: there is a site-independent pdf of the local field. 

\noindent It is well known that only in fully connected systems the MF approximation is exact: we shall call theories of these kind \emph{naive} or \emph{long range} MF theories.
\noindent MF approximation can be improved by including pair correlations between nearest neighbours. The Gibbs measure is factorised as
\begin{eqnarray}
\label{eq:MF_Bethe_Lattice_Factorisation_Gibbs}
    && P(\boldsymbol{S})\,\underset{N\rightarrow\infty}{=}\,\prod_{(ij)}\eta_{ij}(S_i|S_j)\,=\,\prod_{(ij)}\frac{\eta_{ij}(S_i, S_j)}{\eta_i(S_i)\eta_j(S_j)}\prod_k \eta_k(S_k) \\
    && \,=\,\prod_{(ij)}\eta_{ij}(S_i, S_j)\prod_k \eta_k(S_k)^{1-d_k} \nonumber
\end{eqnarray}
where $d_k$ is the degree of node $k$. Normalisation and marginalisation constraints hold
\begin{equation*}
    \underset{S_i}{\Tr}\;\eta_i(S_i)=1\qquad\qquad\underset{S_i\;S_j}{\Tr}\eta_{ij}(S_i, S_j)=1
\end{equation*}
\begin{equation*}
    \underset{S_j}{\Tr}\;\eta_{ij}(S_i, S_j)=\eta_i(S_i)
\end{equation*}
Theories based on this last equation are known as \emph{Bethe-Peierls} MF theories \cite{bethe1935statistical, peierls1936ising}. It can be rigorously proven that they are exact on tree graphs \cite{mezard2009information}. Therefore, in the thermodynamic limit they are exact on any Bethe random graph. 

\subsection{Belief Propagation and Cavity Method}
\label{sec:BP_and_cav}

A way to exploit eq. \eqref{eq:MF_Bethe_Lattice_Factorisation_Gibbs} is to use a probabilistic technique called \emph{Belief Propagation} (BP) \cite{pearl1988probabilistic}, firstly developed in the fields of Information Theory and Artificial Intelligence. Suppose to have a sequence of events labelled with indices $i$ and to assign to each of them a value $x_i$ and a belief $\eta_i(x_i)$ about its probability. If the events are not independent, it is always possible to build a Bayesian network of their mutual causal relations. Let us stick only to the simpler case where events are pairwise related: the probability of event $j$ as a cause of event $i$ is written as an input "message" $\hat{\eta}_{j\rightarrow i}(x_i)$ corresponding to a directed edge on the network. Let us assume that the Bayesian network is sparse and \emph{Markovian}
\begin{equation}
\label{eq:BP_hyp_1}
    \mathbb{P}(x_i|\{x_j\}_{j\neq i})\,=\,\mathbb{P}(x_i|\{x_j\}_{j\in\partial i})
\end{equation}
i.e. that event $i$ depends only on the events directly related to it: $\partial i$ stands for the neighborhood of $i$. In this case, it is possible to substitute cause-effect relations with non-negative functions $\{\psi_{ij}(x_i, x_j)\}$ and express the effect of external sources with non-negative functions $\{\phi_i(x_i)\}$. In a Markovian network one-point and two point marginals satisfy
\begin{eqnarray}
    && \eta_i(x_i)\,=\,\frac{1}{\mathcal{Z}_i}\phi(x_i)\prod_{k\in\partial i}\hat{\eta}_{k\rightarrow i}(x_i) \\
    && \eta_{ij}(x_i, x_j)\,=\,\frac{1}{\mathcal{Z}_{ij}}\phi_i(x_i)\phi_j(x_j)\psi_{ij}(x_i, x_j)\prod_{k\in\partial i/j}\hat{\eta}_{k\rightarrow i}(x_i)\prod_{l\in\partial j/i}\hat{\eta}_{l\rightarrow j}(x_j) \nonumber
\end{eqnarray}
where $\mathcal{Z}_i$ and $\mathcal{Z}_{ij}$ are normalisation constants and $\partial i/j$ is the neighborhood of $i$ minus $j$. The marginalisation condition returns a set of self-consistent equations for input messages
\begin{equation}
    \hat{\eta}_{j\rightarrow i}(x_i)\,=\,\frac{\mathcal{Z}_i}{\mathcal{Z}_{ij}}\underset{x_j}{\Tr}\;\phi_j(x_j)\psi_{ij}(x_i,x_j)\prod_{l\in\partial j/i}\hat{\eta}_{l\rightarrow j}(x_j)\,=\,\underset{x_j}{\Tr}\;\psi_{ij}(x_i,x_j)\eta_{j\rightarrow i}(x_j)
\end{equation}
where we have introduced output "messages"
\begin{equation}
\label{eq:output_messages}
    \eta_{i\rightarrow j}(x_i)\,=\,\frac{\phi_i(x_i)}{\mathcal{Z}_{i\rightarrow j}}\prod_{k\in\partial i/j}\hat{\eta}_{k\rightarrow i}(x_i)
\end{equation}
which satisfy BP equations
\begin{equation}
\label{eq:BP_equations}
    \eta_{i\rightarrow j}(x_i)\,=\,\frac{\phi_i(x_i)}{\mathcal{Z}_{i\rightarrow j}}\prod_{k\in\partial i/j}\underset{x_k}{\Tr}\;\psi_{ik}(x_i,x_k)\eta_{k\rightarrow i}(x_k)
\end{equation}
It is instructive to consider eq.\eqref{eq:BP_equations} as dynamical equations on the network, by including a time index $t+1$ in the r.h.s. and $t$ in the l.h.s.: they describe the flow of information in the network in terms of local updates of the single nodes; the belief of a single node is the balance of input and output messages from and towards neighbors. In a tree graph messages can be updated moving backwards from the leaves to the root. Notice that BP algorithm yields a fixed point in a tree-like graph in a time that grows linearly with the size of the system: it is an exponential speed-up with respect to a naive sampling of configuration space. One and two-point marginal can be computed once a fixed point of BP equations has been reached
\begin{eqnarray}
\label{eq:1p_marginals}
    && \eta_i(x_i)\,=\,\frac{\phi_i(x_i)}{\mathcal{Z}_i}\prod_{k\in\partial i/j}\underset{x_k}{\Tr}\;\psi_{ik}(x_i,x_k)\eta_{k\rightarrow i}^*(x_k) \\
\label{eq:2p_marginals}
    && \eta_{ij}(x_i, x_j)\,=\,\frac{1}{\mathcal{Z}_{ij}}\psi_{ij}(x_i, x_j)\eta_{i\rightarrow j}^*(x_i)\eta_{j\rightarrow i}^*(x_j)
\end{eqnarray}

Let us make a connection with physical systems: if instead of causal relations we consider interactions, then it is natural to set
\begin{equation}
    \psi_{ij}(x_i, x_j)\equiv \exp \beta \mathcal{J}_{ij}(x_i, x_j)\qquad \phi(x_i)\equiv \exp \beta b_i(x_i)
\end{equation}
where $\mathcal{J}_{ij}(x_i, x_j)$ is a pair interaction and $\beta b_i(x_i)$ is the interaction with an external source.  
Indeed, $\psi_{ij}(x_i, x_j)\phi(x_i)\equiv e^{\beta (\mathcal{J}_{ij}(x_i, x_j)+b_i(x_i))}$ is just the number density of $x_i$ given $x_j$. In unfrustrated homogeneous systems one-point and two-points marginals are the same in all the sites. If disordered is added, they become random variables differing from site to site. In the thermodynamic limit finding a fixed point of eqs \eqref{eq:BP_equations} is equivalent to find the pdf of cavity marginals: in fact, one can forget the underlying graph and consider eqs \eqref{eq:BP_equations} as an equivalence in distribution sense; this technique is the \emph{Population Dynamics Algorithm} (PDA) and it will be discussed in greater detail in next section. 
Given a BP fixed point, one can compute the Bethe free-energy
\begin{equation}
\label{eq:Bethe_FE}
    F[\{\eta_{i\rightarrow j}^*\}_{(ij)}]\,=\,-\frac{1}{\beta}(\sum_i\log \mathcal{Z}_i[\{\eta_{i\rightarrow j}^*\}_{(ij)}]-\sum_{(ij)}\mathcal{Z}_{ij}[\{\eta_{i\rightarrow j}^*\}_{(ij)}])
\end{equation}
It consists of the balance between the free energy cost of sites and links. In fact, BP equations are equivalent to what in physics is called \emph{Cavity Method} \cite{mez1987}. If a given node is isolated from the rest of the system or if its degree of freedom is constrained to assume a fixed value, in a tree graph its neighbors get uncorrelated, since the system has only pairwise interactions\footnote{In presence of group interactions, this is not true anymore: if an edge between two nodes is removed, they remain correlated through another node participating to the group interaction. The generalisation of graphs to multi-body interactions are the factor graphs \cite{mez1987}, \cite{zamponi2010mean}.}. The physical interpretation of message $\eta_{i\rightarrow j}(x_j)$ is that of marginal of the system without node $j$, as if this very node is isolated from the system through a cavity: therefore, eq. \eqref{eq:1p_marginals} is also called \emph{cavity marginal}. Eq. \eqref{eq:Bethe_FE} can be straightforwardly interpreted as the balance between the free energy gain of adding new sites and the cost related to the interactions with the nodes they are connected to. Notice that eq. \eqref{eq:BP_hyp_1} can hold if and only if there is no long-range order in the system: hence, at phase transitions the standard BP approach fails. In presence of RSB, eqs. \eqref{eq:BP_equations} do not converge in the SG phase of diluted spin systems. If RS is broken, one has a different pdf of cavity marginals for any different pure state: BP and Cavity method can be extended to RSB systems by correctly taking into account the multiplicity of states \cite{mezard2001bethe}.

\noindent To conclude this section, suppose to expand eqs \eqref{eq:BP_equations} in the couplings: it is not hard to show that in the dense limit $c=O(N)$ and $|J_{ij}|=O(1/\sqrt{N})$ they converge to TAP eqs \eqref{eq:TAP_eqs}. In dense systems cavity method coincide with linear response theory: the system is continuous under addition or subtraction of any spin. Again, this claim is true only in the RS phase: in the RSB phase the system responds non-trivially to any external perturbation, thanks to marginal stability \cite{de1983eigenvalues}.

\chapter{Structural Glasses}

The word \emph{structural glasses} refers to a class of solids whose microscopic spatial structure is disordered. These materials share many properties with spin glasses, such as aging dynamics, memory and an anomalous low temperature response \cite{castellani2005spin, cavagna2009supercooled}. This chapter is devoted to the discussion of the \emph{Random First Order Transition} (RFOT), a mean field theoretical framework that describes the glassy state as the result of a thermodynamic transition: we will do it through the lenses of p-spin models, generalised spin glass models where each spin interacts with the others via $p$-body interactions. 

\section{The glass transition}

\begin{figure}
    \centering
    \includegraphics{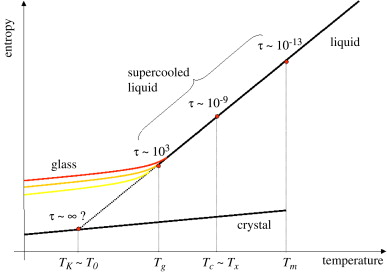}
    \caption{A standard diagram of the liquid-SCL-glass phenomenology. Image taken from \cite{cavagna2009supercooled}.}
    \label{fig:diagram_liquid_pedestrian}
\end{figure}

If a liquid is cooled down to the freezing temperature, it forms a crystal if a long enough time, comparable with the nucleation time scale of crystals, is waited. However, if cooling is sufficiently fast, crystallization can be bypassed: in this situation, the liquid enters a metastable phase and becomes a super-cooled liquid (SCL) \cite{cavagna2009supercooled}. While a liquid shows exponentially fast relaxations, with a unique Arrenhius-like time scale $\tau \propto 1/T$,
the SCL exhibits an increasingly slower relaxation dynamics the lower the temperature, with the appearance of two different relaxation regimes. In figure \ref{fig:diagram_liquid_pedestrian}
we show the typical phase diagram of the liquid-glass transition, featuring the entropy versus the temperature. The glass transition temperature $T_g$ is conventionally defined as the temperature at which the experimental time waited for relaxation is $\tau=10^3 s$. From the freezing (or melting) temperature $T_m$ to $T_g$ the measured relaxation time grows of sixteen orders of magnitude ($\tau=O(10^{-13}) s$ at $T_m$). In the intermediate region between these two temperatures, two different relaxation regimes are found:
\begin{itemize}
    \item $\beta$ or fast relaxation: in this regime, the auto-correlation function of the system develops a plateau
    \begin{equation*}
        C(t)\sim q+c_1(t/\tau_{\beta})^{-a}\qquad 1\ll t \ll \tau_{\beta}
    \end{equation*}
    We can have a physical intuition of this residual correlation with the \emph{cage picture}: the $\beta$ regime is the time scale during which a target particle of the system explores the limited environment defined by its neighborhood. 
    \item $\alpha$ or slow relaxation: this asymptotic regime is related to the diffusion of particles in the system, after they have escaped their cages. The total time needed for such an event is $t\sim \tau_{\alpha}$. As the particle diffuses out of the cage, the auto-correlation function decays from the plateau to zero
    \begin{equation*}
        C(t)\sim \exp(-(t/\tau_{\alpha})^{\delta})\qquad t\simeq\tau_{\alpha}
    \end{equation*}
    following a stretched exponential, $\delta<1$.
\end{itemize}
Between $T_m$ and $T_g$, the SCL undergoes a crossover at $T=T_x$ from a regime where $\tau_{\beta}\sim\tau_{alpha}$ ($T>T_x$) to a regime where $\tau_{\beta}\ll \tau_{\alpha}$ ($T<T_x$). This crossover phenomenon from strongly coupled time-scales to decoupled ones is a signature of the emergence of glassy behavior. This behavior was interpreted in 1969 by Golstein as a crossover for activated dynamics \cite{goldstein1969viscous}. At the temperature $T=T_x$, identified by him as the temperature where experimentally $\tau_{\alpha}\sim 10^{-9} s$, the dynamics of the SCL is driven by hopping processes through potential energy barriers whose height is significantly larger than $k_B T$, the scale of thermal fluctuations. For higher temperatures, jumps are very frequent and the system does not spend much time in a local minimum of the potential energy landscape, so there is no net distinction between the two relaxation regimes; as temperature is lowered, jumps become rarer, the system spends longer times visiting a local minimum and the two time scales separate.
 
The glass transition temperature $T_g$, at variance with the Golstein crossover temperature $T_x$, has less significance being conventional.
It is legit to see how the alpha-relaxation time behaves as temperature is further lowered, by waiting longer experimental times between a cooling step and the next. It is observed experimentally that $\tau_{\alpha}$ becomes exponentially large as temperature is decreased, until the point that the scl appears essentially frozen at antropic time-scales.
It is conjectured that in the limit of quasi-static cooling\footnote{Also called adiabatic, the system is let equilibrate at each step.} the SCL displays a thermodynamic phase transition and freezes in what is called \emph{ideal glass}. The limiting temperature $T=T_K$ is called \emph{Kauzmann temperature} after W. Kauzmann, who claimed of its existence in 1948 \cite{kauzmann1948nature}: he observed that measures of entropy and enthalpy in temperature intercept their respective values on the crystalline state at a finite non-zero temperature. He suggested that at this temperature a phase transition should occur, in order to avoid the paradoxical scenario of a SCL with lower entropy than that of the crystal for $T<T_K$.  

The existence of a thermodynamic phase transition of glasses, in which an amorphous long-range order emerges, is still a debated problem. There are essentially two hypothesis at stake:
\begin{itemize}
    \item The formation of a glass is a purely kynetic process, with no 
    underlying thermodynamic phase transition, but rather a dynamic crossover: in this picture, $\tau_{\alpha}$ grows exponentially as $T\rightarrow 0$. The theoretical framework to describe this phenomenon is \emph{Facilitation theory} \cite{kob1993kinetic, chandler2010dynamics}.
    \item There is an underlying 1RSB thermodynamic phase transition at $T=T_{K}$, and the glass formation process is described by RFOT \cite{kirkpatric87, kirkpatrick1988comparison, kirkpatrick1989scaling, BerthierBirlo2011, biroli2022rfot}: in this picture, $\tau_{\alpha}$ diverges at the Kauzmann temperature, following a super-Arrhenius law driven by the phenomenon of entropy-crisis predicted by Kauzmann. Any glass formed at $T>T_K$ is a meta-stable state\footnote{In finite dimensions there is no universally accepted definition of meta-stable state. Extensions of the RFOT picture to finite dimensions rely on the mosaic picture \cite{kirkpatrick1989scaling} and on the point-to-set correlation length \cite{bouchaud2004adam}.} of the free energy landscape.
\end{itemize}
We devote the rest of this chapter to outline key features of the RFOT, using spherical p-spin models and focusing on static aspects.



\section{Mean field picture of glass formation}

In this section we describe Random First Order Transition (RFOT), the mean field theory of glass transition which is conjectured to exist also in finite dimensions.
At the mean field level, the phenomenology of the liquid-SCL-glass process is essentially captured, with a crucial difference. While in real systems the $\alpha$ relaxation time diverges only at $T_K$, in mean field models it does exist a sharp dynamical phase transition at a higher temperature, which is called \emph{mode coupling temperature} $T=T_{MCT}>T_g$, after the renown \emph{Mode Coupling theory} developed in \cite{gotze1984some}. This mean field theory describes the equilibrium dynamics of a tagged particle with mass $m$ in a liquid. In Fourier space, the transform of the correlation density $F(k, t)$ (it is assumed isotropy) satisfies the self-consistent equations
\begin{equation}
\label{eq:Mcoupl_eq}
    \diffp[2]{F(k, t)}{t}+\frac{k^2 T}{m S(k)}F(k, t)\,=\,-\int_0^t\;ds\;M(k, t-s)F(k, s)
\end{equation}
where $S(k)\,=\,F(k, 0)$ is the structure form factor and $M(k, t-s)$ is a memory kernel between times $s<t$. This theory describes correctly the dynamics of SCL only in the region $T_{x}<T<T_{m}$ where the two modes $\omega_{\alpha}=1/\tau_{\alpha}$ and $\omega_{\beta}=1/\tau_{\beta}$ are coupled.
The solution of eq. \eqref{eq:Mcoupl_eq} predicts a divergence of $\tau_{\alpha}$ at $T_{MCT}\sim T_x$: at the mean-field level barriers are extensive, so the crossover to activation observed in finite dimensions is replaced by a fictitious ergodicity breaking transition.

This undesired mean field result can nevertheless be more useful when a connection to statics is made. This was firstly done by Kirkpatric, Thirumalai and Wolynes \cite{kirkpatrick1989scaling}, who unified the phenomenology of p-spin models to the MCT of Gotze and co-workers. Let us consider the free energy landscape of the system. The anticipated divergence of $\tau_{\alpha}$ is related to the splitting of the Gibbs measure into an exponential number of thermodynamic states, each one separated by extensive barriers. Let us consider the partition function of a thermodynamic system and decompose it in pure states:
\begin{equation}
\label{eq:Z_deco}
    Z=\sum_{\alpha} e^{-\beta F_a}\,=\,\int df e^{-\beta N f} \sum_{\alpha} \delta(f-f_{\alpha})\,=\,\int df e^{-\beta N f}\Omega(f)
\end{equation}
We define as \emph{Complexity} of the free energy level $f$ the quantity
\begin{equation}
\label{eq:Complexity}
    \Sigma(f)\,=\,\lim_{N\rightarrow \infty} \frac{1}{N}\log \Omega(f)
\end{equation}
When $\Sigma(f)$>0, there is an exponential number of metastable states with free energy $f$. Conversely, if the complexity is zero, the number of metastable states with the corresponding free energy is subexponential. Finally, if the complexity is negative, no metastable states exist with that free energy. The dynamical phase $T_K<T<T_{MCT}$ corresponds to a non-negative equilibrium complexity, $\Sigma(f_*)>0$, where $f_*$ is the saddle point in the last r.h.s. of \eqref{eq:Z_deco}. At the dynamical transition the free-energy is analytic, so thermodynamic is untouched.
In the dynamical phase the temperature has a two-fold meaning. Indeed, from the saddle point \eqref{eq:Z_deco} we have
\begin{equation}
\label{eq:def_temp_dyn_phase}
    \diff{\Sigma}{f}(f_*)\equiv \frac{1}{T}
\end{equation}
At short times, when the system is exploring a valley, the temperature is the parameter coupling the system with a thermal bath, according to the usual $\diff{S}{E}(E_*)=1/T$. At later times, when the system has explored multiple basins, the temperature mediates free energy exchanges from one valley to the other through \eqref{eq:def_temp_dyn_phase}. So, in the dynamical phase the total equilibrium entropy reads
\begin{equation}
\label{eq:total_entropy}
    S(T)\,=\,S_0(T)+\Sigma(T)
\end{equation}
where the first addendum in the r.h.s. is the internal entropy of basins and expresses vibrations of particles around their $\beta$ regime configurations, the second is the complexity and accounts for the multiplicity of cages.

What about the Kauzmann temperature? The interpretation is that it is the temperature such that intervalley jumps are not thermodynamically favoured: the complexity of equilibrium states vanishes at $T=T_K$. The equilibrium measure is concentrated on the states at the bottom of the landscape, whose number is subexponential. This is a $1RSB$ thermodynamic transition, since now $f_{eq}$ is non-analytic at $T=T_K$.
From \eqref{eq:total_entropy}, one has
\begin{equation}
    S(T)\,=\,S_0(T).
\end{equation}
which brings to the identification of the condensation of the equilibrium measure in the lowest free energy states in mean field models to the phenomenon of \emph{entropy crisis}.

\section{The pure p-spin model}

We consider a spin glass model with the following Hamiltonian

\begin{equation}
\label{eq:p-spin-Hamiltonian}
    \mathcal{H}_{p}[\boldsymbol{\sigma}]\,=\,-\sum_{(i_1\dots i_p)}J_{i_1\dots i_p}\sigma_{i_1}\dots\sigma_{i_p}
\end{equation}
where $(i_1,\dots, i_p)$ stands for a combination of the $p$ indices, the $\sigma_i$ are real variables satisfying the global spherical constraint
\begin{equation*}
    \sum_{i=1}^N\sigma_i^2\,=\,N
\end{equation*}
and the couplings are quenched normal variables with zero mean and variance
\begin{equation*}
    \mathbb{E}[J_{i_1\dots i_p}^2]\,=\,\frac{p!J^2}{2 N^{p-1}}.
\end{equation*}
As before, we solve the model using replica trick \eqref{eq:Replica_Trick}
\begin{equation*}
    \overline{\log Z}\,=\,\lim_{n\rightarrow 0}\frac{\overline{Z^n}-1}{n}.
\end{equation*}
The computation leads to the Replica Action
\begin{equation}
\label{eq:pSpin_spher_Action}
    A[\mathbb{Q}]\,=\,-\frac{\beta^2J^2}{4}\sum_{ab}Q_{ab}^p-\frac{1}{2}\log \det \mathbb{Q}-n s_{\infty}
\end{equation}
where $s_{\infty}$ is the infinite-temperature entropy.
The free energy evaluated at the RS saddle point is
\begin{equation}
    \label{eq:fe_sphpspin_RS_0}
    f(\beta;q)\,=\,-\frac{\beta J^2}{4}+\frac{\beta J^2}{4}q^p-\frac{1}{2}\log(1-q)-\frac{1}{\beta}s_{\infty}.
\end{equation}
The physical value of $q$ that minimises $f$ is $q_*\,=\,0$, yielding
\begin{equation}
\label{eq:fe_sphpsin_RS_1}
    f(\beta)\,=\,-\frac{J^2\beta}{4}-\frac{1}{\beta}s_{\infty}
\end{equation}
i.e. the paramagnetic free-energy. This solution is stable at all temperatures, but for $T\leq T_K$ the 1RSB saddle point maximises the free energy functional \cite{Crisanti1992}:
\begin{eqnarray}
\label{eq:fe_sphpsin_1RSB}
    && f_{1RSB}(\beta)\,=\,\underset{m, q_1}{\max}\;\Phi_{\beta}(0, q_1, m) \\
\label{eq:functional_sphpspin_1RSB}
    && \Phi_{\beta}(q_0, q_1, m)\,=\,-\frac{\beta J^2}{4}+\frac{\beta J^2}{4}(1-m)q_1^p+\frac{\beta J^2}{4}m\;q_0^p-\frac{1}{\beta}s_{\infty} \nonumber \\
    && -\frac{1}{2m}\log(1-(1-m)q_1-mq_0)+\frac{1-m}{2m}\log\Bigl(1-q_1-\frac{m}{1-m}q_0\Bigr)
\end{eqnarray}
We set $q_0=0$ because there is no external magnetic field. Eq. \eqref{eq:fe_sphpsin_1RSB}, despite not being the optimal solution, is also stable for $T_{K}<T<T_{on}$, where $T=T_{on}$ is the temperature over which only the paramagnetic state exists. At $T=T_K$ the system has a second order thermodynamic phase transition with a jumping order parameter: more precisely, $q_0=q_1=q_*=0$ for $T=T_K^{+}$ and $q_0=0, q_1=q_*>0$ for $T=T_{K}^{-}$.

As expected, the dynamical phase is invisible to any thermodynamic approach. In order to unveil its presence, we need a tool to explore metastable states.

\subsubsection{Monasson Potential}

In \cite{monasson1995structural} R. Monasson develops a technique that allows to explore different metastable states at given temperature $T<T_{on}$.
He proposes the following potential

\begin{eqnarray}
\label{eq:Monasson_Potential}
    && \mathcal{G}_{\beta}(m, q)\,=\,-\frac{1}{\beta}\log{\overline{Z_{\beta}(m, q)}} \\
    && Z_{\beta}(m, q) \,=\, \int d\boldsymbol{\sigma}_1\dots d\boldsymbol{\sigma}_n e^{-\beta \sum_{a=1}^m\mathcal{H}_p[\boldsymbol{\sigma}_a]}\prod_{ab}\delta(N\;q-\boldsymbol{\sigma}_a\cdot\boldsymbol{\sigma}_b) \nonumber 
\end{eqnarray}
Replica of the system are constrained to have fixed mutual overlap: this is equivalent to introduce a coupling between different replicas. In this framework, replica are not introduced as an artifact to perform the computation: their number $m$ is a parameter of the problem, and because of this they are often appointed as \emph{Real Replicas}.

 Consider the standard decomposition in pure states
 \begin{equation*}
    Z=\int df e^{-\beta N\varphi(f)}\qquad \varphi(f)=f-\frac{1}{\beta}\Sigma(f)
 \end{equation*}
 This equation does not allow for a direct computation of the Complexity of meta-stable states. The advantage of \eqref{eq:Monasson_Potential} is that it allows to compute the Complexity by means of a Legendre transform:
\begin{eqnarray}
    && \beta \mathcal{G}(m)\,=\,\beta m f-\Sigma(f) \\
    && \nonumber \\
    && f\equiv \frac{\partial \mathcal{G}}{\partial m}\qquad \Sigma(f)\,=\,\beta\,[m(f)f-\mathcal{G}(m(f))]
\end{eqnarray}
where the overlap is evaluated at the larger of the minima of $\mathcal{G}(m, q)$.
Hence, at a given temperature $T$, varying the parameter $m$ allows the exploration of different families of metastable states. Equilibrium states correspond to the choice $m=1$. 
The expression one obtains from \eqref{eq:Monasson_Potential} is identical to \eqref{eq:functional_sphpspin_1RSB}, with $q_0=0$ 
\begin{eqnarray}
\label{eq:Monasson_Potential_sphpsin_expression}
    && \mathcal{G}_{\beta}(q, m)\,=\,-\frac{\beta J^2}{4}-\frac{\beta J^2}{4}(m-1)q^p \nonumber \\
    && -\frac{1}{2m\beta}\log(1+(m-1)q)-\frac{m-1}{2m\beta}\log(1-q)
\end{eqnarray}
The equation for $q$ reads
\begin{equation}
\label{eq:boh}
    q[q^p+(m-2)q^{p-1}+q^{p-2}-\frac{2}{(\beta J)^2p}]=0.
\end{equation}
In order to study metastable states, one has to select the largest solution $q_*$ of this last equation, which will be a maximum or a minimum depending if $m<1$ or $m>1$. For evaluating the Complexity at equilibrium, one has to consider the solution $q_*$ in the limit $m=1$.
The dynamical temperature is defined as the temperature where the Replicon eigenvalue evaluated at $q=q_*$ vanishes \cite{Crisanti1992}
\begin{equation}
    \lambda_R(T)\,=\,\frac{2}{(1-q_*)^2}-\frac{p(p-1)q_*^{p-2}}{T^2}\Longrightarrow T_d=\sqrt{\frac{p(p-1)q_*^{p-2}}{2}}(1-q_*)
\end{equation}

\subsubsection{Road to $T=0$}

\begin{figure}
    \centering
    \includegraphics[width=0.8\columnwidth]{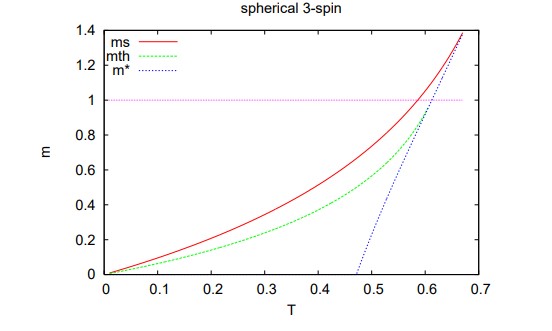}
    \caption{Phase diagram in the plane $m, T$ of the $p=3$ spherical spin glass. Picture taken from \cite{zamponi2010mean}.}
    \label{fig:3spin_spher_phase}
\end{figure}

One can build a phase diagram in the plane $T, m$ for the p-spin model, such that in figure \ref{fig:3spin_spher_phase}.
The tree lines refer to the following:
\begin{itemize}
    \item The line $m_*(T)$ is the spinodal line: it represents the value of $m$ for temperature $T$ such that a $q_*$ solution of \eqref{eq:boh} appears.
    \item The line $m_{th}(T)\geq m_*(T)$ is the stability line: it is identified through the condition of marginal stability $\lambda_R=0$. Solutions with $m<m_{th}(T)$ are unstable, so meta-stable states exist for $m\geq m_{th}$. Thus we identify $f_{th}(T)=f(m_{th}(T))$.
    \item The line $m_s(T)$ is the entropy crisis line, where the configurational entropy vanishes. The free energy and the complexity as functions of $m$ decrease from $m_{th}$ to $m_{s}$. Clearly, $f_{min}(T)=f(m_s(T))$.
\end{itemize}
Notice that in the whole dynamical phase $m_{th}<1<m_{s}$, whereas $m_s<1$ as the glassy phase is entered. 
By inverting $m$ through $f=\partial \mathcal{G}/\partial m$, one can follow TAP states in temperature, down to $T=0$. In pure p-spin models it does exist a perfect mapping between energy levels and TAP levels: in other words, each level $f$ that for some $T$ in the dynamical phase is of equilibrium can be written as $f=f(e(T))$, with $f_{min}=f(e_{min}(T))$ and $f_{th}=f(e_{th}(T))$. This property is a consequence of the homogeneity of the Hamiltonian \eqref{eq:p-spin-Hamiltonian} with respect to the configuration $\boldsymbol{S}$: at finite temperatures, the TAP free energy as well is homogeneous with respect to the magnetisation profile, so thanks to the global spherical constraint a stationary point of the TAP free energy is followed radially in the $N-sphere$ when temperature is changed. In more sophisticated model, such as mixed p-spin models \cite{folena2020mixed}, this property is lost and together with it the perfect matching between $e$ and $f$. Behind this failure there is the so-called \emph{chaos in temperature} \cite{rizzo2003chaos}, the property for which the overlap between any pair of equilibrium states referring to two different temperatures is zero. What happens in mixed models and in the model we will study in chapter 5 as well is that only states that are of equilibrium for some $T_K<T_{SF}<T_d$ can be followed down to zero temperature.
By fixing
\begin{equation}
    y=\frac{m}{T}\equiv \frac{d\Sigma}{df}
\end{equation}
the Monasson free energy $\mathcal{G}_{\beta}(m, q_*)$ has a good limit for $T\rightarrow 0$
\begin{equation}
    \mathcal{G}_{\infty}(y)=-\frac{1}{4}(p\chi_0+y)+\frac{1}{2y}\log\frac{\chi_0}{\chi_0+y}
\end{equation}
where $\chi_0=\lim_{T\rightarrow 0}(1-q_*)/T$. One can compute the Complexity of energy levels from
\begin{equation}
    e=\diff{(y{\mathcal{G}_{\infty}})}{y}\qquad \Sigma(e)=y [e-\mathcal{G}_{\infty}(y(e))].
\end{equation}

\subsubsection{Linear excitations of the pure p-spin}
We conclude this chapter by discussing the properties of the linear excitations of the energy Hessian, evaluated on a point of minimum. The spherical p-spin is the most simple spin glass model featuring continuous spins, so is a ground zero for the study of excitation spectra in spin glasses.

We define a Lagrangian associated to \eqref{eq:p-spin-Hamiltonian}
\begin{equation}
    \mathcal{L}\,=\,\mathcal{H}_p+\frac{1}{2}\mu (|\boldsymbol{S}|^2-N)
\end{equation}
The parameter $\mu$ is a radial force that enforce the spherical constraint. Local minima of the energy landscapes satisfy the stationarity condition:
\begin{equation}
    \mu S_i\,=\,-\frac{\partial \mathcal{H}_p}{\partial S_i}\qquad \mu=p|e|
\end{equation}
The Hessian can be computed with a further derivative
\begin{equation}
\label{eq:Hessian_pspin}
    \frac{\partial\mathcal{L}}{\partial S_i \partial S_j}=\frac{\partial^2 \mathcal{H}_p}{\partial S_i\partial S_j}+p|e|\delta_{ij}
\end{equation}
The Hessian in this last equation is a shifted Wigner random matrix: we will discuss in greater detail these kind of matrices and their properties in section \ref{sec:RMT}. Here we anticipate some of their properties: the spectrum of \eqref{eq:Hessian_pspin} is the interval $[p|e|-2\sqrt{p(p-1)/2},p|e|+2\sqrt{p(p-1)/2}]$, and the eigenvalues are random variables distributed according to a Wigner semi-circular law \cite{wigner1958distribution}
\begin{equation*}
    \rho(\lambda)\,=\,\frac{\sqrt{(\lambda-\lambda_-)(\lambda_+-\lambda)}}{2\pi\sqrt{p(p-1)/2}}
\end{equation*}
with $\lambda_{\pm}(e)=\pm 2\sqrt{p(p-1)/2}+p|e|$. The energy levels can be classified in \cite{folena2020mixed}
\begin{itemize}
    \item Stable minima $e<e_{th}=-2\sqrt{(p-1)/p}$: the spectra of these minima are gapped. Relaxation inside these minima is exponentially fast, with an asymptotic time-scale $\tau\sim 1/\lambda_{-}$.
    \item Marginal minima $e=e_{th}$: the spectra are gapless, with lower edge at $\lambda=0$. Relaxation is only algebraically fast: the exponent of the decay is determined by the concentration of eigenvalues close to the origin: $\rho(\lambda)\sim\lambda^{1/2}\rightarrow C(t)\sim t^{-1/2}$.
    \item Saddles (dominant only for $e>e_{th}$): part of the spectrum is negative. Here it is found $C(t)\sim t^{-2/3}$.
\end{itemize}

When studying random matrices, another important task is the computation of eigenvectors and their statistics. In particular, an interesting property to measure is the degree of localisation, namely how the amplitudes of the components are "spatially" distributed.
We anticipate that in Wigner matrices such as \eqref{eq:Hessian_pspin} eigenvector are completely delocalised and thus featureless. The triviality of eigenvectors is shared by other popular mean-field models of disordered systems, like the perceptron \cite{franz2015universal}. Since in many finite-dimensional systems under specific conditions localised excitations can appear, as we shall see in the first part of next chapter, it is important to have a class of mean field models for disordered systems able to feature less trivial properties in their excitations: in chapters 4, 5, 6 we define a class of mean field spin glass models able to capture non-trivial localisation features and thus to better represent glassy systems at low temperature.

\chapter{Low energy excitations of disordered systems}

The behavior of thermodynamic susceptibilities at low temperatures is a central problem in condensed matter.  
About a century ago, the newborn theory of quanta proved itself to be fundamental for correctly understanding the low temperature regime of solids. As to structural glasses, from early experimental studies it was clear that these systems possessed an anomalous low temperature phenomenology \cite{zeller1971thermal}. The specific heat of many different glassy systems was found to scale linearly in temperature below $10 K$. Since crystals specific heat follows a cubic scaling at low temperature, glassy systems must possess an excess of excitations above Debye prediction. The diverse behavior of glasses at very low temperature was addressed in the first 70s by means of phenomenological tunnelling models \cite{phillips1972tunneling, anderson1972anomalous}.

The Hessian of disordered systems are random matrices \cite{wigner1958distribution, rosenzweig1960repulsion}. In the case of mean field systems, Hessians can be represented by statistical ensembles of random matrices. With the purest spirit of statistical physics, the problem of determining the excitations of a specific physical system is replaced by the problem of diagonalising instances drawn from an assigned measure over the space of matrices. The spectral statistics obey self-averaging properties and thus the full information about the original system in the infinite size limit is equivalent to that of the representative random matrix ensemble.

This chapter is divided in two sections: in the first \label{sec:low_T_solid_exc} we will convey a brief account of the problem of excitations in glassy systems and of recent advancements \cite{lerner2021low}. In section \ref{sec:RMT} we will present fundamental concepts of Random Matrix Theory \cite{potters2020first} and discuss how it helps in the study of the excitations of disordered systems.

\section{Excitations of solids at low temperature}
\label{sec:low_T_solid_exc}

\subsection{Vibrational modes of crystals: Debye-Law}

A crystal is a solid whose constituent molecules are positioned at the vertices of a regular Bravais lattice \cite{ashcroft2022solid}. For most of purposes, electronic degrees of freedom can be decoupled from ionic ones\footnote{This is the Adiabatic Approximation and it is based on the different time scales of electron and ionic dynamics.}, so that the ionic part of the crystal can be conceived as a system of coupled harmonic oscillators\footnote{Anharmonic terms in ionic interactions of crystals are important close to the melting point and in the description of volume thermal dilatation \cite{ashcroft2022solid}.}. 
In the harmonic approximation, the overall dynamical behavior of the system is determined by the \emph{Dynamical Matrix}
\begin{eqnarray*}
   && \mathbb{D}_{R\;R'}=\frac{\partial^2 U}{\partial \Vec{x}_R \partial \Vec{x}_{R'}}(\Vec{x}_i,\dots,\Vec{x}_j) \\
   && \Vec{F}_R \,=\, -\sum_{R'}\mathbb{D}_{RR'}\;\Vec{x}_R \\
   && U(\boldsymbol{x})\,\simeq\,U_0+\frac{1}{2}\sum_{RR'}\;\Vec{x}_R\cdot \mathbb{D}_{RR'}\Vec{x}_{R'}
\end{eqnarray*}
where index $R$ stands for the position of the ion in the lattice, $\{\Vec{x}_R\}_R$ are displacements about equilibrium positions and $\{\Vec{F}_R\}_R$ are forces felt by oscillators. The eigenvalues $\{\lambda_k\}$ of the dynamical matrix define harmonic frequencies $\omega_k\equiv \pm\sqrt{\lambda}/M$, where $M$ is the ionic mass.
Since the system is traslationally invariant, the Dynamical matrix must satisfy $\mathbb{D}_{RR'}\equiv \mathbb{D}_{|R-R'|}$: thus, by Goldstones' theorem eigenvectors are plane waves with wave-vectors $\Vec{k}$ in the Brillouin zone, related to harmonic frequencies through a dispersion relation $\omega\equiv \omega_s(\Vec{k})$, where $s$ is the branch index of the spectrum\footnote{In three dimensions, there are one longitudinal branch and two transverse branches, discriminating vibration modes that are respectively. parallel and orthogonal to displacement vectors.}. In literature they are called \emph{normal modes of vibration}. In the long wavelenght (small frequency) limit they describe the propagation of sound waves in the solid, with velocity $v_s=\omega/k$. 

In the quantum description, necessary at low temperatures, normal modes are substituted by bosonic quasi-particles called \emph{phonons} and the vibrational state of the solid is described in terms of Bose-Einstein density of occupation numbers of phononic levels
\begin{equation*}
    n_s(\Vec{k})\,=\,\frac{1}{e^{\beta\;\hbar\;\omega_s(\Vec{k})}-1}\qquad \langle \mathcal{O}\rangle \equiv \sum_s\int \frac{d^3k}{(2\pi)^{3}}n_s(\Vec{k})\mathcal{O}(\omega_s(\Vec{k}))
\end{equation*}
where the integral is performed in the first Brillouin zone and $\mathcal{O}$ is a physical observable. Since occupation numbers depend on wave-vectors only through phonon frequencies, a fundamental quantity is the \emph{Vibrational Density of States} or \emph{Phonon Level Density}
\begin{equation}
    \label{eq:VDoS_def}
    \mathcal{D}(\omega)\equiv \sum_s\;\int\frac{d^3k}{(2\pi)^{3}}\delta(\omega-\omega_s(\Vec{k}))\qquad \langle \mathcal{O}\rangle=\int d\omega \mathcal{D}(\omega)\;n(\omega)\;\mathcal{O}(\omega)
\end{equation}
At low temperatures, only phonons with energy $\hbar\;\omega\ll k_B T$ are excited and $n(\omega)$ is dominated by them. As a consequence, at low $T$ only the $\omega\rightarrow 0$ behavior of phonon density is relevant: setting $\omega_s(\Vec{k})\,=\,v|\Vec{k}|$ 
\begin{equation}
\label{eq:Debye_Scaling}
    \mathcal{D}(\omega)\sim \frac{3}{2\pi^2}\frac{\omega^2}{v^3}
\end{equation}
Therefore, the lower edge of the vibrational spectrum determines the temperature dependence of physical observables, in particular that of the specific heat. Around a century ago, Debye predicted through \eqref{eq:Debye_Scaling} the behavior $c_v\sim T^3$ for the specific heat: this prediction was the first great success of the quantum theory of solids, 
since classically it was impossible to explain the discrepancy between experimental data of low temperature specific heats and Dulong-Petit law $c_v\,=\,6\;nk_B T$.

Classically, the lowest vibrational modes control the long-time equilibrium dynamics. When the DoS is gapped, the system relaxes exponentially fast, with a characteristic time scale $\tau=1/\omega_{min}$. If instead the spectrum is gapless, relaxation slows down to a power law behavior, controlled by the exponent of the VDoS close to $\omega=0$.


\subsection{Vibrational modes of glasses: the Boson peak}
\label{sec:boson_peak}
    
 The low temperature physics of glasses strongly differs from that of crystals. As the system is cooled down, molecular vibrations are more and more dumped and thus the entropy gets less and less important: the underlying disordered structure of the glass, which determines the potential energy landscape (PEL), becomes relevant. Early experiments \cite{zeller1971thermal} show that the specific heat of many glassy samples as a linear behavior in temperature, at variance with the cubic scaling of crystals: glasses have low frequency excitations in excess. Following these experimental observations, phenomenological theories identified the anomaly in terms of two-level systems \cite{phillips1972tunneling, anderson1972anomalous}: these are localised excitations, consisting in small and localised group of particles that tunnel between two quasi-degenerate mechanically stable configurations. The existence of such a populations of excitations was conjectured for the first in 1962 by Rosenstock \cite{rosenstock1962anomalous}: he claimed that in solids with structural defects non-Debye, short-range acoustic waves could emerge.

At the level of the VDoS, these localised excitations in excess of the Debye spectrum manifest in a peak of the reduced VDoS $\tilde{\mathcal{D}}(\omega)=\mathcal{D}(\omega)/\mathcal{D}_{debye}(\omega)$ at a characteristic frequency $\omega_{BP}$ in the $THz$ range called in literature \emph{boson peak} (\cite{malinovsky1986nature, schirmacher1998harmonic, kirillov1999spatial, taraskin2001origin, gurevich2003anharmonicity, gurevich2005pressure, parshin2007vibrational, shintani2008universal, grigera2003phonon, marruzzo2013heterogeneous, manning2015random, yang2019structural}). The physical origin of this feature is still debated and there is no consensus on a commonly accepted theory.

Another important feature of glassy vibrations is that they seem to possess a high degree of universality. Superimposed to phononic vibrations, the VDoS of localised glassy excitations seems to follow a quartic law at low fequency
 \begin{equation}
\label{eq:non_phononic_VDoS}
     \mathcal{D}_{g}(\omega)\,\sim\, A\;\omega^4
 \end{equation}
 for a variety of models \cite{lerner2016statistics,mizuno2017continuum,lerner2017effect,shimada2018anomalous,kapteijns2018universal,angelani2018probing,wang2019low, Wang2019sound, richard2020universality,bonfanti2020universal,Ji2019,ji2020thermal,ji2021geometry}. The quartic scaling seems to be robust with respect to system type (kind of interactions, symmetries) \cite{richard2020universality, das2020robustness, bonfanti2020universal}, preparation protocol \cite{lerner2016statistics} and physical dimensions \cite{kapteijns2018universal}. The prefactor in \eqref{eq:non_phononic_VDoS} instead depends on all these parameters, including the preparation protocol \cite{ji2020thermal, ji2021geometry}.
 Spatially, these low frequency quasi-localised modes exist in microscopic cores with a diameter of the order of ten average lattice spacings, and their amplitude seems to decay with distance from the core center as $r^{d-1}$, as it happens for the spatial response of elastic media to dipolar perturbations \cite{lerner2021low}. Because of this, these modes are often called \emph{quasi-localised} instead of simply "localised".
 


\subsection{Soft potential model}

Between the 80s and the 90s, new phenomenological theories ascribed the existence of quasi-localised excitations to regions of the glassy sample with anomalously small stiffness \cite{karpov1982atomic,klinger1983atomic,karpov1983theory,buchenau1991anharmonic,buchenau1992interaction,buchenau1992soft,gurevich1993theory}. The theoretical framework resulting from them is known in literature as \emph{soft potential model}. This theory schematise the just mentioned localised groups of soft particles as non-interacting anharmonic oscillators
\begin{eqnarray*}
    && \mathcal{V}\,=\,\sum_i V(x_i) \\
    && V(x)\,=\,V(0)+\sum_{k=1}^{\infty}\frac{a_n}{k!}x^{k}
\end{eqnarray*}
where $x$ is the projection of particle displacements along some assigned direction.
The coefficients of the expansion are assumed to be random variables, with a joint distribution $P(\{a_n\})$ with no zeros and singularities. If one centers the expansion around an equilibrium configuration $\{x_i^{(eq)}\}$, truncating at fourth order one can represent the potential acting on each oscillator as a quartic polynomial. One can show that for a generic minimum, the distribution of its curvature $b_2$ behaves as $P(b_2)\sim b_2$ at small curvatures: this corresponds to $\mathcal{D}(\omega)\sim\omega^3$. With the additional constraint that $\{x_i^{(eq)}\}$ is a global minimum, one has $P(b_2)\sim b_2^{3/2}$, yield $\mathcal{D}(\omega)\sim \omega^4$.

In spite of their ability of predicting the quartic law, this early version of the soft potential model cannot say anything about the degree of localisation of the modes. The next natural step is to include interactions in the effective potential energy
\begin{equation}
    \mathcal{V}\,=\,\sum_i V(x_i)+\frac{1}{2}\sum_{ij}\mathcal{J}(x_i,x_j)
\end{equation}
The first successful phenomenological theory basing on this potential is the Gurevich-Parshin-Schober (GPS) theory \cite{gurevich2003anharmonicity}, formulated almost twenty years ago. Their theory bears some similarities with earlier attempts to introduce interactions between defects \cite{grannan1990low1, grannan1990low2, kuhn1997random}. GPS theory assumes couplings of the form $\mathcal{J}(x_i,x_j)=Jg_{ij}x_ix_j/r_{ij}^{-d}$, with $g_{ij}$ random uniform variables in the interval $[-1/2,1/2]$. They write an Hamiltonian
\begin{equation}
\label{eq:GPS}
    \mathcal{H}_{GPS}=\frac{1}{2}\sum_i k_i x_i^2+\frac{1}{4!}\sum_i x_i^4+\frac{1}{2}\sum_{i,j}\frac{J g_{ij}x_i x_j}{r_{ij}^d}
\end{equation}
with stiffness distribution $P(k)\sim k^{\beta}$ for some $\beta>0$ and $J$ interaction strenght.
Their theory predicts both the quartic law and the presence of the Boson peak, finding for the reduced VDoS the behaviors
\begin{equation}
    \label{eq:glassy_reduced_VDoS_scalings}
    \Tilde{\mathcal{D}}(\omega)\equiv \frac{\mathcal{D}(\omega)}{\omega^2}=
    \begin{cases}
        & A\;\omega^2\qquad \omega\ll \omega_{BP} \\
        & B\;\omega^{-1}\qquad \omega_{BP} \ll \omega \ll \omega_{phon} \\
        & C\qquad \omega\geq\omega_{phon}
    \end{cases}
\end{equation}
They tested their theoretical predictions with numerical simulations, and compared their data with experiments, finding a very good agreement.
Very recently, a mean field theory on a fully-connected graph inspiring from GPS model was proposed \cite{bouchbinder2021low, rainone2021mean, folena2021marginal}. Interactions in the GPS Hamiltonian \eqref{eq:GPS} are replaced by quenched gaussian variables with the usual scaling, and a linear term controlled by an external field $h$ is added. This model has a RS and RSB phase in the plane $h,J$, and in a segment of the critical line separating the two the system features Hessian spectra with quartic VDoS.

So far we did not explicitly mention the relation between glassy excitations and phonons, which in absence of external forces must exist. The frequency $\omega_1$ in eq. \eqref{eq:glassy_reduced_VDoS_scalings} is a crossover frequency beyond which the phonon spectrum is essentially continuous. For lower frequencies, phonons typically gathers in bands of width $\propto N^{-1/2}$, appearing as sharp peaks in the VDoS and thus being more easily distinguishable from non-phononic modes; only for $\omega\ll \omega_{BP}$ phonons are absent. In the region of the Boson peak, there is strong hybridization between phonons and quasi-localised modes: to correctly disentagle them is still an open problem of structural glasses\cite{lerner2021low}. 

\newpage

\section{Random Matrix Theory}

\label{sec:RMT}


\noindent Random Matrix Theory (RMT) is a relatively new branch of probability theory: its history as a well defined research field begins in the late 50s. The theoretical physicist E. Wigner proposed a toy model for excited energy levels of atomic nuclei. At that time, the problem of characterising the energy spectrum of atomic nuclei was very challenging. Popular quantum models for the nucleus consisted in hermitian Hamiltonian operators represented as very big and complicated matrices. Wigner proposed to tackle the problem by considering the entries of this matrix as random gaussian variables \cite{wigner1958distribution}\footnote{This is the same reason why statistical mechanics was introduced: to deal with the problem of many degrees of freedom by means of statistical ensembles.} and to call the resulting operators \emph{Random Matrices}. An important problem was to characterise the typical spacing between atomic energy levels.
Wigner guessed the correct form for the PDF of the spacing $s$ between adjacent levels ($s_0$ is a characteristic scale)
\begin{equation}
\label{eq:Wigner_Surmise}
    P_s(s)\,=\,\frac{\pi s}{2s_0}\;e^{-\frac{\pi s^2}{4s_0^2}}
\end{equation}
which is thus named \emph{Wigner surmise} after him. 
The form of this PDF has very deep physical implications: 
\begin{itemize}
    \item Firstly, since $P_s(0)=0$, arbitrarily close energy levels are suppressed. This means that eigenvalues of random matrices\footnote{Wigner proved this result for Gaussian matrices, but it is actually a general result for the bulk of the spectrum of generic random matrices, as we will show later in this chapter.} repel each other: this effect can happen only if adjacent levels are strongly correlated. Indeed, the form of eq. \eqref{eq:Wigner_Surmise} in the region of small spacings is quite different from the spacing pdf of two adjacent iid variables, for which Poisson statistics holds:
\begin{equation*}
    P_s^{(iid)}(s)\,=\,\frac{1}{s_0}\;e^{-\frac{s}{s_0}}\qquad P_s^{(iid)}(0)=\frac{1}{s_0}>0
\end{equation*}
\item Secondly, the PDF in \eqref{eq:Wigner_Surmise} decay exponentially for large spacings: two adjacent levels cannot be too far, there is a \emph{confinement} effect. 
\end{itemize}
The interplay between repulsion and confinement is the signature feature of the eigenvalues of random matrices. \noindent Another important result obtained by Wigner is the so called \emph{Wigner semi-circle law}\footnote{Ironically, eq. \eqref{eq:semi-circle} is actually a semi-ellipse.} for the eigenvalue probability distribution or \emph{spectral density}

\begin{equation}
    \label{eq:semi-circle}
    \rho(\lambda)\,=\,\frac{1}{2\pi}\sqrt{4-\lambda^2}
\end{equation}

\noindent Wigner's law is an universal law for random matrices in the limit $N\rightarrow\infty$, where $N$ is the rank of the matrix under consideration: it is the limiting eigenvalue distribution for any random matrix whose entries are iid variables with zero mean and finite variance \cite{potters2020first}. Conversely, when fluctuations of the entries are strong or when there are non-negligible correlations between matrix entries, different limiting distributions appear. In the following pages, after introducing some fundamental tools and concepts, we will mostly discuss gaussian matrices and problems related to them.

\subsection{Random matrices: fundamental tools}

We define as Random Matrix Ensemble $\mathcal{E}(V, d\mathcal{M}(N))$ as a set $\{\mathbb{M}_1, \dots, \mathbb{M}_k, \dots\}$ of $N\times N$ matrices drawn from the matrix vectorial space $V$ according to the probability measure
\begin{equation}
    d\mathcal{M}(\mathbb{M})\equiv P_M(\{M_{ij}\})\prod_{ij}dM_{ij}
\end{equation}
We consider matrices with either real, complex or quaternion\footnote{Quaternions are an extension of complex numbers. The set of quaternions obeys a non-commutative algebra, in which each number is represented as a four-dimensional vector $n=a+b\bf{i}+c\bf{j}+d\bf{k}$. The basis vector $\bf{i}, \bf{j}, \bf{k}$ obey the multiplication rule $\bf{i}^2=\bf{j}^2=\bf{k}^2=\bf{i}\bf{j}\bf{k}=-1$. Quaternions in modern physics find many applications: for example, they allow for a simplified descriptions of spinors.} entries: if a random matrix is either symmetric, hermitian or simplectic, then its eigenvalues are random real variables, otherwise the matrix admits a singular values decomposition\footnote{The matrix can be decomposed as $\mathbb{M}=\mathbb{U}\mathbb{S}\mathbb{V}^T$, where $\mathbb{U}$, $\mathbb{V}$ are two orthogonal/unitary/self-similar matrices of orders $N$, $M$ respectively and $\mathbb{S}$ is a $N\times M$ diagonal matrix, whose elements $s_i\equiv \sqrt{\lambda_i}$ are called singular values and are complex numbers. More on \cite{potters2020first}.}. In the following, we will stick to the first case.

\subsubsection{Resolvent Function and Spectral Density}

We define the Resolvent Matrix of matrix $\mathbb{M}$ as
\begin{equation}
\label{eq:Resolvent_def}
    \mathbb{G}_M(z)\,=\,(\mathbb{M}-z\mathbb{I})^{-1}\,=\,\sum_{k=1}^N\frac{\boldsymbol{\psi}(\lambda_k)\boldsymbol{\psi}(\lambda_k)^T}{\lambda_k-z}
\end{equation}
where $\boldsymbol{\psi}(\lambda_k)$ is the $k$-th eigenvector of $\mathbb{M}$. The domain of the resolvent matrix is the resolvent set: it is the complex plane minus first order poles on the eigenvalues of $\mathbb{M}$. Thus, the spectrum of matrix $\mathbb{M}$ is the complementary of the resolvent set.
The resolvent encloses the spectral properties of the system: this can be seen by considering
the \emph{Green Function}
\begin{eqnarray}
    \label{eq:Green_Function_finite_N_def}
    && \mathcal{G}_M^{(N)}(z)\equiv \frac{1}{N}\Tr\mathbb{G}_M(z)\,=\,\frac{1}{N}\sum_{k=1}^N\frac{1}{\lambda_k-z} \\
    \label{eq:Green_Function_def}
    && \mathcal{G}(z)\,=\,\lim_{N\rightarrow\infty}\mathcal{G}_M^{(N)}(z)\,=\,\int d\lambda\frac{\rho(\lambda)}{\lambda-z}.
\end{eqnarray}
In the case of random matrices, the choice to drop the subscript in \eqref{eq:Green_Function_def} is not arbitrary: there is a self-averaging property, which we will explain a few lines below. Before that, let us finish to show the relevance of the Green function for the spectral problem.
The green function is the moment generating function of the spectral density. Indeed, by considering the $|z|\rightarrow\infty$ expansion of \eqref{eq:Green_Function_def} one has
\begin{eqnarray}
    \label{eq:resolvent_expansion}
    && \mathcal{G}(z)\,=\,-\frac{1}{z}\int d\lambda\frac{\rho(\lambda)}{1-\frac{\lambda}{z}}\,=\,\sum_{k=0}^{\infty}(-1)^{k+1}\frac{\overline{\lambda^k}}{z^{k+1}} \\
    && \frac{d}{dw^{k}}\left(\frac{1}{w}\mathcal{G}(w)\right)\Bigl|_{w=0}\equiv (-1)^{k+1}\overline{\lambda^{k}}\qquad w\equiv 1/z
\end{eqnarray} 
The spectral density can be obtained directly throughout the inversion of Steltjes transform, namely \emph{Stokhosky-Plemelj formula}
\begin{equation}
\label{eq:spectral_density_from_resolvent}
    \rho(\lambda)\,=\,\frac{1}{\pi}\lim_{\epsilon\rightarrow 0_+}\Im\mathcal{G}(\lambda+i\epsilon)
\end{equation}
This last formula can be extended to finite matrices by taking the prescription $\epsilon=O(N^{-1})$ for deterministic matrices and $\epsilon=O(N^{-1/2})$ for random ones \cite{potters2020first}.
If one is able to write an equation for the Green function and solve it, one can get the spectral density of the system.
However, when dealing with random matrices, sample fluctuations must be taken into account. It can be rigorously proven that a self-averaging principle holds for the spectral density \cite{mehta2004random}: more specifically, the sample average of the empirical spectral density
\begin{equation}
    \label{eq:spectral_density_empirical}
    \rho_{M}(\lambda)\,=\,\frac{1}{N}\sum_{k=1}^N\;\delta(\lambda-\lambda_k)
\end{equation}
concentrates around $\rho(\lambda)$:
\begin{eqnarray}
\label{eq:spectral_density_self-averaging}
    && \lim_{N\rightarrow\infty} \lVert\overline{\rho_{M}^{(N)}(\lambda)}-\rho(\lambda)\rVert_2\,=\,0 \\
    && \lim_{N\rightarrow\infty}\mathbb{P}\left[\lVert\varphi_{M}(\lambda)-\overline{\rho_{M}(\lambda)}\rVert_2<\epsilon\right]=1\qquad \forall \epsilon>0
\end{eqnarray}
where $\varphi_M(\lambda)$ is the empirical histogram of the eigenvalues of $\mathbb{M}$. Thus, for very large matrices the histogram of a single instance is very close to the asymptotic distribution $\rho(\lambda)$.
Convergence in the bulk of the spectrum is typically fast: large deviations occur typically in vanishing intervals $\lambda_{-}<\lambda<\lambda_{-}(N)$, $\lambda_{+}(N)<\lambda<\lambda_{+}$, i.e. close the edges of the spectrum \cite{potters2020first}
\begin{equation*}
    \label{eq:edges_spectrum}
    \lambda_{-}\equiv \inf \Supp(\rho(\lambda))\qquad \lambda_{+}\equiv \sup \Supp(\rho(\lambda))
\end{equation*}
\noindent Before discussing the statistical properties of eigenvector, it is instructive to discuss two classical random matrices ensemble: the Gaussian and Wishart-Laguerre ensembles.

\subsection{Gaussian Ensembles: GOE, GUE, GSE}


The \emph{Gaussian Orthogonal Ensemble} (GOE) was introduced by E. Wigner to represent Hamiltonian operators of time-reversal invariant systems. An instance of it is a symmetric random matrix $\mathbb{M}$ generated as following

\begin{equation*}
    \mathbb{M}\,=\,\mathbb{X}+\mathbb{X}^T
\end{equation*}
where $\mathbb{X}$ is a random matrix with gaussian real entries with iid zero mean and variance\footnote{The choice of the scaling in $N$ ensures that all elements of the matrix are $O(1)$ in the thermodynamic limit.} $\sigma^2/2N$. The joint pdf of the entries of $\mathbb{M}$ can be readily written

\begin{equation}
\label{eq:joint_pdf_GOE}
    P_{M}(\mathbb{M})\,=\,\prod_{(ij)}\sqrt{\frac{N}{2\pi\sigma^2}}\exp\left(-\frac{N M_{ij}^2}{2\sigma^2}\right)\prod_{i}\sqrt{\frac{N}{4\pi\sigma^2}}\exp\left(-\frac{N M_{ii}^2}{4\sigma^2}\right)
\end{equation}

\noindent The variance of diagonal elements is two times that of off-diagonal ones. This choice has the crucial benefit of making the GOE a \emph{Rotational Invariant} ensemble. 
Indeed, notice that eq. \eqref{eq:joint_pdf_GOE} can be rewritten as
\begin{equation}
\label{eq:joint_pdf_GOE_bis}
    P_{M}(\mathbb{M})\,=\,2^{-N/2}\left(\frac{N}{2\pi\sigma^2}\right)^{N(N+1)/4}\exp\Bigl(-\frac{N}{4\sigma^2}\Tr\mathbb{M}^2\Bigr)
\end{equation}
which implies that
\begin{equation}
    \label{eq:rotational_invariant_jpdf}
    P_{M}(\mathbb{M})\,=\,P_{M}(\mathbb{O}\mathbb{M}\mathbb{O}^T)
\end{equation}
for any orthogonal order $N$ matrix $\mathbb{O}$. Any ensemble such that \eqref{eq:rotational_invariant_jpdf} holds is rotational invariant: we will come back to this later, in section \ref{sec:Harr}. With a similar procedure to that of the GOE, one can define the \emph{Gaussian Unitary Ensemble} (GUE) and the \emph{Gaussian Simplectic Ensemble} (GSE): the first represents quantum hamiltonians in presence of an external magnetic field, whereas the second hamiltonians in presence of a spin-orbit coupling. GUE matrices are hermitian matrices with complex entries, GSE matrices are self-similar matrices with quaternion\footnote{A self-similar matrix with quaternion entries can be represented as the symmetrization of $\mathbb{M}=[A, B; -\operatorname{conj}B, -\operatorname{conj}A]$, where $A$ and $B$ are complex matrices and $\operatorname{conj}$ denotes the conjugation of all matrix entries.} entries.
Let us write down the joint pdf of the eigenvalues of $\mathbb{M}$ in a form valid for GOE, GUE, GSE:
\begin{equation}
    \label{eq:joint_pdf_eigs_Gaussian}
    \rho_M(\boldsymbol{\lambda})\,=\,\frac{1}{\mathcal{Z}_{N,\beta}}\exp\left(-\frac{N}{4\sigma^2}\sum_{k=1}^N\;\lambda_k^2\right)\prod_{(ij)}^{1, N}\;|\lambda_i-\lambda_j|^{\beta}.
\end{equation}
The prefactor enforces the normalisation of the joint distribution of the \emph{unordered} eigenvalues: the distribution of ordered eigenvalues is just \eqref{eq:joint_pdf_eigs_Gaussian} with a rescaled prefactor $\mathcal{Z}_{N,\beta}^{(ord)}\equiv N!\mathcal{Z}_{N,\beta}$.
The squared eigenvalues sum in the exponential clearly stems from the trace term in \eqref{eq:joint_pdf_GOE_bis}. The last term in the r.h.s. is called \emph{Vandermonde} determinant: its role is to suppress statistically very close eigenvalues pairs and it is responsible for the repulsion effect previously discussed, at the beginning of this section about random matrices; it depends by an exponent $\beta=1, 2, 4$ for GOE, GUE, GSE respectively, usually called \emph{Dyson index}.
We will provide a justification of \eqref{eq:joint_pdf_eigs_Gaussian} in section \ref{sec:Harr}. 

\subsubsection{The semi-circle law}
\label{sec:semi-circle}

 The most natural idea to compute $\rho(\lambda)$ is to start from the definition of spectral density in terms of the joint pdf \eqref{eq:joint_pdf_eigs_Gaussian}, writing
\begin{eqnarray}
\label{eq:Stielberg_integral}
    && \rho(\lambda)\,=\,\int\;d\lambda_2\dots d\lambda_N\;\frac{1}{\mathcal{Z}_{\beta}}e^{-\beta N^2\mathcal{V}(\boldsymbol{\lambda})} \\
    && \mathcal{V}(\boldsymbol{\lambda})\,=\,\frac{1}{4\;\beta N\sigma^2}\sum_{k=1}^N\;\lambda_k^2-\frac{1}{N^2}\sum_{(jk)}\log|\lambda_j-\lambda_k|
\end{eqnarray}
From these last expressions it is evident that the eigenvalues of a gaussian matrix can be mapped to a classical one-dimensional electrons gas, confined to be in an interval of width $O(2\sqrt{\beta}\sigma)$ and interacting through the two-dimensional Coulomb potential. In the light of this mapping, notice that sequences of iid variables are equivalent to confined systems of non-interacting particles in one dimension. One can understand from this analogy why correlations between neighboring eigenvalues are responsible for the shape of \eqref{eq:Wigner_Surmise} close to the origin. The integral appearing in \eqref{eq:Stielberg_integral} is a \emph{Stielberg integral}. Its computation is quite laborious, and results in an integral equation for the spectral density
\begin{equation*}
    \Pr\int\;d\lambda'\frac{\rho(\lambda')}{\lambda-\lambda'}\,=\,\lambda
\end{equation*}
Here we prove the semicircular law \eqref{eq:semi-circle} following a different approach. We introduce \emph{Schur complement formula} \footnote{We picked the definition related to a different definition of the resolvent: $\mathbb{G}(z)\equiv (z\mathbb{I}-\mathbb{M})^{-1}$.}
\begin{equation}
\label{eq:Schur_complement}
    \frac{1}{G_{jj}(z)}\,=\,z-M_{jj}-\sum_{k,l\neq j}M_{jk}G_{kl}^{(j)}(z)M_{lj}
\end{equation}
where $\{G_{jj}\}$ are the diagonal elements of the resolvent matrix \eqref{eq:Resolvent_def} and $\{G_{kl}^{(i)}\}$ are the entries of the submatrix of \eqref{eq:Resolvent_def} where row and column $i$ have been removed. Eq. \eqref{eq:Schur_complement} is a very general formula, holding for the resolvent of any invertible symmetric matrix \cite{potters2020first}. Let us trace over the two sides of \eqref{eq:Schur_complement}
\begin{equation*}
    \frac{1}{N}\sum_{j=1}^N\frac{1}{G_{jj}(z)}\,=\,z-\frac{1}{N}\sum_{j=1}^N M_{jj}-\frac{1}{N}\sum_{j=1}^N\sum_{k,l\neq j}M_{jk}G_{kl}^{(j)}(z)M_{lj}
\end{equation*}
\begin{equation*}
    \frac{1}{\mathcal{G}(z)}\,=\,z-\frac{\sigma^2}{N}\sum_{j=1}^N\mathcal{G}^{(j)}(z)+O(N^{-1/2})
\end{equation*}
We have used the fact that the Green function is self-averaging, $\overline{M_{ii}}=0$ and that the elements of $\mathbb{M}$ along row and column $i$ are by construction uncorrelated to the $\{G_{kl}^{(j)}\}$, so that
\begin{equation*}
\sum_{k,l\neq j}M_{jk}G_{kl}^{(j)}(z)M_{lj}\,\simeq \sum_k\;M_{jk}^2G_{kk}^{(j)}\,=\,N\overline{M_{jk}^2G_{kk}^{(j)}(z)}\equiv \sigma^2\beta\overline{G_{kk}^{(j)}}\,=\,\sigma^2\beta \mathcal{G}^{(j)}(z). 
\end{equation*}
 Finally, notice that $\mathcal{G}^{(j)}(z)$ is the green function of a large gaussian matrix of order $N-1$ with variance $(1-1/N)\sigma^2$: it is clear then that $\mathcal{G}^{(j)}(z)=\mathcal{G}(z)+O(N^{-1})$, which allows us to write down a second order algebraic equation for the Green function of gaussian ensembles
\begin{equation}
\label{eq:algebraic_eq_gaussian_resolvent}
    \frac{1}{\mathcal{G}(z)}\,=\,z-\sigma^2\beta\mathcal{G}(z)
\end{equation}
This admits two solutions, of which we choose the one with minus sign, in order to correctly have $\mathcal{G}(z)\sim 1/z$ for large $|z|$:
\begin{equation}
\label{eq:resolvent_eq_gaussian}
    \mathcal{G}(z)\,=\,\frac{z-\sqrt{z-2\beta^{1/2}\sigma}\sqrt{z+2\beta^{1/2}\sigma}}{2\beta\sigma^2}
\end{equation}
Take $z=\lambda+i0+$: when the infinitesimal is removed, the argument of the square root is non-negative for any $|\lambda|>2\sqrt{\beta}\sigma$, so that the Green function in this subset of the real line is real. \emph{Au contraire}, in the interval $|\lambda|<2\sqrt{\beta}\sigma$ there is a non-zero imaginary part. Thus, by using \eqref{eq:spectral_density_from_resolvent} we finally obtain the spectral density of gaussian ensembles
\begin{equation}
    \rho(\lambda)\,=\,\frac{1}{2\pi\beta^{1/2}\sigma^2}\sqrt{4\beta\sigma^2-\lambda}
\end{equation}
which is just \eqref{eq:semi-circle} with $\beta=1$ (GOE case) and $\sigma=1$.

\subsubsection{Levels spacing and Tracy-Widom distribution}

\begin{figure}
    \centering
    \includegraphics[width=0.45\columnwidth]{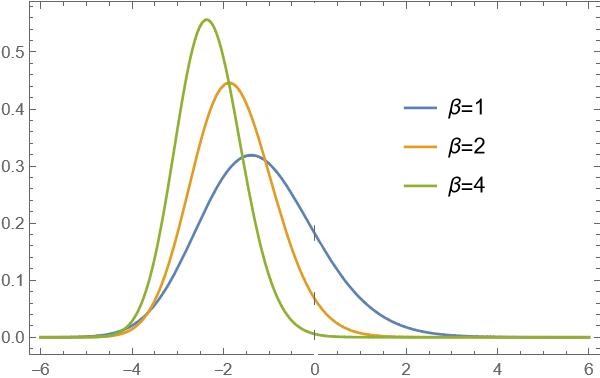}
    \caption{Tracy-Widom distribution for the three cases $\beta=1, 2, 4$.}
    \label{fig:tw}
\end{figure}

Wigner surmise formula \eqref{eq:Wigner_Surmise} is an approximation of the actual level spacing distribution
\begin{equation*}
    P_s(s)\,=\,\int\;d\lambda d\lambda' \rho_{ord}(\lambda, \lambda')\delta(s-(\lambda'-\lambda))
\end{equation*}
Here we will not discuss the exact case\footnote{The two-point eigenvalue correlation is expressed in terms of Airy functions. See \cite{potters2020first} for details.}: we will show how to derive \eqref{eq:Wigner_Surmise} from the study of $2\times 2$ GOE matrices. The characteristic polynomial $\det(\mathbb{M}-\lambda\mathbb{I})$ of a $2\times 2$ matrix is quadratic, so it is easily solvable: the solution is given by the two eigenvalues
\begin{equation*}
    \lambda_{\pm}\,=\,\frac{\Tr \mathbb{M}}{2}\pm\sqrt{\left(\frac{\Tr \mathbb{M}}{2}\right)^2-\det \mathbb{M}}\qquad s\equiv \lambda_+-\lambda_-\,=\,\sqrt{(\Tr \mathbb{M})^2-4\det \mathbb{M}}
\end{equation*}
The distribution of $s$ can be obtained through
\begin{equation*}
    P_s(s)\,=\,\frac{1}{2\pi^2\sigma^4}\int dM_{11}dM_{22}dM_{12}\;e^{-\frac{1}{2\sigma^2}\Tr\mathbb{M}^2}\delta\left(s-\sqrt{(\Tr \mathbb{M})^2-4\det \mathbb{M}}\right)
\end{equation*}
The final results is eq. \eqref{eq:Wigner_Surmise}. The computation for GUE and GSE holds is analogous: one finds $P_s(s)\sim s^{\beta}$ close to the origin. In the light of the Coulomb gas picture, the fact that it suffices the $N=2$ case to grasp the qualitative behavior of neighboring eigenvalues is not surprising, since an eigenvalue "interacts" strongly only with its two neighbors. Taken a pair, the effect of the other eigenvalues can be treated effectively as a boundary condition: eigenvalues in the center of the spectrum are compressed from both sides by other eigenvalues, but as we move towards the edges, the compression effect becomes asymmetric. Then, it is insightful to consider the large deviation problem for the largest eigenvalue of a gaussian matrix:
\begin{equation}
\label{eq:large_deviation_largest_eig}
    \mathbb{P}\left[\lambda_{max}-2\sqrt{\beta}\sigma>x\right]\equiv e^{-N C_{\beta}(x)}
\end{equation}
The resulting density function is the \emph{Tracy-Widom distribution} \cite{tracy1994level}
\begin{equation}
\label{eq:Tracy-Widom}
    \lim_{N\rightarrow\infty} P(N^{2/3}(\lambda_{max}-\lambda_{+}))\underset{x=N^{2/3}(\lambda_{max}-\lambda_{+})}{\equiv} P_{TW}^{(\beta)}(x)\qquad P_{TW}^{(\beta)}(x)\sim e^{-\frac{2\beta}{3}x^{3/2}}\qquad x\gg 1 
\end{equation}
In figures \ref{fig:tw} we show the Tracy-Widom distributions for the three cases $\beta=1, 2, 4$. Notice that they are peaked at negative values of $x$: fluctuations are most likely to be below the upper edge $\lambda_+$. 
The finite size scaling of the largest eigenvalue at leading order is
\begin{equation}
    \lambda_{max}\,=\,\lambda_{+}-C\;N^{-2/3}
\end{equation}
Finite size effects are stronger at the edges than in the bulk: this is a general feature of random matrices. 

\subsection{Haar measure}
\label{sec:Harr}

We begin this section sketching a proof for the eigenvalues joint distribution formula \eqref{eq:joint_PDF_eigs_bis}. For a generic random matrix ensemble, it is always possible to write
\begin{equation}
\label{eq:eigs_eigvecs_measure}
P_{M}(\mathbb{M})\prod_{i=1}^N\;dM_{ii}\prod_{(ij)}\;dM_{ij}\,=\,\rho(\boldsymbol{\lambda},\mathbb{O})\;\prod_{i=1}^N\;d\lambda_i\;\prod_{ij}\;dO_{ij},
\end{equation}
where $\boldsymbol{\lambda}=(\lambda_1,\dots\lambda_N)$ and $\mathbb{O}$ is the eigenvector matrix. Then, the jpdf of eigenvalues reads

\begin{equation}
    \label{eq:joint_PDF_eigs} \rho(\lambda_1,\dots,\lambda_N)\,=\int\mathcal{D}\mathbb{O}\;\rho(\boldsymbol{\lambda},\mathbb{O})=\,\int\mathcal{D}\mathbb{O}\;P_M(\mathbb{M}(\boldsymbol{\lambda},\mathbb{O}))|\mathbb{J}(\mathbb{M}\rightarrow \{\boldsymbol{\lambda},\mathbb{O}\})|
\end{equation}

\noindent The Jacobian in \eqref{eq:joint_PDF_eigs} can be computed explicitly \cite{livan2018introduction} and is equal to Vandermonde determinant \eqref{eq:Vandermonde}: we will show its computation in the case of symmetric matrices. Starting from $\mathbb{M}\equiv \mathbb{O}\mathbb{X}\mathbb{O}^T$, with $\mathbb{X}\equiv \operatorname{diag}(\lambda_1,\dots,\lambda_N)$, one can write
\begin{eqnarray*}
    && \delta \mathbb{M}\,=\,\mathbb{O}\delta\mathbb{X}\mathbb{O}+\delta\mathbb{O}\mathbb{X}\mathbb{O}^T+\mathbb{O}\mathbb{X}\delta\mathbb{O}^T \\
    && \delta(\mathbb{O}\mathbb{O}^T)=0\Longrightarrow\delta\mathbb{O}^T\,=\,-\mathbb{O}^T\delta\mathbb{O}\;\mathbb{O}^T \\
    && \delta\hat{\mathbb{M}}\,=\,\delta\mathbb{X}+\delta\mathbb{W}\,\mathbb{X}-\mathbb{X}\,\delta\mathbb{W}\qquad \delta\hat{\mathbb{M}}\,=\,\mathbb{O}^T\delta\mathbb{M}\mathbb{O}\qquad \delta\mathbb{W}\,=\,\mathbb{O}^T\delta\mathbb{O}
\end{eqnarray*}
Since $\hat{\mathbb{M}}$ and $\mathbb{M}$ are connected by an orthogonal transformation, the original Jacobian is equal to $\delta\hat{\mathbb{M}}/\delta\mathbb{X}\delta\mathbb{W}$, whose elements are
\begin{equation*}
    \frac{\partial\hat{M}_{ij}}{\partial \lambda_k}\,=\,\delta_{ij}\delta_{ik}\qquad \frac{\partial\hat{M}_{ij}}{\partial W_{kl}}\,=\,\delta_{ik}\delta_{jl}(\lambda_j-\lambda_i)
\end{equation*}
This matrix can be made diagonal after a finite number of row switches, so its determinant in absolute value is
\begin{equation}
\label{eq:Vandermonde}
    |\det\partial \hat{\mathbb{M}}/\delta\mathbb{X}\delta\mathbb{W}|\,=\,|\mathcal{J}(\mathbb{M}\rightarrow \{\boldsymbol{\lambda},\mathcal{\mathbb{O}}\})|\,=\,\prod_{(ij)}^{1, N}\;|\lambda_i-\lambda_j|
\end{equation}
which is just eq. \eqref{eq:Vandermonde} in the GOE case. Let us go back to eq. \eqref{eq:joint_PDF_eigs}: after the computation of the Jacobian, it reads

\begin{equation}
\begin{split}
\label{eq:jpdf_eigs_general}
    \rho(\lambda_1,\dots\lambda_N)\,& =\,\prod_{(ij)}^{1, N}\;|\lambda_i-\lambda_j|^{\beta}\int\mathcal{D}\mathbb{O}\;P_M(\mathbb{M}(\boldsymbol{\lambda},\mathbb{O})) \\
    & = \tilde{\rho}(\lambda_1,\dots,\lambda_N)\prod_{(ij)}^{1, N}\;|\lambda_i-\lambda_j|^{\beta}
\end{split}
\end{equation}
This is the most general form for the joint distribution of eigenvalues and in most cases one cannot go any further than it. 
An exception to this is given by rotational invariant ensembles, for which \eqref{eq:jpdf_eigs_general} can be made explicit. In a rotational invariant ensemble, as \eqref{eq:rotational_invariant_jpdf} dictates, any vector basis is statistically equivalent to the other: in particular, the eigenvector basis does not play any special role. In this situation, the joint distribution of eigenvalues and eigenvectors factorises:

\begin{equation}
\label{eq:RMT_ROT-INV_2}
\rho(\boldsymbol{\lambda},\mathbb{O})\;=\;\rho(\boldsymbol{\lambda})\;P_{\text{Haar}}^{(N)}(\mathbb{O}).
\end{equation}

\noindent We introduced \emph{Haar measure}:

\begin{eqnarray}
   & P_{Haar}^{(N)}(\mathbb{O})\,=\,\frac{1}{V_{Haar}(N)}\prod_{i=1}^N\delta(\boldsymbol{O}_i\cdot\boldsymbol{O}_i-1)\prod_{(ij)}^{1,N}\delta(\boldsymbol{O}_i\cdot\boldsymbol{O}_j) \\
   & \nonumber \\
   & V_{Haar}(N)\equiv\int\mathcal{D}\mathbb{O}\,=\,\int\prod_{ij}^{1, N}dO_{ij}\prod_{i=1}^N\delta(\boldsymbol{O}_i\cdot\boldsymbol{O}_i-1)\prod_{(ij)}^{1,N}\delta(\boldsymbol{O}_i\cdot\boldsymbol{O}_j).
\end{eqnarray}

\noindent where $\boldsymbol{O}_i$ is the $i$-th column of matrix $\mathbb{O}$, i.e. the $i$-th eigenvector. In the large $N$ limit, Haar measure is a uniform measure over the $N$/$2N$/$4N$-dimensional spheres of radius one\footnote{In the infinite dimensional limit, the sphere "squeezes" on the equator and the zenit axe, i.e. the scalar product of any two random vectors on it tends to zero.}. Then, the distribution of the rescaled eigenvectors components absolute values $\{N|\psi_k|\equiv \eta_k\}$ is
\begin{eqnarray}
\label{eq:porter_thomas}
    && P_{GOE}(\eta)\,=\,\frac{1}{\sqrt{2\pi\eta}}e^{-\eta} \\
\label{eq:porter_thomas_bis}
    && P_{GUE}(\eta)\,=\,e^{-\eta}.
\end{eqnarray}
The first of these is called \emph{Porter-Thomas distribution} \cite{rosenzweig1960repulsion}: it was introduced in the study of nuclear fission rates by Porter and Thomas. Given the very nature of Haar measure, the normalisation of each eigenvector must be finely distributed among its components, and thus the distributions in \eqref{eq:porter_thomas}.\eqref{eq:porter_thomas_bis} decay exponentially fast for large $\eta$. In this situation, eigenvectors are called \emph{delocalised}: they correspond to wave functions of ergodic physical systems. If the wave-functions of rotational invariant random matrices are extended, one realises that when this symmetry is broken localisation phenomena-i.e. breaking of ergodicity-may happen. This phenomenon, called \emph{localisation transition}, it is at the core of this thesis.

Let us complete the derivation of the joint pdf: a necessary condition for RI is that the distribution $P_M$ satisfies Wyel's Lemma:
\begin{equation}
    P_M(\mathbb{M})\,=\,f(\Tr \mathbb{M},\dots, \Tr \mathbb{M}^N)
\end{equation}
Combining \eqref{eq:joint_PDF_eigs}, \eqref{eq:Vandermonde}, \eqref{eq:RMT_ROT-INV_2}, finally we get

\begin{equation}
    \label{eq:joint_PDF_eigs_bis}
    \rho(\lambda_1,\dots,\lambda_N)\,=\,V_{\text{Haar}}(N)\;f\Bigl(\sum_{i=1}^N\lambda_i,\dots,\sum_{i=1}^N\lambda_i^N\Bigr)\prod_{(ij)}^{1, N}\;|\lambda_i-\lambda_j|^{\beta}.
\end{equation}
This is the general expression for the joint eigenvalues distribution of rotational invariant random matrices. Consider now $f$ factorised as 
\begin{equation*}
    f(s_1,\dots,s_N)\equiv \exp\left(-\frac{N\beta}{2}\sum_{k=1}^N\;a_k s_k\right)\equiv\exp\left(-\frac{N\beta}{2}\sum_{k=1}^N\;a_k\Tr\mathbb{M}^k\right)
\end{equation*}
This is a generalisation of the Coulomb gas picture discussed in section \ref{sec:semi-circle}: a rotational invariant random matrices ensemble, if function $f$ is factorised, is completely identified by its confinement potential (per site)
\begin{equation}
\label{eq:confinement_potential}
    V(x)\,=\,\sum_{k=1}^N a_k x^k
\end{equation}
The gaussian case is $V(x)\,=\,x^2/2$, which is the harmonic potential. To conclude this section, let us make two remarks. First, at short enough scales, the dominant term in the Coulomb gas effective hamiltonian is the repulsive interaction: in this situation the details of the confinement potential are not important, meaning that there is \emph{local universality}. Properties like the spacing between neighboring eigenvalues have a high degree of generality. 
As to the universality of Wigner law \eqref{eq:semi-circle}, at page 48 at the beginning of \ref{sec:RMT} we stated that it holds for any random matrix with centered iid entries with finite variance: if instead one deals with non-trivial\footnote{We mean we neglect the case of a positively shifted spectrum.} random matrices with positive spectrum, it is intuitive from the Coulomb gas picture that the limiting spectral density cannot be Wigner law. If electrons feel an infinite potential barrier at the origin, there is no way their equilibrium configuration density will be \eqref{eq:semi-circle}: this is what happens for instance in ensembles whose matrices have correlated entries.

\subsection{Layman classification of random matrix ensembles}

We concluded last section mentioning a situation in which Wigner law cannot be the limiting distribution. The simplest matrix ensemble satisfying this description is \emph{Wishart-Laguerre ensemble}: an instance of it is a $N\times N$ random matrix
\begin{equation}
    \mathbb{C}\,=\,\frac{1}{N}\mathbb{X}\mathbb{X}^T
\end{equation}
i.e. sample covariance matrices. Here $\mathbb{X}$ is a matrix $N\times T$ matrix of data, containing observations of $N$ quantities at different times: $\{x_i^{(t)}\}$, with $i=1,\dots,N$ and $t=1,\dots,T$ and $\alpha=N/T$ is fixed. Such covariance matrices were studied for the first time by mathematician J. Wishart in the 30s \cite{wishart1933generalised}. Wishart matrices are rotational invariant matrices, non-negative by definition\footnote{For any vector $\boldsymbol{v}$, one has $\boldsymbol{v}^T\mathbb{C}\boldsymbol{v}\equiv \lVert \mathbb{X}^T\boldsymbol{v}\rVert^2\geq 0$.}: their limiting distribution is \emph{Marchenko-Pastur} law
\begin{equation}
\label{eq:Marchenko-Pastur}
    \rho(\lambda)\,=\,\frac{\sqrt{(\lambda_+-\lambda)(\lambda-\lambda_-)}}{2\pi\alpha\lambda}+\frac{\alpha-1}{\alpha}\delta(\lambda)\theta(\alpha-1)
\end{equation}
and their confinement potential per site is
\begin{equation}
    V(x)\,=\,\frac{x}{\alpha}-(1-1/\alpha)\log x
\end{equation}
We introduced Wishart matrices because of \emph{Layman classification} of matrix ensembles \cite{livan2018introduction}. There are essentially two macro-categories
\begin{itemize}
    \item \emph{Random matrix ensembles with independent entries}: the most notable examples are Wigner ensembles, random graphs adjacency matrices \cite{mckay1981expected}, Lévy matrices \cite{cizeau1994theory} and power-law banded matrices. 
    \item \emph{Rotational invariant ensembles}: here we have gaussian ensembles, Wishart-Laguerre ensemble, Jacobi classical ensembles and many other examples. 
\end{itemize}
These two sets, apart from gaussian ensembles\footnote{It is a consequence of a theorem from Rosenzweig-Porter \cite{rosenzweig1960repulsion}.}, are disjoint. If we require rotational symmetry, entries get correlated and eigenvectors become irrelevant. On contrary, if we want to have independent entries, eigenvectors and eigenvalues are not independent.

\subsection{Perturbation of Wigner Matrices}
\label{sec:Perturbation_Wigner_Matrices}

In the first part of this chapter, we discussed and compared the properties of linear excitations for crystal and glassy systems. We then showed how the Hessian matrix of the potential energy of a generic mean field disordered system could be represented by an ensemble of random matrices. This section aims to show this connection: after briefly introducing free probability theory, we discuss the \emph{Rosenzweig-Porter} ensemble and highlight its applications.

\subsubsection{Dyson brownian motion}

In this section we will show how a Wigner matrix is modified by the addition of an another Wigner matrix. In order to do that, let us consider a problem studied by F. Dyson in 1962 \cite{dyson1962brownian}, concerning the study of time-dependent stochastic perturbations of quantum systems:
\begin{equation*}
    \mathbb{M}\,=\,\mathbb{M}_0+\sqrt{dt}\;\mathbb{X}
\end{equation*}
where $\mathbb{M}_0$ is the Wigner initial matrix and $\mathbb{X}$ is another Wigner matrix that is independent of $\mathbb{M}_0$. For our purposes, we will consider it a random matrix problem and take real matrices. The evolution equations of the eigenvalues and eigenvectors can be deduced from perturbation theory and read
\begin{equation}
\label{eq:Dyson_eigvals}
    d\lambda_i\,=\,\sqrt{\frac{2}{\beta N}}d\,B_i+\frac{1}{N}\sum_{j\neq i}\frac{dt}{\lambda_i-\lambda_j}
\end{equation}
\begin{equation}
\label{eq:Dyson_eigvecs}
    d\boldsymbol{v}_i\,=\,\sqrt{\frac{1}{N}}\sum_{j\neq i}\frac{d\,B_{ij}}{\lambda_i-\lambda_j}\boldsymbol{v}_j-\frac{1}{2 N}\sum_{j\neq i}\frac{dt}{(\lambda_i-\lambda_j)^2}\boldsymbol{v}_i
\end{equation}
These equations describe the evolution of a gaussian matrix process
\begin{equation*}
    \mathbb{M}(t)\,=\,\mathbb{M}_0+\mathbb{X}(t)
\end{equation*}
such that at any time $t$ the resulting matrix is a Wigner matrix with variance $\sigma^2\,=\,\sigma_0^2+t$. This is the generalisation of the diffusion process of a particle to random matrices: it shows us that by summing Wigner matrices (respectively, gaussian random variable) up to infinitesimal steps, one still obtains a Wigner matrix (respectively, another gaussian random variable). This stochastic dynamics enforces the Green function to satisfy a Burgers partial differential equation (PDE)
\begin{equation}
\label{eq:Burgers}
    \frac{\partial \mathcal{G}_t(z)}{\partial t}\,=\,-z\frac{\partial\mathcal{G}_t(z)}{\partial z}\qquad \mathcal{G}_0(z)\,=\,\mathcal{G}_{M_0}(z).
\end{equation}
This equation can be solved with the method of characteristics, and its solution
gives us a self-consistent equation for the resolvent
\begin{equation}
    \mathcal{G}_t(z)\,=\,\mathcal{G}_0(z-t\mathcal{G}_t(z))
\end{equation}
Since the perturbation is rotational invariant, we can assume that the initial matrix is diagonal: then we have
\begin{equation}
    \label{eq:resolvent_selfconsist_wigner_pert}
    \mathcal{G}(z, t)\,=\,\int d\lambda_0\frac{\rho(\lambda_0)}{z-t\mathcal{G}_t(z)-\lambda_0}
\end{equation}
If we fix time $t$, what we have obtained is the equation for the Green function of the sum of two different Wigner matrices. But in showing this, we have also shown that as in random scalars or vectors, the stochastic evolution conserves up to continuous times the gaussian measure.
Notice that since we are assuming a rotational invariant perturbation, it is not important for the initial matrix to be Wigner: equation \eqref{eq:resolvent_selfconsist_wigner_pert} would work also for a deterministic initial matrix. The crucial hypothesis is that the eigenvectors of the perturbation are uncorrelated to those of the initial matrix: in next section we define a more precise criterion for the validity of \eqref{eq:resolvent_selfconsist_wigner_pert}.

\subsubsection{Free probability and R-transform}

Let us consider two generic commutative variables $A$ and $B$ and let $\tau(\cdot)$ a scalar function returning their moments, such that the $k$-th moment is $\tau(A^k)$. The two variable are said independent if for any $k, l$ such that the moments exist
\begin{equation}
\label{eq:ind}
    \tau(A^kB^l)\,=\,\tau(A^k)\tau(B^l)
\end{equation}
In particular, this criterion ensures that the probability distribution of their sum is just a convolution of the their respective distributions
\begin{eqnarray*}
    P_{C}(C)\,=\, P_A(A):P_B(C-A) \equiv \int dA P_A(A)P_B(C-A)  
\end{eqnarray*}
The generalisation of this formula for an arbitrary sequence of independent variables satisfies the central limit theorem.
Let us now consider non-commutative variables, such as random matrices: in this case, the function $\tau(\cdot)$ is the rescaled trace: the $k$-th matrix moment is
\begin{equation*}
    \tau(A^k)\,=\,\frac{1}{N}\Tr A^k.
\end{equation*}
Given the existence of Dyson process, one wonders how to extend central limit theorem to random matrices. Unfortunately, the usual independence definition 
\eqref{eq:ind} does not work on non-commutative variables. Indeed, because of the presence of terms like $\tau(ABAB)$, i.e. products of matrices invariant under cyclic permutations i.e. terms that cannot be simplified from the trace, the matrix cumulants of the sum of many random matrices do not converge to those of gaussian matrices. A stronger criterion of independence is \emph{freeness}: two random variables $A$ and $B$ are "free" if for any two sequences of polynomials $p_1,\dots,p_n$, $q_1,\dots, q_n$ for which
\begin{equation*}
    \tau(p_k(A))\,=\,0\qquad \tau(q_k(B))\,=\,0\qquad \forall k
\end{equation*}
it holds
\begin{equation}
\label{eq:}
    \tau\left(\prod_{k=1}^n\;p_k(A)q_k(B)\right)\,=\,0
\end{equation}
This property ensures that all terms like $\tau(ABAB)$ vanish, allowing the generalisation of central limit theorem to random matrices \cite{potters2020first}:
\begin{quote}
    The matrix cumulants of the sum of $N$ free random matrices with zero mean and normalised by $N^{-1/2}$ converge to the matrix cumulants of a Wigner matrix:
    \begin{equation}
        \kappa_{A_1+\dots +A_N}^{(k)}(C)\underset{N\rightarrow\infty}{=} \delta_{k,2}
    \end{equation}
\end{quote}
We conclude this section introducing the \emph{R-transform}, a tool which allows to write a self-consistent equation for the Green function of the sum of two free random matrices:
\begin{equation}
\label{eq:R_transf}
    \mathcal{R}(g)\,=\,\mathcal{B}(g)-\frac{1}{g}
\end{equation}
where $\mathcal{B}(g)$, the \emph{Blue function}, is the inverse of the Green function\footnote{The inversion of the green function is a delicate job: in this thesis we keep a low level of rigor and we assume to be in the conditions such that this inverse exists.}. In the gaussian case, by comparison with eq. \eqref{eq:resolvent_eq_gaussian} one sees that the $R$-transform is the identity (we put $\beta, \sigma$=1)
\begin{equation*}
    \mathcal{R}_{gauss}(g)\,=\,g
\end{equation*}
The $R$-transform is additive under addition of two free matrices:
\begin{equation}
\label{eq:R_transf_additive}
    \mathcal{R}_C(g)\,=\,\mathcal{R}_A(g)+\mathcal{R}_B(g)
\end{equation}
If we have diagonalised one between $A$ and $B$, let say $A$, we can write a self-consistent equation for $\mathcal{G}_c(z)\equiv g(z)$ by exploiting \eqref{eq:R_transf_additive}. Indeed, from definition \eqref{eq:R_transf} we have
\begin{eqnarray*}
    && \mathcal{R}_C(g)\,=\,\mathcal{B}_C(g)-\frac{1}{g}\,=\,\mathcal{B}_A(g)-\frac{1}{g}+\mathcal{R}_B(g), \\
    && \mathcal{B}_A(g)\,=\,z-\mathcal{R}_B(g) \\
\end{eqnarray*}
where we used $\mathcal{B}_c(g)\equiv z$. The final formula, after inverting the last of these two equations, reads
\begin{equation}
\label{eq:resolvent_self_const_free}
    g(z)\,=\,\mathcal{G}_A(z-\mathcal{R}_B(g(z)))\equiv \int d\lambda\;\frac{\rho_A(\lambda)}{z-\mathcal{R}_B(g(z))-\lambda}
\end{equation}
If we know the $R$-transform of $B$, we can solve this equation and compute the green function of the new system. In particular, notice that when $B$ is a gaussian matrix, \eqref{eq:resolvent_self_const_free} becomes \eqref{eq:resolvent_selfconsist_wigner_pert} with $t=1$. Hessians of mean field disordered systems usually figure as the sum of a off-diagonal coupling matrix and a diagonal stiffness matrix, representing respectively structural and on site disorder: typically, the off-diagonal matrix is a Wigner matrix\footnote{In some disordered models the interaction matrix is a Wishart matrix: renown examples are the perceptron \cite{rosenblatt1958perceptron} and Hopfield model \cite{hopfield1982neural}.} and the diagonal elements have a non-gaussian distribution a-priori known:
\begin{equation}
\label{eq:Hessian_mean_field_disordered_systems}
    \mathbb{M}\,=\,\mathbb{J}+diag(d_1,\dots,d_N)
\end{equation}
For finite $N$, interactions and stiffness are correlated to make the spectrum non-negative, and thus the two addenda are not free. However, since interactions are weak and stiffness depend weakly on couplings, in the thermodynamic limit these two contributions are free: hessians of mean field disordered systems can be represented as instances of a random matrix ensemble.

\subsubsection{Rosenzweig-Porter ensemble}

We define as \emph{Rosenzweig-Porter ensemble} \cite{rosenzweig1960repulsion} an ensemble of matrices given by the sum of two free matrices, a Wigner matrix and a diagonal matrix, with iid elements distributed according to some $P_v(v)$:
\begin{equation}
    \label{eq:Rosenzweig-Porter}
    \mathbb{M}\,=\,\mathbb{W}+diag(v_1,\dots,v_N)
\end{equation}
In the standard setting, the diagonal matrix represents a quantum hamiltonian operator represented on its eigenbasis and the Wigner matrix acts as a noisy perturbative term. In the thermodynamic limit, Hessians of mean field disordered systems become instances of \eqref{eq:Rosenzweig-Porter}.
The spectral density of $\mathbb{M}$ can be computed by solving \eqref{eq:resolvent_self_const_free} and computing the imaginary part of the Green function through \eqref{eq:spectral_density_from_resolvent}. Here, we analyse the statistical properties of the eigenvectors of these matrices: since the unperturbed system has fully localised eigenstates, the idea is to understand how these are changed by the action of a Wigner perturbation.
We consider a diagonal matrix with $P_v(v)\,=\,\mathcal{N}(0, \sigma^2)$, and discuss the three cases $\sigma^2=O(N^0)\ll 1$, $\sigma^2=O(N^0)\gg 1$, $\sigma^2=O(N^{\gamma-1})$ for some real $\gamma$, reporting the result of \cite{truong2016eigenvectors}; in order to have the two terms in \eqref{eq:Rosenzweig-Porter} of the same order, the typical spacing between the $\{v_i\}$ must be $O(1/N)$. The last scaling is the one studied in the original Rosenzweig-Porter model in \cite{rosenzweig1960repulsion}. Let us consider local eigenvector moments of the form $I_n^{(q)}(\lambda)\equiv \langle|\psi_n(\lambda)|^{2q}\rangle_W$, where $n$ stands for the component, the argument is the eigenvalue and the average is performed only with respect to the Wigner matrix, i.e. at fixed diagonal. Notice that eigenvector moments can be computed from eq. \eqref{eq:Resolvent_def} through
\begin{equation}
\label{eq:eigvec_moments}
    I^{(q)}(\lambda)\,=\,\lim_{\epsilon\rightarrow 0_+}\epsilon^q\sum_{n=1}^N \overline{|G_{nn}(\lambda+i\epsilon)|^q}^W
\end{equation}
For Rosenzweig-Porter ensemble, the following formula stands \cite{truong2016eigenvectors}
\begin{equation}
\label{eq:local_eigenvector_moment_Ros_Por}
\begin{split}
I_n^{(q)}(\lambda)\,& =\,\frac{1}{N^q}\left[\frac{1}{(\lambda-\tau(\lambda)-v_n)^2+\pi^2\rho(\lambda)^2}\right]^q\Gamma(1+q) \\
& \equiv \lim_{\epsilon\rightarrow 0_+}\epsilon^q \overline{|G_{nn}(\lambda+i\epsilon)|^q}^W
\end{split}
\end{equation}
where $\tau(\lambda)$ and $\rho(\lambda)$ are respectively the real and imaginary part (spectral density) of the Green function.
Let us start from $\sigma=O(N^0)$. For $\sigma$ ranging from zero to infinity, the spectral density of the system interpolates from the semi-circle \eqref{eq:semi-circle} to the gaussian $\mathcal{N}(0, \sigma^2)$. The total eigenvector moments $I_q\equiv \sum_n I_n^{(q)}$ averaged over the diagonal elements have the following asymptotic forms \cite{truong2016eigenvectors}
\begin{equation}
    \overline{I_q(\lambda)}\equiv \mathcal{I}_q(\lambda)\sim \begin{cases}
        \frac{1}{N^{q-1}}\qquad \sigma\ll 1,\quad\forall \lambda \\
        \frac{\sigma^{2q}}{N^{q-1}}\qquad \sigma \gg 1,\quad \forall \lambda
    \end{cases}
\end{equation}
In both cases, the Wigner perturbation has the effect of making eigenstates delocalised, no matter how strong is $\sigma$. However, since $\mathcal{I}_q/\mathcal{I}_q^{(gauss)}\sim \sigma^{2q}$, the degree of delocalisation in the non-perturbative regime is far less than the gaussian one. Indeed, eigenvectors moments give us information about how the normalisation is distributed: in particular the second moment, called \emph{Inverse Participation Ratio}, is a measure of the fraction of components participating to eigenvector normalisation. So, one understands that in the strong $\sigma^2$ regime, the normalisation mass among the rescaled components $\{N\psi_n^2\}$ is not as homogeneous as in the gaussian case. When $\sigma^2\,=\,N^{\gamma-1}$, the statistical properties of eigenvectors become richer. It is found
\begin{equation}
    \mathcal{I}_q(\lambda)\sim \begin{cases}
        N^{1-q}\qquad \gamma <1 \\
        N^{D_q (1-q)}\qquad 1<\gamma<2 \\
        N^0 \qquad \gamma \geq 2
    \end{cases}
\end{equation}
for any $\lambda$. In the interval $1<\gamma<2$, eigenstates become non-ergodic: they are still delocalised, since level spacing statistics follow Wigner-Dyson formula, but their moments scale with system size in a non-trivial way. This regime is called \emph{multi-fractal regime} and $D_q$ are the corresponding fractal dimensions. For Rosenzweig-Porter model, it is found for any $q>1/2$ \cite{truong2016eigenvectors}
\begin{equation*}
    D_q\,=\,2-\gamma
\end{equation*}
For $\gamma \leq 1$, the fractal dimension is unity and eigenvectors are statistically equivalent to gaussian ones.
At $\gamma=2$, eigenvectors are fully localised and two points correlations have a transition from Wigner-Dyson formula to Poisson statistics, meaning that Wigner perturbation does not change the nature of initial states. This is true also for $\gamma>2$, but the relative importance of the two terms in \eqref{eq:Rosenzweig-Porter} changes, with the Wigner matrix becoming dominant with respect to the diagonal, implying that the problem should be studied using a different setting.

\subsection{Sparse matrices}
\label{sec:sparse_matrices}

A $N\times N$ matrix is said \emph{sparse} when the number of non-zero entries per row is finite for $N$ going to infinity. The adjacency matrices of networks and graphs are sparse: they appear naturally in the study of Hessian matrices in statistical physics \cite{Doro2003, Reka2002} and Markovian transition matrices in stochastic processes in physics and information theory \cite{lovasz1993random}. Historically, the first physical problem studied in a sparse random matrix setting is Anderson localisation \cite{And58}: this phenomenon consists in the transition of a metal to an insulator when disorder is sufficiently strong. The electron gas wave-functions, which in a metal are extended, can become localised around specific lattice sites in presence of disorder. Anderson localisation has been observed in semiconductors with insulating impurities or defects. Sparse random matrices have been studied also in the field of disordered systems, from Replica method \cite{edwards1976eigenvalue} to semi-numerical techniques, like Population Dynamics for the resolution of cavity self-consistent equations \cite{abou1973selfconsistent}. In this section, we discuss notable spectral properties of adjacency matrices and the relevance of cavity method in this context.

\subsubsection{Kesten-Mc Kay distribution}

We recall for reader convenience the definition of adjacency matrix of a graph \eqref{eq:adjacency}:
\begin{equation*}
    A_{ij}=\begin{cases}
        & 1\qquad (ij)\in\mathcal{E}\quad i\text{,}j\in\mathcal{V}\equiv \{1,\dots, N\} \\
        & 0\qquad \text{otherwise}
    \end{cases}
\end{equation*}
The adjacency matrix of a random regular graph (RRG) has a fixed number $c$ of non-zero elements per row, where $c$ is the connectivity of the graph. The spectral density of random matrices $\mathbb{A}$ corresponding to the adjacency of RRG is named \emph{Kesten-Mc Kay} distribution \cite{kesten1959symmetric, mckay1981expected}. Its expression reads
\begin{equation}
\label{eq:Kesten-Mc Kay}
    \rho(\lambda)\,=\,\frac{c\sqrt{4(c-1)-\lambda^2}}{2\pi(c^2-\lambda^2)}\qquad |\lambda|\leq 2\sqrt{c-1}
\end{equation}
In addition to \eqref{eq:Kesten-Mc Kay}, which describes the continuous spectrum, there is for any $N$ an isolated eigenvalue located at $\lambda=c$: indeed, the constant vector is always an eigenvector of the adjacency matrix. If one considers a random matrix whose entries are $\Tilde{A}_{ij}\,=\,\sigma_{ij} A_{ij}$, with $\sigma_{ij}\,=\,\pm 1$ with probability $1/2$, the isolated eigenvalue disappears and the continuous spectrum retains \eqref{eq:Kesten-Mc Kay} as its spectral density. Eigenvectors of the RRG are all delocalised: intuitively, since each node is perfectly equivalent, there is no way the system can exhibit localisation, which usually is driven by extreme values of site or bond quantities. In figure (ref) we show a plot of Kesten-Mc Kay formula \eqref{eq:Kesten-Mc Kay} for growing values of $c$, showing how it approaches Wigner law \eqref{eq:semi-circle} in the dense limit.

\subsubsection{Introducing disorder}

Let us consider sparse matrices of the form
\begin{equation}
\label{eq:dirty_RRG}
    M_{ij}\,=\,A_{ij}\sigma_{ij}+V_i\delta_{ij}
\end{equation}
where $\{A_{ij}\}$ is the connectivity matrix of a RRG and $\sigma_{ij}$ and $V_i$ are non correlated\footnote{The reader may have noticed that \eqref{eq:dirty_RRG} has the standard form of Hessian matrices appearing in physical systems. Unfortunately, in the sparse case usually it is not possible to define proper random matrix ensembles to describe Hessians: this happens because, in order for the spectrum to be positive, stiffness and interactions are strongly correlated also in the thermodynamic limit.} random variables with given probability distributions. When disorder is introduced\footnote{We mean that we deal with sparse random matrices with non homogeneous entries.}, localised states appear in the spectrum, close to the lower and upper edges. When disorder is sufficiently strong, all delocalised states disappear. Let us make a specific example, considering \eqref{eq:dirty_RRG} with only diagonal disorder $\{V_i\}$ distributed uniformly in $[-W/2, W/2]$, for some $W>0$. This case is the original model studied by Anderson, whose adaption to random graphs topology has been studied in \cite{abou1973selfconsistent} and then more recently in \cite{biroli2010anderson}. For non-zero $W$, no matter how small, the system develops a band of localised states, close to the edges $\lambda_{\pm}\,=\,2\sqrt{c-1}$: indeed, the edges remain the same as in \eqref{eq:Kesten-Mc Kay}. At a critical value $W_c\approx 17.4$, localised states extend up to $\lambda=0$, and extended ones disappear. An interesting phenomenon occurring in these models is that of \emph{Lifshitz tails} \cite{lifshitz1964energy, bapst2011lifshitz}: the spectral density close to the edges has a singular behavior of the form
\begin{equation}
\label{eq:Lifshitz_tails}
    \rho(\lambda)\sim \exp\left(-k_1\;\exp\left(\frac{k_2}{\delta_{\pm}^{1/2}}\right)\right)\qquad \delta_{\pm}=|\lambda_{\pm}-\lambda|
\end{equation}
with $k_1$ and $k_2$ constant depending on $W$. The spectral density decays extremely fast as the edges are approached, making its numerical estimation close to them quite hard. The behavior in \eqref{eq:Lifshitz_tails} can be understood through an argument made by Lifshitz \cite{lifshitz1964energy} for finite dimensional systems, which we adapt in the case of RRG. Consider the node around which a wave-function with energy close to the band-edge localises, and suppose that in a bubble of radius $n$ around this node the $V_i$ assume values in $[-W/2,-W/2+\delta]$: the probability of this event is $(\delta/W)^{\#_n}$, with $\#_n\propto (c-1)^n$ the number of nodes in the bubble. The radius of the bubble must be chosen as $n\propto \delta^{-1/2}$, and so one finds back eq. \eqref{eq:Lifshitz_tails}. Lifshitz tails like \eqref{eq:Lifshitz_tails} are not present for arbitrary forms of the $V_i$ distribution: for instance, if one chooses a gaussian, the spectrum will extend to the whole real line and the spectral density will have a different decay \cite{bapst2011lifshitz}.

\subsubsection{Spectral Cavity Method and Population Dynamics Algorithm}

Let us consider the following self-consistent equations
\begin{equation}
\label{eq:spectral_cavity}
    \mathcal{G}_{i\rightarrow j}(z)\,=\,\frac{1}{z-M_{ii}-\sum_{k\in\partial i/j}M_{ij}^2 \mathcal{G}_{k\rightarrow i}(z)}
\end{equation}
where $z$ is a complex number, $M$ is a generic sparse matrix of the form \eqref{eq:dirty_RRG} and $\mathcal{G}_{i\rightarrow j}(z)$ are the entries of the resolvent matrix in a system where the link $(i, j)$ has been removed. These equations were firstly derived in \cite{abou1973selfconsistent} through perturbation theory: they can be derived rigorously through Schur-Complement formula \eqref{eq:Schur_complement}. Once a fixed point of eq. \eqref{eq:spectral_cavity} is found, the diagonal entries of the resolvent, the spectral density and eigenvectors moments can be computed through
\begin{eqnarray}
\label{eq:spectral_cavity_solution_given_sample}
    && \mathcal{G}_{ii}(z)\,=\,\frac{1}{z-M_{ii}-\sum_{k\in\partial i}M_{ij}^2 \mathcal{G}_{k\rightarrow i}^*(z)} \\
    && \rho(\lambda)\,=\,\frac{1}{\pi}\Im \sum_{k}\mathcal{G}_{kk}(\lambda+i\epsilon) \\
    && I_q^{(n)}(\lambda)\,=\,\epsilon^q \sum_k |\mathcal{G}_{kk}(\lambda+i\epsilon)|^q
\end{eqnarray}
Eqs. \eqref{eq:spectral_cavity}, \eqref{eq:spectral_cavity_solution_given_sample} yield the solution for a given sample, so a full solution should be an average over many different solutions \eqref{eq:spectral_cavity_solution_given_sample}. 

\part{Vector spin glasses on fully connected graphs}

\chapter{The fully connected vector spin glass}

In this chapter, we introduce and study in detail a random field vector spin glass, defined on a fully connected graph: as discussed in chapter 1, these models are generalisations to multi-dimensional spins of the renown SK model. 
After an initial general presentation of the model, we focus on the problem of linear excitations of energy minima. The motivation behind this choice is to find a toy model for excitations of stable glassy minima.

The chapter is organised as follows: in section \ref{sec:the_model_chap4} we study the general case, performing a Replica Computation of the free energy of the model and discussing the properties of the solution both in the RS and in the RSB phase. After that, in section \ref{sec:TAP_equations_vector_model} we re-derive the TAP equations of the model and discuss linear excitations around TAP solutions.

Finally, in section we present our results about the problem of the linear excitations around minima of the energy landscape. This part of the chapter is a re-elaboration of our work \cite{franz2021delocalization}.
We show that energy minima in the RS phase feature soft localised excitations at the edge and that the onset of the Spin Glass transition is related to a delocalisation transition of the softest modes of the Energy Hessian spectrum, providing a non trivial connection between the static response properties of the system at criticality and Random Matrix Theory. 

\section{The model}
\label{sec:the_model_chap4}

We consider a spin glass model with the following Hamiltonian

\begin{eqnarray}
\label{eq:Model}
    && \mathcal{H}[{\bf S}]\,=\,\mathcal{H}_0[{\bf S}]-\sum_{i=1}^N\;\Vec{S}_i\cdot\Vec{b}_i \\
\label{eq:Model_H0}
    && \mathcal{H}_0[{\bf S}]\,=\,-\sum_{i j}\;J_{ij}\Vec{S}_i\cdot\Vec{S}_j
\end{eqnarray}
As usual, we consider $m$-dimensional vector spins with unit norm. The disorder is site-site uncorrelated and quenched, where
\begin{itemize}
    \item Couplings $J_{ij}$ are uncorrelated gaussians with zero mean and variance
    \begin{equation}
    \label{eq:couplings_variance}
        \overline{J_{ij}J_{kl}}\,=\,\delta_{(ij),(kl)}\frac{J^2}{N}\qquad J_{ii}=0
    \end{equation}
    \item External fields $\{\Vec{b}_i\}$ are uncorrelated random vectors, distributed with an unspecified $P_b(\Vec{b})$ 
    satisfying
    \begin{equation}
    \label{eq:isotropy_ext_fields}
        \overline{b_i^{(\alpha)}}=0\qquad \overline{b_i^{(\alpha)}b_j^{(\beta)}}=\delta_{ij}\delta^{\alpha\beta}H^2.
    \end{equation}
\end{itemize}
The gaussianity of the couplings is not strictly necessary: since couplings are weak for large $N$, their distribution has no impact on the thermodynamics, provided that it has finite variance. In this case, for the central limit theorem the Hamiltonian \eqref{eq:Model_H0} is a gaussian random function with zero mean and covariance
\begin{eqnarray}
    \label{eq:covariance_H0}
    \overline{ \mathcal{H}_0[{\bf S}] \mathcal{H}_0[{\bf S}']}\,& =&\,\frac{1}{N}\sum_{\alpha\beta}^{1, m}\left(\sum_{i=1}^N S_i^{(\alpha)}S_i^{(\beta)}\right)^2 \\
    &\equiv& N\;\sum_{\alpha\beta}^{1, m} g(q_{\alpha\beta}({\bf S}, {\bf S}^{'}))
\end{eqnarray}
where
\begin{equation}
    q_{\alpha\beta}({\bf S}, {\bf S}^{'})\,=\,\frac{1}{N}\sum_{i=1}^N S_i^{(\alpha)} S_i^{(\beta)}\qquad g(x)\,=\,\frac{x^2}{2}
\end{equation}
Notice that if one considers a different Hamiltonian for the interactions, like the isotropic p-spin Hamiltonian
\begin{equation*}
    \mathcal{H}_0[{\bf S}]\,=\,-\frac{1}{p!}\sum_{i_1\dots i_p}J_{i_1\dots i_p}\sum_{\alpha}S_{i_1}^{\alpha}\cdots S_{i_p}^{\alpha}
\end{equation*}
with couplings properly normalised, all the dependence on the kind of interaction will be bound to the function $g(q_{\alpha\beta})$: for this reason, this function is called \emph{characteristic function} of the spin glass model.

At variance with couplings, the distribution of external fields is relevant in the thermodynamic limit. On one hand, it affects quantitatively the values of physical quantities. Consider for instance a $m$-dimensional gaussian distribution and a uniform distribution on the $m$-sphere of radius $H$: 
\begin{equation}
\label{eq:ext_field_pdfs}
    P_b^{(1)}(\Vec{b})\,=\,\frac{1}{(2\pi H^2)^{m/2}}e^{-|\Vec{b}|^2/2H^2}\qquad P_b^{(2)}(\Vec{b})\,=\,\frac{\delta(|\Vec{b}|-H)}{S_m(H)}
\end{equation}
In the first case, external fields can be arbitrarily small in magnitude: this clearly affects all relevant physical observables, since they are computed in terms of averages over the distribution of the local fields
\begin{equation*}
    \Vec{\mu}_i\equiv \sum_{j} J_{ij}\Vec{S}_j+\Vec{b}_i
\end{equation*}
For instance, the two distributions \eqref{eq:ext_field_pdfs} lead to very different values of the magnetic susceptibility for a given value of $H$, and to a different value of the critical point.

On the other hand, the choice of the distribution does not change qualitatively the physical behavior of the system at $T=0$, provided has zero mean and finite variance, which eqs. \eqref{eq:isotropy_ext_fields} ensure. As pointed out in \cite{sharma2010almeida},
the crucial property to preserve the dAT picture is to have randomly oriented external fields.

\subsection{Replica computation of the free-energy}

In this section we derive the free energy of the model \eqref{eq:Model}, characterising the RS solution in all ranges of temperatures and fields where it is stable and making considerations regarding the fRSB solution.

From the gaussianity of the Hamiltonian, we can steadily write
\begin{eqnarray*}
    && \overline{\exp\left(-\beta\sum_{a=1}^n \mathcal{H}_0[{\bf S}_a]\right)}\,=\,\exp\left(\frac{\beta^2}{2}\sum_{ab}\overline{\mathcal{H}_0[{\bf S}_a]\mathcal{H}_0[{\bf S}_b]}\right) \\
    \,&=&\,\exp\left(\frac{\beta^2 N}{2}\sum_{ab}\sum_{\alpha\beta}g(Q_{ab}^{\alpha\beta})\right)\int D\mathbb{Q}\;\prod_{\alpha\beta}\prod_{ab}\delta\left(\sum_{i}S_{i,a}^{\alpha}S_{i, b}^{\beta}-N Q_{ab}^{\alpha\beta}\right)
\end{eqnarray*}
After replacing the deltas with their Laplace representation and neglecting unrelevant prefactors of the integral, the replicated partition function at fixed $\{\Vec{b}_i\}$ reads
\begin{eqnarray}
  \label{eq:replica_Z}
  &&\underset{J}{\mathbb{E}}[Z^n]\,=\,\int D\mathbb{Q}D\mathbb{\hat{Q}} \exp\left\{\frac{\beta^2 N}{2}\sum_{a,b,\alpha,\beta}
  \Bigl[g(Q_{ab}^{\alpha\beta}) -{\hat Q}_{ab}^{\alpha\beta} {
  Q}_{ab}^{\alpha\beta}\Bigr]\right\} \\ 
  && \times\int D{\bf S}_1\dots D{\bf S}_n\exp\left(\sum_{a,b}\sum_{\alpha,\beta}\frac{\beta^2{\hat Q}_{ab}^{\alpha\beta}}{2}\sum_i S_{i, a}^{\alpha} S_{i, b}^{\beta}+\beta\sum_{a=1}^n\sum_{i=1}^N\Vec{b}_i\cdot \Vec{S}_{i, a}\right) \nonumber 
  \nonumber
  \end{eqnarray}
  where we set
  \begin{equation}
      D{\bf S}_a\equiv\prod_{i=1}^N\prod_{a=1}^n d^mS_{i, a}.
  \end{equation}
  We absorbed the constraints on the norms of the spin into the constraints of the overlaps, in order to have the summations involving the latter running over all indices.
  Introducing the average over the external fields with the symbol $\underset{b}{\mathbb{E}}[\cdot]$, the replica action assumes the standard form
  \begin{eqnarray}
    \label{eq:replica_A}
     & A[\mathbb{Q},\mathbb{\hat{Q}}]\,=\,-\frac{\beta^2}{2}\sum_{a,b}\sum_{\alpha,\beta}\Bigl[g(Q_{ab}^{\alpha\beta})-{\hat Q}_{ab}^{\alpha\beta}{Q}_{ab}^{\alpha\beta}\Bigr]-\log\underset{b}{\mathbb{E}}[W_b[\mathbb{\hat{Q}}]] \\
     & \nonumber \\
     \label{eq:W_0}
     & W_b[\mathbb{\hat{Q}}]\,=\,\int d^mS_1\cdots d^mS_n \exp\left\{\frac{\beta^2}{2} \sum_{a b}^{1, n}\sum_{\alpha \beta}^{1, m}{\hat Q}_{ab}^{\alpha\beta}S_a^{\alpha}S_b^{\beta}+\beta\sum_a \Vec{S}_a\cdot\Vec{b}\right\}
  \end{eqnarray}
  These conjugated parameters $\{\hat{Q}_{ab}^{\alpha\beta}\}$ can be eliminated through saddle point equations
  \begin{equation}
      \label{eq:hatQ_SP}
      \frac{\partial A}{\partial Q_{ab}^{\alpha\beta}}=0\Longrightarrow \hat{Q}_{ab}^{\alpha\beta}\,=\,g'(Q^{\alpha\beta}_{ab}).
  \end{equation}
  Moreover, since the system is statistically isotropic, we shall restrict to order parameter matrices $\mathbb{Q}$ of the form
  \begin{equation}
      \label{eq:isotropy_overlaps}
      Q^{\alpha\beta}_{ab}\,=\,\delta_{\alpha\beta}Q_{ab}.
  \end{equation}
Let us consider first the RS solution and then the fRSB solution.

\subsubsection{RS solution}

If we plug the RS ansatz
\begin{eqnarray*}
   & Q_{ab}=q,\qquad a\neq b \\
   & Q_{aa}=1/m.
\end{eqnarray*}
into \eqref{eq:W_0}, we can apply a Hubbard-Stratonovich transform and rewrite it as
\begin{equation*}
    W_b(q)\,=\,e^{\frac{\beta}{2}g'(1/m)-\frac{\beta}{2}g'(q)-\frac{1}{2}b^2}\int d\mathcal{G}_{g'(q)}^{(m)}(h)\exp\left(\frac{\Vec{h}\cdot\Vec{b}}{g'(q)}\right)\mathcal{K}_m(\beta h).
\end{equation*}
Once we plug it into \eqref{eq:replica_A}, the free energy density reads
\begin{eqnarray}
\label{eq:f-RS}
   && f(\beta, H)=\underset{q}{\max}\;\Phi_{RS}(q) \\
   && \nonumber \\
   && \Phi_{RS}(q)=\frac{\beta m}{2}\Bigl[g(q)-g'(q)q\Bigr]-\frac{\beta m}{2}[g(1/m)-g'(1/m)/m\Bigr] \\
   && \nonumber \\
   && -\frac{\beta}{2}(g'(1/m)-g'(q))-\frac{1}{\beta}\int d\mathcal{G}_{g'(q)}^{(m)}(h)\underset{b}{\mathbb{E}}\left[\exp\left(-\frac{b^2+2\Vec{h}\cdot\Vec{b}}{2g'(q)}\right)\right]\log \mathcal{K}_m(\beta h). \nonumber
\end{eqnarray}
where $d\mathcal{G}_{g'(q)}^{(m)}$ is a $m$-dimensional isotropic gaussian measure
\begin{equation}
    d\mathcal{G}_{g'(q)}^{(m)}\,=\,\frac{d^m h}{(2\pi g'(q))^{m/2}}\exp\left(-\frac{|\Vec{h}|^2}{2g'(q)}\right)
\end{equation}
and $\mathcal{K}_m$ is the usual rescaled modified Bessel function, which we report here for reader convenience
\begin{equation*}
    \mathcal{K}_m(x)\equiv (2\pi)^{m/2}\frac{I_{m/2-1}(x)}{x^{m/2-1}}.
\end{equation*}
Equation \eqref{eq:f-RS} is just a generalisation to random external fields of the free energy derived by de Almeida and Thouless (eq. \eqref{eq:fe_para_mVector_noextfield}) in \cite{de1978infinite} for the zero external field case.
By noticing that 
\begin{equation*}
    \int d\mathcal{G}_{g'(q)}^{(m)}(h)\underset{b}{\mathbb{E}}\left[\exp\left(-\frac{b^2+2\Vec{h}\cdot\Vec{b}}{2g'(q)}\right)\right]\,=\,1
\end{equation*}
we deduce that the distribution of cavity fields is
\begin{equation}
\label{eq:cav_field_pdf}
    P_h(\Vec{h})\,=\,\mathcal{G}_{g'(q)}^{(m)}(h)\underset{b}{\mathbb{E}}\left[\exp\left(-\frac{b^2+2\Vec{h}\cdot\Vec{b}}{2g'(q)}\right)\right]
\end{equation}
For example, in the case of gaussian and uniform external fields we have
\begin{equation}
    \label{eq:cav_fields_pdf_gaussian_ext_fields}
    P_h(\Vec{h})\,=\,\mathcal{G}_{g'(q)+H^2}^{(m)}(h)\qquad \text{Gaussian}
\end{equation}
\begin{equation}
    \label{eq:cav_fields_pdf_uniform_ext_fields}
    P_h(\Vec{h})\,=\,\frac{\mathcal{G}_{g'(q)}^{(m)}(h)\;\mathcal{K}_m\left(\frac{hH}{g'(q)}\right)}{S_m(H)\;e^{H^2/2g'(q)}}\qquad \text{Uniform}
\end{equation}
The guassian case is trivial, since it is the distribution of the sum of two independent gaussian vectors, the internal fields $\Vec{h}_i^{(0)}\,=\,-\partial_{i}\mathcal{H}_0$ and the external fields $\Vec{b}_i$. In the uniform case we find a less trivial distribution. We make two remarks about \eqref{eq:cav_fields_pdf_uniform_ext_fields}:
\begin{itemize}
    \item The scaling of $P_h(|\Vec{h}|)$ in the pseudo-gap region ($h\rightarrow 0$) does not depend on the particular choice of external fields distribution. Indeed, the term inside the average over the external field in \eqref{eq:cav_field_pdf} is a constant for $|\Vec{h}|\rightarrow 0$. Thus, it holds
    \begin{equation}
        \label{eq:psuedo_gap_scaling_cavity_pdf}
        P_h(h)\sim h^{m-1}\qquad \forall\;\;\text{isotropic}\;\;P_b(\Vec{b}),
    \end{equation}
    the abundance of small fields only depends on the dimensions of the spins. Conversely, the tail of the distributions at large $h$ depends on the statistics of the external fields: in the uniform case for instance
    \begin{equation*}
        P_h(h)\sim e^{-h^2/2g'(q)+hH/g'(q)}
    \end{equation*}
    the decay is slower than in the gaussian case. 
   \item The normalisation of the distributions can have very different behaviors as a function of $H$. Indeed, in the gaussian and uniform cases we find respectively
   \begin{eqnarray}
       && Z_m^{gauss}(H)\,=\,(2\pi)^{m/2}[g'(q)+H^2]^{m/2} \\
       && \nonumber \\
       && Z_m^{unif}(H)\,=\,[2\pi g'(q)]^{m/2}S_m(1) H^{m-1}e^{H^2/2g'(q)},
   \end{eqnarray} 
    meaning that the normalisation in presence of random uniform fields is exponentially larger than that of the case with gaussian fields. This result is easily understandable: in the uniform case the only way that a cavity field can be small is that the internal field balances the external field, whereas in the gaussian case external fields can be arbitrarily small, so there are many more ways to make $|\Vec{h}_i|$ small\footnote{The abundance of small fields is $\int_0^{h} h'^{m-1}/Z_m\propto h^m/Z_m$, thus the larger the normalisation the more depleted are small fields.}. In numerical simulations it seems more convenient to deal with gaussian external fields, so from now on we will only consider gaussian external fields.
   
\end{itemize}

The overlap as usual satisfies the self-consistent equation
\begin{equation}
    \label{eq:overlap_mVector_RS_self-cons}
    q\,=\,\frac{1}{m}\int d^mh\;P_h(h)\left[\frac{I_{m/2}(\beta h)}{I_{m/2-1}(\beta h)}\right]^2
\end{equation}
from which the magnetic susceptibility $\chi=\beta (1-q)$ follows.


In next section we expand observables in the paramagnetic phase around $T=0$.

\subsubsection{Low temperature expansion of observables}

 We begin with the expansion of the overlap \eqref{eq:overlap_mVector_RS_self-cons} up to order $T^2$: the result is
\begin{eqnarray}
\label{eq:overlap_RS_low_T}
    && q\,=\,\frac{1}{m}-\chi_{LR}(0)T-\left[\frac{\chi_{LR}(0)^2}{2}+\frac{(m-1)(m-5)A}{8}\right]T^2+O(T^3) \\
\label{eq:zero_T_chiLR_A}
    && \chi_{LR}(0)\,=\,(1-1/m)\left\langle\frac{1}{h}\right\rangle\qquad A\,=\,(1-1/m)\left\langle\frac{1}{h^2}\right\rangle
\end{eqnarray}
and is obtainable by considering the expansions of Bessel functions in Appendix \ref{sec:Bessel}.
For $m>1$, the expansion in powers of $T$ terminates at $O(T^{m-1})$ because the coefficient of next term contains $\langle 1/h^m\rangle$ and thus is divergent, remembering that $P_h(h)\sim h^{m-1}$: so in general the overlap can be written as the sum of a polynomial in $T$ of order $m-1$ and a singular function with leading behavior
\begin{equation}
    q_{sing}\sim \langle 1/h^m \rangle T^m \sim T^m\log(1/T)
\end{equation}
When $m>3$, if $H>H_c(0)$ the coefficients of the regular part of the expansion are all $O(1)$ in $H$, but in the cases $m=1, 2$ one has to scale $H$ with temperature, since the dAT line is divergent (the critica field is infinite): indeed, $\chi_{LR}(0)=O(H^{-1})$ and $A=O(H^{-2})$, and one has to plug in this expressions the divergence of $H$ in $T$. We are not interested in discussing this case, since we want to study the paramagnetic phase at $T=0$, so in other words the $m>2$ case.

The expansions of thermodynamic observables can be readily obtained once \eqref{eq:overlap_RS_low_T} is substituted. Let us consider the internal energy and the specific heat:
\begin{eqnarray}
\label{eq:internal_energy_AND_specific_heat_RS}
    U(T)\,&=&\,-\langle h \rangle_h+c_v(0)T+O(T^{2}) \\ 
    c_v(0)\,&=&\,\frac{m-1}{2}+\frac{m}{2(1/m+H^2)^{1/2}}\left[\frac{\sqrt{2}}{m}\frac{\Gamma(m/2+1/2)}{\Gamma(m/2)}\right]^2
\end{eqnarray}
For large $H$ spins do not interact and the specific heat satisfies equipartition theorem\footnote{Each spin has $m-1$ degrees of freedom, since their moduli are fixed.}, as it should be. Thus the second term in the expression of the specific heat is a correction coming from interactions. Notice that coherently with continuous classical theories, the positive specific heat implies a logarithmic divergence of the entropy.
At the transition, this RS expression \eqref{eq:internal_energy_AND_specific_heat_RS} does not exhibit any singularity, at variance with the RS $H=0$ model whose susceptibility has a cusp at the transition \cite{sherrington1975solvable, edwards1975theory, edwards1976theory}. 
The dAT line expanded close to $T=0$ for $m>2$ reads
\begin{equation}
    H_c(T)\,=\,\frac{1}{\sqrt{m(m-2)}}-\sum_{k=1}^{m-3}c_k\;T^k-k_{sing}T^{m-2}\log(1/T)
\end{equation}
so for $m=3$ one has the singular leading behavior $\Delta H_c\propto T\log(1/T)$. 

\subsubsection{fRSB solution}

Let us go back to \eqref{eq:replica_A}, \eqref{eq:W_0} and let us consider the isotropic saddle point \eqref{eq:isotropy_overlaps}. We average $W_b$ over gaussian external fields
\begin{equation}
\label{eq:W_2_avg}
    \underset{b}{\mathbb{E}}[W_b[\mathbb{Q}]]\,=\,\int d^m S_1\cdots d^m S_n \exp\left\{\frac{\beta^2}{2} \sum_{a b}^{1, n}[g'(Q_{ab})+H^2]\Vec{S}_a\cdot \Vec{S}_b\right\}
\end{equation}
The logarithm of \eqref{eq:W_2_avg} can be written exactly as the integral in \eqref{eq:kRSB_fe_functional}, except for the final condition:
\begin{equation}
\label{eq:final_condition_Parisi_eq_vector}
    \log I(q_{ea}, h)\equiv \varphi(q_{ea}, h)\,=\,\log \mathcal{K}_m(\beta h).
\end{equation}
The first term in \eqref{eq:replica_A} is trivial to write in the fRSB formalism:
\begin{equation*}
    \frac{\beta^2}{2}\sum_{a,b}\sum_{\alpha,\beta}\Bigl[g(Q_{ab}^{\alpha\beta})-{\hat Q}_{ab}^{\alpha\beta}{Q}_{ab}^{\alpha\beta}\Bigr]\,=\,\frac{m\beta}{2}\left\{g(1/m)-g'(1/m)/m-\int_{0}^{1/m} dq\;P(q)[g(q)-g'(q)q]\right\}
\end{equation*}
We write the fRSB free energy functional following a variational approached used in \cite{sommers1984distribution}:
\begin{eqnarray}
\label{eq:fRSB_fe_functional_vector}
&&\Phi[x(q)]=-\frac{m\beta}{2}\left\{g(1/m)-g'(1/m)/m-\int_{q_0}^{q_{ea}} dq\;P(q)[g(q)-g'(q)q]\right\} \\
&&-\frac{\beta(g'(1/m)-g'(q_{ea}))}{2}-\frac{1}{\beta}\int d^m h P(q_0, h)\varphi(q_0,h)\nonumber \\
&&-\frac{1}{\beta}\int_{0}^{1/m}dqP(q)\int d^m{h}\; P(q,h)\left\{\diffp{\varphi}{q}+\frac 1 2 g''(q)\left[\diffp[2]{\varphi}{h}+\frac{m-1}{h}\diffp{\varphi}{h} +x(q)\,\left|\diffp{\varphi}{h}\right|^2\right]\right\}
\nonumber \\
&&+\frac{1}{\beta}\int d^m{h}\; P(q_{ea}, h)[\varphi(q_{ea}, h)-\log \mathcal{K}_m(\beta h)]. \nonumber
\end{eqnarray}
We introduced the functional Lagrange multipliers $P(q, h)$ to enforce Parisi equations
\begin{equation}
\label{eq:Parisi_eqs}
    \diffp{\varphi}{q}\,=\,-\frac{1}{2}g''(q)\left[\diffp[2]{\varphi}{h}+\frac{m-1}{h}\diffp{\varphi}{h}+x(q)\left|\diffp{\varphi}{h}\right|^2\right]
\end{equation}
and their boundary condition, eq. \eqref{eq:final_condition_Parisi_eq_vector}. The physical meaning of $P(q, h)$ is the following: they are the probability density of observing a cavity field $\Vec{h}$ when measuring the system in a cluster of the phase space with overlap $q$. Dynamically, this corresponds to measure the cavity field acting on a spin for a time interval covering all dynamical epochs from $q_{ini}=q_{ea}$ to $q_{fin}=q$, following the interpretation of RSB in \cite{Sompolinsky1981}.
The $P(q,h)$ satisfy the pde
\begin{eqnarray}
  \label{eq:pde_pdf_cavity_fRSB}
  \diffp{P}{q}=\frac 1 2 g''(q)\left[\diffp[2]{P}{h}+\frac{m-1}{h}\diffp{P}{h}-2 x\,\diffp{}{h}\cdot\left(P\diffp{\varphi}{h}\right)\right] 
\end{eqnarray}
with initial condition
\begin{eqnarray}
  \label{eq:initial_condition_pdf_cavity_fRSB}
  P(q_0, h)=\mathcal{G}_{g'(q_0)+H^2}^{(m)}(h).
\end{eqnarray}
We can derive a self-consistent equation for the overlap if we impose that $\varphi(q, h)$ and $P(q, h)$ satisfy their pde:
\begin{equation}
\label{eq:overlap_mVector_self_consist}
    \frac{\delta \Phi}{\delta x(q)}=0\quad\Longrightarrow\quad q\,=\,\frac{1}{m}\int d^m h\;P(q, h)\;T^2\left(\diffp{\varphi}{h}\right)^2
\end{equation}
In the RS limit one finds back eq. \eqref{eq:overlap_mVector_RS_self-cons}.
By deriving again $\Phi$ and setting the result to zero, one finds the condition of marginal stability ($\Lambda$=0) of the spin glass phase
\begin{equation}
\label{eq:replicon_RSB}
    1\,=\,\int d^m h P(q, h)\left[\frac{m-1}{m}\left(\frac{M}{h}\right)^2+\frac{1}{m}\left(\diffp{M}{h}\right)^2\right].
\end{equation}
where we introduced $M(q, h)\,=\,T\partial_h\varphi(q, h)$: this is the RSB magnetisation corresponding to a cavity field $\Vec{h}$. The functional found in \eqref{eq:fRSB_fe_functional_vector}, constraints aside, is just a special case of that found by D. Panchenko in \cite{panchenko2018free}. There, the author vector spin variables constraint to generic compact subsets of $\mathbb{R}^m$, whereas we analyse only the case of spins constraint to live on the hypersphere.

\subsubsection{Zero temperature limit}

Eqs. \eqref{eq:fRSB_fe_functional_vector}, \eqref{eq:pde_pdf_cavity_fRSB}, \eqref{eq:overlap_mVector_self_consist} and \eqref{eq:replicon_RSB} can be straightforwardly extended to $T=0$ if the following scaling laws are assumed
\begin{itemize}
    \item $\varphi(q, h)\,=\,\beta\,\phi(q, h)=O(\beta)$
    \item $P(q, h)\,=\,\,P_0(q, h)=O(1)$
    \item $\beta x(q)\,=\,y(q)=O(1)$
    \item $\beta\,(1/m-q_{ea})\,=\,\chi_{LR}=O(1)$
\end{itemize}
The first assumption is motivated by observing that the boundary condition function \eqref{eq:final_condition_Parisi_eq_vector} behaves as
\begin{equation*}
    \varphi(q_{ea}, h)\,=\,\log \mathcal{K}_m(\beta h)\,=\,\beta h+O(\log \beta)\qquad \beta\rightarrow\infty
\end{equation*}
as one can verify by checking the asymptotic expansions of Bessel functions in Appendix \ref{sec:Bessel}. The third condition is just a consequence of the first, see eq. \eqref{eq:Parisi_eqs} for confirmation. The second condition ensures that the consistency equation \eqref{eq:overlap_mVector_self_consist}
\begin{equation*}
    q\,=\,\frac{1}{m}\int d^m h\;P(q, h)\;T^2\left(\diffp{\varphi}{h}\right)^2\,=\,\frac{1}{m}\int d^m h\;P_0(q, h)\;\left(\diffp{\phi}{h}\right)^2
\end{equation*}
has a good $T=0$ limit. It is known fact that the smallest overlap $q_0$ depends very weakly on temperature \cite{mez1987}, so that it can be considered temperature independent: thus we have that $P(q_0, h)\,=\,P_0(q_0, h)$.
The fourth assumption ensures the stability of the zero temperature solution in the RSB phase \cite{palmer1979internal}. At variance with the SK model ($m=1$), we expect the local fields $\{\mu_i^{(\alpha)}\}$ inside any pure state $\alpha$ to be gapped at $T=0$, and that
\begin{equation*}
    \min \mu_i^{(\alpha)}\,=\,\chi_{LR}\equiv (1-1/m)\int d^m h\frac{P(q_{ea}, h)}{h}.
\end{equation*}
Note that this would also ensure continuity at the critical field.
The fact that $\chi_{LR}(0)\neq 0$ at $H=0$ follows from the continuity of the vector spins, thus the existence of arbitrarily small excitations. Note that the isotropy of the model implies that the cavity field distributions should have the form
\begin{equation}
    P(h, q)=h^{m-1}p(h, q)
\end{equation}
In the $m=1$ case it is known that the $p(h, q)$ starting from $q=q_0$ should progressive go from a gaussian behavior at small fields to the behavior at $q=q_{ea}$ predicted in equation \eqref{eq:TAP_mods}, $P(h, q_{ea})\sim h$. As it was previously discussed, this scaling implies $q_{ea}\sim 1-aT^2$: this low temperature behavior of Edwards-Anderson overlap is equivalent to the absence of low energy excitations, $\chi_{LR}=0$. It is not yet clear to us if for $m>1$ the function $p(h, q)$ should change close to the origin similarly to what happens in the $m=1$ case. 

With these assumptions we can write the equation for the ground state energy in the RSB phase:

\begin{eqnarray}
\label{eq:GS_RSB_1}
    E(H)\,=\,\underset{y(q), q_0}{\max}\Bigl\{\frac{m}{2}\,\int_{q_0}^{1/m}\,dq\,q\,y(q)-
    \int d^m h\,P_0(q_0,h)\,\phi(q_0,h)\Bigr\}.
\end{eqnarray}

The ground state can be computed numerically from \eqref{eq:Parisi_eqs}, \eqref{eq:pde_pdf_cavity_fRSB} at $T=0$ and \eqref{eq:GS_RSB_1} by adapting to the vector spin glass the strategy in \cite{oppermann2008universality, schmidt2008method} where the authors maximise \eqref{eq:fe_SK_fRSB} with respect to $q(x)$, or that in \cite{alaoui2020algorithmic}, where instead the equations are solved by maximising $x(q)$.
Parisi solution of vector systems has been explored mainly close to the critical line \cite{elderfield1982parisi}. As a future development of our work, it would be interesting to solve the RSB phase of vector models in full generality.

\section{TAP equations}
\label{sec:TAP_equations_vector_model}

In this section we derive the TAP equations of the model, by means of a \emph{Georges-Yepidia-Plefka expansion} \cite{plefka1982convergence, georges1991expand} of the Gibbs potential. The TAP equations of vectorial model were derived for the first time in \cite{bray1980metastable}: we find back their result, modulo a different normalisation of the spins. The derivation of the equations and in particular of the TAP free energy is preparatory for the discussion we wish to do about linear excitations around TAP solutions in next section. 
We expand at high temperatures the free-energy functional
\begin{equation}
\label{eq:Adef}
    -\beta F[{\bf m}]\equiv A[{\bf m};\beta] = \log{\int D{\bf S}\exp[{-\beta \mathcal{H}[{\bf S}]+\sum_{i}{\Vec\lambda_i^{\beta}}\cdot({\Vec S_i}-{\Vec m_i})}]}
\end{equation}
where the Lagrange multipliers $\{\bf{\lambda_i^{\beta}}=\beta\bf{\eta_i}\}$ are external fields that enforce magnetizations to their ensemble averages:
\begin{equation}
    \label{eq:mEq}
    \partial_{\Vec{\lambda_i^{\beta}}}A[{\bf m};\beta] = 0\quad\Longrightarrow\quad\Vec{m_i} = {\langle\Vec{S_i}\rangle}_{\lambda}.
\end{equation}
This self-consistent equation returns physical magnetisations when the Lagrange fields are set to zero. Solutions of \eqref{eq:mEq} define metastable states of the system. Thus, the function \eqref{eq:Adef} is generally non-convex: one can obtain the Gibbs potential through its convex-envelope.

In this section and in the next we use the letter "m" for magnetisations and "d" for spin dimension.
We expand \eqref{eq:Adef} up to order $\beta^2$:
\begin{equation}
\label{eq:GY}
    A[m;\beta]=A[m;0]+\partial_{\beta}A[m;0]\beta+\frac{1}{2}\partial_{\beta}^2A[m;0]\beta^2+O(\beta^3).
\end{equation}
For fully connected models it is well known that the high-temperature expansion truncates after a finite number of terms \cite{thouless1977solution}. In the case of disordered models such those studied hereby, in the thermodynamic limit all terms $o(\beta^2)$ vanish.
Consider the following observable
\begin{equation}
    \label{eq:Udef}
    U_{\beta}=\mathcal{H}-{\langle \mathcal{H}\rangle}-\sum_i\partial_{\beta}\Vec{\lambda_i^\beta}\cdot(\Vec{S_i}-\Vec{m_i}).
\end{equation}
It's not hard to show\footnote{To retrieve \eqref{eqs:U-A_rel}, use that $\langle U\rangle=0$, $\frac{d}{d\beta}\langle O\rangle=\langle\frac{dO}{d\beta}\rangle-\langle OU\rangle$ and finally that $\frac{d^n}{d\beta}\bold{\lambda_i^{\beta}}=-\frac{\partial}{\partial\bold{m_i}}\frac{\partial^nA[m;\beta]}{\partial\beta^n}$.} that
\begin{subequations}
    \label{eqs:U-A_rel}
    \begin{align}
        \partial_{\beta}A_{\beta} & = -{\langle \mathcal{H}\rangle}_{\beta} \\
        \partial_{\beta}^2A_{\beta} & = {\langle U_{\beta}^2\rangle}
    \end{align}
\end{subequations}
Let's compute the order zero term of \eqref{eq:GY}
\begin{equation}
    \label{eq:A0_def}
    A_0 = \log{\int D{\bf S} \exp\sum_{i}\Vec{\lambda_i^0}\cdot(\Vec{S_i}-\Vec{m_i})}=\sum_i[\mathcal{K}_d(\lambda_i^{0})-\Vec{\lambda}_i^{0}\cdot\Vec{m}_i]
\end{equation}
where as usual
\begin{equation*}
\int d\Omega_d e^{\boldsymbol{\lambda_i^0}\cdot\hat{\bold{z}}} = S_d(1)\frac{\int_0^{\pi} d\theta (\sin{\theta})^{d-2}e^{\lambda_i^0\cos{\theta}}}{\int_0^{\pi} d\theta (\sin{\theta})^{d-2}}\,=\,(2\pi)^{d/2}\frac{I_{d/2-1}(\lambda_i^0)}{(\lambda_i^0)^{d/2-1}} 
\end{equation*}
Note that at $\beta=0$ equation \eqref{eq:mEq} can be rewritten as
\begin{equation}
\label{eq:mEqBeta0}
m_i^{\mu}\equiv{\langle S_i^{\mu}\rangle}_0=\frac{d}{d\lambda_i^0}[\log{\mathcal{K}_d(\lambda_i^0)}]\frac{\lambda_i^{\mu}(0)}{\lambda_i^0}\equiv g_n(\lambda_i^0)\frac{\lambda_i^{\mu}(0)}{\lambda_i^0}    
\end{equation}
where
\begin{equation}
\label{eq:gDef}
g_d(\lambda)=\frac{I_{d/2}(\lambda)}{I_{d/2-1}(\lambda)}.
\end{equation}
This is the generalisation to $d$-dimensional spins of eq. \eqref{eq:TAP_eqs} found in \cite{thouless1977solution} for the $d=1$ case. Function \eqref{eq:gDef} in the $d=3$ case is the well known Langevin function describing the polarisation of electric or magnetic dipoles under the effect of an external electrostatic or magnetostatic field:
\begin{equation*}
    g_3(\lambda)\equiv\coth(\lambda)-\frac{1}{\lambda}.
\end{equation*}
The properties of functions $g_d$ for general $d$ can be found in Appendix \ref{sec:Bessel}.
Having obtained the relation between magnetisations and fields, in order to close them we need to find the expression of the $\{\lambda_i\}$ in terms of the $\{m_i\}$.
Let us now compute the second order term of \eqref{eq:GY} using the second relation of \eqref{eqs:U-A_rel}, at $\beta=0$: firstly, we note that
\begin{equation*}
(\partial_{\beta}\Vec{\lambda_i^{\beta}})(0)=[\partial_{\beta}(-\partial_{\Vec{m_i}} A[m;\beta])](0)=-(\partial_{\Vec{m_i}}\partial_{\beta}A[m;\beta])(0)=\partial_{\Vec{m_i}}{\langle \mathcal{H}\rangle}_0
\end{equation*}
and thanks to this we can write the square of \eqref{eq:Udef} at $\beta=0$ in a very compact way:
\begin{equation}
\label{eq:Usq}
(U_0)^2=\frac{1}{4}\sum_{i\neq j}\sum_{k\neq l}J_{ij}J_{kl}(\Vec{S_i}-\Vec{m_i})\cdot(\Vec{S_j}-\Vec{m_j})(\Vec{S_k}-\Vec{m_k})\cdot(\Vec{S_l}-\Vec{m_l})
\end{equation}
At $\beta=0$, which corresponds to infinite temperature, the spins are non interacting: all the connected correlation functions are strictly null in the Gibbs state, and this allows us to factorize all the thermal averages. In particular, in the thermal average of \eqref{eq:Usq} only the $i=k, j=l$ or $i=l$, $j=k$ terms contribute, thanks to condition \eqref{eq:mEq}.
Thus, we have
\begin{equation*}
{\langle U_0^2\rangle}_0=\frac{1}{2}\sum_{i\neq j}J_{ij}^2[{\langle(\Vec{S_i}\cdot\Vec{S_j})^2\rangle}_0-{\langle(\Vec{S_i}\cdot\Vec{m_j})^2\rangle}_0-{\langle(\Vec{m_i}\cdot\Vec{S_j})^2\rangle}_0-(\Vec{m_i}\cdot\Vec{m_j})^2]
\end{equation*}
We can further simplify this last expression: writing the scalar products explicitly, factorizing the thermal averages in single site factors and remembering that (derive once again \eqref{eq:mEqBeta0})
\begin{equation*}
 {\langle S_i^{\mu}S_i^{\nu}\rangle}_0=\partial_{\lambda_i^{\mu}(0)}m_i^{\nu}+m_i^{\mu}m_j^{\nu}
\end{equation*}
with some manipulations we finally get
\begin{eqnarray*}
 \partial_{\beta}^2A[m;0]=\frac{1}{2}\sum_{i\neq j}J_{ij}^2\sum_{\mu}\nabla_{\boldsymbol{\lambda_i^0}}m_i^{\mu}\cdot\nabla_{\boldsymbol{\lambda_j^0}}m_j^{\mu}=\frac{1}{2}\sum_{i\neq j}J_{ij}^2\sum_{\mu}(\tilde{\boldsymbol{\chi}}_{ii}\tilde{\boldsymbol{\chi}}_{jj})_{\mu\mu} =\frac{N}{2}\Tr (\overline{\mathbf{\tilde{\chi}}}^2).
\end{eqnarray*}
where in the last equality we used self-averaging.
We introduced the tensor
\begin{equation}
\label{eq:chiTildeDef}
\tilde{\chi}_{ii}^{\mu\nu}=\frac{\partial m_i^{\mu}}{\partial_{\lambda_i^{\nu}(0)}}.
\end{equation}
which is related to the magnetic susceptibility tensor by $\chi_{ii}^{\mu\nu}\,=\,\beta \tilde{\chi}_{ii}^{\mu\nu}$.
Finally, the free energy functional in power of $\beta$ is
\begin{equation}
\label{eq:TapFreeEn_vector}
F_{TAP}[{\bf m}]=-\frac{A_0}{\beta}-\left[\frac{1}{2}\sum_{i\neq j}J_{ij}\Vec{m}_i\cdot\Vec{m}_j+\sum_i\Vec{b}_i\cdot\Vec{m}_i\right]-\frac{N}{4\beta}\Tr (\overline{\mathbf{\chi}}^2)
\end{equation}
Let us comment the three terms at the r.h.s. of this last expression: the first term is the infinite temperature entropy of the system; the second term is the familiar 'mean field' energy; finally 
the third term is the energetic contribution of Onsager reaction field, so it is a signature of the linear response that follows the addition of a novel spin. For the systems we are studying are statistically isotropic (random external fields), Onsager term could be further simplified
\begin{equation*}
    \Tr \overline{\mathbf{\chi}}^2\,=\,d\;\chi^2\qquad \chi\,=\,\overline{\chi_{ii}^{\mu\mu}}
\end{equation*}
In deriving the TAP free energy, we neglected higher order terms because they yield subextensive contribution. One can show that in the RSB phase these neglected terms yield a divergent contribution. The stability condition that one can find for the high-temperature expansion at a given $T, H$ is exactly the positivity of the Replicon eigenvalue \cite{plefka1982convergence}.

TAP equations are straightforwardly obtained from TAP free energy \eqref{eq:TapFreeEn_vector} with a derivation with respect to $\Vec{m}_i$, finding

\begin{equation}
\label{eq:tapLocField_i}    
\Vec{h}_i = \sum_j J_{ij}\Vec{m}_j+\Vec{b}_i-\chi\Vec{m}_i\qquad \Vec{m}_i\,=\,g_d(\beta |\Vec{h}_i|)\frac{\Vec{h}_i}{h_i}
\end{equation}
 A solution of \eqref{eq:tapLocField_i} is a \emph{TAP state}: in the RS high temperature phase, it represents a local minimum of the free-energy landscape. In 1RSB systems TAP states are equivalent to pure states, but in fRSB systems this is false in general \cite{mez1987}.
 TAP equations \eqref{eq:tapLocField_i} lend themselves to an iterative algorithm to compute magnetisations corresponding to local minima of the free energy. In the spin glass phase, simple iterative approaches hardly converge \cite{bray1979evidence}; an improved scheme, based on the minimisation of $|m_i-\tanh \beta h_i|$, is proposed in \cite{nemoto1985tap}. A rigorous proof of the convergence of Eqs. \eqref{eq:tapLocField_i} in the RS phase is provided in \cite{bolthausen2014iterative}, where the author considers the SK model ($m=1$). The proof is based on the behavior of the function
 \begin{equation*}
     \psi(t)\,=\,\underset{Z, Z', Z''}{\mathbb{E}}[\tanh(b+\sqrt{t}Z+\sqrt{q-t}Z')\tanh(b+\sqrt{t}Z+\sqrt{q-t}Z'')]
 \end{equation*}
 defined in $[0, q]$, which corresponds to the overlap between the on-site magnetisation at different times. Indeed, one finds that $\psi(t)\sim q-(1-\Lambda)^t$, where $\Lambda$ is the Replicon eigenvalue. Thus, $|m_i(t)-m_i(t-1)|^2=2(q-\psi(t))\rightarrow 0$ and \eqref{eq:tapLocField_i} admit a fixed point. The proof could be straightforwardly extended to the family of vector models hereby studied, since the functions $\mathcal{g}_m(x)$ relating magnetisations and cavity fields are qualitatively very similar to the hyperbolic tangent.
 

\subsection{Fluctuations around TAP states: the Hessian}

In this section we compute the Hessian matrix related to TAP free energy, eq. \eqref{eq:TapFreeEn_vector}. To do the computation, we define the inverse function of $g_d(x)$:
\begin{equation*}
    h_d(y)\equiv g_d^{-1}(y)
\end{equation*}
With this definition, one has
\begin{equation*}
    \frac{\partial F_{TAP}}{\partial m_i^{\mu}}\,=\,\frac{h_d(m_i)}{m_i}m_i^{\mu}+\chi m_i^{\mu}-\sum_{j}J_{ij}m_j^{\mu}-b_i^{\mu}
\end{equation*}
and deriving one more time, one obtains the Hessian matrix, expressed in terms of its matrix $i, j$ elements:
\begin{eqnarray}
\label{eq:TAP_Hessian}
    && \mathbb{M}_{ij}\,=\,\mathbb{M}_{ij}^{(L)}+\mathbb{M}_{ij}^{(T)} \\
    && \mathbb{M}_{ij}^{(L)}\,=\,\mathbb{P}_i\left[-J_{ij}-\frac{2\beta d}{N}m_i m_j+\left(\chi+\diff{h_d}{m}(m_i)\right)\delta_{ij}\right]\mathbb{P}_j \\
    &&\mathbb{M}_{ij}^{(T)}\,=\,\mathbb{P}_i^{(\perp)}\left[-J_{ij}+\left(\chi+\frac{h_d(m_i)}{m_i}\right)\delta_{ij}\right]\mathbb{P}_j^{(\perp)} \\
    && \mathbb{P}_i\,=\,\frac{\Vec{m}_i\Vec{m}_i^{\intercal}}{m_i^2}\qquad\mathbb{P}_i^{(\perp)}\,=\,\mathbb{I}-\frac{\Vec{m}_i\Vec{m}_i^{\intercal}}{m_i^2}
\end{eqnarray}
The Hessian is naturally decomposed in two contributions: the first accounts for \emph{longitudinal} fluctuations $\Vec{m}_i\rightarrow \alpha_i \Vec{m}_i$ for some random $\alpha_i$, the second for \emph{transverse} ones $\Vec{m}_i\rightarrow \mathbb{R}_i \Vec{m}_i$ for some random local rotation $\mathbb{R}_i$. The eigenvectors of the Hessian thus split into two orthogonal subspaces: longitudinal excitations in the $N$-dimensional space spanned by ${\bf m}$, whereas transverse modes span the orthogonal space ${\bf m}^{\perp}$ with dimension $N\times (m-1)$. As a first check, the Hessian evaluated at the $H=0$ paramagnetic solution $\boldsymbol{m}$ reads
\begin{equation}
\label{eq:HessianF_zeroField1}
\frac{\partial^2F_{TAP}}{\partial m_l^{\nu}\partial m_k^{\mu}}\Bigl|_{[\bold{m}_i]=0} = \delta^{\mu\nu}\Bigl[\Bigl(\frac{d}{\beta}+\frac{\beta}{d}\Bigr)\delta_{lk}-J_{lk}\Bigr]
\end{equation}
from which we can easily\footnote{The maximal eigenvalue of the interactions matrix is $2$, and this shall be equated to the diagonal element.} see that $T_c\,=\,1/d$. When one wants to study the model in the infinite $d$ limit (spherical limit), it is convenient then to rescale all spins to have norm $\sqrt{d}$.

The eigenvalue spectrum of \eqref{eq:TAP_Hessian} was studied in \cite{yeo2004complexity} considering the isotropic case $H=0$ in the spin glass phase. The authors find a spectrum made of $m$ null eigenvalues and a continuum band with strictly positive lower edge for any $0<T<T_c$. Exactly $m-1$ null eigenvalues are related to the global $O(m)$ symmetry; the remainder null eigenvalue is non-trivial and stems from the projector term $-2\beta \Vec{m}_i\Vec{m}_j^{\intercal}/N$ in the longitudinal Hessian \eqref{eq:TAP_Hessian}. The null eigenvalue is found in the case of Ising spins as well \cite{aspelmeier2004complexity}. It was shown in \cite{parisi2004supersymmetry} in the case $m=1$ that the nullity of this isolated eigenvalue is related to the breaking of super-symmetry (SUSY) of the Action appearing in the computation of the Bray-Moore complexity \cite{bray1980metastable}. The physical meaning of the isolated eigenvalue is the marginal stability of the spin glass phase. TAP states of spin glass models with pairwise interactions cannot trap the system forever, thus a remanent magnetization is asymptotically lost with relaxation: this is at variance with p-spin models, where instead in the glassy phase there are stable TAP solutions.

Let us restrict to the paramagnetic phase.
The longitudinal and the transverse Hessian matrix for large $N$ are both instances of a Rosenzweig-Porter ensemble. In both cases the Hessian is the sum of a Wigner matrix and a diagonal matrix, whose entries distribution differs from that of the couplings. In order to acquire some physical insight, let us consider the diagonal elements in both cases. The uniform shift provided by $\chi$ ensures the stability of the solution, coherently with \cite{palmer1979internal}. The residual terms are related to the inverse of the local susceptibilities: from \eqref{eq:tapLocField_i}, by deriving the second equation with respect to $\Vec{h}_i$ one finds for the susceptibility tensor
\begin{equation}
\label{eq:susceptibility_tensor}
    {\boldsymbol \chi}_{ii}\,=\,\beta g_d'(\beta h_i)\mathbf{P}_i+\frac{g_d(\beta h_i)}{h_i}\mathbf{P}_i^{(\perp)}\equiv \frac{1}{h_d'(m_i)}\mathbf{P}_i+\frac{m_i}{h_d(m_i)}\mathbf{P}_i^{(\perp)}
\end{equation}
Therefore, the smallest residual values over the stability offset in the diagonal of both matrices are related to the most susceptible spins. As a check, let us verify that longitudinal excitations disappear at zero temperature and how temperature affects transverse ones.
It is convenient to write $\{h_d(m_i)\}$ as functions of the cavity fields, and introduce random variables $\{\kappa_i\}, \{\ell_i\}$
\begin{equation}
    h_d(m_i)\equiv \beta h_i\quad\Longrightarrow\quad
    \begin{cases}
        \diff{h_d}{m}(m_i)\equiv \frac{1}{\beta\;g_d'(\beta h_i)}=\frac{d}{\beta}+\kappa_i \\
         \frac{h_d(m_i)}{m_i}\equiv \frac{h_i}{g_d(\beta h_i)}\,=\,\frac{d}{\beta}+\ell_i.
    \end{cases}
\end{equation}
where we used $g_d(x)\,\sim\,x/d$ for small $x$. The term $d/\beta$ is just the energy provided to each degree of freedom by the thermal bath. In the zero temperature limit, variables $\kappa_i$ are of order $\beta$ for any $h_i \gg 1/\beta$, whereas $\ell_i$ are equal to $h_i$ in the same situation: as it is natural to expect, when temperature is decreased longitudinal excitations are increasingly harder to trigger than transverse ones; at exactly $T=0$, longitudinal modes are absent, as it should be. More specifically, for $d>1$ we have ($\alpha_i\,=\,\beta h_i$) the following behaviors
\begin{eqnarray}
    && \kappa_i\,\sim\,\frac{2\alpha_i h_i}{d-1}\qquad \alpha_i \gg 1 \\
    && \kappa_i\,\sim\,\frac{3\alpha_i h_i}{2(d+2)}\qquad \alpha_i \ll 1 \\
    && \ell_i\,\sim\,\left[1-\frac{d+1}{2\alpha_i}\right]h_i\qquad \alpha_i \gg 1 \\
    && \ell_i\,\sim\,\left[\frac{\alpha_i d}{2(d+2)}\right]h_i\qquad \alpha_i \ll 1
\end{eqnarray}
The temperature affects transverse excitations only in a region $O(T)$ close to the stability edge: if we treat a small temperature as a perturbative effect, it is clear that its effect would affect strongly only cavity fields of the same order of the effect. 
A detailed study of the TAP Hessian in presence of an external random field would be an interesting development of our current research. In the following, we begin to study the $T=0$ limit of \eqref{eq:TAP_Hessian}, the Hessian of the Hamiltonian, whose spectrum describes the properties of linear excitations around energy minima.

\section{Linear excitations of inherent structures}

We finally begin to study the stability properties of energy minima. 
In RS phases, or in simple glassy models such as the spherical p-spin, energy minima can be obtained by following TAP minima down to zero temperature. In the generic RSB case this is no possible, either because of \emph{temperature chaos} \cite{rizzo2003chaos}, the phenomenon for which energy levels cross each other following tiny variations of temperatures, or for the renown Gardner transition \cite{Gardner1985}, the phenomenon observed in the p-spin Ising spin glass for which a stable state becomes marginal.

The energy landscape of the SK model and its vector version was firstly characterised in \cite{bray1980metastable, bray1981metastable, bray1981metastable2}. In these works Bray and Moore compute the complexity of stationary points of the TAP free energy and the Hamiltonian. They identify a critical energy level such that for lower energies the annealed and the quenched complexities ($\frac{1}{N}\log \overline{\mathcal{N}}$ and $\frac{1}{N}\overline{\log\mathcal{N}}$) are equal, but for higher energies the quenched complexity undergoes a RSB transition and it becomes lower than the annealed one. Later, it was shown that Bray-Moore complexity breaks a SUSY symmetry \cite{kurchan1991replica}.



A practical approach to study the properties of energy minima is to define a numerical minimisation protocol for the Hamiltonian.
The problem of reaching fixed points of \eqref{eq:Hamiltonian_Vector_Models_Random_Fields} is a constrained optimization problem: we want to find a configuration of spins ${\bf S}\equiv (\Vec{S}_1,\dots,\Vec{S}_N)$ in the subset $\mathcal{S}_m(1)^{N}$ of $\mathbb{R}^N$, where $\mathcal{S}_m(1)$ is the $m$-dimensional unit sphere, that is a local minimum of \eqref{eq:Hamiltonian_Vector_Models_Random_Fields}. 
In order to take account of constraints, we introduce the Lagrangian

\begin{equation}
\label{eq:Model_again}
    \mathcal{L}[{\bf S}, {\boldsymbol{\mu}}]\,=\,-\frac{1}{2}\sum_{ij}\;J_{ij}\Vec{S}_i\cdot\Vec{S}_j-\sum_{i}\Vec{b}_i\cdot\Vec{S}_i+\sum_{i}\mu_i\,(|\Vec{S_i}|-1).
\end{equation}
A stationary point of the Hamiltonian satisfying the local spherical constraints is a solution of the $N\times m+N$ equations
\begin{eqnarray}
\label{eq:fixed_points_energy_0}
   && \frac{\partial \mathcal{L}}{\partial \Vec{S}_i}\,=\,0 \\
   && \nonumber \\
   && \frac{\partial \mathcal{L}}{\partial \mu_i}\,=\,0
\end{eqnarray}
One can easily show by combining these last two equations that stationary points of \eqref{eq:Model_again} are configurations ${\bf S}^{(*)}$ such that each spin is aligned to its local field
\begin{eqnarray}
\label{eq:fixed_points_energy_1}
    && \mu_i\Vec{S}_i^{(*)}\equiv\Vec{\mu}_i\,=\,\sum_j\;J_{ij}\Vec{S}_j^{(*)}+\Vec{b}_i \\
    && \Vec{S}_i^{(*)}\,=\,\frac{\Vec{\mu_i}}{\mu_i}
\end{eqnarray}
The local field counts two contributions: the first is the internal field $\sum_j J_{ij}\Vec{S}_j$ generated by its neighbors, the second is the random external field. While the first is responsible for mutual correlations which result in frustrated ordering, the second can be regarded as a "spatial" noise term, which reduces the correlations among neighboring spins. At $H=0$ the system is in the RSB phase: as the standard deviation of the external field is raised, spins alignments are perturbed by the external fields, to the extent the latter become the dominant contribution in \eqref{eq:fixed_points_energy_1} and the system becomes RS. This scenario describes the zero temperature phase transition in spin glass systems. As it has been already pointed out in section \ref{sec:vector_models_chap1}, in fully-connected models this transition does not occur for spins with less than $m=3$ components. Only in sparse models one can study a $T=0$ SG transition for Ising and XY spins \cite{parisi2014diluted}, \cite{lupo2017critical}.




\subsection{Numerical minimisation of the energy}

The fixed point condition \eqref{eq:fixed_points_energy_1} invites for the use of a renown gradient descent protocol, named \emph{Greedy Coordinate Descent} (GCD) algorithm, also known as \emph{Gauss-Seidel} procedure: a gradient descent move consists in aligning at each step a spin to its instantaneous local field. Introducing a discrete time step $t$, we can write an equation for the update move
\begin{eqnarray}
    && (\dots, \Vec{S}_{i-1}, \Vec{S}_i, \Vec{S}_{i+1}, \dots)\Longrightarrow(\dots, \Vec{S}_{i-1}, \Vec{S'}_i, \Vec{S}_{i+1}, \dots) \\
    && \nonumber \\
    \label{eq:GD_update_move}
    && \Vec{S}_i(t)\,=\,\frac{\sum_{j}J_{ij}\Vec{S}_j(t-1)+\Vec{b}_i}{|\sum_{j}J_{ij}\Vec{S}_j(t-1)+\Vec{b}_i|}
\end{eqnarray}
We propose code \ref{alg:single_upd} as an algorithm for the single update step.

\begin{algorithm}[t]
\caption{Update move of the GCD algorithm.}
\label{alg:single_upd}
\begin{algorithmic}[1]

\State Read spin label $label\leftarrow i$.
\State
\State Compute the local field on site $i$: $\Vec{\mu}_i\leftarrow \sum_{j} J_{ij}\Vec{S}_j+\Vec{b}_i$
\State
\State Align spin $i$ to local field: $\Vec{S}_i\leftarrow \frac{\Vec{\mu}_i}{|\Vec{\mu}_i|}$
\State
\State Return $\Vec{S}_i$.
\end{algorithmic}
\end{algorithm}

We compute the local field acting on spin $i$ at time $t-1$, we align $\Vec{S}_i(t)$ to it and then we select another spin and repeat the procedure. Every time $N$ spins have been update, we count a sweep time and measure the overlap of the global configuration with the configuration at the previous sweep time: 
\begin{equation}
\label{eq:GD_check_overlap_at_sweep}
    q(t_{sw})\,=\,\frac{1}{N}{\bf S}(t_{sw})\cdot {\bf S}(t_{sw}-1)
\end{equation}
If this quantity is closer to unity than a user input threshold $\epsilon$, the algorithm reaches convergence. The full algorithm is reported in code \ref{alg:gcd_algo}: the algorithmic complexity to complete a sweep is clearly $O(N^2)$, since for any spin we have to compute the instantaneous local field. 
\begin{algorithm}
\caption{GCD algorithm.}
\label{alg:gcd_algo}
    \begin{algorithmic}[1]
        \While{$1-q(t_{sw})\geq \epsilon$}
            \State
            \State Save previous spin configuration ${\bm S}_{old}(t_{sw})\leftarrow {\bm S}(t_{sw}-1)$
            \State
            \For{$1\leq i \leq N$}
                \State Update Spin $i$ with code \ref{alg:single_upd}.
            \EndFor
            \State
            \State Compute overlap with previous configuration: $q(t_{sw})\leftarrow \frac{1}{N}{\bf S}(t_{sw}){\bf S}_{old}(t_{sw})$.
            \State
        \EndWhile
    \end{algorithmic}
\end{algorithm}
The total algorithmic time complexity at reached convergence is $O(N^2 t_{conv}(N, H))$, where $t_{conv}(N, H)$ is the typical convergence time at size $N$ and field width $H$: we will measure later in this section the scaling with $N$ of $t_{conv}(N, H)$ in the Heisenberg case ($m=3$).

How do we choose the next spin after an update? There are two sensible options: we can either select at each time step a spin at random, or update each spin sequentially with its index $i=1,\dots, N$. The first case corresponds to a $T=0$ Monte-Carlo algorithm, the second is a deterministic gradient descent protocol. The first choice allows to study different trajectories and stationary points obtained by starting the descent dynamics from a fixed initial condition, whereas the second approach yields a deterministic descent dynamics, such that for each initial condition there is a unique trajectory and final stationary point. We chose to adhere to the second protocol, since our goal in this chapter is only to study excitation spectra of energy minima.

An important question we did not raised so far is that of initialisation: what is the most convenient initialisation?
A random initialisation, obtained by drawing uniformly\footnote{We can obtain these by extracting each spin component from a gaussian and then to normalise to unity each spin.} a configuration on the manifold $S_m(1)^N$, is equivalent to a quench from infinite temperature. A local energy minimum reached after performing a gradient descent from a quenched initial condition is called in gergon \emph{inherent structure}. In phases that exhibit RSB, inherent structures related to infinite-temperature quenches usually are not very good in energy. 
In models of structural glasses, such as p-spin models, under the right conditions\footnote{In the case of generalised glasses, to consider states that can be followed in temperature.} one can improve the quality of inherent structures by considering a quench from states that are of equilibrium for some preparation temperature in the dynamical phase \cite{folena2020mixed}. This technique is not helpful in fRSB models, because of temperature chaos and level crossing phenomena. In this case, a standard approach to numerical simulations is given by Monte-Carlo simulated tempering \cite{marinari1992simulated}.

 Since we focused mainly on the paramagnetic phase, we chose to use a random initial condition for our minimisation. Even though in finite size systems one does encounter structured landscapes with multiple energy minima close to criticality, one can still modify the GCD algorithm \eqref{eq:GD_update_move} in order to avoid bad high energy minima, even with a random initial condition. 

\subsubsection{Over-Relaxation}

In finite size systems one can have more than an energy minimum also in the RS phase, up to very large sizes in the vicinity of a critical point. Since in numerical simulations one always deal with finite systems, it is crucial to have a minimisation algorithm able to avoid high-energy minima. With a random initial condition, unfortunately, when there are many local minima there is a high probability to fall in a high-energy minimum through a simple gradient descent algorithm like the GCD. Intuitively, this happens because as $H$ is decreased toward the critical field, the landscape loses convexity and new basins in the energy landscape start to form.

In order to deal with this unfortunate situation, we adopt the trick of \emph{Over-Relaxation} (OR). We modify equation \eqref{eq:GD_update_move} as follows
\begin{eqnarray}
\label{eq:GD_update_move_OR}
    \Vec{S}_i(t)\,&=&\,\frac{\sum_{j}J_{ij}\Vec{S}_j(t-1)+\Vec{b}_i+O\;\Vec{S}_i^{(R)}(t-1)}{|\sum_{j}J_{ij}\Vec{S}_j(t-1)+\Vec{b}_i+O\;\Vec{S}_i^{(R)}(t-1)|} \\
    && \nonumber \\
    \Vec{S}_i^{(R)}\,&=&\,\Vec{S}_i^{(\parallel)}-\Vec{S}_i^{(\perp)}\,=\,2[\Vec{S}_i\cdot\hat{\mu_i}]\hat{\mu_i}-\Vec{S}_i
\end{eqnarray}
The new update move consists in the linear combination of two vectors: the instantaneous local field and the OR term with weight $O>0$. The vector $\Vec{S}_i^{(R)}(t)$ is a reflection of the spin $\Vec{S}_i$ about the local field axis, and is weighted trough the over-relaxation parameter $O$. Since the OR vector has the same angle with the local field as the spin it refers, the OR term preserves the energy. In particular, the reflection move maximises the distance from the starting point in the constant energy manifold. 
The OR descent can be visualised as a spiralling descent in the landscape.
The bigger $O$, the more "spiralling" is the descent in the energy landscape.
The OR algorithm was used with success in previous studies of vector spin glass models, like for instance in \cite{baity2015soft}.

\subsubsection{Study of algorithmic performances in energy and sweep time}

In this section we analyse the performance in energy and convergence time of the OR algorithm, with threshold $\delta=10^{-12}$. We modify the single update move \ref{alg:single_upd} by replacing the true local field with the OR one \eqref{eq:GD_update_move_OR}
We restrict our study to the Heisenberg $m=3$ system and consider the five values $H/H_c=0.5,0.87,1.04,1.7,5.0$. These values, for the sizes $N=128, 256, 512, 1024$ considered by us, refer respectively to the case of a complex landscape ($H=0.5H_c$, deep spin glass phase), a weakly complex landscape ($H=0.87H_c$, not deep spin glass phase), the emergence of complexity ($H=1.04H_c$, close to critical point), the non-convex paramagnetic phase ($H=1.7H_c$) and the trivial paramagnetic phase ($H=5.0 H_c$). For each size and external field of this test, we simulated $N_s=100$ samples; for each sample, we performed $N_{run}=10$ runs for different values $O=0,\dots,10$, starting from a random initial condition: in total, we performed $N_s\times N_{run}\times 10$ GD-OR, yielding as much measures of $E$ and $t_{conv}$.

The main conclusions of this brief study are the following:
\begin{itemize}
    \item The OR algorithm performs largely better in convergence time than the simple gradient descent for $H\sim H_c$ and below. The improvement holds both for fixed value of $N$ and for growing $N$: for a given range of sizes, the OR parameter $O$ can be chosen in order to have computational complexity $CC=O(N^2)$, namely to have $t_{conv}$ independent of $N$. Conversely, the simple GD performs with $CC\gg O(N^2)$, $t_{conv}$ is always an increasing function of $N$.
    \item In the paramagnetic phase, the gain in energy of the OR is marginal if compared to that in convergence time. In the spin glass phase, the average energy gain of the OR algorithm is a slowly increasing function of the OR parameter: the greater its value, the deeper in the landscape.
\end{itemize}

In order to measure the energy gain with respect to the simple GD algorithm, $O=0$, we consider for each different run the following quantity
\begin{equation}
    \Delta(O)\,=\,\frac{E(0)-E(O)}{|E(0)|}.
\end{equation}
We consider then its average with respect to the different runs within each sample and finally the sample average of the run-averaged gains. We found that in the spin glass phase the fluctuations of $\Delta(O)$ between runs are comparable with those of $\langle \Delta(O)\rangle_{run}$ between different samples, whereas in the case $H=1.04 H_c$ run-to-run fluctuations are negligible. In the cases $H/H_c=1.7, 3.0$ we did not found any complex landscape. As to the convergence time, for all $H, N$ we found that run-to-run fluctuations are similar to sample-to-sample fluctuations, so we consider the average of $t_{conv}$ as a function of $O$ over $N_{s}\times N_{run}$ values.

\begin{figure}
    \centering
    \includegraphics[width=0.45\columnwidth]{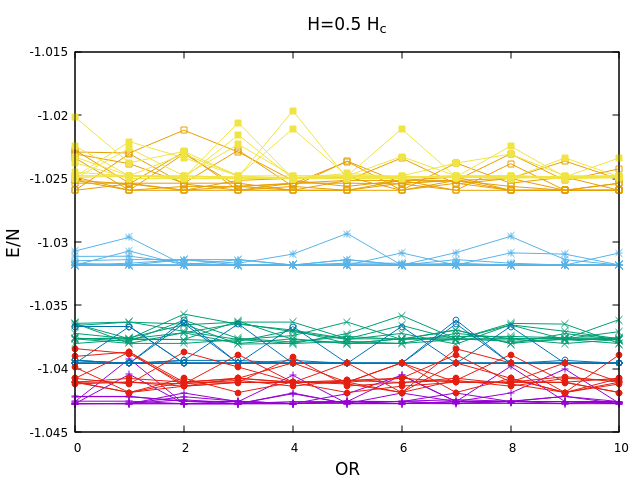}
    \includegraphics[width=0.45\columnwidth]{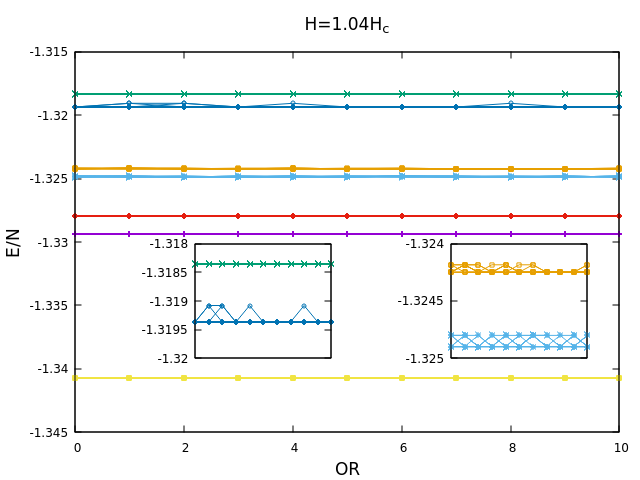}
    \includegraphics[width=0.45\columnwidth]{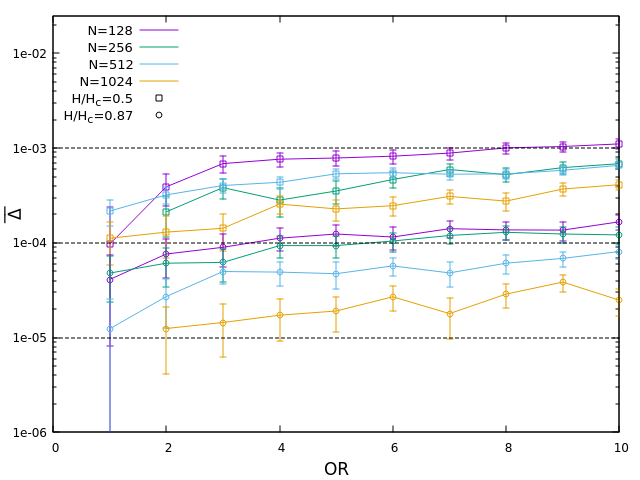}
    \includegraphics[width=0.45\columnwidth]{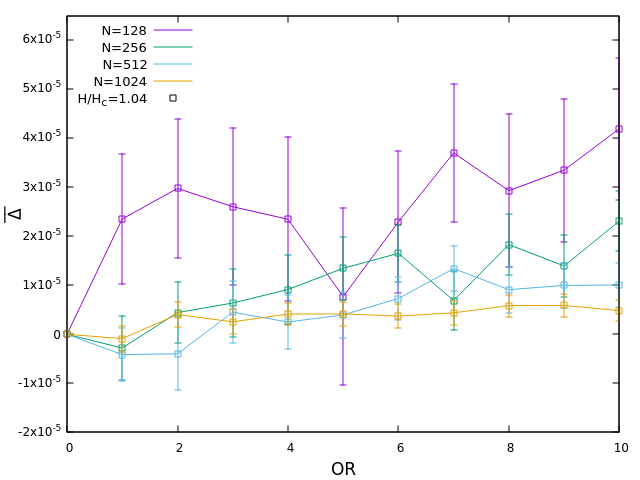}
    \caption{\textbf{Top}: comparisons between energies reached in different samples, for the largest size $N=1024$. Each color is a different sample, each curve is a sequence of runs from the same initial condition, where each point of the curve is the energy reached with a value of $O$ reported on the $x$-axis. The figure on the left shows fluctuations in the spin glass phase, on the right in the paramagnetic phase, close to the critical point. The inset is a zoom on specific samples that in the case $H=1.04H_c$ exhibit a multi-level structure.\newline
    \textbf{Bottom}: On the left, average and typical energy gains of the OR algorithm, for $H/H_c=0.5, 0.87$ and all sizes simulated. On the right, same for $H/H_c=1.04$.
    }
    \label{fig:energies_testopt}
\end{figure}

Let us begin by showing a comparison between sample-to-sample fluctuations of the energies reached and fluctuations within samples, or to be more precise between levels of the same energy landscape. In figures \ref{fig:energies_testopt} we show how they looks in the spin glass phase ($H=0.5 H_c$, left figure) and close to criticality ($H=1.04 H_c$, right figure), for the largest size $N=1024$ and some samples. While in the spin glass phase the difference is less marked, close to criticality sample-to-sample fluctuations are dominant. In the inset of the right figure in \ref{fig:energies_testopt} we highlight the presence of non-trivial landscapes, by zooming on samples with a two-levels structure.

On the bottom of the same figure, we show the sample-averaged relative energy gain of the OR algorithm for $H/H_c=0.5, 0.87$ (left) and for $H/H_c=1.04$ (right), considering all sizes simulated. The relative energy gain ranges between $10^{-4}$ and $10^{-3}$ for $H/H_c=0.5$, between $10^{-5}$ and $10^{-4}$ for $H/H_c=0.87$ and between $10^{-6}$ and $10^{-5}$ for $H/H_c=1.04$. The energy gain of the OR algorithm is the better the deeper in the spin glass phase. Moreover, for the values $H<H_c$ the gain appears to decrease with increasing size: the curves $\Delta(O)$ versus $O$ seem to approach the $x$-axis. This observation is consistent with the theoretical prediction that in the spin glass phase levels of the energy density are typically split by $O(1/N)$.

\begin{figure}
    \centering
    \includegraphics[width=0.8\columnwidth]{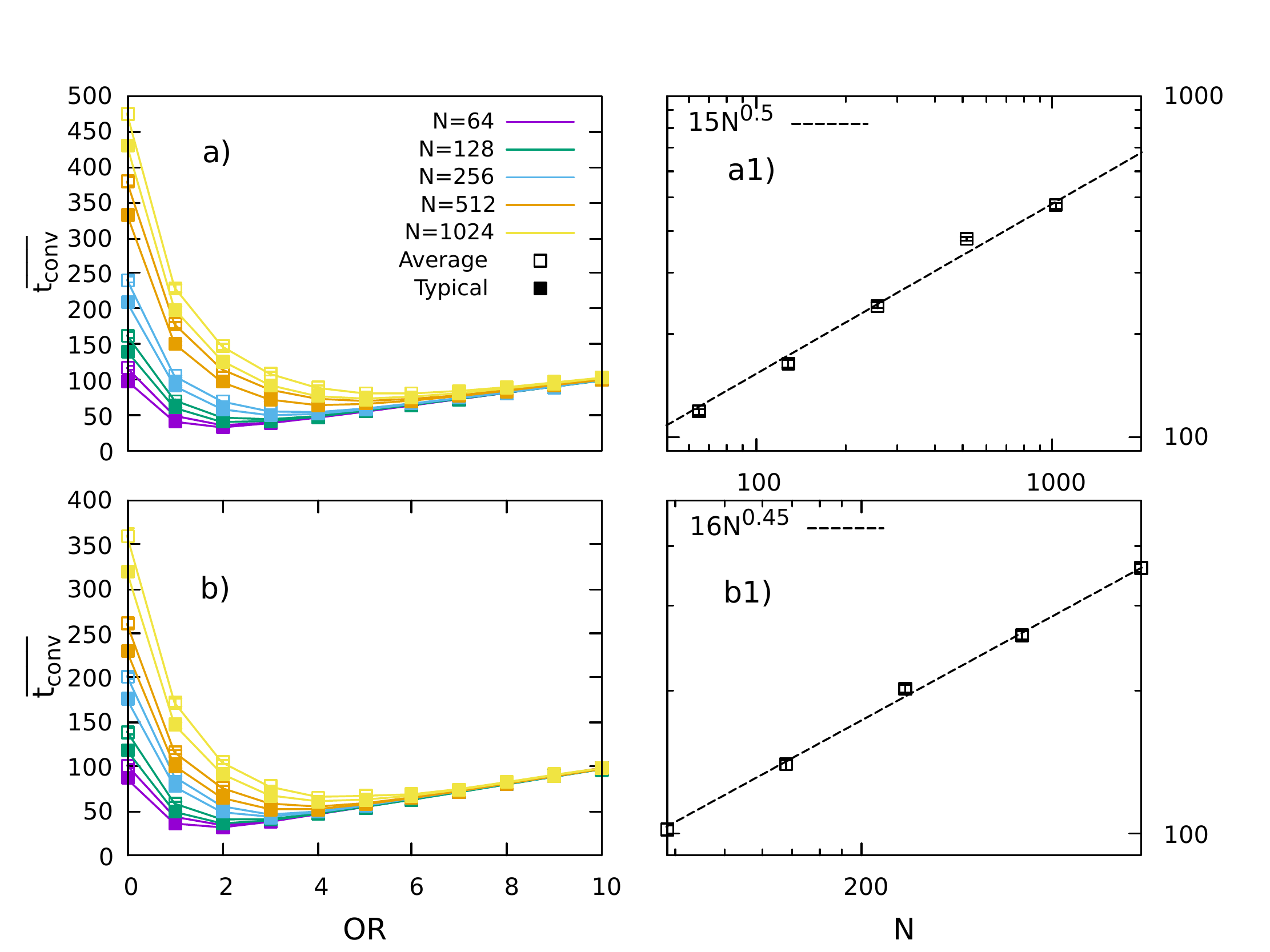}
    \includegraphics[width=0.8\columnwidth]{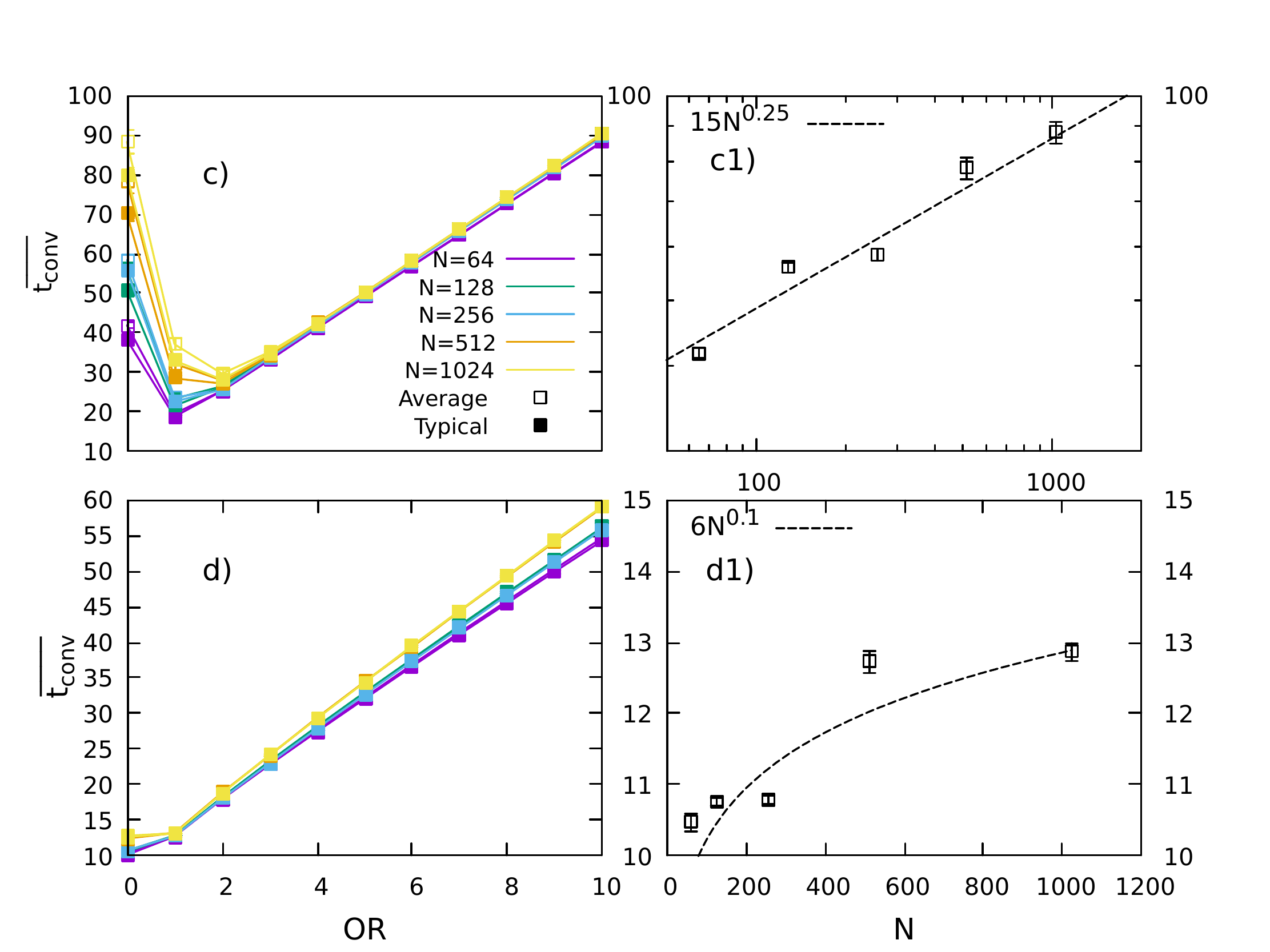}
    \caption{Measures of the averaged convergence time. Figures a-b): measures for $H=0.5 H_c$ (a) and $H=1.04 H_c$ (b). Figures a1-b1): finite size scaling of the convergence time of the GCD algorithm ($O=0$). Figures c-d): measures for $H=1.7 H_c$ (c) and $H=3.00 H_c$ (d). Figures c1-d1): finite size scaling of the convergence time of the GCD algorithm ($O=0$)}
    \label{fig:t_conv}
\end{figure}

Our measures of convergence time show that also for apparently large values of $H$, like $H/H_c=1.7$, the GD-OR algorithm offers a notable improvement in time performance, when compared to the simple GCD algorithm. We show this in figures \ref{fig:t_conv}, where we show the sample-run averaged convergence time as a function of the OR weight $O$ for the values $H/H_c=0.5, 1.04, 1.7, 5.0$. It appears that for the range of sizes of our simulations, the convergence time can be minimised by choosing $0<O<10$, only in the trivial case $H/H_c=5.0$ the GCD algorithm outperforms GD-OR. We also studied the finite size scaling of the averaged convergence time of the GCD algorithm ($O=0$), as represented in figures a1,b1,c1,d1 in \ref{fig:t_conv}. The average convergence time of the GCD algorithm grows with size
\begin{equation*}
    t_{conv}(N, H, O=0)\sim N^a.
\end{equation*}
The exponent $a$ is very small for large $H$ and seems to be roughly $a\approx 0.5$ in the spin glass phase: we provide its values for the values of $H$ studied in table \ref{tab:a_exponents}. Therefore, the time complexity of the GCD algorithm is roughly $O(N^{2.5})$ for $H<H_c$ and tends to $O(N^2)$ for large $H$. As to the GD-OR algorithm ($O>0$), note that for any $H$ one can choose $O$ large enough in order to have $a=0$, at least for the range of sizes we simulated, which are those typically accessible in numerical simulations of dense systems. Thus, over-relaxation brings two time advantages: on one side, it sensitively reduces the convergence time for any given $N$ and $H$ not too large; on the other side, it weakens the dependence of the convergence time on the size of the system, reducing the computational time complexity of the minimisation algorithm.

\begin{table}[]
    \centering
    \begin{tabular}{c|c}
     $H/H_c$ & $a$  \\
     \hline
     $0.50$ & $0.50$ \\
     $0.87$ & $0.50$ \\
     $1.04$ & $0.45$ \\
     $1.70$ & $0.25$ \\
     $3.00$ & $0.10$  
\end{tabular}
    \caption{The size-scaling exponent of the converge time in absence of over-relaxation, $t_{conv}\sim N^a$, as a function of $H$. In the spin glass phase it seems to be independent of $H$ and equal to $a=0.5$, and to decrease as a function of $H$ in the paramagnetic phase.}
    \label{tab:a_exponents}
\end{table}

The GD-OR algorithm impacts more significantly time performances than energy ones in the non-trivial paramagnetic phase: our data suggest that up to the critical point the algorithm ensures for $O$ large enough to go deep in the energy landscape, if not directly into the ground state. In the spin glass phase, for sufficiently large sizes this is clearly impossible. Nevertheless, the GD-OR algorithm is far better than the naive GCD, yielding deeper energy minima the larger the over-relaxation.

\subsection{The Hessian}

Having discussed how to compute numerically configurations of minimum of the energy function, it is time to discuss how to compute the spectrum of the Hessian matrix and the related eigenvectors.

The Hessian matrix we wish to diagonalise is the zero temperature limit of the transverse Hessian \eqref{eq:TAP_Hessian}, namely
\begin{eqnarray}
\label{eq:Hessian_energy}
    && \mathbb{M}_{ij}[{\bf S}]\,=\,\mathbb{P}^{(\perp)}(\Vec{S}_i)(-J_{ij}+|\Vec{\mu}_i[{\bf S}]|\delta_{ij})\mathbb{P}^{(\perp)}(\Vec{S}_j) \\
    && \nonumber \\
    &&
\label{eq:Hessian_energy_fields}
    \Vec{\mu}_i[{\bf S}]\,=\,\sum_{j}\;J_{ij}\Vec{S}_j+\Vec{b}_i \\
    && \nonumber \\
    && \mathbb{P}^{(\perp)}(\Vec{S})\,=\,\mathbb{I}-\Vec{S}\Vec{S}^{\intercal}
\end{eqnarray}
The Hessian, when evaluated in a configuration of minimum, is positive definite. In order to achieve that, correlations among the entries must exist: the form of these correlations in the general case is arbitrarily complicated, making a description in terms of RMT hard. In the specific case of Hessians of disordered mean field systems, correlations among the entries are weak. Indeed, in all theories described by an Hamiltonian which is linear in its disorder parameter, diagonal entries are correlated with the off-diagonal ones through functions of linear combinations of them, such as the $\{\mu_i\}$ in \eqref{eq:Hessian_energy_fields}. In order to ensure the existence of the thermodynamic limit, couplings must scale as $N^{-(p-1)/2}$, being $p$ the degree of interaction, and as a direct consequence the diagonal elements, the local forces, are $O(1)$. When $N$ grows, the correlation of any $\mu_i$ with single values of the couplings becomes increasingly weaker. So, correlations among interactions (off-diagonal entries) and fields (diagonal entries) must vanish in the thermodynamic limit). As to the correlations between off-diagonal entries with each other, in this theory for finite $N$ there are correlations between entries $(i,j), (k, l)$, with $i\neq j\neq k\neq l$, enforced by the orthogonal constraint on the spins. These correlations vanish because in a fully connected system spins orientations tend to be random for $N\rightarrow\infty$. Similarly, diagonal entries become uncorrelated.
Therefore, for $N\rightarrow\infty$, the Hessian matrix \eqref{eq:Hessian_energy} is an instance of a Rosenzweig-Porter or Deformed Wigner ensemble
\begin{eqnarray}
\label{eq:Hessian_as_RWPR}
    && \mathbb{M}\sim \text{RP}_N(\mathbb{W},\operatorname{blockdiag}(\{{\mathbb V}\})) \\
\label{eq:off-diag}
    && \mathbb{W}_{ij}\,=\,-J_{ij}\mathbb{P}_i^{(\perp)}\mathbb{P}_j^{(\perp)}\qquad \mathbb{W}\sim \text{Wig}(N) \\
\label{eq:diag}
    && \mathbb{V}_i\,=\,\mu_i\mathbb{P}_i^{(\perp)}\qquad \mu_i-\chi_0\sim \text{Chi}_m(1/m+H^2) 
\end{eqnarray}
where $\text{Chi}_m(\sigma^2)$ is the chi-distribution with $m$ degrees of freedom and scale $\sigma$
\begin{equation}
\label{eq:cav_field_pdf}
   \text{Chi}_m(x;\sigma^2)\,=\,\frac{x^{m-1}e^{-x^2/2\sigma^2}}{2^{m/2-1}\Gamma(m/2)\sigma^m} 
\end{equation}
that it is the distribution of the norm of a gaussian vector with $m$ components. Recall that thanks to cavity method in the thermodynamic limit the cavity fields
\begin{equation*}
    \Vec{h}_i\,\overset{N\rightarrow\infty}{\,=\,}\,\sum_j\;J_{ij}\Vec{S}_j+\Vec{b}_i-\chi_0\Vec{S}_i
\end{equation*}
are gaussian vectors with zero mean and covariance matrix $(1/m+H^2)\mathbb{I}$. Notice that since in a configuration of minimum spins and local fields are aligned, cavity fields and spins are also aligned, thus one can also sum their norms $\mu_i\,=\,h_i+\chi_0$.
In the infinite volume limit, Hessians of disordered systems can be represented by RM ensembles. However, for finite sizes there is a finite probability of measuring a negative eigenvalue in the spectrum of a RM drawn from \eqref{eq:Hessian_as_RWPR}. We shall write equations for the resolvent function that give us the infinite size theory.

\subsubsection{Derivation of resolvent equations through the cavity method}

We learnt in section \ref{sec:Perturbation_Wigner_Matrices} that the equation for the resolvent or Green function of the sum of two free matrices yield by the R-transform is \eqref{eq:resolvent_self_const_free}, which we rewrite for reader convenience
\begin{equation}
    g(z)\,=\,\mathcal{G}_A(z-\mathcal{R}_B(g(z)))\equiv \int d\lambda\;\frac{\rho_A(\lambda)}{z-\mathcal{R}_B(g(z))-\lambda}
\end{equation}
This equation adapted for the resolvent of \eqref{eq:Hessian_as_RWPR} reads (we change convention on the resolvent and consider $\mathbb{G}(z)\,=\,(\mathbb{M}-z\mathbb{I})^{-1}$)
\begin{equation}
\label{eq:resolvent_eq_Hessian_SG}
    \mathcal{G}(z)\,=\,\left(1-\frac{1}{m}\right)\int_0^{\infty} dh\;\frac{P_{h}(h)}{h+\chi_0-z-\mathcal{G}(z)}
\end{equation}
where the prefactor comes from the projectors $\mathbb{P}_i$ in \eqref{eq:diag}. Indeed, since projectors are idempotent, all moments of the random variables $\mu_i\mathbb{P}_i$ are just rescaled of a factor $\Tr \mathbb{P}/m\,=\,(m-1)/m$. Notice that \eqref{eq:resolvent_eq_Hessian_SG} implies
\begin{equation}
\label{eq:chi0_is_G0}
    \mathcal{G}(0)\,=\,\left(1-\frac{1}{m}\right)\int_0^{\infty} dh\;\frac{P_{h}(h)}{h}
\end{equation}
so we identify $G(0)\,=\,\chi_0$. This is just a consistency check, since we know by definition that the susceptibility tensor is the inverse of the Hessian, or equivalently, the resolvent matrix \eqref{eq:Resolvent_def} in $z=0$. 

We want to provide a different derivation of \eqref{eq:resolvent_eq_Hessian_SG}, using the cavity method. To this purpose, we write the eigenvalue equations for the eigenpairs $\{\lambda,\boldsymbol{\psi}(\lambda)\}$ in presence of a small external 
source $\epsilon_i^{\alpha}$ (we consider the form \eqref{eq:Hessian_energy} of the Hessian)
\begin{eqnarray}
  \label{eq:eigenvalue_equation}
-\sum_j J_{ij} (\mathbb{P}_i^{(\perp)}\Vec{\psi}_j)^\alpha +\mu_i \psi_i^\alpha-\lambda \psi_i^\alpha=(\mathbb{P}_i^{(\perp)}\epsilon_i)^{\alpha}\;,
\end{eqnarray}
where the index $\alpha$ runs over the $m$ components, the eigenvector with $mN$ components can be written as $\psi(\lambda)=(\Vec{\psi}_1,\ldots,\Vec{\psi}_N)$ and each $\Vec{\psi}_i$ is orthogonal to the corresponding $\Vec{S}_i$, i.e.\ $\sum_{\alpha=1}^m \psi_i^\alpha S_i^\alpha=0$. 
Since we are considering a perturbation of the inverse of the resolvent in \eqref{eq:eigenvalue_equation}, a small imaginary part in $\lambda$ is implicitly assumed to insure invertibility. 

We single out a site $i$ and compare the solution of the full system (\ref{eq:eigenvalue_equation}) to the one where the site $i$ is removed.
Defining $\Vec{\psi}_j^{(i)}$ the solution of Eq.~(\ref{eq:eigenvalue_equation}) in absence of spin $i$ and assuming continuity, we can write
\begin{eqnarray}
  \label{eq:eigenvalue_equation_cavity}
  -\sum_j J_{ij} \mathbb{P}_i^{(\perp)}\Vec{\psi}_j^{(i)} -\mathcal{G}(\lambda) \Vec{\psi}_i +\mu_i \Vec{\psi}_i -\lambda \Vec{\psi}_i = \mathbb{P}_i^{(\perp)}\Vec{\epsilon}_i\;,
 \end{eqnarray}
where 
\begin{equation*}
    \mathcal{G}(\lambda)\,=\,\frac{1}{N m}\Tr (\mathbb{M}-\lambda\mathbb{I})^{-1}
\end{equation*}
is the Green function of our Hessian. The continuity assumption, necessary to use cavity method, does not work in the spin glass phase because of marginal stability. Knowing that resolvent entries are eigenvector susceptibilities $\mathbb{G}_{ij}(\lambda)\equiv d\Vec{\psi}_i(\lambda)/d\Vec{e}_j$, by deriving \eqref{eq:eigenvalue_equation_cavity} with respect to $\Vec{\epsilon}_i$
and tracing over the $m$ components, we get an equation for the local resolvent\footnote{The $\Vec{\psi}_j^{(i)}$ do not depend on the small perturbation by construction.}
\begin{eqnarray}
  \label{eq:resolvent_local_self_consistent}
  G_{ii}(\lambda)=(1-1/m)\left(h_i+\mathcal{G}_0-\lambda -\mathcal{G}(\lambda)\right)^{-1}\;,
\end{eqnarray}
from which we get the spectral density by the usual limit
\begin{equation*}
\rho(\lambda)=\lim_{\eta\to 0} \frac {m}{\pi (m-1)} \Im(\mathcal{G}(\lambda+i\eta))\;.
\end{equation*}
The prefactor $m/(m-1)$ takes into account that fluctuations are restricted to the directions orthogonal to the spins.
Finally, by averaging over $i$, we arrive to \eqref{eq:resolvent_eq_Hessian_SG}, the self-consistent equation for $\mathcal{G}(\lambda)$. Equations of the form of \eqref{eq:resolvent_eq_Hessian_SG}, \eqref{eq:resolvent_local_self_consistent} were already derived in \cite{bray1981metastable, bray1982eigenvalue, bray1982spin}: in the last paper in particular, the authors show a connection between the asymptotic decay of the spin autocorrelation function and the abundance of eigenvalues close to $\lambda=0$. Under the hypothesis that temperature is sufficiently small in order to exclude barrier hopping and that the initial perturbation is sufficiently small so that it does not trigger a change of state, they consider the linearised dynamical equations with a white noise term to model the thermal bath
\begin{equation*}
    \frac{d\Vec{S}_i}{dt}\,=\,\mathbb{P}_{i}^{(\perp)}\Vec{\mu}_i+2k_BT\eta
\end{equation*}
that describe the relaxation dynamics at low temperatures and in a range of perturbations where linear response theory holds.
In this context, if $\rho(\lambda)\sim\lambda^x$, then $C(t)\sim C(\infty)-\frac{c}{t^x}$, according to
\begin{equation*}
    C(t)\,=\,1-(m-1)k_B T\int_0^{\infty}\frac{d\lambda}{\lambda}\rho(\lambda)[1-\exp(-\lambda t)].
\end{equation*}
The smallest eigenvalue refer to the longest time-scales of relaxation dynamics, if linear response theory is valid. In the thermodynamic limit, we know that linear response theory should fail in the spin glass phase of fRSB because of marginality.

If one knows the cavity field distribution $P_m(h)$, Eq.~(\ref{eq:resolvent_eq_Hessian_SG}) can be solved numerically and analysed analytically for small $\lambda$. 
giving us access the spectral density, once we separate the real, $\Re \mathcal{G}(\lambda)$, and imaginary, $\Im \mathcal{G}(\lambda)$, parts of $\mathcal{G}(\lambda)$.

\subsubsection{Numerical simulations}

The numerical diagonalisation of the Hessian has been carried out through Lapack libraries package \url{https://netlib.org/lapack/}. We used the \emph{dsyev} subroutine which returns the complete set of eigenvalues and eigenvectors of a symmetric matrix. This is achieved by reducing the symmetric matrix in a tridiagonal form and then using the Pal-Walker-Kahan variant of the QR algorithm \cite{francis1961qr, francis1962qr}, an improvement of the original LR transformation \cite{rutishauser1958solution}, to compute eigenvalues. The QR algorithm works as follows: the original matrix $\mathbb{A}$, given in input is associated a sequence of matrices $\{\mathbb{A}_k\}$ satisfying
\begin{eqnarray*}
    \mathbb{A}_{1}\,&&=\,\mathbb{A} \\
    \mathbb{A}_{k}\,&&=\,\mathbb{Q}_k\mathbb{R}_k \\
    \mathbb{A}_{k+1}\,&&=\,\mathbb{R}_k\mathbb{Q}_k\qquad k\geq 1
\end{eqnarray*}
where $\{\mathbb{Q}_k\}$ are orthogonal matrices and $\{\mathbb{R}\}_k$ are upper triangular matrices. In the limit $k\rightarrow\infty$ the sequence converges to the Schur form of the original matrix, which is a tridiagonal matrix having the eigenvalues on the diagonal. The computational complexity of this algorithm is $O(N^3)$, with $N$ being the size of the square matrix. Since in addition a dense matrix occupies $O(N^2)$ of memory, in all practical applications Lapack can be used efficiently with dense matrices of size $N=O(10^3)$. On a machines cluster one can also deal with $N=O(10^4)$, but the computational complexity, to which computation time is proportional, in this situation is very demanding being $O(10^{12})$.

We performed numerical simulations combining the OR version of code \ref{alg:gcd_algo} and the diagonalisation of the Hessian evaluated on the computed spin configuration. The Hessian was given in the following form
\begin{equation}
\label{eq:Hessian_energy_form2}
    M_{ij}^{ab}\,=\,(\hat{e}_i^{a}\cdot \hat{e}_j^{b})(-J_{ij}+\mu_i\delta_{ij})
\end{equation}
where $\{\hat{e}_i^{a}\}_a$, $a=1,\dots,m-1$, is a orthonormal basis of the space orthogonal to spin $\Vec{S}_i$. In this form the Hessian is full rank, that is $N\times (m-1)$: this choice reduces the RAM requirement for the Hessian. If the resulting spectrum was strictly positive, we saved eigenvalues and the IPR of the related eigenvectors. In some cases, we saved the full eigenvectors and higher order eigenvectors moments \eqref{eq:eigvec_moments}.

We chose external scales $H$ in the paramagnetic phase ranging in $H\,=\,H_c,\ldots,10 H_c$, and a few simulations in the spin glass phase, considering $H\,=\,0.87 H_c$ and $H\,=\,0.5 H_c$. For each value of $H$, we measured system sizes $N=2^k\times 64$, with $k=0,\dots, 5$, with a number of samples per size such that $N\;N_s\geq 3\times 10^6$. We report in \ref{tab:data_1}, \ref{tab:data_2} two detailed tables with the parameters of our simulations, reporting also the average convergence time and energy density. Data reported are for the $m=3$ system.

\begin{table}[]
    \centering
    \begin{tabular}{c|c|c|c|c|c|c|c}
        \hline
        $H/H_c$ & $N$ & $N_s$ & $OR$ & $\overline{t_{conv}}$ & $\overline{E}/N$ & Theory ($N=\infty$) \\
        \hline
        10.4 & 64 & $4\times 10^4$ & 0 & $5.425(2)$ & $-9.617(3)$ & $-9.61884$ \\
        10.4 & 128 & $2\times 10^4$ & 0 & $5.578(4)$ & $-9.626(3)$ & "" \\
        10.4 & 256 & $10^4$ & 0 & $5.758(6)$ & $-9.621(3)$ & "" \\
        10.4 & 512 & $5\times 10^3$ & 0 & $5.92(1)$ & $-9.620(2)$ & "" \\
        10.4 & 1024 & $2\times 10^3$ & 0 & $6.06(2)$ & $-9.627(4)$ & "" \\
        10.4 & 2048 & $10^3$ & 0 & $6.18(4)$ & $-9.621(2)$ & "" \\
        \hline 
        5.0 & 64 &$4\times 10^4$ & 0 & $10.05(1)$ & $-4.712(3)$ & $-4.718550$ \\
        5.0 & 128  & $2.5\times 10^4$ & 0 & $11.09(2)$ & $-4.718(1)$ & "" \\
        5.0 & 256  & $2.0\times 10^4$ & 0 & $11.83(2)$ & $-4.717(1)$ & "" \\
        5.0 & 512  & $10^4$ & 0 & $12.61(4)$ & $-4.717(1)$ & "" \\
        5.0 & 1024  & $2\times 10^3$ & 0 & $13.46(7)$ & $-4.720(1)$ & "" \\
        5.0 & 2048  & $10^3$ & 0 & $14.5(1)$ & $-4.719(1)$ & "" \\
        \hline
        3.1 & 64 & $4\times 10^4$ & 1 & $7.21(1)$ & $-3.0110(5)$ & $-3.016524$ \\
        3.1 & 128 & $2.5\times 10^4$ & 1 & $8.44(4)$ & $-3.0124(2)$ & "" \\
        3.1 & 256 & $2\times 10^4$ & 1 & $8.91(1)$ & $-3.0145(4)$ & "" \\
        3.1 & 512 & $10^4$ & 1 & $9.06(2)$ & $-3.0152(7)$ & "" \\
        3.1 & 1024 & $2\times 10^3$ & 1 & $9.39(2)$ & $-3.0159(8)$ & "" \\
        3.1 & 1500 & $1.2\times 10^3$ & 1 & $9.40(2)$ & $-3.016(1)$ & "" \\
        \hline
        2.08 & 64 & $4\times 10^4$ & 2 & $14.6(3)$ & $-2.1215(5)$ & $-2.125031$ \\
        2.08 & 128 & $2.5\times 10^4$ & 2 & $15.02(3)$ & $-2.1227(5)$ & "" \\
        2.08 & 256 & $2.0\times 10^4$ & 2 & $15.81(2)$ & $-2.1240(3)$ & "" \\
        2.08 & 512 & $10^4$ & 2 & $16.40(2)$ & $-2.1240(4)$ & "" \\
        2.08 & 1024 & $2\times 10^3$ & 2 & $16.82(4)$ & $-2.1246(5)$ & "" \\
        2.08 & 2048 & $10^3$ & 2 & $16.98(4)$ & $-2.1251(4)$ & "" \\
        \hline
        1.73 & 64 & $4\times 10^4$ & 2 & $19.2(3)$ & $-1.8466(4)$ & $-1.842636$ \\
        1.73 & 128 & $2\times 10^4$ & 2 & $19.6(2)$ & $-1.8446(6)$ & "" \\
        1.73 & 256 & $10^4$ & 2 & $20.1(4)$ & $-1.8442(2)$ & "" \\
        1.73 & 512 & $5\times 10^3$ & 2 & $20.5(2)$ & $-1.8439(2)$ & "" \\
        1.73 & 1024 & $10^3$ & 3 & $20.98(6)$ & $-1.8436(6)$ & "" \\
        1.73 & 2048 & $7.5\times 10^2$ & 2 & $18.9(3)$ & $-1.8431(5)$ & "" \\
        \hline
        1.47 & 64 & $4\times 10^4$ & 3 & $19.04(1)$ & $-1.6320(5)$ & $-1.639713$ \\
        1.47 & 128 & $2\times 10^4$ & 3 & $19.18(2)$ & $-1.6361(4)$ & "" \\
        1.47 & 256 & $2\times 10^4$ & 3 & $19.9(1)$ & $-1.6380(5)$ & ""  \\
        1.47 & 512 & $10^4$ & 3 & $23.3(2)$ & $-1.6391(3)$ & ""  \\
        1.47 & 1024 & $3\times 10^3$ & 3 & $26.0(3)$ & $-1.6397(3)$ & "" \\
        1.47 & 1500 & $7\times 10^2$ & 3 & $28.0(5)$ & $-1.6390(6)$ & "" \\
        \hline
    \end{tabular}
    \caption{Table with the parameters of our simulations and average energies and convergence times.}
    \label{tab:data_1}
\end{table}

\begin{table}[]
    \centering
    \begin{tabular}{c|c|c|c|c|c|c|c}
        $H/H_c$ & $N$ & $N_s$ & $O$ & $\overline{t_{conv}}$ & $\overline{E}/N$ & Theory ($N=\infty$)  \\
        \hline
        1.30 & 64 & $4\times 10^4$ & 3 & $19.77(2)$ & $-1.5007(4)$ & $-1.510371$ \\
        1.30 & 128 & $2\times 10^4$ & 3 & $20.07(2)$ & $-1.5059(3)$ & "" \\
        1.30 & 256 & $6\times 10^3$ & 3 & $25.8(2)$ & $-1.5079(4)$ & "" \\
        1.30 & 512 & $6\times 10^3$ & 3 & $30.9(2)$ & $-1.5090(3)$ & "" \\
        1.30 & 1024 & $3\times 10^3$ & 3 & $36.7(4)$ & $-1.5092(3)$ & "" \\
        1.30 & 1500 & $7\times 10^2$ & 3 & $39.7(7)$ & $-1.5099(4)$ & "" \\
        \hline
        1.13 & 64 & $4\times 10^4$ & 3 & $18.1(4)$ & $-1.3810(2)$ & $-1.387340$ \\
        1.13 & 128 & $2\times 10^4$ & 3 & $21.4$ & $-1.3825(1)$ & "" \\
        1.13 & 256 & $10^4$ & 3 & $26.2(1)$ & $-1.3840(3)$ & "" \\
        1.13 & 512 & $3\times 10^3$ & 3 & $29.6(3)$ & $-1.3860(4)$ & "" \\
        1.13 & 1024 & $1.5\times 10^3$ & 3 & $34.2(4)$ & $-1.3870(3)$ & "" \\
        1.13 & 1500 & $7\times 10^2$ & 3 & $39.0(7)$ & $-1.3870(4)$ & "" \\
        \hline
        1.04 & 64 & $2\times 10^4$ & 3 & $21.1(7)$ & $-1.3211(4)$ & $-1.328743$ \\
        1.04 & 128 & $2\times 10^4$ & 3 & $24.4(5)$ & $-1.3225(3)$ & "" \\
        1.04 & 256 & $10^4$ & 3 & $28.3(2)$ & $-1.3249(3)$ & "" \\
        1.04 & 512 & $5\times 10^3$ & 3 & $33.8(3)$ & $-1.3269(3)$ & "" \\
        1.04 & 1024 & $1.5\times 10^3$ & 3 & $42.0(6)$ & $-1.3275(3)$ & "" \\
        1.04 & 1500 & $7\times 10^2$ & 3 & $50(1)$ & $-1.3278(4)$ & "" \\
        \hline
        1.0 & 64 & $4\times 10^4$ & 5 & $52.1(5)$ & $-1.2932(4)$ & $-1.302940$ \\
        1.0 & 128 & $2.5\times 10^4$ & 5 & $56.04(7)$ & $-1.2957(2)$ & "" \\
        1.0 & 256 & $2\times 10^4$ & 5 & $58.0(1)$ & $-1.2989(2)$ & "" \\
        1.0 & 512 & $10^4$ & 5 & $62.1(3)$ & $-1.3011(2)$ & "" \\
        1.0 & 1024 & $2.5\times 10^3$ & 5 & $73.8(9)$ & $-1.3023(2)$ & "" \\
        1.0 & 2048 & $1.5\times 10^3$ & 5 & $98(2)$ & $-1.3027(2)$ & "" \\
        \hline
        0.87 & 64 & $4\times 10^4$ & 5 & $54.3(2)$ & $-1.2054(2)$ & X \\
        0.87 & 128 & $2.5\times 10^4$ & 5 & $57.52(7)$ & $-1.2086(2)$ & "" \\
        0.87 & 256 & $2\times 10^4$ & 5 & $61.1(2)$ & $-1.2135(1)$ & "" \\
        0.87 & 512 & $10^4$ & 5 & $69.1(4)$ & $-1.2161(1)$ & "" \\
        0.87 & 1024 & $2.5\times 10^3$ & 5 & $88(1)$ & $-1.2172(2)$ & "" \\
        0.87 & 2048 & $1.5\times 10^3$ & 5 & $125(2)$ & $-1.2181(2)$ & "" \\
        \hline
        0.52 & 64 & $4\times 10^4$ & 10 & $95.2(1)$ & $-1.0121(2)$ & X \\
        0.52 & 128 & $2.5\times 10^4$ & 10 & $103.93(5)$ & $-1.0166(1)$ & "" \\
        0.52 & 256 & $2\times 10^4$ & 10 & $105.1(1)$ & $-1.0257(1)$ & "" \\
        0.52 & 512 & $10^4$ & 10 & $110.4(4)$ & $-1.0311(1)$ & "" \\
        0.52 & 1024 & $1.5\times 10^3$ & 10 & $124(1)$ & $-1.0346(7)$ & "" \\
        0.52 & 2048 & $5\times 10^2$ & 10 & $163(4)$ & $-1.0356(2)$ & "" \\
        \hline
    \end{tabular}
    \caption{Second table with the parameters of our simulations and average energies and convergence times.}
    \label{tab:data_2}
\end{table}

\subsection{The spectral density}
\label{sec:spectral_density}

The spectral density can be compute numerically by solving \eqref{eq:resolvent_eq_Hessian_SG}, written in terms of its real and imaginary part. Defining the quantity $\mathit{x}(\lambda)=\mathcal{G}(0)-\mathcal{G}(\lambda)-\lambda$ and calling its real and imaginary part $x'$ and $x''$ respectively, we rewrite equation \eqref{eq:resolvent_eq_Hessian_SG} as
\begin{eqnarray}
  \label{eq:resolvent_eq_Hessian_SG_re_and_im}
&&  x'(\lambda)=\mathcal{G}(0)-\lambda-\left(1-\frac{1}{m}\right)\int dh\; P_h(h) \frac {h+x'}{(h+x')^2+(x'')^2}\\
&&\nonumber \\
&&  1=\left(1-\frac{1}{m}\right)\int dh\; \frac {P_h(h)}{(h+x')^2+(x'')^2}
\end{eqnarray}
In \cite{sharma2016metastable}, by studying the spectrum of metastable energy minima of Heisenberg spin glasses, the authors claim that as the paramagnetic phase is entered a spectral gap appears. In this chapter we shall revisit this result: indeed, also in the paramagnetic phase, the lower band edge of the spectrum is $\lambda=0$. In order to see this, consider \eqref{eq:resolvent_eq_Hessian_SG} and set to zero the imaginary part of the Green function, $\Im \mathcal{G}(\lambda)\equiv -x''$. In the paramagnetic phase $H>H_c$ one can expand $x'$ close to $\lambda=0$ one has
\begin{equation}
    x'(\lambda)\,\simeq\,-\lambda-\diff{G}{\lambda}(0)\lambda\,=\,-\frac{\lambda}{\Lambda}\qquad \Lambda\,=\,1-\left(1-\frac{1}{m}\right)\left\langle\frac{1}{h^2}\right\rangle
\end{equation}
where $\Lambda$ is the zero temperature limit of the Replicon eigenvalue of the RS saddle point of the replica action, so $\Lambda>0$ if $H>H_c$. Since in addition $P_h(h)$ is gapless, the denominator of \eqref{eq:resolvent_eq_Hessian_SG} close to zero eigenvalue would be $(h-\lambda/\Lambda)$: if $x''=0$, the integral is divergent. So, it is necessary that $|x''|=\pi\rho(\lambda)\neq 0$ for any positive $\lambda$ if the equation defining the resolvent exists.

\begin{figure}
    \centering
    \includegraphics[width=0.8\columnwidth]{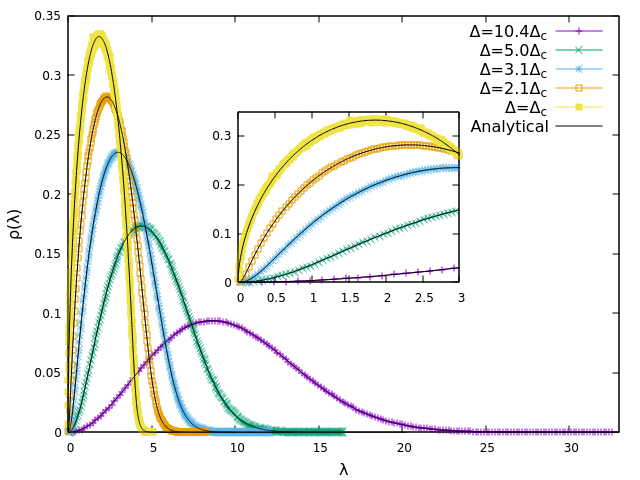}
    \includegraphics[width=0.8\columnwidth]{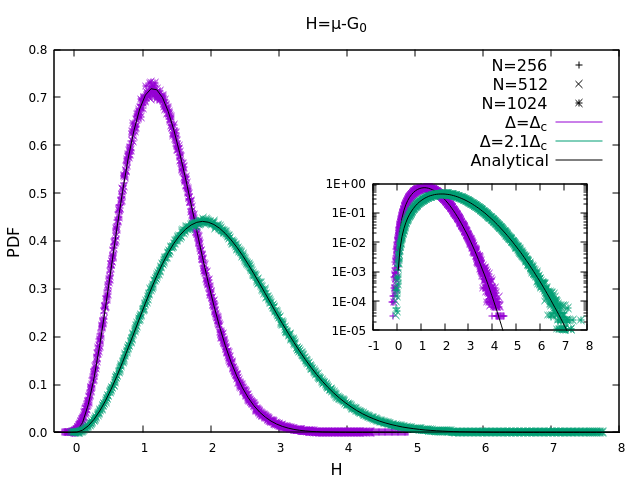}
    \caption{{\textbf{Top}}: The full spectrum $\rho(\lambda)$ for $N=1024$ and several $H$ values. The inset is a zoom on the lower edge. The pseudo-gap is clearly visible for large $H$ values. Approaching the critical point the pseudo-gap region shrinks and the curves progressively approach the critical density. \textbf{Bottom}: The distribution of cavity fields, estimated as $h=\mu-\chi_0$.}
    \label{fig:spec_density_0_and_cav_fields_pdf}
\end{figure}

Equations \eqref{eq:resolvent_eq_Hessian_SG_re_and_im} are closed in the variables $x'$ and $x''$, so we can solve them numerically. In \ref{fig:spec_density_0_and_cav_fields_pdf} we show some of the analytical curves we obtained for the spectral density and compare them with eigenvalues histograms from our numerical data for size $N=2^{10}=1024$. Thanks to the self-averaging property \eqref{eq:spectral_density_self-averaging}, we can compute the histogram by considering for fixed $N, H$ the eigenvalues of the different samples all together. We see from this figure a excellent agreement with the theoretical prediction in a wide range of values of $H$. In addition, we show in the same panel histograms of the cavity field norms for $H$ in the paramagnetic phase and $H_c$. We estimated cavity fields through the asymptotic cavity relation
\begin{equation}
\label{eq:cav}
    h_i\,=\,\mu_i-\chi_0
\end{equation}
As already discussed, the gap of the local fields norm $\mu$ is the zero $T$ susceptibility only at $N=\infty$. Then, it may happen in some simulations that with \eqref{eq:cav} the smallest estimated cavity fields are negative. This effect is stronger the smaller the size. We observed $\mu_i<\chi_0$ for reasonably large sizes only very close to criticality, where $\chi_0$ is higher. Clearly \eqref{eq:cav} brings systematic finite size effects, that will be analysed in section \ref{sec:finite_size_effects}. Having said this, we observe anyway that histograms shapes agree very well with the theoretical distribution \eqref{eq:cav_field_pdf}.

\subsubsection{Expansions close to the edges}

The asymptotic behavior of the spectral density in the vicinity of the band edges can be compute from equations \eqref{eq:resolvent_eq_Hessian_SG}. In Appendix \ref{sec:Expansion_rho_band_edges} we convey the full calculation. Here, we report the final. In the paramagnetic phase $H>H_c$
\begin{equation}
\label{eq:edges_expansion_spectral_density}
    \rho(\lambda)\sim\begin{cases}
        \frac{1}{\Lambda}P_h\left(\frac{\lambda}{\Lambda}\right)\simeq \frac{\lambda^{m-1}}{\Lambda^m Z_m}\qquad \lambda\rightarrow 0_+ \\
        \\
        P_h(\lambda)\,=\,\frac{\lambda^{m-1}}{Z_m}\exp\left(-\frac{\lambda^2}{2(1/m+H^2)}\right)\qquad\lambda\rightarrow\infty
    \end{cases}
\end{equation}
where again $\Lambda$ is the Replicon eigenvalue. We write its explicit expression as a function of the distance $w=H-H_c>0$ to the critical field
\begin{eqnarray}
\label{eq:Replicon_T0_explicit_expression}
    \Lambda\,=\,1-A\,=\,1-\left(1-\frac{1}{m}\right)\left\langle\frac{1}{h^2}\right\rangle\,=\,\frac{(2H_c+w)w}{(m-1)H_c^2+(2H_c+w)w}
\end{eqnarray}
As expected from mean field theory, $\Lambda\propto w$ close to the critical point and thus $\chi_{SG}\propto 1/\Lambda \propto w^{-1}$.
Close to the band edges, the spectral density follows the distribution of cavity fields moduli, i.e. the non-trivial part of the diagonal of the Hessian matrix. This suggests that edges eigenvalues are strongly correlated to edges cavity fields. This statement has been rigorously proven recently in \cite{lee2016extremal}: we will come back to the consequences of this correlation later at the end of section \ref{sec:Eigenvectors}.

\begin{figure}
    \centering
    \includegraphics[width=0.8\columnwidth]{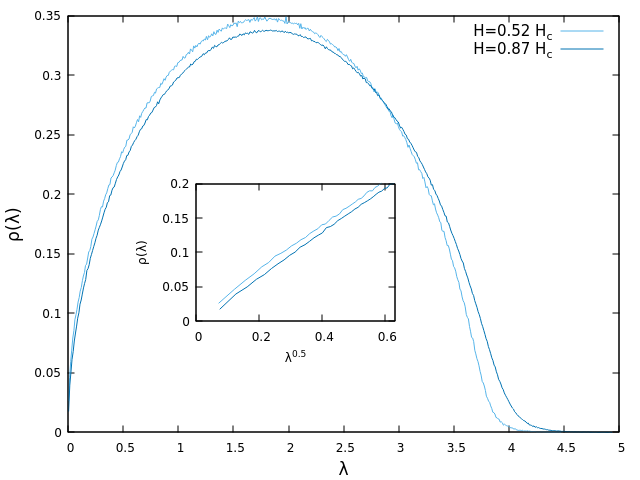}
    \caption{Spectral densities in the spin glass phase.}
    \label{fig:rho_lambda_SG}
\end{figure}

Notice that while the second expansion in \eqref{eq:edges_expansion_spectral_density} is true for any $H>0$, at the lower edge the expansion in \eqref{eq:edges_expansion_spectral_density} cannot hold at the critical point, where $\Lambda=0$. The computation of the expansions in Appendix \ref{sec:Expansion_rho_band_edges} holds on the hypothesis $|x''|\ll |x'|$ approaching the band edges. Whenever this hypothesis is verified, the related spectrum possesses localised excitation states \cite{potters2020first}. So, at the critical point localised states disappear: in this situation, one has that the inequality reverses and $|x''|\gg |x'|$. The spectral density at the critical point has the lower edge behavior (details in \ref{sec:Expansion_rho_band_edges})
\begin{equation}
\label{eq:low_edge_expansion_rho_critical}
    \rho(\lambda)\sim\begin{cases}
        \frac{1}{\pi}\sqrt{\frac{\lambda}{J}}\qquad m>3 \\
        \\ \nonumber
        \sqrt{\frac{Z_3}{\pi}}\sqrt{\frac{\lambda}{|\log\lambda|}}\qquad m=3
    \end{cases}
\end{equation}
the square root behavior typical of the edges of Wigner matrices. We claim that this scaling, pre-factor aside, holds as it is in the whole spin glass phase. Indeed, the square-root behavior was observed deep in the spin glass phase in \cite{sharma2016metastable}. Moreover, in \cite{bray1982spin} the scaling exponent of the autocorrelation is found to be $1/2$ for mean field vector spin glasses, for a low temperature spin glass with no external field. In addition, in \cite{bray1981metastable} it is found that the typical energy minimum\footnote{With typical we mean the level with the largest complexity.} for $m\geq 2$ at $T=0$ and $H=0$ has a cavity field distribution exhibiting an essential singular behavior close to the origin:
\begin{equation}
\label{eq:cavity_field_pdf_metastable_states_RSB}
    P_h(h)\sim h^{m-1}e^{-\frac{m-1}{h}}
\end{equation}
With this distribution in the asymptotic expressions in \ref{sec:Expansion_rho_band_edges}, one still finds \eqref{eq:low_edge_expansion_rho_critical}. What makes a crucial difference is that $\Lambda=0$, or equally, $\chi_{SG}=\infty$ for any $H\leq H_c$: it is this result that implies the scaling \eqref{eq:low_edge_expansion_rho_critical} of the spectral density. As a check of these claims, in figure \ref{fig:rho_lambda_SG} we show numerical measures of the spectral density for $H=0.5, 0.87 H_c$ and $N=2048$ spins. In the inset we zoom on the lower edge, showing the square-root behavior.

\subsubsection{Statistics of edge eigenvalues}

In many situations one could be interested not only in determining the bulk distribution of a set of iid random variables, but also the distribution of the smallest or the biggest elements of such a set, which is a large deviation problem. It is known that there are three universality classes for the distributions of extremes: Gumbel, Fréchet, Weibull, whose cumulative functions read (we consider the case $x>0$)
\begin{eqnarray}
\label{eq:three_classes_univ}
    && C(x)\,=\,e^{-e^{-\frac{x-\mu}{\nu}}}\qquad \text{Gumbel} \\
    && C(x)\,=\,e^{-x^{-\alpha}}\qquad \text{Fréchet} \\
    && C(x)\,=\,1-e^{-\left(\frac{x}{\nu}\right)^{\alpha}}\qquad \text{Weibull}
\end{eqnarray}
We are interested in Weibull distribution, since it is the distribution of the smallest eigenvalue. 
In Appendix \ref{sec:stat_eigs_edges} we show that in the paramagnetic phase the smallest eigenvalue is distributed with a Weibull distribution with shape parameter $\alpha=m$ and scale $\nu\,=\,(m Z_m)^{1/m}\;\Lambda/N^{1/m}$ in the infinite size limit
\begin{equation}
    \label{eq:Weibull_smallest_eig}
    \mathcal{P}(\lambda_1|\lambda_1<\lambda_2<\cdots<\lambda_{(m-1)N})\,\sim\,1-\exp\left(-\frac{N\lambda_1^{m}}{m Z_m \Lambda^m}\right)
\end{equation}
It is not hard to show that also the smallest cavity field is a Weibull variable, with
\begin{equation}
    \label{eq:Weibull_smallest_cav_field}
    \mathcal{P}(h_1|h_1<h_2<\cdots<h_{(m-1)N})\,\sim\,1-\exp\left(-\frac{N h_1^{m}}{m Z_m}\right)
\end{equation}
Upon comparing these two last equations we see that
\begin{equation}
\label{eq:correlation_eigmin_cavmin}
    \lambda_1\overset{d}{=}\Lambda h_1\qquad N\rightarrow\infty
\end{equation}
where $\overset{d}{=}$ stands for "equal in distribution". Eq. \eqref{eq:correlation_eigmin_cavmin} expresses the correlation between the smallest eigenvalue and the smallest cavity field. If one generalises these statements to the distribution of the smallest $k$-th eigenvalue, with $k=O(1)$ for $N\rightarrow\infty$, one can see that it follows Weibull order statistics: this statement has been proven rigorously in \cite{lee2016extremal}. The lower edge eigenvalues, considered all together, are a Poisson point process, the spatial distribution on the real line of uncorrelated random variables. This implies tha the distribution of their spacings is exponential. Thus, in the paramagnetic phase, the correlation between smallest eigenvalues and smallest cavity fields is so strong that repulsion between neighbors eigenvalues (see chapter 3) is overcome. 
For eigenvalues with rank $k=O(1)$ it holds asymptotically (notice that equation below could be guessed also from equation \eqref{eq:edges_expansion_spectral_density})
\begin{equation}
\label{eq:correlation_smallesteigs_smallestcavs}
    \lambda_k\overset{d}{=}\Lambda h_k\qquad N\rightarrow\infty
\end{equation}
We will come back to equation \eqref{eq:correlation_smallesteigs_smallestcavs} later in the last subsection of \ref{sec:Eigenvectors}, where we show the stronger proposition that \eqref{eq:correlation_smallesteigs_smallestcavs} is a full identity.

\begin{figure}
    \centering
    \includegraphics[width=0.8\columnwidth]{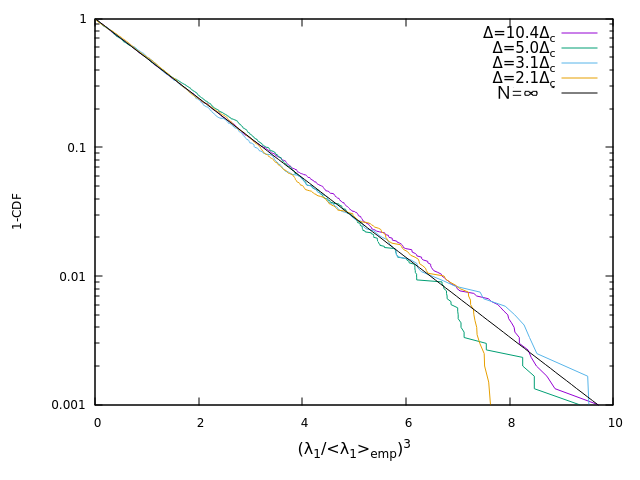}
    \includegraphics[width=0.8\columnwidth]{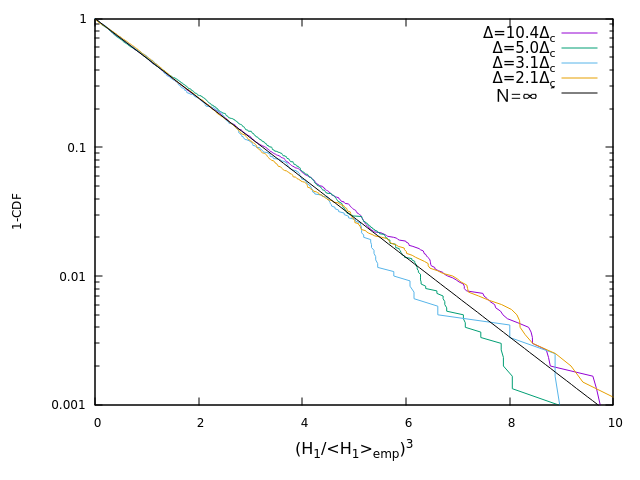}
    \caption{Cumulative distributions for the lowest eigenvalue and lowest cavity field, measured at different $\Delta$ values, follow nicely the theoretical prediction ($N=\infty$). We considered high values of external field because of the strong finite size effects emerging at criticality, see section \ref{sec:finite_size_effects}.}
    \label{fig:cumul}
\end{figure}

In figure \ref{fig:cumul} the empirical CDF of the smallest eigenvalue for some values of $H$ in the paramagnetic phase and the same for the smallest cavity field. We scaled both the variables in the x-axis with the empirical mean, rather than the asymptotic expectations values
\begin{equation}
    \label{eq:eig1_h1_expect_value_Weib}
    \langle h_1 \rangle_W = (m Z_m)^{1/m}\Gamma\left(1+\frac{1}{m}\right)\qquad \langle \lambda_1\rangle_W\,=\,\Lambda\langle h_1 \rangle_W
\end{equation}
This choice seems to reduce finite size effects for the lowest values of the external field.
Having specified that, it seems that plots in \ref{fig:cumul} seem to fully confirm the validity of Weibull statistics and of \eqref{eq:correlation_eigmin_cavmin}.

\subsubsection{Computation of the pseudo-gap width}

Let us estimate the interval $0<\lambda<\lambda_*$ such that \eqref{eq:edges_expansion_spectral_density} holds. In order to do this, we consider \eqref{eq:resolvent_eq_Hessian_SG} and rewrite it in the following form:
\begin{eqnarray}
\label{eq:clog}
  -\lambda - \Lambda x =\left(1-\frac{1}{m}\right)x^2\int dh \frac{P(h)}{h^2(h+x)}. 
\end{eqnarray}
 If $m>3$, the integral appearing in the r.h.s. of (\ref{eq:clog}) is convergent for $x\to 0$, 
 in our region of interest 
 we can therefore estimate it simply as $B=(1-1/m)\int dh P(h)/h^3$.
 We obtain 
\begin{eqnarray}
\label{eq:pseudo_gap_eq}
  \lambda=(1-A)x+Bx^2
\end{eqnarray}
with $B=(1-1/m)\int dh P(h)/h^3$ and $A$ defined in \eqref{eq:Replicon_T0_explicit_expression}. We actually cannot remove $x$ from the integral for arbitrary small value of $\lambda$ (see \ref{sec:Expansion_rho_band_edges}): close to the lower edge $|x''|\ll|x'|$ and the integral in \eqref{eq:clog} develops a divergence. Whenever $|x''|\gg|x'|$, we can correctly approximate the integral with $B$ to leading order in small $\lambda$.

\begin{figure}
    \centering
    \includegraphics[width=0.8\columnwidth]{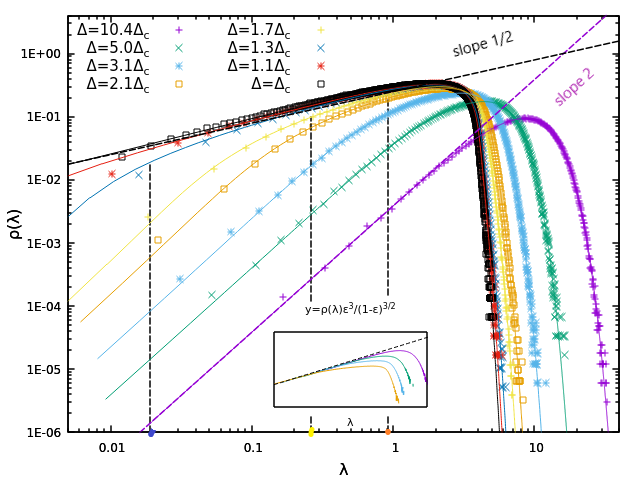}
    \caption{The log-log plot of the Hessian spectrum $\rho(\lambda)$ for $m=3$ and $N=1024$ clearly shows the crossover in the behavior at the lower band edge: from $\lambda^2$ at large fields to $\sqrt\lambda$ at the critical field. The continuous lines are the analytical spectral densities computed in the large $N$ limit. The dashed vertical lines mark the cross-over values $\lambda^*$ in the curves corresponding to $H/H_c=1.3,1.7,2.1$. The inset is a scaling plot with the dependence of the coefficient \eqref{eq:edges_expansion_spectral_density} (first equation) on the replicon eigenvalue made full explicit.}
    \label{fig:fig01}
\end{figure}

The width of the pseudo-gap is the value of $\lambda$ that cancels the discriminant of \eqref{eq:pseudo_gap_eq}, namely
\begin{equation}
\label{eq:pseudo-gap_m>3}
    \lambda_*\,=\,\frac{\Lambda^2}{4B}
\end{equation}
In the $m=3$ case, $B$ is divergent: we estimate its singularity as in section \ref{sec:Expansion_rho_band_edges}, obtaining $B\sim (1-1/m)\log(1/\Lambda)/Z_3$, so that
\begin{equation}
\label{eq:psuedo_gap_m=3}
    \lambda_*\,=\,\frac{3Z_3\Lambda^2}{8|\log\Lambda|}
\end{equation}
Close to the apparent lower edge, the spectral density can be written as
\begin{equation}
\label{eq:rho_close_to_pseudo_gap}
    \rho(\lambda)\sim\frac{\lambda_*}{\pi\Lambda}\sqrt{\frac{\lambda}{\lambda_*}-1}\qquad \lambda\overset{>}{\sim}\lambda_*
\end{equation}
In figure \ref{fig:fig01} we show for $m=3$ the spectral density for several values of fields and $N=1024$ in log-scale, showing for some values of $H$ the position of $\lambda_*$.

The scenario of the lower edge is as follows: in the paramagnetic phase, $H>H_c$, the spectral edge has a crossover from the Wigner form in \eqref{eq:rho_close_to_pseudo_gap} to the power law $\lambda^{m-1}$ in \eqref{eq:edges_expansion_spectral_density} at an eigenvalue proportional to the square of the Replicon eigenvalue. Therefore, the region of the psuedo-gap shrinks as $(H-H_c)^2$ when the critical field is approached from above, as equation \eqref{eq:Replicon_T0_explicit_expression} suggests. At the critical point and below it, the Replicon eigenvalue is identically zero and no pseudo-gap exists. This crossover occurring a criticality is actually a random matrix delocalisation transition, as we are going to show in next section, where we study the eigenmodes of our model. 

\subsection{Eigenvectors}
\label{sec:Eigenvectors}

The study of eigenvector statistics give us important information on the localisation properties of excitation modes. In the fully connected model object of this chapter, there is no spatial structure, so the concept of localisation has to be interpreted in spin space: what are the sites where the eigenvectors has maximal amplitudes? We learnt in chapter 3 that this information is enclosed in eigenvector moments \eqref{eq:eigvec_moments}. Let us local consider eigenvector moments of Rosenzweig-Porter ensemble at fixed diagonal disorder, equation \eqref{eq:local_eigenvector_moment_Ros_Por}, and the total moments: for our model, these can be written as
\begin{equation}
\label{eq:local_eigvec_moments_SG}
I_{n,\alpha}^{(q)}(\lambda)\,=\,\frac{\Gamma(1+q)P^{(\perp)}(n)_{\alpha\alpha}}{m N^{q}|h_n+x(\lambda)|^{2q}}
\end{equation}
\begin{equation}
\label{eq:eigvec_moments_SG}
    I_q(\lambda)\,=\,\frac{1}{m}\sum_{{n,\alpha}}I_{n,\alpha}^{(q)}(\lambda)\,=\,\frac{(1-1/m)\Gamma(1+q)}{m N^{q-1}}\int dh\;\frac{P_h(h)}{|h+x(\lambda)|^{2q}}
\end{equation}
where index $n$ is a site index and $\alpha$ a cartesian component. These moments are not rotational invariant in the space of spins: however, from a choice of basis to the other the only difference is a geometrical pre-factor, whose computation we show in Appendix \ref{sec:change_base_eigvec}. The rotational invariant eigenvector moments
\begin{equation}
\label{eq:eigvec_moments_rot_inv}
    \Tilde{I}_q(\lambda)\equiv \sum_{n}\;|\Vec{\psi}_n(\lambda)|^{2q}
\end{equation}
differ from equations \eqref{eq:eigvec_moments_SG} of a constant $m$-dependent prefactor, as $N$ goes to infinity.

Let us comment on equations \eqref{eq:local_eigvec_moments_SG}, \eqref{eq:eigvec_moments_SG}. The $q=1$ moment is the normalisation condition and reads
\begin{equation}
\label{eq:normalisation_condition_eigvecs}
    1\,=\,(1-1/m)\int dh\;\frac{P_h(h)}{|h+x(\lambda)|^2}
\end{equation}
which is no others than the second equation in \eqref{eq:resolvent_eq_Hessian_SG_re_and_im}, the imaginary part of \eqref{eq:resolvent_eq_Hessian_SG}. The $q=2$ moment is the Inverse Participation Ratio (IPR)
\begin{equation}
\label{eq:bulk_IPR_Hessian_SG}
    I_2(\lambda)\,=\,\frac{(1-1/m)}{N}\int dh\;\frac{P_h(h)}{|h+x(\lambda)|^4}\,=\,\frac{i_2(\lambda)}{N}.
\end{equation}
The IPR measures how the normalisation of the eigenvector is distributed among its components. Whenever the integral in \eqref{eq:bulk_IPR_Hessian_SG} is finite, the IPR scales as the inverse of system size: in this situation, the normalisation is distributed equally among $O(N)$ components and we call the eigenmode \emph{delocalised}. Higher-order $q$ moments
\begin{equation}
\label{eq:higher_than_IPR_hessian_SG}
    I_q(\lambda)\equiv \frac{i_q(\lambda)}{N}
\end{equation}
give more and more grained information on localisation properties, since with increasing $q$ eigenvector components are more and more suppressed in magnitude: for instance, moment $I_4$ gives us information about the degree of localisation within the cluster of components identified by $I_2$. Usually, in dense mean field systems localisation can only occur at the edges when $i_2(\lambda)\rightarrow \infty$ faster than $N$ as $\lambda\rightarrow 0$. The only exception is given by the ensemble of \emph{Levy Matrices}\footnote{It is an ensemble of random matrices having entries with heavy tail statistics (like the Cauchy distribution).} \cite{cizeau1994theory}: the spectra of these matrices can have localised bands, similarly to many problems involving sparse random matrices.

What about fractional moments? The only interesting case is $q<1/2$. Since for $q<1/2$ the importance of components with small amplitudes is enhanced, these moments allow to probe the scattering of normalisation in the bulk: they are particularly useful in systems defined on random graphs, as they can highlight the presence of strong multi-fractal behaviors even within localised phases \cite{GarciaMata2020}.

\subsubsection{Edge modes in the paramagnetic phase}

\begin{figure}
    \centering
    \includegraphics[width=0.8\columnwidth]{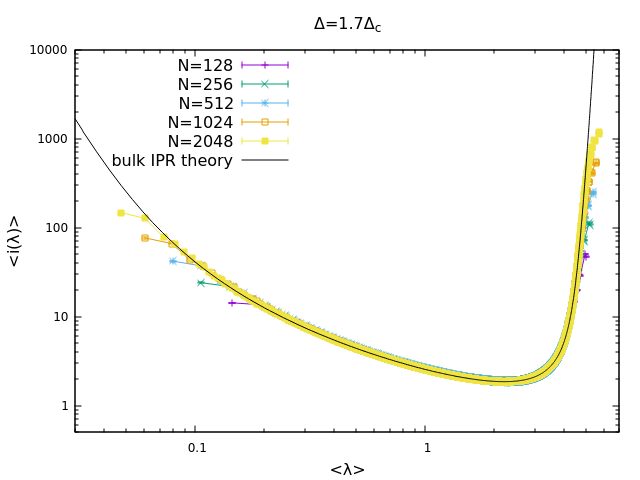}
    \caption{Plot of the sample average of $i(\lambda)$ versus the sample average of $\lambda$, for $H=1.0\simeq\,1.7\Delta_c$.}
    \label{fig:ipr_para_rescaled_plots_1}
\end{figure}

The eigenvector moments of our model are finite for any finite $\lambda$, being the integrals in $i_q(\lambda)$ well defined in this situation. In the limits $\lambda\rightarrow 0$ and $\lambda\rightarrow\infty$, the definition \eqref{eq:eigvec_moments_SG}, based on the assumption
\begin{equation}
\label{eq:bulk_condition}
    \frac{1}{|h+x(\lambda)|}=O(1)\qquad N\rightarrow\infty
\end{equation}
brings to absurdity when $H>H_c$ and thus $\Lambda>0$. Indeed, we find the behaviors
\begin{equation}
\label{eq:incosistent_scalings}
i_q(\lambda)\sim\begin{cases}
   \Lambda^{2q-1}\left(\frac{\lambda}{\Lambda}\right)^{-2(m-1)(q-1)}\qquad \lambda\rightarrow 0 \\
   \\
    P_h(\lambda)^{-3(q-1)}\qquad \lambda\rightarrow\infty
\end{cases}
\end{equation}
Since $\lambda\sim \Lambda N^{-1/m}$, we find $I_q(\lambda)\sim \lambda^{-(m-2)(q-1)}$ close to the lower edge, whereas close to the upper edge $\lambda\sim \log(N)$ and $I_q(\lambda)\sim P_h(\lambda)^{3(q-1)}$. These predictions are non-sense for any $q$: for $q>1$ they predict diverging moments at the edges, which is impossible because these moments are all upper bounded by the normalisation, the $q=1$ moment
\begin{equation*}
    \sum_{n,\alpha}|\psi_{n,\alpha}(\lambda)|^{2q}<\sum_{n,\alpha}|\psi_{n,\alpha}(\lambda)|^2=1.
\end{equation*}
For $q<1$ they are wrong as well, since they predict vanishing $I_q(\lambda)$, but $q<1$ moments are by definition all greater than unity. Finally, for $q=1$ we find that in the limits $\lambda=0$ and $\lambda=\infty$ limit the normalisation condition is violated
\begin{equation}
\label{eq:violation_normalisation_low}
    I_1(0)\,=\,(1-1/m)\left\langle\frac{1}{h^2}\right\rangle\,=\,1-\Lambda<1\qquad \text{if}\quad \Lambda>0
\end{equation}
\begin{equation}
\label{eq:violation_normalisation_up}
    I_1(\lambda)\sim \frac{1}{\lambda^2}\qquad \lambda\rightarrow\infty
\end{equation}
It is evident that hypothesis \eqref{eq:bulk_condition} must be revisited. We should choose a scaling of \eqref{eq:bulk_condition} in $N$ for which all expressions of eigenvector moments are consistent up to the edges. We need first of all to fix  equations \eqref{eq:violation_normalisation_low}, \eqref{eq:violation_normalisation_up}. Looking at \eqref{eq:eigvec_moments_SG}, we understand that the only way to do that is to set
\begin{equation}
\label{eq:edges_condition}
    \frac{1}{|h+x(\lambda)|}=O(N^{1/2})
\end{equation}
as lower and upper edges are approached. This last condition implies the existence of a condensate, a finite set of components that give a finite contribution to the normalisation of edge modes in the thermodynamic limit.
We expect bulk condition\eqref{eq:bulk_condition} to hold only for $\lambda>\lambda_*$, the crossover computed in section \ref{sec:spectral_density} separating the pseudo-gap band from the bulk, and to interpolate from \eqref{eq:bulk_condition} to \eqref{eq:edges_condition} as $\lambda<\lambda_*$.
With \eqref{eq:edges_condition} normalisation is restored
\begin{equation}
\label{eq:normalisation_restored_lower}
    I_1(0)\,=\,\frac{1}{N}\sum_{i=1}^N\frac{(1-1/m)}{h_i^2}\,=\,\Psi_L^2+(1-\Lambda)\Longrightarrow\Psi_L^2\,=\,\Lambda
\end{equation}
\begin{equation}
\label{eq:normalisation_restored_upper}
    I_1(\lambda)\,=\,\Psi_U^2+(1-1/m)\int dh\;\frac{P_h(h)}{|h+x(\lambda)|^2}\overset{\lambda\longrightarrow\infty}{\sim} \Psi_U^2+O(\lambda^{-2})\Longrightarrow \Psi_U^2\,=\,1
\end{equation}
and higher order moments $I_q(\lambda)\propto \Lambda^{q}$ for $\lambda$ going to zero.
In the lower edge the condensate is the Replicon eigenvalue itself: given that $\Lambda<1$, the normalisation of the modes is split between the condensate and the bulk. We refer to these as \emph{localised states}.
On the contrary, in the upper edge the condensate yields all the normalisation. We call the related states \emph{fully localised}. We show in \ref{fig:ipr_para_rescaled_plots_1} the sample-averaged rescaled IPRs $i_2\equiv N I_2$ with ranks $k$ versus the related sample averaged eigenvalues, for a value $H=1.7H_c$ and several sizes in the paramagnetic phase. This measure shows clearly that approaching the edges the bulk prediction \eqref{eq:eigvec_moments_SG} fails.

\begin{figure}
    \centering
    \includegraphics[width=0.8\columnwidth]{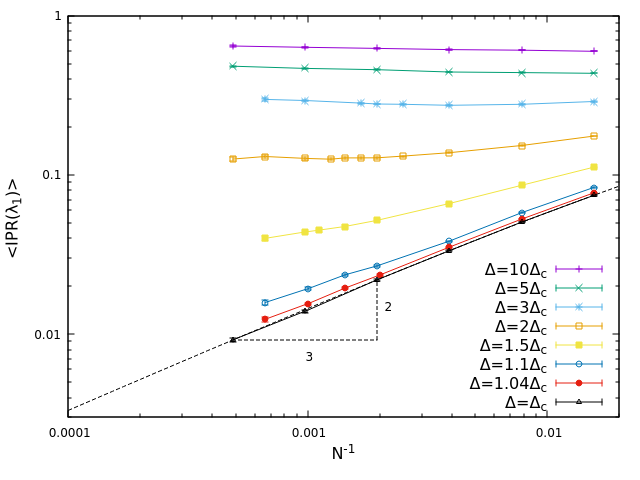}
    \caption{The $IPR$ of the lowest eigenvector versus $N^{-1}$ for several values of $H$. For $H>H_c$ the IPR converges to a finite value, signalling a localization on sites with the smallest external field. At the critical point a delocalization transition takes place, and the $IPR$ decays to zero as $N^{-2/3}$.}
    \label{fig:ipr_scaling_sev_H}
\end{figure}

The physical interpretation in our model of localisation is straightforward: if the random external field is sufficiently strong, the vast majority of spins is locked by strong local fields, responding weakly to small perturbations. Only on a finite number of sites the local field is sufficiently week so that spins can be appreciably excited. Together with this, sites with the strongest fields, whose number again is finite, are practically frozen. Upon lowering $H$, the number of soft spots in the system increases until the condensate disappears. In \ref{fig:ipr_scaling_sev_H} we show the sample averaged IPR of the smallest eigenvalue versus the inverse of the size $1/N$, for the $m=3$ case. It is evident that as $H$ is lowered the IPR crosses over from finite values to the critical scaling $N^{-2/3}$, which we will discuss in next subsection.

Since $\Lambda=0$ at the critical point $H=H_c$, we find the result anticipated at the end of section \ref{sec:spectral_density}: soft modes delocalise entering the SG phase, or conversely, soft modes exhibit a condensation transition entering the paramagnetic phase. In fact, equations \eqref{eq:normalisation_restored_lower}, \eqref{eq:normalisation_restored_upper} have the mathematical form of a Bose-Einstein condensation, an ubiquitous phenomenon in Bosonic systems. However, in order to fully comply with this analogy, the condensation should occur on a single component. We will show that this is indeed the case soon. Before discussing this interesting matter, we first consider equations \eqref{eq:eigvec_moments_SG} in the $\Lambda=0$ case. 

\subsubsection{Edge modes at the critical point}

Eigenvector moments are consistent up to the lower edge whenever $\Lambda=0$, whereas modes at the upper edge are always fully localised. Thus, let us focus on the non trivial case $\lambda\rightarrow 0$. When the system is critical, close to the lower edge rescaled eigenvector moments $i_q(\lambda)\equiv N^{q-1}I_q(\lambda)$ behave as
\begin{eqnarray}
\label{eq:critical_eigvec_modes}
    && i_q(\lambda)\sim\begin{cases}
        \Gamma(1+q)(1-1/m)C_1^{(q)}\rho(\lambda)^{-2(q-m/2)}\sim \lambda^{-(q-m/2)}\qquad q>m/2 \\
        \\
        C_1^{(q)}\,=\,\frac{1}{Z_m}\int_0^{\infty}\frac{h^{m-1}}{(1+h^2)^{q}}dh
    \end{cases} \\
    && \nonumber \\
\label{eq:critical_eigvec_modes1}
    && i_q(\lambda)\sim\begin{cases}
        \Gamma(1+q)(1-1/m)C_2^{(q)}\sim\text{const}\qquad 1<q<m/2 \\
        \\
        C_2^{(q)}\,=\,\frac{1}{Z_m}\int_0^{\infty}h^{m-1-2q}e^{-\frac{h^2}{2(1/m+H_c^2)}}\;dh
    \end{cases} \\
    && \nonumber \\
\label{eq:critical_eigvec_modes2}
    && i_q(\lambda)\sim
        \frac{\Gamma(1+q)(1-1/m)}{Z_m}|\log\rho(\lambda)|\sim |\log\lambda|\qquad q=m/2
\end{eqnarray}
Moments with $1<q<m/2$ scale regularly as $N^{q-1}$, whereas for $q>m/2$ a multi-fractal regime appears (see equations \eqref{eq:multi_frac_IPR_SG}, \eqref{eq:fractal_dimension_SG_crit}). In order to make it manifest, we use the scaling of the spectral density close to the lower edge: since $\rho(\lambda)\sim (1/\pi)\sqrt{\lambda/J}$ (let us ignore the log correction at $m=3$)
\begin{equation*}
    \int_0^{\lambda}\rho(\lambda')d\lambda'\,=\,\frac{1}{N}\Longrightarrow \lambda_N \sim \left(\frac{3}{2}\frac{\sqrt{J}}{N}\right)^{2/3}
\end{equation*}
and we can use this to get
\begin{equation}
\label{eq:multi_frac_IPR_SG}
    I_q(\lambda)\,=\,\frac{i_q(\lambda)}{N^{q-1}}\propto N^{-(q-1)}\rho(N^{-2/3})^{-2(q-m/2)}\propto N^{-(q-1)D(q)}
\end{equation}
with a fractal dimension
\begin{equation}
\label{eq:fractal_dimension_SG_crit}
    D(q)\,=\,\frac{1}{3}\frac{q+m-3}{q-1}
\end{equation}
for $q>m/2$. In particular, the IPR shows the behaviors
\begin{equation}
\label{eq:critical_IPRs}
    I_2\propto\begin{cases}
        N^{-2/3}\qquad m=3 \\
        \frac{\log N}{N}\qquad m=4 \\
        \frac{1}{N}\qquad m>4
    \end{cases}
\end{equation}
This implies that in the $m=3$ case the dominant cluster grows to infinity but it is still subextensive $O(N^{2/3})$ (there is actually an irrelevant $(\log N)^{1/3}$ factor coming from the log previously ignored), whereas for $m>3$ the dominant cluster is extensive (modulo the log term for $m=4$). In this last case, multi-fractality emerges only when probing the hierarchy of normalisation weights within the dominant cluster. Remarkably, multi-fractality is a signature feature of the extended states of Anderson model in presence of non-zero disorder \cite{de2014anderson}.

We expect this scenario to hold deep within the spin glass phase: nevertheless, one has to pay attention to correctly evaluate the prefactors $C^{(q)}$ in eq. \eqref{eq:critical_eigvec_modes}. One could compute them using the distribution found by Bray and Moore in \cite{bray1981metastable} for typical energy minima in the $H=0$ case
\begin{equation*}
    P_h(h)\,=\,\frac{1}{Z_m}h^{m-1}\exp\left(-\frac{(m-1)\Delta_0^2}{h}-\frac{m(h-\Delta_0)^2}{2}\right)
\end{equation*}
where $\Delta_0>0$ is an order parameter, and adapt it to the $H>0$ case.

\begin{figure}
    \centering
    \includegraphics[width=0.8\columnwidth]{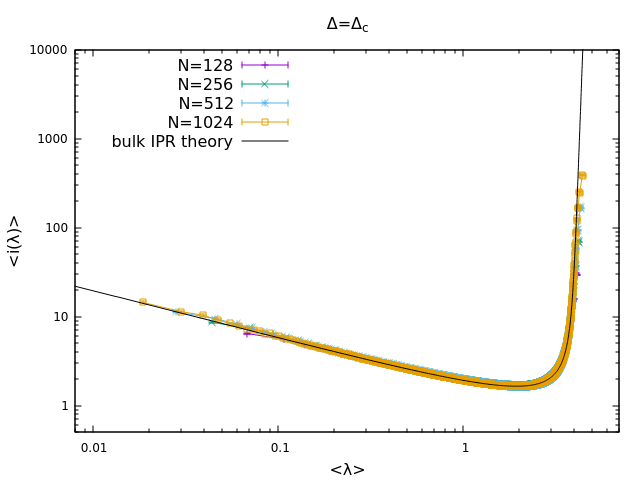}
    \caption{The sample-averaged IPR versus the sample-averaged eigenvalue, at the critical point in the $m=3$ case.}
    \label{fig:ipr_crit}
\end{figure}

In \ref{fig:ipr_crit} we show the sample averaged rescaled IPR at criticality, for several sizes in the $m=3$ case, versus the respective sample averaged eigenvalue. 
We see that our numerical data at the critical point agree perfectly with the bulk prediction \eqref{eq:eigvec_moments_SG}.

\subsubsection{Condensation theorems}

Let us come back to the condensation phenomenon. We found the following results, for $H>H_c$, $\lambda\rightarrow 0$ and $N\rightarrow\infty$
\begin{equation}
\label{eq:condensation_1}
    \lambda\overset{d}{=} \Lambda h
\end{equation}
\begin{equation}
\label{eq:condensation_2}
    \tilde{I}_q(\lambda)\,=\,\Lambda^q.
\end{equation}
If one considers the non-rotational invariant moments, $I_q$, a factor $(1-1/m)$ should be introduced.
Relations \eqref{eq:condensation_1}, \eqref{eq:condensation_2} are actually stronger, as proven recently in \cite{lee2016extremal}. Consider a random matrix ensemble
\begin{equation*}
    \mathbb{M}\,=\,\mathbb{W}+\operatorname{diag}(v_1,\dots,v_N)
\end{equation*}
where $\mathbb{W}$ is a Wigner matrix and the $\{v\}_i$ are iid random variables drawn from a $P_v(v)\sim v^{\alpha}$ close to $v=0$, with $\alpha>1$. Suppose that the diagonal elements are in ascending order \footnote{Notice that if the diagonal elements are in ascending order, also the eigenvalues are so.}.
Lee-Schnelli condensation theorems say that, for $N$ going to infinity, the following statements hold true for the $k=O(1)$ smallest eigenvalues:
\begin{itemize}
    \item The order statistics of the smallest eigenvalues is the same as that of the smallest diagonal elements, a Poisson order statistics. In particular, the distribution of the smallest eigenvalue is a Weibull distribution.
    \item With high probability\footnote{They establish conditions for likely events based on large deviations relations on the elements of their matrix. Details in \cite{lee2016extremal}.}
    \begin{equation}
    \label{eq:LS_eigs}
        N^{1/(1+\alpha)}|\lambda_k-(1-\langle 1/v^2 \rangle_v)\;v_k|\,\leq\,\frac{c_1}{N^{1/2-1/(1+\alpha)}}+\frac{c_2(\log N)^2}{N^{1/(1+\alpha)}}
    \end{equation}
    which means that \eqref{eq:condensation_1} is an exact identity, or stated in another way, the distribution of the difference of the r.h.s. and l.h.s. of \eqref{eq:condensation_1} tends to a delta function.
    \item With high probability
    \begin{equation}
    \label{eq:LS_vectors}
    |\psi_k(\lambda_k)^2-(1-\langle 1/v^2 \rangle_v)|\leq \frac{d_1}{N^{1/2-1/(1+\alpha)}}+\frac{d_2}{N^{1/(1+\alpha)}}
    \end{equation}
    which means that the condensate set is given, for mode $k$, by a single element, corresponding to the site with the $k$-th smallest diagonal element. A single component of the eigenvector yields a finite contribution to its normalisation. So, the delocalisation transition we discussed so far is a genuine Bose-Einstein condensation for random matrices. A more general study of this kind of problem has been recently carried out in \cite{ikeda2023bose}. See \cite{grad2017, evans2008condensation} for condensation phenomena in extreme value statistics related problem.
\end{itemize}
It is time to compare our numerical data with these theoretical predictions.

\subsection{Finite size effects}
\label{sec:finite_size_effects}

To fully characterise localised states in finite size system, we shall study finite size effects on condensation relations \eqref{eq:LS_eigs}, \eqref{eq:LS_vectors}. In order to take into account the vector nature of our degrees of freedom, the correct way to write these relations, showing explicitly Lee-Schnelli leading finite size corrections, is as follows
\begin{eqnarray}
\label{eq:condensation_relation_correct_eig_m}
    N^{1/m}|\lambda_i^{a}-\Lambda h_i|=O(\max(c_1N^{-1/2+1/m},c_2N^{-2/m}))
\end{eqnarray}
\begin{eqnarray}
\label{eq:condensation_relations_vector_m}
    ||\Vec{\psi}_i(\lambda_i^{a})|^{2}-\Lambda|=O(\max(c_1N^{-1/2+1/m},c_2N^{-2/m}))
\end{eqnarray}
where $a=1,\dots,m-1$ and $i=O(1)$ but in general $i=1,\dots, N$. Eigenvalues are divided in multiplets of $m-1$, which in the termodynamic limit form a degenerate space. This is an echo of the infinite $H$ limit: in this limit, the Hessian is given only by the diagonal part. The $m-1$ degeneracy comes from the fact that in \eqref{eq:Hessian_energy_form2} each diagonal element $\mu_i$ is repeated $m-1$ times.

The leading order of \eqref{eq:condensation_relation_correct_eig_m} can be deduced within our formalism. First, assuming $|x''|<|x'|$, we write (remember $x(\lambda)=\mathcal{G}(0)-\mathcal{G}(\lambda)-\lambda$)
\begin{equation}
\label{eq:aaa}
    |h_i+x(\lambda_i^a)|\,=\,|h_i+x_{i,a}'|\sqrt{1+\frac{x_{i,a}''^2}{(h_i+x_{i,a}')^2}}\simeq |h_i+x_{i,a}'|+\frac{1}{2}\frac{x_{i,a}''^2}{|h_i+x_{i,a}'|}
\end{equation}
Then, remembering condition \eqref{eq:edges_condition}, we set $|h_i+x_{i,a}|=k_{i,a}\;N^{-1/2}$ , with $a>0$. Solving the quadratic equation in $|h_i+x(\lambda_i)|$ in \eqref{eq:aaa}, we have the solutions
\begin{equation*}
    |h_i+x_{i,a}'|\,=\,\frac{k_{i,a}N^{-1/2}}{2}\left[1\pm\sqrt{1-2a^2x_{i,a}''^2N}\right]
\end{equation*}
We select the '$+$' solution because is the only one consistent with condensation condition. We have
\begin{equation*}
    |h_i+x_{i,a}'|\,=\,k_{i,a}N^{-1/2}-\frac{a^3N^{1/2}x_{i,a}''^2}{2}.
\end{equation*}
Next, we expand $x_{i,a}'$ up to second order
\begin{equation}
\label{eq:Rex_2nd_order}
   x_{i,a}'\simeq x_{i,a}'(0)+\frac{dx_{i,a}'}{d\lambda_i}\Bigl|_{\lambda=0}\lambda_i^a+\frac{1}{2}\frac{d^2x_{i,a}'}{d\lambda_i^2}\Bigl|_{\lambda=0}(\lambda_i^a)^2+\dots\,=\,-\frac{\lambda_i^a}{\Lambda}+\frac{1}{2}\Delta_2(\lambda_i^a)^2.
\end{equation}
Noticing that $N^{1/2}x_{i,a}''^2=O(N^{-2+\frac{2}{m}+\frac{1}{2}})\ll N^{-1/2}$, we deduce that $h_i+x_i>0$. Finally, we have after writing $(\lambda_i^a)^2$ at leading order in $N$
\begin{equation}
\label{eq:condensation_correct_from_SG}
    \Lambda h_i-\lambda_i^a\,=\,\Lambda k_{i,a}N^{-1/2}+\frac{1}{2}|\Delta_2|\Lambda^3(mZ_m)^{1/m}N^{-2/m}+\dots
\end{equation}
The coefficient $\Delta_2$ is negative: it is equal to minus the second derivative of the real part of $\mathcal{G}$ evaluated in zero, which is a response function and so non-negative by definition. Its expression is
\begin{equation*}
    \Delta_2\,=\,\begin{cases}
        -\frac{3(1-1/m)}{\Lambda^3}\left\langle\frac{1}{h^3}\right\rangle\qquad m>3 \\
        -\frac{2}{Z_3\Lambda^3}\log(1/\lambda)+O(1).\qquad m=3
    \end{cases}
\end{equation*}
We could not yield a prediction of coefficient $k_{i,a}$, since there are finite size corrections to $\mathcal{G}(0)$ (the lower edge of the local fields) we cannot evaluate. We know that they scale as $N^{-1/2}$ \cite{lee2016extremal}, but we do not know their coefficient, so we absorb them into $k_{i,a}$.

The finite size effects we just estimated are valid for eigenvalues sufficiently small to ensure $|x''|\ll|x'|$. In our model this condition holds only for $\lambda\ll \lambda_*$, eqs. \eqref{eq:pseudo-gap_m>3}, \eqref{eq:psuedo_gap_m=3}. For $\lambda\sim\lambda_*$ eigenvalues are in a crossover region and for $\lambda>\lambda_*$ they are effectively critical. We are able to estimate the scaling in $H-H_c$ of the crossover size $N_*$ such that, for $N\gg N_*$, estimates \eqref{eq:condensation_correct_from_SG} are correct. Indeed, combining eqs. \eqref{eq:pseudo-gap_m>3}, \eqref{eq:psuedo_gap_m=3} with $\lambda\sim \Lambda (m Z_m)^{1/m} N^{-1/m}$, we get
\begin{equation}
\label{eq:crossover_size}
    N_*\sim \frac{B^m m Z_m}{\Lambda^m}\propto (H-H_c)^{-m}
\end{equation}
modulo a log correction in the $m=3$ case. Close to criticality, it is impossible in numerical simulations to observe condensation.


In what follows, we will study finite size effects directly on our numerical data. We will consider
\begin{itemize}
    \item Finite size effects on the local fields, quantifying the systematic error we make by using $\mathcal{G}(0)$ instead of $\mathcal{G}_N(0)$.
    \item Finite size effects on spectral quantity, to test \eqref{eq:condensation_correct_from_SG} and the related prediction on the condensate.
\end{itemize}
We considered only the $m=3$ case.

\subsubsection{Finite size effects on the local fields}

Cavity fields norms in our analysis have been evaluated from the numerically evaluated local fields, through the asymptotic relation
\begin{equation}
\label{eq:cavvv}
    \mu_i\,=\,h_i+\chi_0\,=\,|\sum_j J_{ij}\Vec{S}_j+\Vec{b}_i+\chi_0\Vec{S}_i|
\end{equation}
where as usual $\chi_0$ the $N=\infty$ is the zero temperature susceptibility:
\begin{equation}
\label{eq:chi0}
    \chi_0(H)\,=\,(1-1/m)\sqrt{\frac{2}{\pi(1/m+H^2)}}
\end{equation}
An improvement would be to use
\begin{equation}
\label{eq:cavity_fields_improved}
    h_i\,=\,\mu_i-\sum_j J_{ij}^2\chi_{jj}
\end{equation}
with
\begin{equation*}
    \chi_{ii}\,=\,\frac{1}{\mu_i-G_N(0)}
\end{equation*}
\begin{equation*}
    G_N(0)\,=\,\frac{1}{N}\sum_i \frac{1}{\mu_i-G_N(0)}
\end{equation*}
Finding a good initial condition for an iterative equation built on this last one is not easy. Indeed, one can verify with euristic attempts that there is a chance already at the second step to develop a destabilizing term: this happens anytime one of local fields verify $\mu_i-G_N^{(2)}(0)\ll 1/N$, which can happen easily at intermediate range of sizes. In any case, the problem related to the usage of \eqref{eq:cavvv} concerning for the finite size effects on the smallest elements is relevant only in an intermediate range between $H\sim H_c$ and $H\gg H_c$. Indeed, in the first case for the range of sizes we were able to simulate we are always in the critical region $N\ll N_*$: in this situation, even though the choice \eqref{eq:cavvv} gives a systematic error that adds up to finite size effects, the trend would be a critical $\lambda\sim N^{-2/3}$ decay. In the second case $\chi_0\propto H^{-1}$ is small and the probability $\mu_i<\chi_0$ even at finite sizes is basically zero. In the cases $H=5:10 H_c$ we never observed a sample where such a thing happened.

\begin{figure}[h!]
    \centering
    \includegraphics[width=0.8\columnwidth]{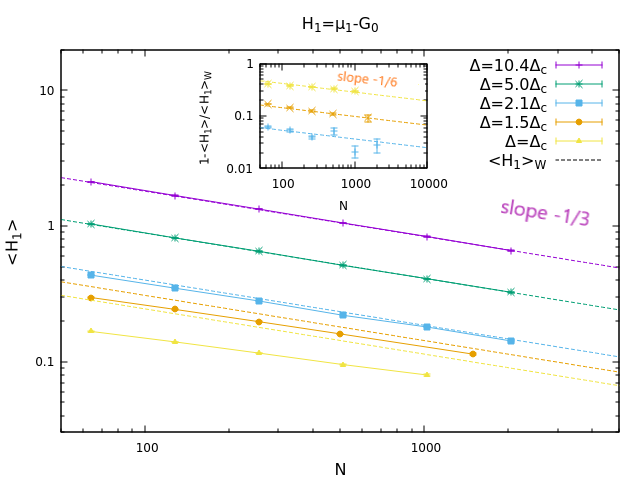}
    \caption{The size dependence of the smallest cavity field. In the inset we show the deviation from the asymptotic values, highlighting finite size effects.}
    \label{fig:smallest_cavity_field_VS_N}
\end{figure}

From estimates in \cite{lee2016extremal}, we expect finite size effects with the leading behavior
\begin{equation}
\label{eq:fse_T0_susc}
    |G(0)-G_N(0)|=O(N^{-1/2})
\end{equation}
In order to verify that, in figure \ref{fig:smallest_cavity_field_VS_N} we show the smallest cavity field versus the size $N$, for several values of the external field and for $m=3$. At high values of $H$ there is perfect agreement with the expectation value from a Weibull distribution
\begin{equation*}
    \langle h_1 \rangle_W = (m Z_m)^{1/m}\Gamma\left(1+\frac{1}{m}\right)N^{-1/m}
\end{equation*}
but already at $H=2 H_c$ we can detect significant deviations. In the inset, we show the quantity $1-\langle h\rangle_{emp}/\langle h\rangle_W$. This quantity is a measure of the difference $G(0)-G_N(0)$: calling $\tilde{h}\,=\,\mu-G_N(0)$ the exact cavity field, we have
\begin{equation}
\label{eq:mmmm}
    1-\frac{\langle h \rangle_{emp}}{\langle h \rangle_W}\,=\,1-\frac{\langle \tilde{h} \rangle_{emp}}{\langle h \rangle_W}+\frac{G(0)-G_N(0)}{\langle h \rangle_W}
\end{equation}
In order to comply with prediction \eqref{eq:fse_T0_susc}, the last quantity on the r.h.s. should scale as $N^{-1/2+1/m}$. This is checked in the inset, where we show that the quantity just described scale as $N^{-1/6}$, in agreement with theoretical expectation. Notice that the first term in the r.h.s. of \eqref{eq:mmmm}, which represents finite size effects on the empirical mean of a Weibull variable, is smaller than $O(N^{-1/2+1/m})$.

\subsubsection{Measure of finite size effects of spectral quantities}

Let us recap the finite size effects we expect for the smallest eigenvalues $\lambda_i^{a}$: for $N\gg N_*$
\begin{eqnarray}
\label{eq:recap_corrections_eigs}
    \lambda_i^{a}&&\sim \Lambda h_i \\
    \Lambda h_i-\lambda_i^{a}\,&&=\,\max\left(\Lambda k_{i,a}N^{-1/2},\frac{1}{2}|\Delta_2|\Lambda^3(mZ_m)^{1/m}N^{-2/m}\right).\nonumber
\end{eqnarray}
For $N\ll N_*$, $\lambda_i^{a}\sim N^{-2/3}$. In the crossover region $N\sim N_*$ terms with different scaling in $N$ are very similar, so we expect $\Lambda h_i-\lambda_i^a$ to feature a maximum for these sizes.
For $m=3$, the leading correction is $O(N^{-1/6})$, in the $m=4$ case the two corrections in the second of \eqref{eq:recap_corrections_eigs} are of the same order, for $m>4$ the correction $O(N^{-2/m})$ is dominant.

\begin{figure}[h!]
    \centering
    \includegraphics[width=0.8\columnwidth]{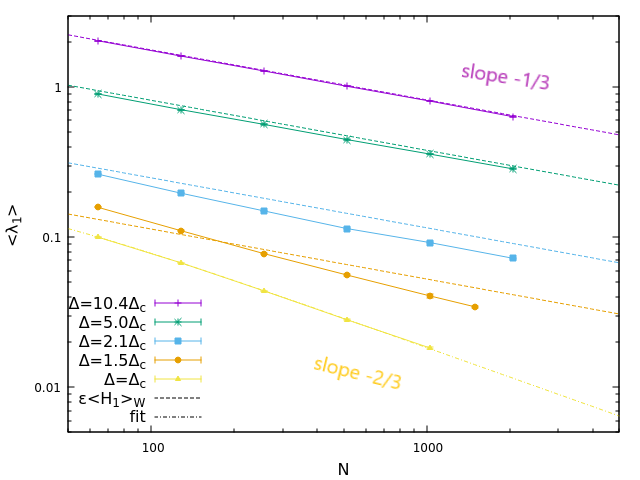}
    \caption{The size-scaling behavior of the sample-averaged smallest eigenvalue, for $H/H_c=10.4, 5.0, 2.1, 1.5, 1.0$. Data in the paramagnetic phase are compared with Weibull prediction, data at the critical point are fitted with $N^{-2/3}$.}
    \label{fig:smallest_eig}
\end{figure}

In figure \ref{fig:smallest_eig}. we show a plot of the smallest eigenvalue versus the size $N$, comparing it to Weibull prediction. While for high $H$ there is very good agreement, approaching the critical field finite size effects emerge in all their strength.

\begin{figure}[h!]
    \centering
    \includegraphics[width=0.5\columnwidth]{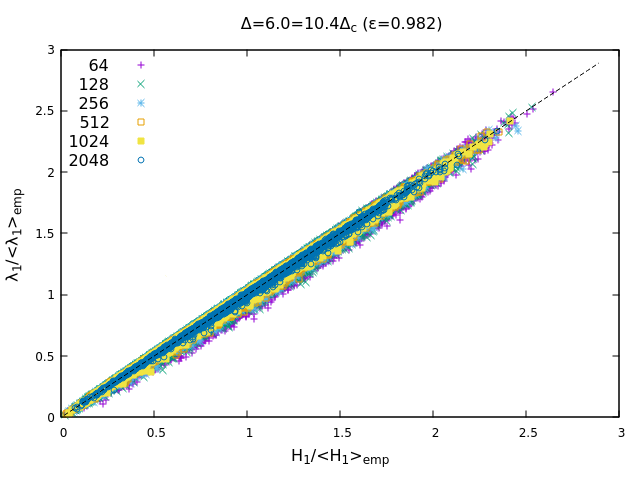}
    \includegraphics[width=0.5\columnwidth]{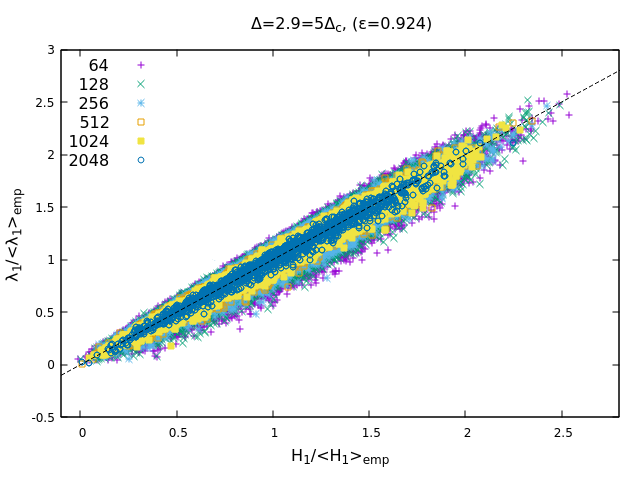}
    \includegraphics[width=0.5\columnwidth]{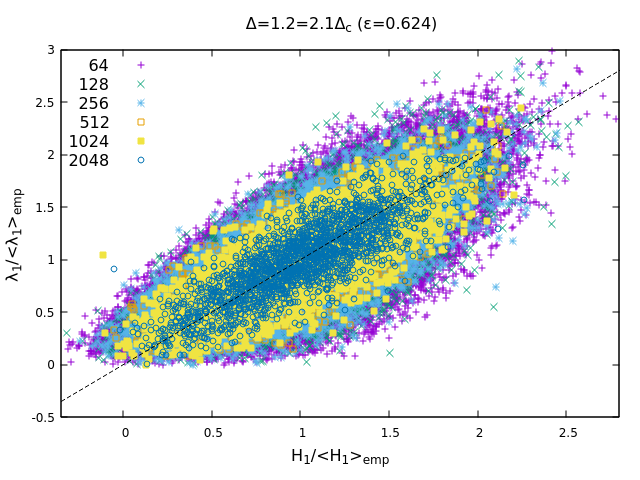}
    \caption{Scatter plots of the smallest eigenvalue versus the smallest cavity field for all sizes simulated, normalised with their empirical averages. The top figure is $H=10.4 H_c$, the central figure is $H=5 H_c$ and the bottom one is $H=2.1 H_c$.}
    \label{fig:scatter_plots}
\end{figure}

In figures \ref{fig:scatter_plots}, we show scatter plots of the smallest eigenvalue, rescaled with its empirical mean, versus the smallest cavity field rescaled with its empirical mean. In the thermodynamic limit all points should align along the bisector line. We see that at high field (top and central figure), where $N_*$ is small, the correlation between the two quantities is very strong, but already at $H=2H_c$ the correlation is way weaker. All this is in perfect agreement with the considerations made so far.

\begin{figure}[h!]
    \centering
    \includegraphics[width=0.8\columnwidth]{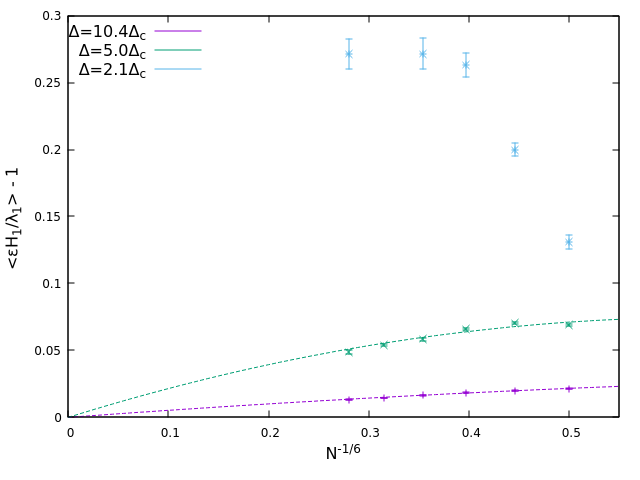}
    \includegraphics[width=0.8\columnwidth]{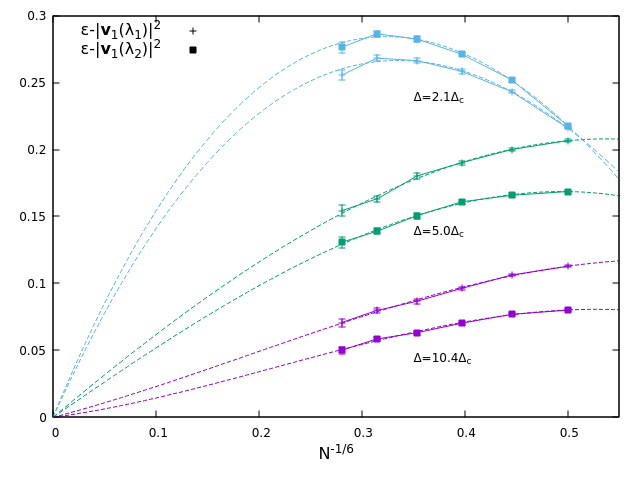}
    \caption{Plots of the relative differences between empirical and asymptotic average quantities: the relation $\langle\Lambda H_1/\lambda_1\rangle-1$ (top) and the largest component of the smallest eigenvectors $\epsilon - \langle |\Vec{\psi}_1(\lambda_1)|^2 \rangle$ and $\Lambda - \langle |\Vec{\psi}_1(\lambda_2)|^2 \rangle$ (bottom). 
    We plot these differences as a function of $N^{-1/6}$ which is the expected leading finite size correction. All data are compatible with a zero value in the large $N$ limit, but finite size effects become very severe as $H$ is lowered, showing a non-monotonic dependence of the curves with respect to the size $N$. Dashed lines are quadratic interpolations to the data.}
    \label{fig:FSE}
\end{figure}

To corroborate this assessment, in \ref{fig:FSE} we show a plot of $\Lambda\langle h_1/\lambda_1\rangle_{emp}-1$ versus $N^{-1/6}$. While at $H=5.0, 10.0 H_c$ our data are consistent with a curve decaying as $N^{-1/6}$, in the $H=2.0H_c$ case we are far from the asymptotic behavior. Naively, one could think that data are converging to a finite value. It is actually the maximum of $\Lambda h_i-\lambda_i^a$ that appears at $N=O(N_*)$. To give further evidence to this claim, in the plot below in the same figure we show the difference between the asymptotic value of the condensate and the measured squared larger component $|\psi_k(\lambda_k^a)|^2$ for the first two eigenvalues. It is more evident in these plots that the $H=2H_c$ curve is starting to fall down from a peak approximately at $N=1024$.

\begin{figure}[h!]
    \centering
    \includegraphics[width=0.8\columnwidth]{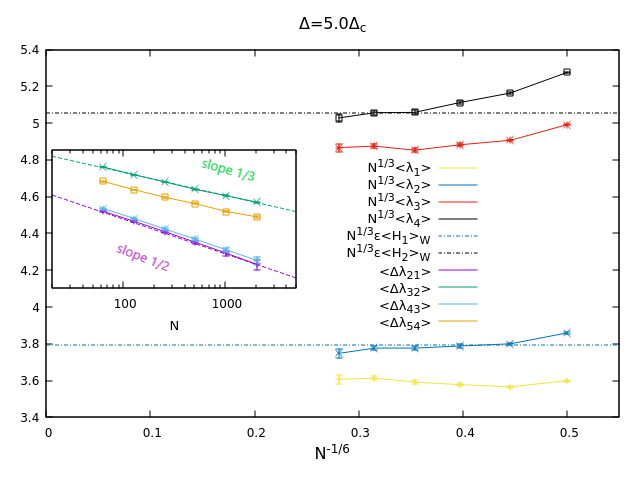}
    \caption{The sample means of the four smallest rescaled eigenvalues $N^{1/3}\lambda_i$ converge to a constant in the large $N$ limit, as they should. The differences within a pair scale as $N^{-1/2}$, while differences between pairs scale as $N^{-1/3}$ (see the inset), so each pair converges to a unique value with $N^{-1/6}$ corrections (hence the horizontal scale).}
    \label{fig:scaledLambda}
\end{figure}

\begin{figure}
    \centering
    \includegraphics[width=0.8\columnwidth]{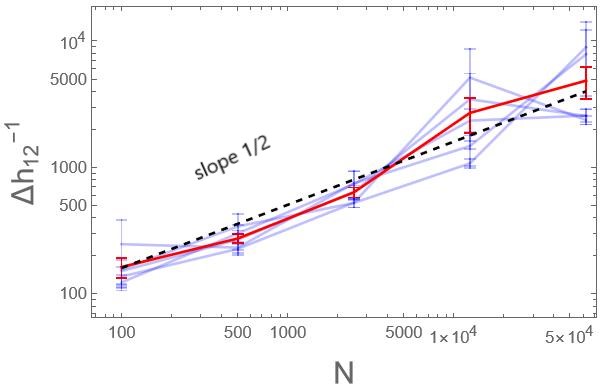}
    \caption{The divergence with size of the $1/(h_1-h_2)$ term appearing in the perturbative approach given by eq. \eqref{eq:pert_approach}.}
    \label{fig:divergence_inverse_spacings}
\end{figure}

To conclude this section, we show how finite size effects impact degenerate $m-1$ multiplets. Our theory predicts $|\Delta\lambda_{i}^{ab}|=O(\max(N^{-1/2}),\max(N^{-2/m}))$, with $a, b=1,\dots, m-1$. We see it easily thanks to triangular inequalities 
\begin{equation*}
|\Lambda h_i-\lambda_i^{a}|-|\Lambda h_i-\lambda_i^{b}|\,\leq\,|\lambda_i^{a}-\lambda_i^{b}|\leq |\Lambda h_i-\lambda_i^{a}|+|\Lambda h_i-\lambda_i^{b}|.
\end{equation*}
In figure \ref{fig:scaledLambda} we show our measures of the splittings of the two smallest duplets in the $m=3$ case, for $H=5H_c$. For lower fields, we could not observe duplets: it is a feature of the asymptotic $N\gg N_*$ system.
On the $y$-axis of the main plot, we show rescaled eigenvalues $N\lambda_i^a$, in the inset the splitting between adjacent eigenvalues. Our measures met theoretical predictions: splittings between eigenvalues of the same pair scale as $N^{-1/2}$, those between different pairs as $N^{-1/3}$.
Our measures show that within a pair the largest eigenvalue is closer to the asymptotic behavior. 
This results can be interpreted more easily with the help of perturbation theory: the degenerate pair of eigenvalues is split at finite sizes in two distinct eigenvalues, with spacing $\propto 1/\sqrt{N}$, which is the magnitude of the perturbation, coming from the interaction matrix. More generally, one can consider the large $H$ limit: since the replicon $\Lambda\rightarrow 1$ for $H\rightarrow\infty$, one can define $\delta=1-\Lambda$ and after proper rescaling consider the interaction matrix a perturbation proportional to $\sqrt{\delta/N}$. Then, by defining the rescaled matrices (we use the notation of eq. \eqref{eq:Hessian_as_RWPR} for the Hessian, the interaction matrix and the diagonal matrix) $\mathbb{\Tilde{M}}\,=\,\sqrt{\delta}\mathbb{M}$, $\mathbb{\Tilde{V}}\,=\,\sqrt{\delta}\mathbb{V}$, $\mathbb{\Tilde{W}}\,=\,\sqrt{N}\mathbb{W}$, one can rewrite the Hessian as
\begin{equation}
    \mathbb{\Tilde{M}}=\sqrt{\frac{\delta}{N}}\mathbb{\Tilde{W}}+\mathbb{\Tilde{V}}.
\end{equation}
The perturbative expansion of the smallest eigenvalues pair up to second order formally reads
\begin{equation}
\label{eq:pert_eigs}
    \tilde{\lambda}_1^{\pm}\,=\,\tilde{h}_1+\delta\;\tilde{\mathcal{G}}(0)\pm 2\sqrt{\frac{\delta}{N}}+\frac{2\delta}{3N}\sum_{j=1}^N\frac{1}{\tilde{h}_1-\tilde{h}_j}+O\left(\frac{\delta^{3/2}}{N^{3/2}}\right)
\end{equation}
The leading perturbative order is given by the splitting of the degenerate pair triggered by the perturbation.
As to next order, we can divide the summation in two parts: the sum over $j>2$ and the term with $j=2$. We thus have
\begin{eqnarray*}
    && \frac{2\delta}{3N}\sum_{j=1}^N\frac{1}{\tilde{h}_1-\tilde{h}_j}\simeq -\frac{2\delta}{3N}\frac{1}{\tilde{h}_2-\tilde{h}_1}-\frac{2\delta}{3N}\sum_{j\geq 3}\frac{1}{\tilde{h}_j-\tilde{h}_1} \\
    && \simeq -\frac{2\delta}{3N}\frac{1}{\tilde{h}_2-\tilde{h}_1}-\frac{2\delta}{3}\left\langle \frac{1}{\tilde{h}}\right\rangle -\frac{2\delta}{3}h_i\left\langle\frac{1}{\tilde{h}^2}\right\rangle \\
    && = -\frac{2\delta}{3N}\frac{1}{\tilde{h}_2-\tilde{h}_1}-\delta\;\mathcal{G}(0)-\delta\;\tilde{h}_i
\end{eqnarray*}
so that we can rewrite \eqref{eq:pert_eigs} as
\begin{equation}
\label{eq:pert_approach}
    \tilde{\lambda}_1^{\pm}\,\simeq \,\Lambda\tilde{h}_1\pm 2\sqrt{\frac{\delta}{N}}-\frac{2\delta}{3N}\frac{1}{\tilde{h}_2-\tilde{h}_1}+O\left(\frac{\delta^{3/2}}{N^{3/2}}\right)
\end{equation}
For $N$ going to infinity, the average of the coefficient of the $O(\delta/N)$ term is divergent: indeed, the pdf of the differences of two iid random variables is finite in zero, so the average value of $1/\Delta h$ is infinite. However, if one considers finite $N$ sequences, one finds that the average value of the splitting $\Delta h$ scale as $N^{1/2}$. We show this in figure \ref{fig:divergence_inverse_spacings}, where the averages of $1/(h_{2}-h_{1})$ 
for five distinct runs of $N_s=10^4$ samples are computed for growing $N$. We considered sequences of $N$ iid random variables $h_i$ sampled by a Chi distribution with two degrees of freedom (the pdf of the cavity fields in the thermodynamic limit).

We conclude that the separations between eigenvalues in each pairs observed in figure \ref{fig:scaledLambda} are the results of two contributions: the first comes from the splitting of degenerated eigenvalues in perturbation theory, the second from the divergence of the quantities $1/\Delta h$. Physically, these are related to the finite size effects of the unperturbed resolvent function 
\begin{equation*}
    \mathcal{G}^{(0)}_{N}(z)\,=\,\frac{2}{3N}\sum_{k=1}^N\frac{1}{h_k-z}
\end{equation*}
In fact, by setting $z=h_1+i\epsilon$ and $\epsilon=O(N^{-1/2})$, we get
\begin{equation*}
    \mathcal{G}^{(0)}_N(h_1+i\epsilon)\,=\,\frac{2i}{3N\epsilon}+\frac{2}{3N(h_2-h_1)}+\mathcal{G}^{(0)}(h_k+i\epsilon)\,=\,\mathcal{G}^{(0)}(h_k+i\epsilon)+O(N^{-1/2}).
\end{equation*}

\section{Conclusions}

In this chapter we studied a fully connected model of vector spin glasses. We began with a replica computation of the free energy of the model, discussing both the RS and RSB phase, and then we focused on the low temperature phase, studying the problem of the linear excitations of energy minima. We showed that this mean field model features localised modes at the lower edge, a feature observed in real glasses. This result is at variance with what is often a common belief, that mean field models with Hessian represented by dense matrix ensembles cannot have localised modes. In addition to that, we showed that the spin glass transition, by the point of view of linear excitations, is a delocalisation transition. We believe that these results enrich the picture of the zero temperature spin glass phase transition.

\chapter{The fully connected vector p-spin}

In previous chapter we studied the spin glass transition at $T=0$ of a generalisation of the original SK model to vector spins with random external fields. In the RS phase, where linear response theory is exact, energy minima possess localised excitations down to $\lambda=0$, with the spectral density featuring a pseudo-gap $\rho(\lambda)\sim\lambda^{m-1}$. At the onset of the transition, the number of soft excitations grows rapidly as $\rho(\lambda)\sim\sqrt{\lambda}$ and excitations delocalise. This random matrix transition is connected to the onset of marginal stability, i. e. full replica symmetri breaking.
It is natural to consider as a natural continuation of our work in the previous chapter a p-spin model with vector spins. These models have been extensively studied in the annhealed and quenched case in \cite{taucher1992annealedn, taucher1993quenchedn}: the authors however focused on the $m\rightarrow \infty$ limit.

As discussed in Chapter 2, p-spin models are toy models of the glass transition, according to the 1RSB-RFOT scheme. While in Ising p-spin models the low-temperature phase is fRSB \cite{Gardner1985}, spherical p-spin models feature at zero temperature an energy band with stable minima, i.e., inside each basin linear response theory is satisfied \cite{Crisanti1992}. Actually, the minima of spherical p-spin model are ultra-stable: the spectrum is gapped in the whole stable band, becoming gapless only at the threshold level. As to vector p-spin models, the results from our previous work hint the existence of a stable band of energy minima with gapless spectra. This is what we will show in this chapter, which shall be thought of as an application of our results in Chapter four to mean-field models of structural glasses.

The chapter is organised as follows: in Section \ref{sec:model_chap_5} we introduce the model, discussing its general features. In Section \ref{sec:dyn_phase_vector_pspin}, we briefly discuss the dynamical phase, computing the complexity through the Monasson method and the Replicon eigenvalue from the Hessian of the replicated free-energy.
In section \ref{sec:stable_glasses} we perform the $T=0$ limit and study eigenvalue spectra as functions of the energy level of the 1RSB landscape. We show how our results of Chapter 4 extend to these minima. Finally, in section \ref{sec:ultra_stable_glasses} we show the most interesting result of this chapter: the existence of rare ultra-stable energy minima among the exponentially many gapless ones. These minima are interpreted as mean-field equivalent of ultra-stable glasses \cite{rodriguez2022ultrastable}.

The content of this chapter is based on our work \cite{franz2022linear}.

\section{The model}
\label{sec:model_chap_5}

The model is defined by the following p-spin Hamiltonian
\begin{eqnarray}
  \label{eq:Hamiltonian_vector_pspin}
 {\mathcal H}[{\bf S}]=-\sum_{p=3}^{\infty} a_p \sum_{{\boldsymbol i},{\boldsymbol \alpha}}J_{i_1,\dots,i_p}^{\alpha_1,\dots,\alpha_p}S_{i_1}^{\alpha_1}\cdots S_{i_p}^{\alpha_p}
\end{eqnarray}
where in the second summation ${\bf i}=(i_1,\dots,i_p)$ is a disposition of the $p$ indices without repetitions, $i_1\neq i_2\neq \dots i_{p-1}\neq i_p$, whence ${\bf \alpha}=\{\alpha_1\dots \alpha_p\}$.
The couplings $J_{i_1,\dots,i_p}^{\alpha_1,\dots,\alpha_p}$ are
Gaussian variables symmetric over all the indexes but otherwise
independent, with zero mean and variance
$\overline{(J_{i_1,\dots,i_p}^{\alpha_1,\dots,\alpha_p})^2}=\frac{p!}{2}N^{-(p-1)}$. The model generalizes to $m$-components spins the mixed p-spin model usually considered for Ising or spherical variables. It differs from the model considered by Panchenko in \cite{panchenko2018free} by the fact that here all the spin components interact with each others, while in that model only components with the same label interact. This is a minor difference that does not affect physics, and it is only for notational simplicity that we choose the present version.

The model defined in \eqref{eq:Hamiltonian_vector_pspin} represents a mixed p-spin model, where contributions with different interactions are weighted by $a_p$. Note that one must have $\sum_p a_p^2<\infty$. As we learnt in section \ref{sec:the_model_chap4} of Chapter 4, the Hamiltonian of spin glass models is a Gaussian function with covariance
\begin{eqnarray}
  \label{eq:covariance_Hamiltonian_pspin}
\overline{{\mathcal H}[{\bf S}]{\mathcal H}[{\bf S'}] } = N g(q({\bf S},{\bf S'}))
\end{eqnarray}
where $q({\bf S},{\bf S'})$ is the overlap 
\begin{eqnarray}
\label{eq:overlap_in_chap_5}
  q({\bf S},{\bf S'})=\frac{1}{N}\sum_{i=1}^N {\bf S}_i\cdot {\bf S'}_i
\end{eqnarray}
and the function $g$ is
\begin{eqnarray}
  \label{eq:characteristic_function}
  g(q)=\frac 1 2 \sum_{p} a_p^2 q^p. 
\end{eqnarray}
Notice that here, because the couplings are nonisotropic\footnote{However, since they have zero mean and an isotropic covariance matrix, the physical solution is isotropic}, the natural definition of the overlap \eqref{eq:overlap_in_chap_5} returns a value in the interval $[0, 1]$ for any $m$. This different definition will result in a minor change from $(1-1/m)$ factors to $(m-1)$ ones in all equations related to transverse excitations.
In this paper we concentrate on the cases $m>2$ and the pure monomial case where a single $a_p$ with $p>2$ does not vanish.

\section{The dynamical phase}
\label{sec:dyn_phase_vector_pspin}

The dynamical phase can be unveiled by replica method exploiting the smart Monasson method \cite{monasson1995structural}
 described in chapter 2. The complexity of meta-stable states is the Legendre transform of minus the Monasson potential, evaluated at the largest $q=q_{*}$ among those extremising it:
\begin{eqnarray}
&& \Phi(n,\beta, q)=\frac{1}{nN}\log\overline{\int \boldsymbol{dS}\exp\left(-\beta\sum_{a=1}^n \mathcal{H}[\boldsymbol{S}_a]\right)\prod_{a,b}\delta(\boldsymbol{S}_a\cdot\boldsymbol{S}_b-N q)} \\
&& \Sigma(f, \beta)\,=\,\beta n f+ \Phi(n,\beta, q_{*})\qquad \left. f\,=\,-\frac{1}{\beta}\diffp{\Phi}{n}\right|_{q=q_{*}}
\end{eqnarray}
The complexity of equilibrium states is achieved at $n=1$, while $n\neq 1$ allows for the exploration of different families of metastable states at a certain temperature $T$. The equilibrium free energy is obtained at $n=1$ by evaluating the Monasson potential at the trivial $q=0$ solution
\begin{equation}
    \left. f_{eq}(\beta)\,=\,-\frac{1}{\beta}\diffp{\Phi}{n}\right|_{n=1,q=0}
\end{equation}
and is equal to the paramagnetic free energy $f_{para}=\frac{g(1)}{2}+\log S_m(1)$. As usual, the dynamical phase is defined as the interval $[T_K,T_d]$ such that $\Sigma_{eq}(T)>0$ and the replicon $\Lambda>0$. The complexity of the equilibrium states vanishes at the glass transition, identified by the Kautzmann temperature $T_K$, whereas the Replicon eigenvalue vanishes at the dynamical transition, identified by $T_d$. 

\subsection{The complexity}

\begin{figure}
    \centering
    \includegraphics[width=0.8\columnwidth]{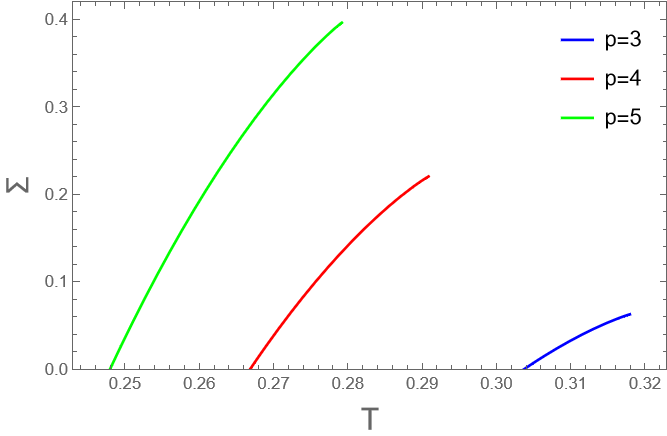}
    \caption{\textbf{Left}: Equilibrium complexity in the dynamical phase, for $p=3, 4, 5$. The complexity is non zero only in the interval $[T_K,T_d]$. The maximal complexity at $T=T_d$ is $\Sigma_{d}=0.0627787$ ($p=3$), $\Sigma_{d}=0.220444$ ($p=4$), and $\Sigma_{d}=0.396359$ ($p=5$).}
    \label{fig:Sigma_eq_dyn_phase}
\end{figure}

In our model, the Monasson potential reads (computation in Appendix \ref{sec:der_mon_pot})
\begin{align}
  \label{eq:Monasson_potential_vector_spin}
\Phi(n,\beta, q)=&\frac{n \beta^2}{2}\left\{  g(1) + (n - 1) [g(q) -  q g'(q)] -  g'(q) \right\} \nonumber\\
 &+\log\left(\int_0^\infty \frac{dh}{Z_m^{(0)}}\; h^{m-1} e^{-\frac{h^2}{2g'(q)}}\mathcal{K}_m(\beta h)^n
 \right)\;,
\\
 &\mathcal{K}_m(u)= (2\,\pi)^{m/2}\frac{I_{\frac{m-2}{2}}(u)}{u^{\frac{m-2}{2}}}\qquad Z_m^{(0)}=\int_0^\infty dh\; h^{m-1} e^{-\frac{h^2}{2g'(q)}}\;,\nonumber
\end{align}
The equation for the overlap is obtained by extremising \eqref{eq:Monasson_potential_vector_spin} with respect to $q$, and is equal to
\begin{eqnarray}
  \label{eq:overlap_vector_pspin}
  && q=\frac{\int_0^\infty dh\; h^{m-1} \exp\left[-\frac{h^2}{2g'(q)}\right] \mathcal{K}_m(\beta
  h)^{n}\mathcal{g}_m(\beta h)^2}{\int_0^\infty dh\; h^{m-1} \exp\left[-\frac{h^2}{2g'(q)}\right]\mathcal{K}_m(\beta h)^n}.\quad \\
  && \mathcal{g}_m(x)\,=\,\diff{\log \mathcal{K}_m}{x}
\end{eqnarray}
Notice that from this last equation we read the distribution of cavity fields as a function of $n$ and $\beta$
\begin{eqnarray}
  \label{eq:cavity_fields_pdf_pspin_vector}
  P_h(h;n,\beta)=\frac{ h^{m-1} \exp\left[-\frac{h^2}{2g'(q)}\right] \mathcal{K}_m(\beta
  h)^n}{\int_0^\infty dh\; h^{m-1} \exp\left[-\frac{h^2}{2g'(q)}\right] \mathcal{K}_m(\beta
  h)^n}
\end{eqnarray}
where $q=q_{*}$ to have a nontrivial result. Eq. \eqref{eq:cavity_fields_pdf_pspin_vector} prescribes the statistics of cavity fields for different families of meta-stable states. The Complexity reads
\begin{eqnarray}
    && \Sigma(n,\beta)=-\frac{n^2\beta^2}{2}[g(q)-q g'(q)]+\log\zeta-n\langle\;\log Y(\beta h)\;\rangle_{n} \\
    && \zeta=\int_0^\infty \frac{dh}{Z_m^{(0)}}\; h^{m-1} e^{-\frac{h^2}{2g'(q)}}\mathcal{K}_m(\beta h)^n
\end{eqnarray}
where $\langle\cdot\rangle_n$ is an average over \eqref{eq:cavity_fields_pdf_pspin_vector} and $q=q_{*}$. In figure \ref{fig:Sigma_eq_dyn_phase}, we show the equilibrium Complexity in the dynamical phase, for the pure p-spin models with $p=3, 4, 5$ and $m=4$ components. We see that the maximum complexity increases with $p$, as expected since the number of ways to obtain a stationary configuration grows when the coupling involves increasingly larger groups of spins. 

\subsection{The Replicon}

\begin{figure}
    \centering
    \includegraphics[width=0.8\columnwidth]{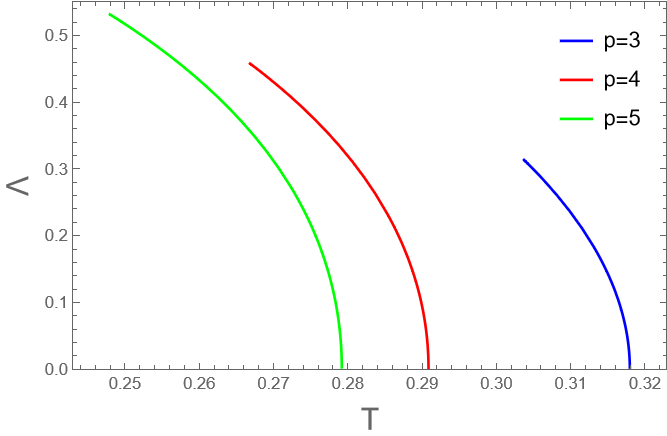}
    \caption{The replicon eigenvalue $\Lambda$ for the pure models with $m=4$ and $p=3$ (blue), $p=4$ (red) and $p=5$ (green). The replicon eigenvalue vanishes at $T_d$ as $(T_d-T)^{1/2}$.}
    \label{fig:Replicon_T}
\end{figure}

The stability of TAP states within the dynamical phase can be studied by means of the Replicon eigenvalue of the Hessian of the Replicated Action related to \eqref{eq:Hamiltonian_vector_pspin}. The elements of its Hessian read
\begin{equation*}
    D_{(ab)(cd)}\,=\,-\frac{\beta^2 g''(Q_{ab})}{2}[\delta_{(ab)(cd)}-g''(Q_{cd})(\langle\Vec{S}_a\cdot\Vec{S}_b\;\Vec{S}_c\cdot\Vec{S}_d\rangle-\langle\Vec{S}_a\cdot\Vec{S}_b\rangle\langle\Vec{S}_c\cdot\Vec{S}_d\rangle)].
\end{equation*}
To study the stability of the dynamical phase, we have to evaluate the Hessian on 1RSB saddle points with $Q_{ab}=q$ on the block diagonals and $Q_{ab}=0$ otherwise; in order to study the stability of equilibrium states in the dynamical phase, the dimension $x$ of the diagonal blocks must be sent to $1$.
The resulting Replicon eigenvalue is equal to
\begin{equation}
\label{eq:Replicon_pspin}
    \Lambda(\beta)\,=\,1-\left.\beta^2g''(q)\left[(m-1)\left\langle\frac{\mathcal{g}_m^2(\beta h)}{(\beta h)^2}\right\rangle+\left\langle(\mathcal{g}_m'(\beta h))^2\right\rangle\right]\right|_{q=q_{*}}
\end{equation}
where as usual $\mathcal{g}_m(x)=d\log\mathcal{K}_m(x)/dx$. We plot the Replicon eigenvalue in Figure \ref{fig:Replicon_T}, for $m=4$ and $p=3, 4, 5$. The Replicon eigenvalue has a square-root singularity close to the dynamical temperature. 

\section{Stable glasses}
\label{sec:stable_glasses}

In 1RSB-RFOT mean-field models with continuous variables the energy landscape usually features a low-lying band dominated by stable gapped minima and a high-energy band dominated by marginal gapless minima. In the pure p-spin spherical model, the marginal band corresponds to a single level $E=E_{mg}$, with saddles dominating the landscape for larger energies \cite{Ros2020saddles}. In the spherical mixed p-spin model, there is a wide band of marginal energy minima in an energy interval centred around the threshold energy. In this model, the $T=0$ relaxation dynamics in the marginal phase is richer in comparison to that of the pure p-sin model, featuring both memoryless and memorious relaxations according to the temperature at which the system has been prepared before the quench to zero temperature \cite{folena2020mixed}.

In this model we focus on the stable band of the energy landscape: at variance with spherical models, we show that the energy minima of our vector model represent mean field gapless stable glasses, meaning that within these basins linear response theory holds but at the same time excitation modes related to arbitrary small energies exist.

\subsubsection{Zero temperature limit of Monasson Potential}

To take the zero temperature limit, as usual, we fix $y=n/T$ for $n\rightarrow 0$ and $T\rightarrow 0$. This corresponds to fixing the slope of the Complexity of meta-stable states:
\begin{equation}
    \diff{\Sigma}{f}\,=\,\frac{n(f)}{T}=y
\end{equation}
which allows to have the Legendre relation between $\Sigma$ and $\Phi_0$ weel defined in the $T=0$ limit. This condition is satisfied for all equilibrium states that can be followed down to $T=0$, that is, those equilibrated in $T_K<T<T_{SF}$.

\begin{figure}
    \centering
    \includegraphics[width=0.8\columnwidth]{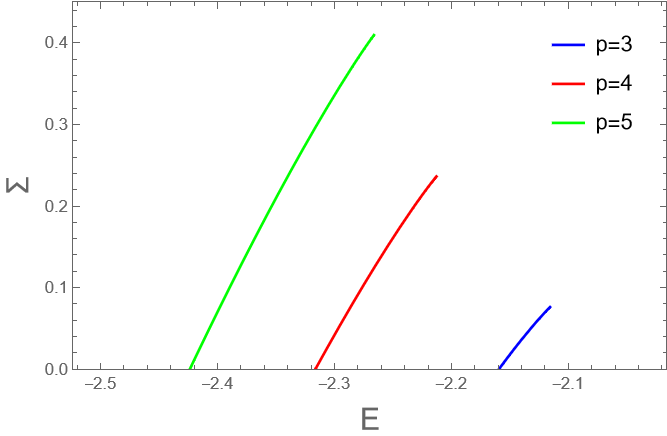}
    \caption{The complexity of the energy minima for the pure models with $m=4$ and $p=3$ (blue), $p=4$ (red) and $p=5$ (green). The maximum complexity is $\Sigma_{max}=0.0760961$ ($p=3$), $\Sigma_{max}=0.236176$ ($p=4$) and $\Sigma_{max}=0.409372$ ($p=5$). The number of stable minima is considerably larger that the number of states at $T_d$}
    \label{fig:complexity_energy_stable}
\end{figure}

The Monasson potential \eqref{eq:Monasson_Potential}, evaluated at $q=q_{*}$, in the $T=0$ limit becomes
\begin{equation}
  \label{eq:zero_T_Monasson_Potential}
\Phi_0(y)= \frac{1}{2} y^2 \left[g(1)-g'(1)\right]+\log \left[ \frac{\int_0^{\infty}dh\,h^{m-1}\exp \left(-\frac{h^2}{2 g'(1)}+y h\right)}{\int_0^{\infty}dh\,h^{m-1}\exp \left(-\frac{h^2}{2 g'(1)}\right)}\right]
\end{equation}
and the Complexity of energy minima is evaluated through a Legendre transform of minus \eqref{eq:zero_T_Monasson_Potential}:
\begin{equation}
\label{eq:Complexity_energy_minima}
    \Sigma(E)\,=\,y(E)E+\Phi_0(y(E))\qquad E(y)\,=\,-\diffp{\Phi_{0}}{y}
\end{equation}
The figure in \ref{fig:complexity_energy_stable} is the Complexity of pure p-spins with $p=3, 4, 5$. The complexity grows from the ground state $E=E_{gs}$, where it is zero, until the level $E=E_{mg}$ where marginal minima become dominant. The computation of the dominant branch of the Complexity in this regime requires RSB \cite{montanari2003nature, rizzo2013replica}. The complexity at $E=E_{mg}$ is larger than that of the states at $T=T_{d}$, consistently with the impossibility of following these states down to zero temperature.

\begin{figure}
    \centering
    \includegraphics[width=0.8\columnwidth]{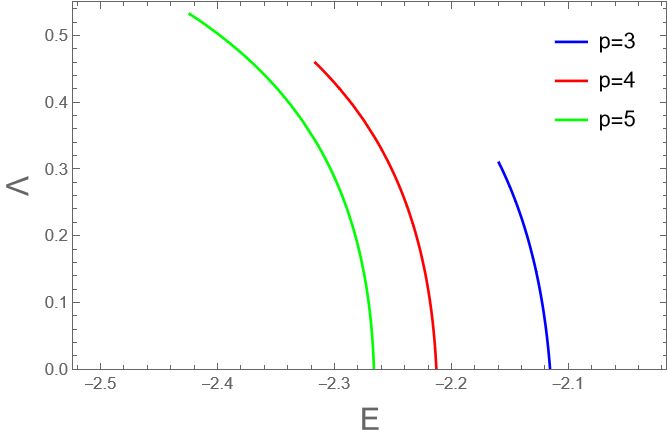}
    \caption{The replicon eigenvalue in the energy minima for the pure models with $m=4$ and $p=3$ (blue), $p=4$ (red) and $p=5$ (green). Notice that here the replicon eigenvalue vanishes as $E_{mg}-E$, although the slope is very large: we have $|\Lambda'(E_{mg})|\simeq 23, 82, 212$ respectively for $p=3, 4, 5$}
    \label{fig:Replicon_T0}
\end{figure}

The Replicon eigenvalue in the $T=0$ limit assumes the simpler form
\begin{eqnarray}
  \label{eq:Replicon_T0}
  \Lambda=1-(m-1)g''(1)\left\langle \frac{1}{h^2}\right\rangle
\end{eqnarray}
where the distribution of cavity field reads
\begin{eqnarray}
  \label{eq:cavity_field_pdf_zero_T_pspin}
  && P(h;y)=\frac{h^{m-1}}{Z_0(y)} \exp\left(-\frac{h^2}{2f'(1)}+y h\right) \\
  && Z^{(0)}(y)=\int_0^{\infty}dh\,h^{m-1}\,e^{-\frac{h^2}{2 f'(1)}+y h}.
\end{eqnarray}
The value of $y$ corresponding to $E=E_{gs}$ returns the Parisi distribution $P_0(1/m, h)$ inside the ground states, whereas trivially $P_0(0, h)=\delta(\Vec{h})/(2\pi)^{m/2}$. The parameter $y$ when manipulated in the interval $[y_{mg}, y_{gs}]$ corresponding to $[E_{gs}, E_{mg}]$ allows one to explore the statistics of different energy minima of the 1RSB part of the landscape, where $E(y)$ is a decreasing function of $y$. The energy level at which marginal states become dominant is evaluated from the nullity condition of the Replicon eigenvalue:
\begin{eqnarray}
  \label{eq:condition_Emg}
  \left. 1=(m-1)g''(1)\left\langle \frac{1}{h^2}\right\rangle\right|_{y=y_{mg}}
\end{eqnarray}
We show the Replicon eigenvalue as a function of the energy $E$ in figure \ref{fig:Replicon_T0} The Replicon eigenvalue is linear at the left of $E=E_{mg}$, at variance with the square root behavior found for the Replicon eigenvalue close to $T=T_d$.

For $E>E_{mg}$, the RSB theory is required to describe the distribution of cavity fields and thus study the stability.
In the stable phase, we see that $P_h(h)\propto h^{m-1}$ in the pseudo-gap for any level $E$. The dependence on the specific level in the pseudo-gap is all enclosed in the prefactor $1/Z_0(y)$. A weaker dependence is found at the tail of the distribution, where the linear term in the exponential in \eqref{eq:cavity_field_pdf_zero_T_pspin} modulates the decay of the distribution.
It is enough to show that the Hessian of Hamiltonian \eqref{eq:Hamiltonian_vector_pspin} is of the Rosenzweig-Porter ensemble as \eqref{eq:Hamiltonian_Vector_Models_Random_Fields} to conclude that our results found in chapter 4 extend to the excitations of the stable glasses of this model.

\subsection{Excitation spectra of stable energy minima}

The Hessian of the Hamiltonian \eqref{eq:Hamiltonian_vector_pspin} with spins constrained to have unit norm is given by
\begin{eqnarray}
\label{eq:Hessian_vector_pspin}
&& \mathbb{M}_{ij}=\mathbb{P}_i^{(\perp)}(\partial\partial\mathcal{\mathbb{H}}_{ij}+\mu_i\delta_{ij}\mathbb{I})\mathbb{P}_j^{(\perp)} \\
&& \nonumber \\
&& \partial\partial\mathcal{H}_{ij}^{\alpha\beta}\equiv \frac{\partial^2\mathcal{H}}{\partial S_{i}^{\alpha}\partial S_{j}^{\beta}} \\
&& \nonumber \\
&& \Vec{\mu}_i\,=\,-\frac{\partial\mathcal{H}}{\partial \Vec{S}_i}\qquad \mu_i=h_i+g''(1)\chi\qquad \chi=(m-1)\left\langle\frac{1}{h}\right\rangle
\end{eqnarray}
The off-diagonal part, containing interaction, now depends explicitly on the configuration: this dependence, however, has no effect. Indeed, the matrix $\partial\partial\mathcal{\mathbb{H}}$ in p-spin models is a Wigner matrix \cite{cavagna1998stationary, auffinger2013random}. To prove this, we have to evaluate the covariance functions
\begin{equation}
    W_{ijkl}^{\alpha\beta\gamma\delta}({\bf S},{\bf S'})\,=\,\overline{\frac{\partial^2\mathcal{H}}{\partial S_{i}^{\alpha}\partial S_{j}^{\beta}}}\overline{\frac{\partial^2\mathcal{H}}{\partial {S'}_{k}^{\gamma}\partial {S'}_{l}^{\delta}}}
\end{equation}
Clearly, the average value of each entry of $\partial\partial\mathcal{\mathbb{H}}$ is zero since it is a linear function of the couplings. With some effort, one finds the following covariance functions for the first and second derivatives of the Hamiltonian:
\begin{eqnarray}
    && \overline{\partial\mathcal{H}_{i}^{\alpha}\partial\mathcal{H}_{j}^{\beta}}\,=\,\delta_{ij}\delta_{\alpha\beta}g'(q) \\
    && \overline{\partial\partial\mathcal{H}_{ij}^{\alpha\beta}\partial\partial\mathcal{H}_{kl}^{\gamma\delta}}\,=\,\delta_{(ij)(kl)}\delta_{(\alpha\beta)(\gamma\delta)}\frac{g''(q)}{N}
\end{eqnarray}
Therefore, given that the entries of the interaction matrix have zero mean and finite variance $O(1/N)$, for $N\rightarrow\infty$ the off-diagonal matrix is a Wigner matrix with semi-circular spectrum. Thus, the Hessian of the vector p-spin Hamiltonian is a Rosenzweig-Porter matrix. In the stable phase, since we know the statistics of the strengths of the local fields $\mu_i$, we are able to analytically solve the spectrum, following the same methods as in Chapter 4. We compute the spectral density as usual from the solution of the self-consistent equation for the resolvent, written in its real and imaginary part
\begin{eqnarray}
    && \Re \mathcal{G}(\lambda)\,=\,(m-1)\left\langle\frac{h+\Re x(\lambda)}{(h+\Re x(\lambda))^2+\Im x(\lambda)^2}\right\rangle \\
    && 1\,=\,g''(1)(m-1)\left\langle\frac{1}{(h+\Re x(\lambda))^2+\Im x(\lambda)^2}\right\rangle
\end{eqnarray}
with $x(\lambda)=g''(1)[\mathcal{G}(0)-\mathcal{G}(\lambda)]-\lambda$ and $\rho(\lambda)=\frac{\pi}{g''(1)(m-1)}|\Im x|$. The only difference is that the distribution of cavity fields is controlled not by an external control parameter (the external field) but by an internal one (the energy level we consider). 

\subsection{Localised soft excitations in the whole stable phase}

\begin{figure}
    \centering
    \includegraphics[width=0.5\columnwidth]{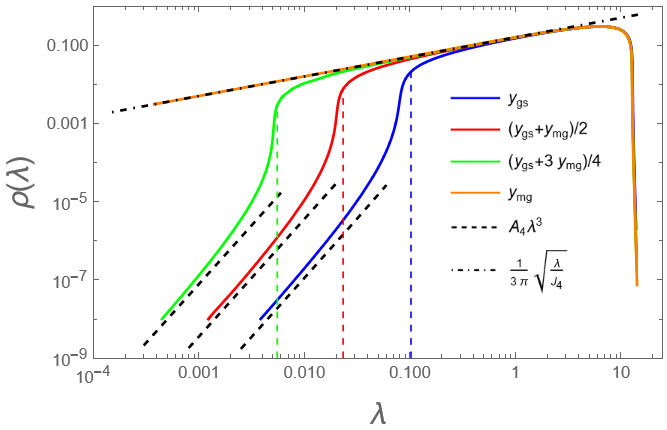}
    \includegraphics[width=0.5\columnwidth]{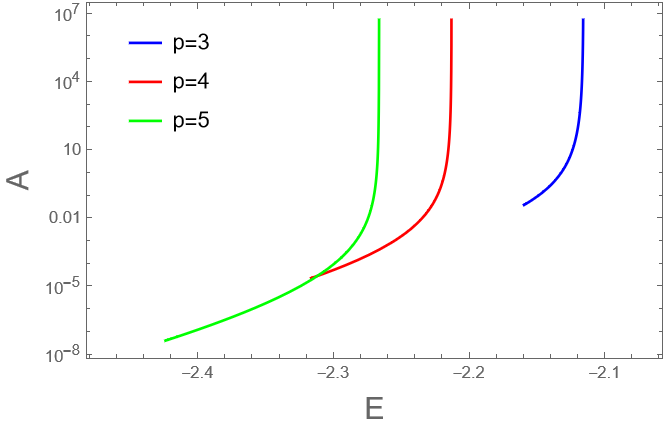}
    \includegraphics[width=0.5\columnwidth]{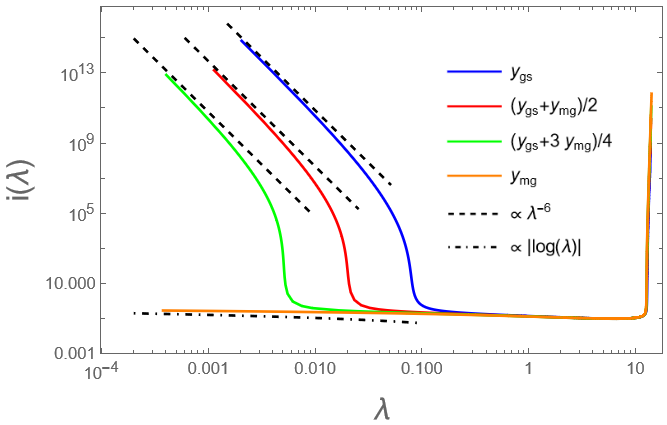}
    \caption{\textbf{Top}: The spectrum of the Hessian in log-log scale for $m=4$ and $p=3$. The curves for $y<y_{mg}$ cross-over from a $\lambda^3$ behavior to a $\sqrt{\lambda}$ behavior at $\lambda_*$ marked by coloured vertical dashed lines. In the bulk of the spectrum, the spectral density does not depend on $y$. \\
    \textbf{Center}: The prefactor $A_4$ of stable glassy minima is smaller for better optimized glasses. The dependence on the energy level $E$ is very strong for high values of $p$: even far from $E_{mg}$ this quantity varies by several order of magnitudes. \\
    \textbf{Bottom}: The scaled bulk inverse participation ratio $i(\lambda)$ as a function of $\lambda$ for $m=4$ and $p=3$ on a log-log scale. Notice the different behavior between the stable minima and the marginal one. The curve at $y_{mg}$ diverges logarithmically, while the other curves behave as $\lambda^{-6}$ for $\lambda\to 0$.}
    \label{fig:spectrum_stable_phase}
\end{figure}

Using the results of Chapter 4, we can straightforwardly write the behaviour of the spectral density close to the lower edge
 \begin{equation}
 \label{eq:res_eqs_pspin}
\begin{split}
    &\rho(\lambda)\simeq\begin{cases}\,&A(E)\lambda^{m-1}\qquad\,\lambda\ll\lambda_* \\
        & \\
        & \sqrt{\lambda-\lambda_*(E)}\qquad\,\lambda\sim\lambda_*
    \end{cases}
        \nonumber \\
        & \nonumber \\
        & \lambda_*(E)\,\sim\,\Lambda^2\propto (E_{mg}-E)^2 \nonumber \\
         &
    \nonumber
\end{split}
\end{equation}
The spectral density behaves as a power law in the pseudo-gap $\lambda\ll\lambda_*$. At the marginal energy level $E=E_{mg}$ the spectral density changes behavior to a Wigner law, and the Hessian undergoes a delocalisation transition. In figures \ref{fig:spectrum_stable_phase} we show the spectral densities of the $m=4$ and $p=3$ system corresponding to values of $y$ in $y_{mg}\leq y\leq y_{gs}$ (top), the coefficient of the spectral density for $m=4$ and $p=3, 4, 5$ (center) and the rescaled IPR $i(\lambda)=NI_2(\lambda)$ (from eq. \eqref{eq:eigvec_moments}) for $m=4$, $p=3$ and $y$ in $y_{mg}\leq y\leq y_{gs}$ (bottom). The curves shown are theoretical curves obtained by numerically solving the resolvent equations \eqref{eq:res_eqs_pspin}. Note that the theoretical curves of $i(\lambda)$ report the bulk prediction \eqref{eq:bulk_IPR_Hessian_SG}: for any $N$, at sufficiently low $\lambda\ll \lambda_*$ localisation effects take place.
We remark that the spectral density and the IPRs seem to be independent of $y$ in the bulk. This property is probably related to the fact that the variance of the $P_h(h)$ has a weak dependence on $y$ in the physical interval $[y_{mg},y_{gs}]$. The dependence on the energy level is mostly in the lower edge, through the pre-factor 
\begin{equation}
\label{eq:pre-factor-rho}
    A(E)=\frac{1}{Z^{(0)}(E)\Lambda^m(E)}
\end{equation}
where $Z^{(0)}(E)$ is the normalisation of the cavity pdf in \eqref{eq:cavity_fields_pdf_pspin_vector}, written as a function of $E$. In the central figure of \ref{fig:spectrum_stable_phase}, we show the dependence of $A$ on the energy for $p=3, 4, 5$ and $m=4$. The prefactor is an increasing function of the energy, implying that low-lying minima in the landscape are more stable and with modes in the pseudo-gap increasingly localised. This property has been observed in numerical simulations of three-dimensional computer glasses: better optimised samples, corresponding to low energy minima of the landscape, are depleted of soft excitations, in corrispondence of a lower value of the prefactor \cite{ji2020thermal, ji2021geometry}. 

\section{Ultra-stable glasses}
\label{sec:ultra_stable_glasses}

Ultra-stable glasses are non-crystalline materials which have enhanced thermal and mechanical stability properties \cite{swallen2007organic, kearns2007influence, perez2014suppression, yu2015suppression, luttich2018anti, rodriguez2022ultrastable}, with respect to ordinary glasses obtained by annhilation from the super-cooled liquid state. An ultra-stable glass is usually prepared through a physical vapour deposition process: inside a vacuum chamber, the original system is deposited at the gaseous state over a thin film of a different material, which is in contact with a substrate material that acts as thermal bath.
Discovered in 2007 by Swallen et al. \cite{swallen2007organic, kearns2007influence}, the study of these materials has more recently led to successful numerical studies \cite{singh2013ultrastable, fullerton2017density, Kaptjeins2019fastgenultrast, Parmar2020}, confirming the hypothesis that these glasses correspond to deep minima of the potential energy landscape. In particular, numerical glassy minima obtained through a generalisation of Swap Monte-Carlo techniques \cite{BerthierSwapHS2016, BerthierNinnarelloSwap2017} in \cite{Kaptjeins2019fastgenultrast} show that the DoS of ultra-stable glasses features a gap, whose amplitude is the greater the more polydisperse the glass former.

In this section, we provide mean-field representatives of ultra-stable glasses. We consider energy minima with holed cavity field distributions: we introduce a \emph{cavity gap} $h_0$
\begin{equation}
\label{eq:cavity_gap}
    \mu_i\,=\,h_i+g''(1)\chi+h_0.
\end{equation}
This quantity enhances the stability of the system: indeed, as we learnt so far, in stable phases low energy excitations are driven by the statistics of small local fields, therefore a cavity gap induces a gap in the spectrum of linear excitations. Last but not least, the linear response of the system in these states will necessarily be smaller than that of gapless minima. As we will show in next section, our model features a non-zero complexity of ultra-stable minima, which however is always lower than that of gapless minima.

\subsection{Complexity of energy minima}

In order to confirm these predictions, we have to compute the complexity of energy minima satisfying the constraint \eqref{eq:cavity_gap}. We do this using the Kac-Rice formula \cite{berzin2022kac}, a standard technique for complexity calculation, used for the first time in the context of spin glasses in \cite{bray1980metastable}. We define
\begin{eqnarray}
\label{eq:Mon_Pot_cav_gap_Kac_Rice}
  \Phi_0(y, h_0)=\log\overline{\int_{h_i>h_0} d{\bf S} d {\boldsymbol{\mu}}\;e^{-y\mathcal{H}}
  \prod_{i,\alpha}\delta\left(
\partial\mathcal{H}_i^\alpha-\mu_i S_i^\alpha\right)\left| \det\left( \partial\partial\mathbb{\mathcal{H}}-\operatorname{diag}(\boldsymbol{\mu})\right)\right|}
\end{eqnarray}
This 'free energy' is computed through a partition function that weights the stationary points of the Hamiltonian with the parameter $y$: the ground state is given by 'low temperatures' $y\rightarrow y_{gs}^{-}$, whence marginal states are approached in the opposite limit $y\rightarrow y_{mg}^{+}$. The delta functions enforce the configurations and the local field to satisfy the stationary configuration condition $\partial\mathcal{H}_i^\alpha=\mu_i S_i^\alpha$, the modulus of the Hessian determinant is the usual Jacobi volume factor that accompanies the constraint. While this formula is generally valid, we added an additional constraint on the integration, $h_i>h_0$ for any $i=1,\dots,N$, in order to unveil the presence of ultra-stable minima. The use of the annhealed approximation is justified by the 1RSB form of the energy landscape.

In order to compute a complexity of energy minima, we shall consider solutions with a strictly positive determinant of the Hessian. If we remove the modulus from \eqref{eq:Mon_Pot_cav_gap_Kac_Rice}, we can straightforwardly carry on the computation, following the steps of Bray and Moore \cite{bray1980metastable}.
The computation with the signed determinant, however, is equivalent to computing the number of solutions weighted with their Morse index\footnote{The number of negative directions of the eigenspace: stationary points with an even number of these feature a positive determinant, the others negative.}, including also saddles in the calculations. Luckily, previous results on p-spin models ensure us that there are no saddles in the stable band, so we can steadily proceed with the computation. We commence with the determinant: by exploiting self-averageness, we can evaluate its disorder average separately, writing
\begin{eqnarray}
   & \overline{\left|\det\left( \partial\partial\mathbb{\mathcal{H}}-\text{diag}(\mu)
  \right)\right|}\,=\,\overline{\det\left( -\partial\partial\mathbb{\mathcal{H}}+\text{diag}(\boldsymbol{\mu})
  \right)}\nonumber\\
  & =\left\{\overline{\int \frac{d\boldsymbol{X}}{(2 \pi )^{\frac{N\,(m-1)}{2}}}\,\exp\left[-\frac{1}{2}\boldsymbol{X}^T\cdot(-\partial\partial\mathcal{H}+\text{diag}(\mu))\boldsymbol{X}\right]}\right\}^{-2}.\nonumber
\end{eqnarray}
where we introduced the gaussian representation of the determinant. By averaging over the couplings, performing a Hubbard-Stratonovich transformation and subsequently applying the saddle point method, we end up with
\begin{eqnarray}
\label{eq:det}
  \overline{\left|\det\left( \partial\partial\mathcal{H}-\text{diag}(\mu)
  \right)\right|}=e^{\frac{Ng''(1) w^2}{2 }}\prod_i [\mu_i-g''(1) w]^{m-1}
\end{eqnarray}
where $w$ is given by the solution of the saddle point equation 
\begin{eqnarray}
\label{eq:chih0}
  w=\frac 1 N \overset{N}{\underset{i=1}{\sum}} \,\frac{(m-1)}{\mu_i-g''(1)w}
\end{eqnarray}
Notice that by setting $\mu_i\,=\,h_i+g''(1)w$, this last equation can be rewritten for $N\rightarrow\infty$ as
\begin{eqnarray}
\label{eq:w}
    w\,=\,(m-1)\,\int_{h_0}^{\infty}\,dh\,\frac{P_{h_0}(h)}{h},
\end{eqnarray}
so that $w$ is just a 'cut' version of the $h_0=0$ susceptibility. To distinguish it from this quantity, from now on we will write \eqref{eq:w} $w\equiv \chi_{h_0}$, introducing the cut susceptibility.

The disorder average over the delta function times the 'Boltzmann weight' in \eqref{eq:Mon_Pot_cav_gap_Kac_Rice} is performed following standard paths
\begin{eqnarray}
    & \overline{\int\,d\boldsymbol{S}\,e^{-y\mathcal{H}}\prod_{k, \alpha}\,\delta(\partial\mathcal{H}_k^{\alpha}+\mu_k\,S_k)}\,=\nonumber \\
    & \int\,d\boldsymbol{S}d\boldsymbol{\hat{S}}\,\exp{-i\sum_{k=1}^N\,\mu_k\,\Vec{S}_k\cdot\Vec{\hat{S}}_k}\nonumber \\
    &\nonumber \\
    &\times\overline{\exp\left[{-\left(i\sum_{k,\alpha}\,\hat{S}_k^{\alpha}\frac{\partial}{\partial S_k^{\alpha}}+y\right)\mathcal{H}}\right]}\nonumber
\end{eqnarray}
where the $\boldsymbol{\hat{S}}$ are Lagrange multipliers introduced by the Fourier Representation of the delta function. After the average over the disorder and one Hubbard-Stratonovich transformation this last expression becomes
\begin{eqnarray}
\label{eq:fromDelta}
&&\left[
  \frac{1}{\Gamma(m/2)g'(1)^{m/2}}
  \right]^N\exp\Bigg[\frac 1 2 Ny^2g(1)-N\frac { g''(1) u^2}{2} \\
  &&-\sum_i\frac {1}{2g'(1)}[yg'(1)+g''(1) u-\mu_i]^2\Bigg]\nonumber
\end{eqnarray}
with $u$ given by the saddle point equation 
\begin{eqnarray}
\label{eq:u}
  u=\frac{1}{g'(1)N}\overset{N}{\underset{i=1}{\sum}} [\mu_i-yg'(1)-g''(1) u]\Longrightarrow u\,=\,\frac{\overline{\mu}-g'(1)y}{g'(1)+g''(1)}.
\end{eqnarray}

\begin{figure}
     \centering
     \includegraphics[width=0.45\columnwidth]{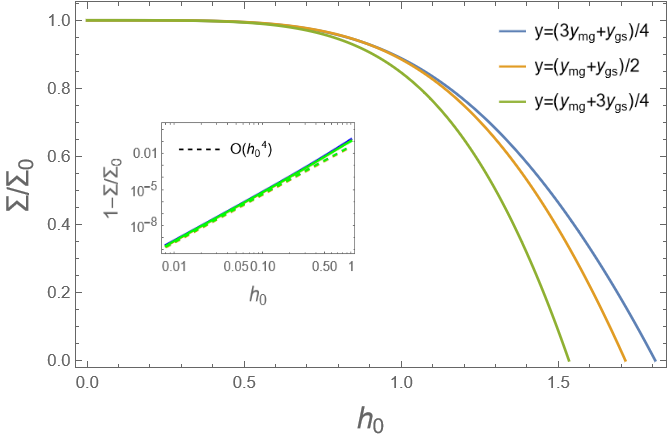}
     \includegraphics[width=0.45\columnwidth]{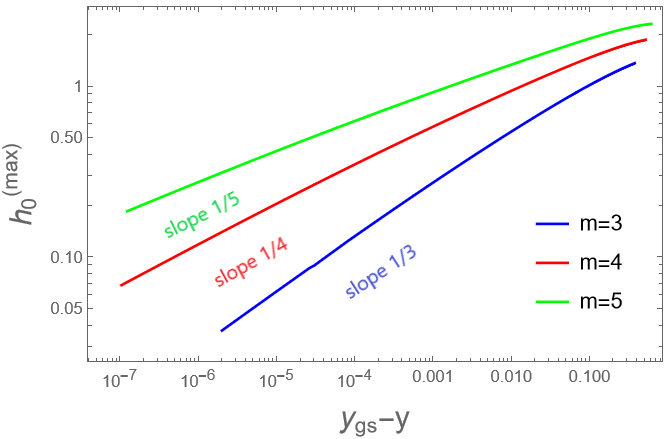}
     \includegraphics[width=0.45\columnwidth]{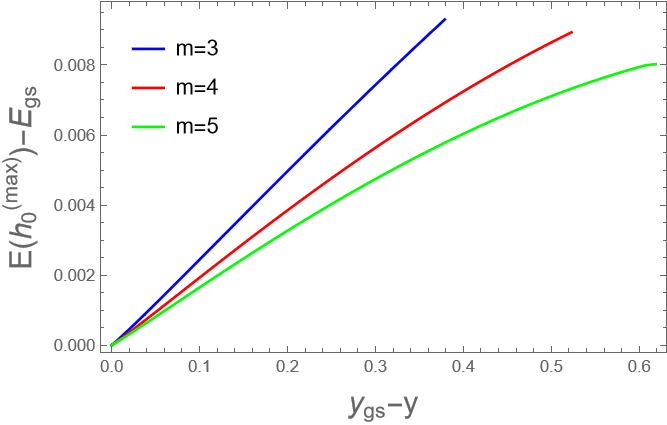}
     \caption{\textbf{Top-left}: the Complexity of gapped minima, normalised with that of ungapped minima, for the values of $y$ reported in legend. \textbf{Top-right}: the maximal cavity gap as a function of the distance from the ground state. This quantity close to this point is singular. \textbf{Bottom}: the energy of the gapped minima as a function of the position $y$ in the landscape. The are no gapped minima at the ground state level.}
     \label{fig:Sigma_Tzero_gapped}
 \end{figure}

Putting \eqref{eq:det} and \eqref{eq:fromDelta} together and setting $h_i=\mu_i-g''(1)\chi_{h_0}$ we finally write
\begin{eqnarray}
\label{eq:g0}
&&{\Phi_0(y; h_0)= \frac{y^2}{2}[g(1)-g'(1)]-\frac{g''(1)}{2 g'(1)}(\chi_{h_0}-u)^2} \nonumber\\
&& {-g''(1) y (u-\chi_{h_0})-\frac{g''(1)}{2}(u^2-\chi_{h_0}^2)}+\ln I(y; h_0) \\
&& \nonumber \\
&& {I = \frac{\int_{h_0}^{\infty}dh\,h^{m-1}e^{-\frac{h^2}{2 g'(1)}+\frac{h}{g'(1)}[g''(1)(u-\chi_{h_0})+yg'(1)]}}{\int_0^{\infty}\,dh\,h^{m-1}e^{-\frac{h^2}{2g'(1)}}}} \nonumber
  \nonumber
\end{eqnarray}
Notice that the cavity field probability distribution
\begin{equation}
\label{eq:cavFpdfh0}
    P_{h_0}(h)\,=\,\frac{\theta(h-h_0)}{Z(y; h_0)}h^{m-1}e^{-\frac{h^2}{2 f'(1)}+\left[y+f''(1)\frac{(u-\chi_{h_0})}{f'(1)}\right]h}
\end{equation}
for $h_0>0$ has a finite cut on the lower edge, that is, $P_{h_0}(h_0)>0$, and is reweighted in the exponential through the coefficient $y(h_0)\,=\,y+\frac{f''(1)(u-\chi_{h_0})}{f'(1)}$. Different families of ultra-stable minima can be studied by varying $y$ and $h_0$. The energy level and the Complexity at a given $y$ and $h_0$ are obtained from \eqref{eq:Complexity_energy_minima} adapted to the presence of a cavity gap:
\begin{equation}
\label{eq:Complexity_energy_minima_with_cav_gap}
    \Sigma(y, h_0)\,=\,y E(y, h_0)+\Phi_0(y; h_0)\qquad E(y; h_0)\,=\,-\diffp{\Phi_{0}(y, h_0)}{y}
\end{equation}
Let us begin by considering the complexity in the small $h_0$ limit.
The Complexity of our model has a quite lengthy expression (see Appendix \ref{sec:compl_us_minima_app}), which however simplifies a lot when expanded at small $h_0$. The result of the expansion is
\begin{equation}
 \label{eq:SigmaGapSmallh0}
     \Sigma=\,\Sigma_0-\left[\frac{1+y\,\langle h\rangle_0}{m\,Z_0}\right]h_0^m+O(h_0^{m+1})
 \end{equation}
 which shows that the complexity of ultra-stable minima is a decreasing function of $h_0$. The difference between the Complexity of dominant gapless minima and gapped ones is small for small cavity gaps, $\Delta\Sigma=O(h_0^{m})$. In figure \ref{fig:Sigma_Tzero_gapped} we report on the left the Complexity of the $p=3$, $m=4$ system for three different values of $y$ in $[y_{mg},y_{gs}]$. In the inset, a zoom in on the small $h_0$ region, showing the $O(h_0^4)$ behaviour.

 The plot on the right is the value of $h_0$ at which the complexity, for the chosen level $y$ vanishes. It is shown that this maximal cavity gap behaves singularly as $h_0^{(max)}(y)\sim (y_{gs}-y)^{1/m}$ close to the ground state level.
 This singularity can be estimated from \eqref{eq:SigmaGapSmallh0}, by setting $\Sigma=0$ on the l.h.s. and expanding $\Sigma_0$ close to $y-y_{gs}$. Since the gapless minima complexity is linear on the left of $y_{gs}$, we get the estimate
 \begin{eqnarray}
    &\left[\frac{1+y\,\langle h\rangle_0}{m\,Z_0}\right](h_0^{max})^m\simeq \frac{d\Sigma_0}{dy}(y_{gs})(y_{gs}-y)\nonumber\\
    & h_0^{(max)}\simeq A\,(y_{gs}-y)^{1/m} \\
    & A\,=\,\left[(m\,Z_0)\frac{\Sigma_0'(y_)}{1+y\langle h \rangle_0}\right]^{1/m}\Bigl |_{y=y_{gs}}.
\end{eqnarray}
We conclude this subsection by showing in the bottom figure of \ref{fig:Sigma_Tzero_gapped} the energy at the maximal cavity gap as a function of $y$. The energy is equal to the ground state at $y=y_{gs}$, as it should. This self-consistently checks the validity of our results.

\subsection{Response function of ultra-stable glasses}

In complexity computations \emph{à la Bray-Moore} usually one of the order parameters of the problem is the susceptibility inside a solution, when this is positive definite. In our problem, we have two candidates: the cut susceptibility $\chi_{h_0}$ and the order parameter $u$, given by \eqref{eq:u}. When $h_0=0$, by integrating by parts \eqref{eq:u}, one immediately realises that these two quantities are the same, $u=\chi$. However, when $h_0>0$ the true susceptibility is the function $u$. Eq. \eqref{eq:u}, after an integration by parts, reads
\begin{equation}
\label{eq:u_bis}
    u\,=\,\chi_{h_0}+P_{h_0}(h_0)
\end{equation}
which is a self-consistent equation since $P_{h_0}(h)$ depends on $u$. Note that the existence of the cut $P_{h_0}(h_0)$ for $h_0>0$ implies that the coefficient of the linear term in \eqref{eq:cavFpdfh0} is strictly positive: $u>\chi_{h_0}$ from \eqref{eq:u_bis}. This is equivalent to consider a shifted level $y(h_0)=y+(u-\chi_{h_0})/g'(1)>y$, corresponding to a lower energy in the landscape and thus to increased stability.

One can see that $u$ is the true response function by direct computation of this quantity. Suppose to perturb the system with an external field $\Vec{\epsilon}_i$ on each site: the static linear response function is given by
\begin{eqnarray}
\label{eq:response}
    & \mathcal{R}\,=\,\frac{1}{N}\sum_{i,\alpha}R_{ii}^{\alpha\alpha} \\
    & R_{i j}^{\alpha \beta}\,=\,\frac{\partial \overline{\langle S_i^{\alpha}  \rangle}}{\partial \epsilon_j^{\beta}}\Bigl|_{\epsilon=0}
\end{eqnarray}
Here $\langle\cdot\rangle$ is an average according to Kac-Rice-Moore measure:
\begin{eqnarray}
\label{eq:KRM}
    P_{KRM}\,\propto\,e^{-y{\cal H}}
  \prod_{i,\alpha}\delta\left(
{\cal H}_i^{\alpha'}-\mu_i S_i^\alpha
\right)\left| \det\left( H''-\text{diag}(\mu)\right)\right|.
  \nonumber
\end{eqnarray}
The full computation of \eqref{eq:response}, where we show that $\mathcal{R}=u$, can be found in Appendix \ref{sec:resp_US_comp}.
In figure \ref{fig:u_VS_h0} we show the response function $u$ of the $p=3$, $m=4$ system as a function of $h_0$, for some values of $y$. At small $h_0$, its expression is
\begin{equation}
\label{eq:u_small_h0}
    u\,=\,\chi-\frac{1}{Z_0}[(m-1)(m-2)-1]h_0^{m-1}+O(h_0^m).
\end{equation}

\begin{figure}
    \centering
    \includegraphics[width=0.6\columnwidth]{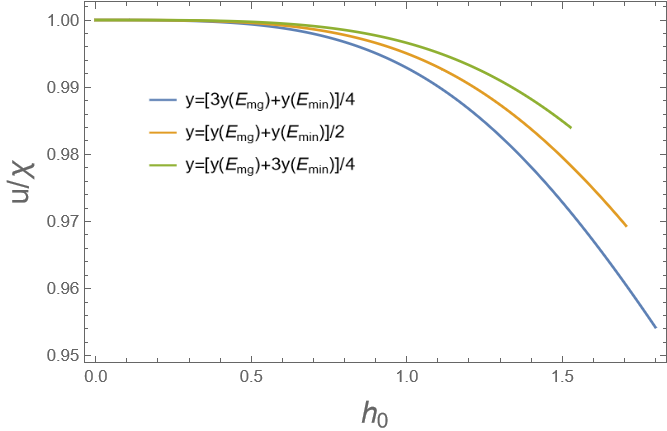}
    \caption{The response function of ultra-stable minima, normalised with the susceptibility of gapless minima $\chi(y)$, for the $m=4$, $p=3$ system and values of $y$ in legend.}
    \label{fig:u_VS_h0}
\end{figure}

\subsection{Spectra of ultra-stable glasses}

Ultra-stable minima have gapped spectra. The spectral gap $\lambda_0$ can be calculated from the resolvent equations \eqref{eq:resolvent_eq_Hessian_SG_re_and_im} in the limit $\lambda\rightarrow\lambda_0^{+}$
\begin{eqnarray}
\label{eq:gapSpeceqs_1}
& 1\,=(m-1)g''(1)\,\int_{h_0}^{\infty}dh\,\frac{P_{h_0}(h)}{[h+\Re x(\lambda_0)]^2} \\
& \nonumber \\
\label{eq:gapSpeceqs_2}
& \lambda_0\,=\,(m-1)(\Re x(\lambda_0))^2\,\int_{h_0}^{\infty}dh\,\frac{P_{h_0}(h)}{h\,[h+\Re x(\lambda_0)]^2}
\end{eqnarray}
where we remind that $x(\lambda)=g''(1)[\mathcal{G}(0)-\mathcal{G}(\lambda)]-\lambda$. We solve Eq. \eqref{eq:gapSpeceqs_1} and plug its solution into \eqref{eq:gapSpeceqs_2}. Notice that \eqref{eq:gapSpeceqs_2} in this form can be obtained with few manipulations of the equation of the real part in \eqref{eq:resolvent_eq_Hessian_SG_re_and_im}.

\begin{figure}
    \centering
    \includegraphics[width=0.5\columnwidth]{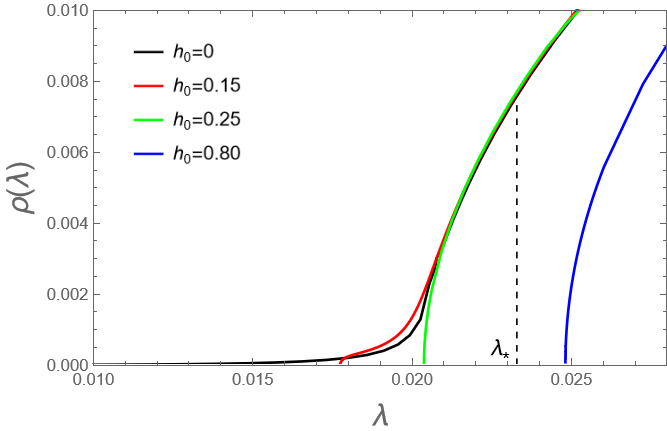}
    \includegraphics[width=0.5\columnwidth]{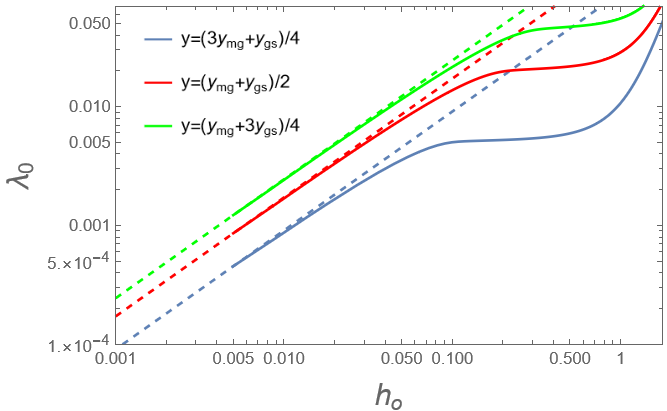}
    \includegraphics[width=0.5\columnwidth]{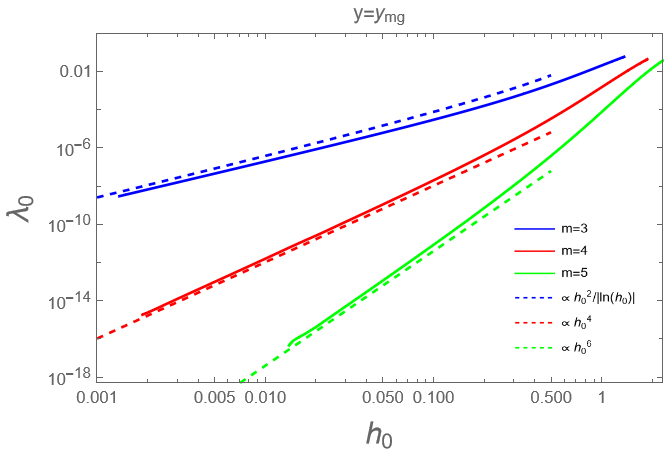}
    \caption{\textbf{Top}: Spectral properties in presence of a cavity gap $h_0$, for the $m=4$ and $p=3$ pure p-spin at $y=(y_{gs}+y_{mg})/2$. The spectral density of gapless minima is compared to that of minima with cavity gaps $h_0=0.15, 0.25, 0.8$. The dashed vertical line marks the position of the crossover $\lambda_*\equiv \frac{\Lambda^2}{4\langle 1/h^3\rangle}$.
     \newline
     \textbf{Center}: The relation between the spectral gap and the cavity gap for the three values of $y\in[y_{mg}, y_{gs}]$, the dotted lines are $\Lambda(y)\,h_0$. 
     \newline
     \textbf{Bottom}: The spectral gap at the critical point $y=y_{mg}$ for $m=3, 4, 5$: the scaling provided by eq. \eqref{eq:gap_spec} is verified. Marginal minima develop extremely small gaps in a broad range of values of $h_0$.}
    \label{fig:spectra_US_glasses}
\end{figure}

The spectral gap for small $h_0$ behaves as
\begin{eqnarray}
 \label{eq:gap_spec}
   \lambda_0=
   \begin{cases}
   \Lambda\,h_0+O(h_0^2)\qquad\,y>y_{mg} \\
   \\
   \left[\frac{g''(1)(m-1)}{4\, (m-2)^2\,Z_0^2\,\langle 1/h^3 \rangle_0}\right] h_0^{2(m-2)}+O(h_0^{2(m-1)})\qquad\,m>3,\quad y=y_{mg} \\
   \\
   \left[\frac{g''(1)}{2\, \,Z_0}\right] \frac{h_0^{2}}{|\ln h_0|}+O(h_0^{4})\qquad m=3,\quad y=y_{mg}.
   \end{cases} \nonumber
 \end{eqnarray}
The result at $y>y_{mg}$ tells us that when $h_0$ is small, the spectrum of ultra-stable minima is obtained by cutting off excitations $O(h_0)$ from the gapless spectrum. The prefactor is the expected one since for $h_0$ small the spectral gap is small and \eqref{eq:condensation_relation_correct_eig_m} holds. We show in the central figure in \ref{fig:spectra_US_glasses} the spectral gap for the $m=4$, $p=3$ system and some values of $y$ as shown in legend. The spectral gap has a linear behavior at small cavity gap and saturates for higher values of $h$, being roughly constant in a wide interval before having a final steep growth close to $h_0=h_0^{(max)}(y)$. In general, it appears that the spectral gaps are numerically very small if compared with the cavity gap: for instance, for $y=(y_{mg}+y_{gs})/2$ we have $h_0^{(max)}\simeq 1.71$ and $\lambda_0^{(max)}\simeq 0.07$. The inertia to develop a spectral gap is even more pronounced in the vicinity of the marginal level: in the bottom figure of \ref{fig:spectra_US_glasses} we show the spectral gap as a function of $h_0$, comparing it with the predictions in \eqref{eq:gap_spec} for $m=3, 4, 5$. The different order of magnitudes spanned by the spectral gaps in the $y<y_{mg}$ and $y=y_{mg}$ cases is remarkable.

The spectral density of ultra-stable minima is Wigner-like sufficiently close to $\lambda_0$
\begin{equation}
\label{eq:spectral_density_us}
    \rho(\lambda)\sim\sqrt{\lambda-\lambda_0}
\end{equation}
Indeed, the existence of the spectral gap necessarily requires (see \eqref{eq:gapSpeceqs_1} and remember that $P_{h_0}(h_0)>0$)
$|h_0+\Re x(\lambda_0)|>0$ as a necessary condition, and as a consequence no condensation can occur in ultra-stable minima. Indeed, in order to observe a crossover to a power-law behaviour, one should see $|h+x|=O(N^{-1/2})$ in the limit $\lambda\rightarrow \lambda_0$. We show the spectral density close to the lower edge for $m=4$, $p=3$, $y=(y_{mg}+y_{gs})/2$ and $h_0=0.0, 0.15, 0.25, 0.80$ in the upper figure of \ref{fig:spectra_US_glasses}. When $\lambda_0\ll \lambda_*$, the crossover eigenvalue of the $h_0=0$ system, the spectral density of the ultra-stable minima tends to follow the spectral density of the related gapless minimum, departing from it only in the immediate surroundings of $\lambda_0$.
The excitations of ultra-stable minima close to the lower edge are completely delocalised, $I_2(\lambda_0)\propto 1/N$ with no multi-fractal behavior. Through formula
\begin{equation}
    I_2(\lambda)\,=\,\frac{3(m-1)g''(1)}{N}\int \frac{P_{h_0}(h)}{|h+x(\lambda)|^2}
\end{equation}
one can show that for small $h_0$ the rescaled IPR $i(\lambda)=N I_2(\lambda)$ mimes the divergence of the bulk IPR of gapless minima close to $\lambda=0$. Indeed, we find the following.
\begin{equation}
   \begin{cases}
   i_2(\lambda)\sim h_0^{-2\,(m-1)},\,\,\,y>y_{mg} \\
   i_2(\lambda)\sim 1/\sqrt{h_0},\,\,\,y=y_{mg},\,\,\,m=3 \\
   i_2(\lambda)\sim |\ln h_0|,\,\,\,y=y_{mg},\,\,\,m=4 \\
   i_2(\lambda)\sim const,\,\,\,y=y_{mg},\,\,\,m\geq 5.
   \end{cases} 
\end{equation}
to be compared with equations \eqref{eq:incosistent_scalings}, \eqref{eq:critical_eigvec_modes}, \eqref{eq:critical_eigvec_modes1}, \eqref{eq:critical_eigvec_modes2} with $q=2$. The fact that excitations are completely delocalised is a direct consequence of the cut on the distribution of cavity fields, $P_{h_0}(h_0)>0$, which implies that in ultra-stable minima there are always finite fractions of spins in the thermodynamic limit feeling cavity fields $O(h_0)$.

\section{Conclusions}

In this chapter, we studied a toy model of structural glasses, the vector p-spin model. We consider the problem of linear excitations of energy minima belonging to the stable energy band of the landscape. We found that typical minima feature gapless spectra with localised excitations, following the condensation mechanism described in Chapter 4. At the top of stable band marginal minima become dominant and excitations delocalise. In the stable band, in addition to stable gapless energy minima, there are rare stable and gapped energy minima, named by us in this chapter ultra-stable energy minima. These minima constitute a mean field representative of ultra-stable glasses, disordered materials obtained by circumventing the usual supercooling through a physical vapour deposition process.
We believe that this mean-field model constitutes an improvement with respect to spherical p-spin models, for it is able to reproduce localisation mechanisms which are present in real glassy system and absent in spherical models.

An interesting path for future research would be to study its dynamics, extending the studies made in \cite{folena2020mixed} for mixed p-spin models.

\part{Vector spin glasses on sparse random graphs}

\chapter{The sparse vector spin-glass}

We learnt from the last two chapters that it is possible to observe in fully connected first-principle mean field models low energy excitation modes with localisation features. The natural development of our previous work is to consider mean-field models defined in finitely connected random graphs. The study of magnetic systems in the Bethe lattice began over eighty years ago: Peierls studied the Ising ferromagnet \cite{peierls1936ising}, weiss the Heisenberg ferromagnet \cite{weiss1948application}, while Ziman and Li the Ising and Heisenberg antiferromagnets respectively \cite{ziman1951antiferromagnetism, li1951application}. For long time, Bethe lattice theories were considered to be just a refinement of naive mean field theories for quantitative predictions. The tipping point is due to the work of Anderson, Thouless and Abou-Chakra \cite{abou1973selfconsistent}, who pioneered the study Anderson transition on the Bethe lattice.
With years passing, the interest in network models in physics has flourished.

In particular, in the field of spin glasses in the last decade, much effort has been put into understanding the Bethe lattice SK model \cite{viana1985phase, mezard2001bethe, parisi2014diluted, parisi2017marginally, Perrupato2022, Angelini2022, angelini2023real}. One of the most fascinating idea is that the $T=0$ Bethe lattice theory of SK model is the correct mean field limit of the finite dimensional model \cite{Angelini2022}. In general, understanding the properties of the $T=0$ random field critical point is crucial \cite{angelini2023real}.

In this chapter, we consider again the problem of linear excitation modes. We will study a vector spin glass model. In the first section, a replica computation is considered to show how in diluted systems a new order parameter emerges. After that, we will introduce the BP equations \eqref{eq:BP_equations} for vector systems. We will develop an algorithm to solve them at finite temperature and estimate the location of the dAT line of our model, for the Heisenberg case $m=3$. Finally, in the last section we consider linear excitations of inherent structures: after conveying a general picture about the qualitative behaviour of the system in different parametric regions in $H$, we will focus on the "gapless" region, where low-energy quasi-localised modes featuring a quartic law in their DoS are observed. Note that the results discussed in this chapter are unpublished, therefore they may be limited and incomplete in some aspects.

\section{The model}

The Hamiltonian of the model is 
\begin{equation}
\label{eq:Hamiltonian_sparse}
    \mathcal{H}[{\bf S}]\,=\,-\sum_{[ij]\in\mathcal{E}}J_{ij}\Vec{S}_i\cdot\Vec{S}_j-H\sum_{i=1}^N \Vec{b}_i\cdot\Vec{S}_i
\end{equation}
where as usual the spins are with unit norm and $m$ components. The system is defined on a RRG with connectivity $c$, $[i,j]$ is a link and $\mathcal{E}$ is the edge set.
We consider homogeneous disorder
\begin{itemize}
    \item Rademacher uncorrelated couplings
    \begin{equation}
        P_J(J)\,=\,\frac{1}{2}\delta(J-1)+\frac{1}{2}\delta(J+1)
    \end{equation}
    \item Spherical uniform uncorrelated external fields
    \begin{equation}
        P_b(\Vec{b}_i)\,=\,\frac{\delta(|\Vec{b}_i|-1)}{S_m(1)}
    \end{equation}
\end{itemize}
At variance with chapter 4, here we prefer to work with uniform spherical fields. Given the sparse nature of the model, if one chose gaussian couplings or gaussian fields the disorder would be strongly heterogeneous, and linear excitations would be strongly affected by weakly interacting neighbourhoods or neighbourhoods with very small external fields. In the dense case due to the extensive number of neighbours, the effect of heterogeneity could not affect the localisation properties of the system. The reason we want a homogeneous disorder is because we are interested in studying how the properties of linear modes change with respect to the interplay between the internal field and the external field. While the former enforces correlation between neighboring spins, the external field can be regarded as a spatial (in the sense of random graphs) white noise term. It is interesting then to study the spectral properties of this system as one approaches the spin glass phase and compare the results with the dense model studied in Chapter 4.

Before studying linear modes of inherent structures, however, we prefer to outline some analytical results. We want to show how the sparse model contains a richer degree of information, by showing through replica method the emergence of a new order parameter.


\section{Replica computation for diluted systems}

In this section, we explain how to use Replica method in diluted system. Let us consider the Viana-Bray \cite{viana1985phase} version of \eqref{eq:Hamiltonian_sparse}, corresponding to substituting the RRG graph with an ERG graph with average connectivity $c$: setting $J_{ij}\,=\,\eta_{ij}K_{ij}$
\begin{eqnarray}
    && P_J(J)\,=\,P_{\eta}(\eta)P_K(K) \\
    && P_{\eta}(\eta)\,=\,\frac{c}{N}\delta(\eta-1)+\left(1-\frac{c}{N}\right)\delta(\eta) \\
    && P_K(K)\,=\,\frac{1}{2}\delta(K-\kappa)+\frac{1}{2}\delta(K+\kappa)
\end{eqnarray}
We choose this setting because the computation is simpler than the RRG. Clearly, the linear excitation properties change on an ERG graph, but in this section we are interested in discussing the order parameter that emerges in diluted systems and the properties of the solution. We will come back to the original problem later in this chapter. We also set the absolute value of the couplings to $\kappa$, to show later how the solution we find converges to that of the dense model if for $c\rightarrow\infty$ $\kappa\,=\,1/\sqrt{c-1}$ is chosen.

The computation of the free energy proceeds as usual: after averaging out the couplings and the connectivity factors, we are left with
\begin{equation}
\label{eq:mlml}
    \overline{\mathcal{Z}^n}\,=\,\underset{{\bf S}_1}{\Tr}\cdots \underset{{\bf S}_1}{\Tr}\;\exp\left(\frac{c}{2N}\sum_{ij}\left[\cosh\left(\beta\kappa\sum_a \Vec{S}_i^a\cdot\Vec{S}_j^a\right)-1\right]\right).
\end{equation}
For the moment, we do not consider the external field and set $H=0$.
We need to find an order parameter that allows us to decouple spins and apply the saddle point method.  

\subsection{The order parameter}

The order parameter of diluted systems is
\begin{equation}
\label{eq:order_parameter_diluted}
    \rho(\boldsymbol{\sigma})\,=\,\frac{1}{N}\sum_{k=1}^N\prod_{a=1}^n \delta^m(\Vec{\sigma}_a-\Vec{S}_k^a)
\end{equation}
The physical meaning of this order parameter is the following: while the Hamiltonian of dense systems is a gaussian function whose covariance matrix is a function of the sole overlaps (second order spin correlations), in the diluted case the Hamiltonian is not gaussian, and its properties are determined by the whole set of spin correlations. Physically, this corresponds to the strong degree of correlation between a spin and its finite neighbourhood.
So, the order parameter \eqref{eq:order_parameter_diluted} is exactly the probability density function of observing $\boldsymbol{\sigma}^{\intercal}\,=\,(\Vec{\sigma}_1,\dots,\Vec{\sigma}_n)$ for the $n$ replica of a single spin. 

The necessity of such an order parameter can be deduced by the Mc-Laurin expansion of hyperbolic cosine in \eqref{eq:mlml}
\begin{eqnarray*}
   && \cosh\left(\beta\kappa\sum_a \Vec{S}_i^a\cdot\Vec{S}_j^a\right)\,=\,1+\frac{\beta^2\kappa^2}{2}\sum_{ab}(\Vec{S}_i^a\cdot\Vec{S}_j^a)(\Vec{S}_i^b\cdot\Vec{S}_j^b )\\
   && +\frac{\beta^4\kappa^4}{4!}\sum_{abcd}(\Vec{S}_i^a\cdot\Vec{S}_j^a)(\Vec{S}_i^b\cdot\Vec{S}_j^b)(\Vec{S}_i^c\cdot\Vec{S}_j^c)(\Vec{S}_i^d\cdot\Vec{S}_j^d)+\dots
\end{eqnarray*}
If $\kappa=O(1)$, we have to keep all terms of the expansion, whereas if $\kappa\,=\,1/\sqrt{N}$ (dense limit) all terms of order greater than two are irrelevant in the thermodynamic limit and the order parameter is just the overlap matrix. 

\subsection{A Bray-Rodgers equation}

The order parameter is inserted in \eqref{eq:mlml} through a functional delta
\begin{equation*}
    1\,=\,N\int \mathcal{D}\rho\;\delta\left[N\rho(\boldsymbol{\sigma})-\sum_{k=1}^N\prod_{a=1}^n \delta^m(\Vec{\sigma}_a-\Vec{S}_k^a)\right]
\end{equation*}
After this, by following standard steps one arrives to
\begin{equation}
    \overline{\mathcal{Z}^n}\,\propto\,\int \mathcal{D}\rho\mathcal{D}\hat{\rho}\;e^{-N\mathcal{S}[\rho,\hat{\rho}]}
\end{equation}
where $\mathcal{S}[\rho,\hat{\rho}]\,=\,\mathcal{S}_1[\rho,\hat{\rho}]+\mathcal{S}_2[\rho,\hat{\rho}]+\mathcal{S}_3[\hat{\rho}]$, $\hat{\rho}$ is the conjugated of the order parameter and
\begin{eqnarray}
    && \mathcal{S}_1[\rho,\hat{\rho}]\,=\,i\int_{S_m(1)^n}d\boldsymbol{\sigma}\hat{\rho}(\boldsymbol{\sigma})\rho(\boldsymbol{\sigma}) \\
    && \mathcal{S}_2[\rho,\hat{\rho}]\,=\,-\frac{c}{2}\int_{S_m(1)^n}\int_{S_m(1)^n}d\boldsymbol{\sigma}d\boldsymbol{\sigma}'\rho(\boldsymbol{\sigma})\rho(\boldsymbol{\sigma}')[\cosh(\beta\kappa \boldsymbol{\sigma}\cdot \boldsymbol{\sigma}')-1] \\
    && \mathcal{S}_3[\hat{\rho}]\,=\,-\log \int_{S_m(1)^n}d\boldsymbol{\sigma}\exp(i\hat{\rho}(\boldsymbol{\sigma})).
\end{eqnarray}
From the saddle point equations we get
\begin{equation}
   \rho_*(\boldsymbol{\sigma})\,=\,\frac{e^{i\hat{\rho}(\boldsymbol{\sigma})}}{\int_{S_m(1)^n}d\boldsymbol{\sigma}e^{i\hat{\rho(\boldsymbol{\sigma})}}} 
\end{equation}
\begin{equation}
    i\hat{\rho}_*(\boldsymbol{\sigma})\,=\,c\int_{S_m(1)^n}d\boldsymbol{\sigma}'\rho(\boldsymbol{\sigma}')[\cosh(\beta\;\boldsymbol{\sigma}\cdot \boldsymbol{\sigma}')-1].
\end{equation}
These equations combined together yield \emph{Bray-Rodgers equation}
\begin{eqnarray}
\label{eq:Bray-Rodgers}
       && \rho_*(\boldsymbol{\sigma})\,=\,\frac{\exp(g[\boldsymbol{\sigma};\rho_*])}{\int_{S_m(1)^n}d\boldsymbol{\sigma}'\exp(g[\boldsymbol{\sigma};\rho_*)])} \\
       && \nonumber \\
       && g[\boldsymbol{\sigma};\rho]\,=\,c\int_{S_m(1)^n}d\boldsymbol{\sigma}'\rho(\boldsymbol{\sigma}')[\cosh(\beta\kappa\boldsymbol{\sigma}\cdot \boldsymbol{\sigma}')-1]
\end{eqnarray}
which is a functional self-consistent equation for the saddle point value of the order parameter. It was introduced by A. J. Bray and G. J. Rodgers in \cite{rodgers1988density}, where the authors study the random matrix problem of determining the density of states of the interaction matrix of the Viana-Bray model.

The free energy density of the system is computed as usual
\begin{eqnarray}
\label{eq:fe_diluted}
    && f(\beta)\,=\,\lim_{n\rightarrow 0}\frac{1}{n\beta}\mathcal{S}[\rho_*] \\
    && \mathcal{S}[\rho]\,=\,\frac{1}{2}\int_{S_m(1)^n}d\boldsymbol{\sigma}g(\boldsymbol{\sigma};\rho)\rho(\boldsymbol{\sigma})-\log \int_{S_m(1)^n}d\boldsymbol{\sigma}'\exp(g[\boldsymbol{\sigma};\rho])
\end{eqnarray}
In the following, we show how to deal with \eqref{eq:Bray-Rodgers},\eqref{eq:fe_diluted} in the paramagnetic case, in absence and in presence of an external random field.

\subsection{Paramagnetic solution at zero external field}

To solve \eqref{eq:Bray-Rodgers} in the paramagnetic case, we have to define an RS ansatz for the order parameter $\rho(\boldsymbol{\sigma})$. The simplest choice is to consider isotropic functions in the replica space. Given the spherical constraint on each of the replicas $n$, this amounts to considering the uniform solution.
\begin{equation}
\label{eq:para_H0_sol}
    \rho_{0}(|\boldsymbol{\sigma}|)\,=\,\frac{1}{S_m(1)^n}.
\end{equation}
This solution is the correct one for the $H=0$ paramagnetic phase: it tells us that within any single replica each spin value is equally probable, so it reproduces all features of the high-temperature phase. With \eqref{eq:para_H0_sol}, also the Bray-Rodgers function is constant
\begin{equation*}
    g(|\boldsymbol{\sigma}|,\rho_0)\,=\,\frac{c}{S_m(1)^n}\int d\boldsymbol{\sigma}'[\cosh(\beta\kappa\boldsymbol{\sigma}\cdot \boldsymbol{\sigma}')-1]\,=\,c\;[\mathcal{K}_m(\kappa\beta)^n-1]
\end{equation*}
where $\mathcal{K}_m(\beta)$ is as usual defined by \eqref{eq:mVector_K}. After performing the $n=0$ limit, the free energy density reads
\begin{equation}
\label{eq:fe_para_H0_diluted_explicit}
    f(\beta)\,=\,\left(\frac{c}{2}-1\right)\frac{\log S_m(1)}{\beta}-\frac{c}{2\beta}\log\mathcal{K}_m(\beta\kappa)
\end{equation}
while the internal energy and the entropy densities are
\begin{equation}
    u(\beta)\,=\,\diff{(\beta f)}{\beta}\,=\,-\frac{c\kappa}{2}\frac{I_{m/2}(\beta \kappa)}{I_{m/2-1}(\beta \kappa)}
\end{equation}
\begin{equation}
    s(\beta)\,=\,\beta(u-f)\,=\,-\frac{c\kappa\beta}{2}\frac{I_{m/2}(\beta \kappa)}{I_{m/2-1}(\beta \kappa)}-\left(\frac{c}{2}-1\right)\log S_m(1)+\frac{c}{2}\log \mathcal{K}_m(\beta\kappa).
\end{equation}
In the high temperature limit, we correctly find $s(\beta)\approx \log S_m(1)$, since $\mathcal{K}_m(0)\,=\,S_m(1)$. If we set $\kappa\,=\,J/\sqrt{c}$, in the limit $c\rightarrow\infty$ we retrieve the solution of the dense system \eqref{eq:fe_para_mVector_noextfield}.

\begin{figure}
    \centering
    \includegraphics[width=0.8\columnwidth]{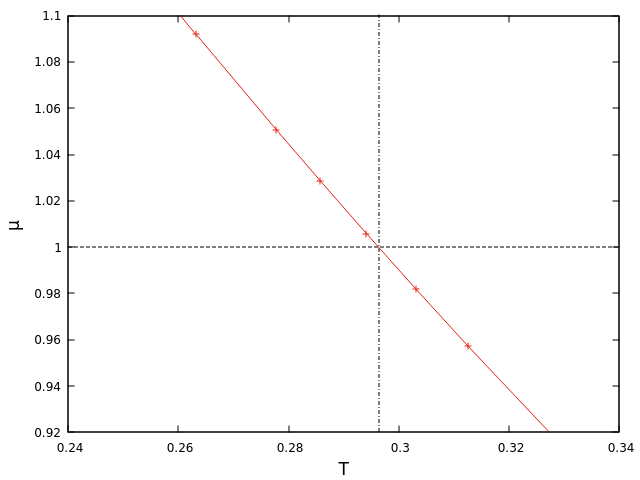}
    \caption{The largest eigenvalue of the Heisenberg kernel \eqref{eq:kernel_heis}. The zero field spin glass transition for the $c=3$ RRG and for unit couplings $\kappa=1$ is at $T_c(0)=0.296(2)$. We represented the Kernel with a $500\times 500$ matrix: we observed that for this graining discretisation errors were neglibile.}
    \label{fig:largest_eig_Kernel}
\end{figure}

The location of the critical temperature can be determined by studying the stability of the uniform solution under perturbations. The condition for criticality is determined by solving the following kernel eigenvalue equation \cite{coolen2005finitely}:
\begin{eqnarray}
\label{eq:crit_eq_Tc_general}
    && \iint_{S_m(1)}d^m\tau_1d^m\tau_2\;e^{\beta\kappa (\Vec{\sigma}_1\cdot\Vec{\tau}_1+\Vec{\sigma}_2\cdot\Vec{\tau}_2)}\psi(\Vec{\tau}_1\cdot\Vec{\tau}_2)\,=\,\frac{\mathcal{K}_m^2(\beta\kappa)}{c}\psi(\Vec{\sigma}_1\cdot\Vec{\sigma}_2) \\
    && \\
    && \int_{S_m(1)}d^m\tau\psi(\tau_1)\,=\,0
\end{eqnarray}
where the constraint stems from the requirement that the perturbed probability measure is still normalised. One has to find the temperature $T$ at which the largest eigenvalue of the Kernel operator becomes equal to unity.
In the cases $m=1, 2$, the last equation can be solved explicitly.
The $m=1$ case has been studied extensively in the last decades, starting from the initial papers \cite{viana1985phase, Kanter1987, mezard1987mean, dominicis1987replica}. The condition for criticality is found easily from \eqref{eq:crit_eq_Tc_general} by remembering that $\int_{S_1(1)}\equiv \sum_{\sigma=\pm 1}$ and $\mathcal{K}_1(x)\,=\,2\cosh(x)$. The result is
\begin{equation*}
    c\tanh^2(\beta_c\kappa)\,=\,1
\end{equation*}
For XY spins ($m=2$), \eqref{eq:crit_eq_Tc_general} is solved with Fourier modes \cite{skantzos2005cavity, coolen2005finitely, lupo2017critical}. The solution is given by the lowest-order mode and reads
\begin{equation*}
    c\left[\frac{I_1(\beta_c\kappa)}{I_0(\beta_c\kappa)}\right]^2\,=\,1
\end{equation*}
For general $m$, the solution of \eqref{eq:crit_eq_Tc_general} is more involving and, unfortunately, does not lead to explicit equations. However, in the Heisenberg case, the problem can be solved by considering eigenvectors $\psi_k(\varphi, \theta)\,=\,f_k(\cos \theta) e^{i k\varphi}$. This parametrisation is similar to that of spherical harmonics. The original equation simplifies and the solution is obtained by diagonalising the following symmetric Kernel\cite{coolen2005finitely} ($x\equiv \tau_1\cdot\tau_2\equiv \cos\theta$, the overlap between two different replica vectors)
\begin{eqnarray}
\label{eq:kernel_heis}
    & \mathcal{M}(x, y)\,=\,\frac{1}{4\pi}\int_{0}^1\int_0^1 ds\;dt I_0\left(\beta \kappa\sqrt{1-x^2}\sqrt{1-s^2}\right)e^{\beta\kappa (sx+t)}\nonumber \\
    & \times\frac{\theta((1-s^2)(1-t^2)-(y-st)^2)}{\sqrt{(1-s^2)(1-t^2)-(y-st)^2}} \frac{(\beta\kappa)^2c}{\sinh(\beta\kappa)^2} 
\end{eqnarray}
We found that its largest eigenvalue in the case $c=3$, $\kappa=1$ studied by us becomes equal to unity at $T_c(0)=0.296(2)$: in figure \ref{fig:largest_eig_Kernel} we show the largest eigenvalue of the kernel operator as a function of temperature. The temperature we found, when rescaled by $\sqrt{c-1}$, is equal to $T_c(0)'=0.209(1)$: this value is less than the dense limit $T_c^{(\infty)}(0)=1/3$, as expected.
We replaced the $c$ factor in \eqref{eq:kernel_heis} with $c-1$, since we work on RRG: this takes into account the different branching factor between the two graphs, $c$ in ERG and $c-1$ in RRG respectively.


\subsection{General RS solution}

When a random external field is added, the solution \eqref{eq:fe_diluted} is modified via Bray-Rodgers function which becomes
\begin{equation}
    g[\boldsymbol{\sigma};\rho]\,=\,c\int_{S_m(1)^n}d\boldsymbol{\sigma}'\rho(\boldsymbol{\sigma}')[\cosh(\beta\kappa\boldsymbol{\sigma}\cdot \boldsymbol{\sigma}')-1]\langle e^{\Vec{b}\cdot\sum_a\Vec{\sigma}_a}\rangle_b
\end{equation}
Clearly, now the uniform solution is wrong. The most general RS ansatz for $\rho(\boldsymbol{\sigma})$ must take into account all the possible combinations of the $n$ replica that are invariant under permutations. Excluding those RS combinations that couple replica, like $\sigma_1^{\mu}\cdots\sigma_n^{\mu}$, we have that the order parameter must be a function of any possible sum of the form $\sum_{a,\mu}(\sigma_a^{\mu})^k$, for $k>0$ integer\footnote{In the $m=1$ case only $\sum_a \sigma_a$ terms are present: in this situation the order parameter is represented in terms of a distribution of cavity fields, rather than a functional distribution like in the vector case \eqref{eq:RS_general_parametrisation_sparse}.}. This is accomplished by parametrising the order parameter as follows \cite{coolen2005finitely}
\begin{equation}
\label{eq:RS_general_parametrisation_sparse}
    \rho(\boldsymbol{\sigma})\,=\,\int \mathcal{D}p\;W[p]\prod_{a=1}^n\;p(\Vec{\sigma}_a)
\end{equation}
where we introduced the functional measure $\mathcal{D}pW[p]$ which, together with the factorisation in the single replica of the remainder of the integrand accounts for all the permutation invariant terms described above. Notice that the previously used uniform solution consists in the choice $W[p]\,=\,\delta[p(\Vec{\sigma})-1/S_m(1)]$. 

The functional $W[p]$ is the true order parameter of the sparse spin glass. When the system is RS, $W$ is unique and its stationary value describes the Gibbs state. When RS breaks, the stationary point should be given by a set of functionals $\{W_k[p]\}$ that represent spin statistics within different clusters in the phase space, following the ultrametric scheme. For instance, in \cite{mezard2003cavity} M. Mézard and G. Parisi consider the sparse Ising spin glass at zero temperature, in the 1RSB approximation.

Once \eqref{eq:RS_general_parametrisation_sparse} is imposed, with some effort, the $n=0$ limit can be performed, giving the RS free energy density. Without specifying the distribution of couplings and random fields and the underlying random graph, its expression reads in terms of the functional $W$ reads \cite{coolen2005finitely}
\begin{eqnarray}
\label{eq:RS_fe_general_sparse_vector}
    && f_{RS}\,=\,\frac{c}{2\beta}\int \mathcal{D}p_1\mathcal{D}p_2W[p_1]W[p_2]\int dK\;P_K(K)\log \mathcal{Z}_{\beta}^{(2)}[p_1, p_2; K]  \\
    && -\frac{1}{\beta}\sum_{l=0}^{\infty}P_c(l)\int \prod_{i=1}^l \mathcal{D}p_i W[p_i] dK_iP_K(K_i) d^mb P_b(\Vec{b})\log \mathcal{Z}_{\beta, H}^{(1)}[\{p_l\}; K, \Vec{b}] \nonumber \\
    && \nonumber \\
    && \mathcal{Z}_{\beta, H}^{(1)}[\{p_l\}; K, \Vec{b}]\,=\, \int d^m\sigma e^{\beta \Vec{\sigma}\cdot\Vec{b}}\prod_{i=1}^l \int d^m\sigma' p_i(\Vec{\sigma}^{'})e^{\beta K \Vec{\sigma}\cdot\Vec{\sigma}'} \\
    && \nonumber \\
    && \mathcal{Z}_{\beta}^{(2)}(p_1, p_2; K)\,=\,\int d^m\sigma d^m\sigma' p_1(\Vec{\sigma}) p_2(\Vec{\sigma'}) e^{\beta K \Vec{\sigma}\cdot\Vec{\sigma}'}
\end{eqnarray}
where $P_c(l)$ is the probability of having connectivity $l$: in a RRG with connectivity $c$ one has $P_c(l)\,=\,\delta(l-c)$, in an ERG $P_c(l)\,=\,c^l e^{-c}/l!$, the Poisson distribution.
In this expression we can read an average over probability densities, connectivity, couplings, and fields of the Bethe free energy \eqref{eq:Bethe_FE}, derived in the context of Belief Propagation or Cavity Method in Section \ref{sec:BP_and_cav}. We identify the single site distribution $p(\Vec{\sigma})$ as an instance of the spin cavity marginal, eq. \eqref{eq:BP_equations} defined in section \ref{sec:BP_and_cav}. Then, 
the functionals $\mathcal{Z}_{\beta, H}^{(1)}[\{p_l\}; K, \Vec{b}]$ and $\mathcal{Z}_{\beta}^{(2)}(p_1, p_2; K)$ are the site and link partition functions respectively, written in a form suitable for population dynamics algorithms (PDA). All this proves the equivalence between Replica and Cavity approaches in the diluted case.

\section{Cavity method}
\label{sec:Cavity_Method_sparse}

The RS solution of the sparse case for any $H\neq 0$ is given in terms of a nontrivial stationary functional density $W_*[p]$ of spin cavity marginals. This statement is equivalent to considering the BP equations, which yield the set of cavity marginals for a given graph, in a distributional sense. Then, if one is not interested in the properties of a particular instance of the graph ensemble, as we learnt in Section \ref{sec:BP_and_cav} the most profitable approach is to consider a PDA algorithm. 

\subsection{BP equations and Discretization}

Let us write equations \eqref{eq:BP_equations} explicitly for the problem we are interested in (we switch back to the notation of section \ref{sec:BP_and_cav})
\begin{equation}
\label{eq:BP_equations_vector}
    \eta_{i\rightarrow j}(\Vec{S}_i)\,=\,e^{\beta \Vec{b}_i\cdot \Vec{S}_i}\prod_{k\in\partial i/j}\int_{S_m(1)}d^mS_k e^{\beta J_{ij}\Vec{S}_i\cdot \Vec{S}_j}\eta_{k\rightarrow i}(\Vec{S}_k)
\end{equation}
The PDA equivalent of this last equation reads
\begin{equation}
\label{eq:BP_equations_vector_pda}
    \eta_c(\Vec{S})\,=\,e^{\beta \Vec{b}\cdot \Vec{S}}\prod_{k=1}^{c-1}\int_{S_m(1)}d^mS' e^{\beta J\Vec{S}\cdot \Vec{S}'}\eta_{c}^{(k)}(\Vec{S}')
\end{equation}
where $\Vec{b}$ and $J$ are drawn from the respective PDFs. As usual, \eqref{eq:BP_equations_vector_pda} can be implemented numerically by representing the target density functional with a population $\{\eta_c^{(p)}(\Vec{S})\}_{p=1}^{N_p}$. Each member of the population should be discretized on the $m$ sphere:
\begin{equation}
\label{eq:discretize_cav_marginals_vector}
    \eta_c^{(p)}(\Vec{S})\Longrightarrow (\eta_c^{(p)}(\Vec{S_1}),\dots, \eta_c^{(p)}(\Vec{S}_{N_d}))
\end{equation}
In the $m=1$ case, cavity marginals can be parametrised in terms of a single parameter $u$, a cavity field, as
\begin{equation}
    \eta_c(S)\,=\,\frac{e^{u S}}{2\cosh(u)}
\end{equation}
The discretisation over the scalar variable $u$ is straightforward, and in the RS phase the pda fixed point is the cavity fields distribution $P_u(u)$. 

In the $m=2$ case, spins are continuous variables, and our fixed point is a probability density functional over spin marginals $W[\eta_c]$. The discretisation of XY models has been studied thoroughly in \cite{lupo2017approximating}, and is based on the so-called \emph{clock model} \cite{Nob1986, Nob1989, Ilk13, Ilk14}. The circumference is uniformly discretised in $Q$ notches $\phi_q\,=\,(2\pi/Q)(q-1)$, with $q\,=\,1,\dots, Q$. For high temperatures, physical observables of the clock model converge exponentially fast to the observables of the XY model, whereas at very low $T$ convergence is only algebraically fast \cite{lupo2017approximating, lupo2019random}.

For $m>2$, the problem concerning the discretization of the $m$-sphere is much harder. In 1904 J. J. Thomson\footnote{He is the discoverer of the electron!} studied the problem of determining the ground state configuration of classical electrons interacting with a repulsive Coulomb potential on a sphere, the first atomic model in history \cite{thomson1913structure}. The problem generalised to an arbitrary repulsive potential\footnote{Some interesting physical realizations regard multi-electron bubbles \cite{leiderer1995ions} and surface ordering of liquid metal drops \cite{davis1997history}.} is a century-old mathematical puzzle \cite{saff1997distributing}.
The optimal arrangement of particles on a sphere can be regarded as well as a Voronoi tiling of the sphere \cite{saff1997distributing}. The sphere is covered in terms of non-overlapping Voronoi-Dirichlet cells: 
\begin{equation}
    \mathcal{D}_i\,=\,\{\Vec{x}\in S_3(1)\;:\;|\Vec{x}-\Vec{x}_i|\,=\,\min |\Vec{x}-\Vec{x}_k|\}\qquad \bigcup_{k=1}^N \mathcal{D}_k\,=\,S_3(1)
\end{equation}
The Voronoi-Dirichlet cell of particle with label $i$ is the set of points of the sphere closer\footnote{Using the Euclidean distance in $\mathbb{R}^3$.} to $i$ than to any other particle. Extensive numerical studies with large number of particles have shown that all but exactly $12$ of the Dirichlet cells of an optimal tiling configuration are hexagonal, the exceptional cells are pentagons \cite{saff1997distributing}. With a relatively small number of particles, $N\leq 150$, the $12$ exceptional particles are at the vertices of an icosahedron \cite{erber1997complex}. The presence of exactly twelve topological defects (disclinations) is a consequence of a Euler theorem, stating that the number of topological defects for the optimal tessellation of a surface is $6\varepsilon$, where $\varepsilon$ is the Euler characteristic ($\mathit{f}$ is the number of faces, $e$ the number of edges, $v$ the number of vertices)
\begin{equation*}
    \varepsilon\,=\,\mathit{f}-e+v.
\end{equation*}
The characteristic of the sphere is $\varepsilon=2$, that of the plane is $\varepsilon=0$: indeed, the hexagonal tiling is an optimal planar tessellation.

The potential energy landscape in the generalised Thomson problem is complex: it is estimated that the number of local minimum is exponentially large in $N$ \cite{erber1995comment}. So, determining an optimal configuration is a non-convex optimisation problem. First numerical attempts relied on building so-called icosadeltahedral configurations, i.e. arrangements with icosahedral symmetry. The number of particles of such configurations satisfies the equation $N\,=\,10(h+k)^2-10hk+2$, with $h, k$ integer. It has been conjectured in \cite{altschuler1997possible} that ground states configuration of the Thomson problem possess this symmetry. However, in \cite{perez1997influence, perez1997comment} it was pointed out that there are configurations with five-fold and seven-fold dislocations\footnote{A dislocation is a translation topological defect of a crystal: it can be obtained for instance through an external shear.} with lower energy.
Later, in \cite{perez1999symmetric} it was shown that there exist configurations with icosahedral symmetry and dislocation defects.

Finding a good discretization of the sphere is crucial for solving \eqref{eq:BP_equations_vector_pda} at low temperature. Even in the RS phase, when temperature is low we shall expect that typical fixed point marginals of the populations to polarise in regions with small solid angle. Therefore, the systematic error related to discretization in the small temperature region is enhanced. A poor tiling has a significant impact also on the factor with the external field in \eqref{eq:BP_equations_vector_pda}. When the external field is drawn in a badly covered region (large Voronoi cell), the scalar product $\Vec{b}\cdot\Vec{S}$ is biased to be with relatively high probability less than $\Vec{b}\cdot\Vec{S}_*$, where $\Vec{S}$ is the spin representing the cell. These events bias the population dynamics as if the standard deviation of the external field $H$ had a lower value.

\subsection{Random grid algorithm}

We introduce an algorithm to solve eqs \eqref{eq:BP_equations_vector_pda} in the $m=3$ case. We consider random uniform tessellations of the unit sphere
\begin{equation}
\label{eq:random_grid_generator_uniform}
    S(N_p)=\{\Vec{r}_1,\dots,\Vec{r}_{N_p}\}\qquad \Vec{r}_i\overset{d}{=}\operatorname{Unif}(S_m(1))
\end{equation}
and solve BP equations \eqref{eq:BP_equations_vector_pda} written in discretised form (let us use the symbol $\nu$ for cavity marginals)
\begin{equation}
\label{eq:BP_equations_vector_pda_discr}
    \nu(\Vec{r})\,=\,e^{\beta \Vec{b}\cdot \Vec{r}}\prod_{k=1}^{c-1}\sum_{\Vec{r'}\in S(N_p)} e^{\beta J\Vec{r}\cdot \Vec{r}'}\nu^{(k)}(\Vec{r}').
\end{equation}
We kept random uniform external field, but we chose to use anti-ferromagnetic couplings $P_J(J)=\delta(J+1)$. The RS phase of the antiferromagnet is the same as that of the $J\,=\,\pm 1$ spin glass, and the location of the instability line is also the same. 

\begin{figure}
    \centering
    \includegraphics[width=0.35\columnwidth]{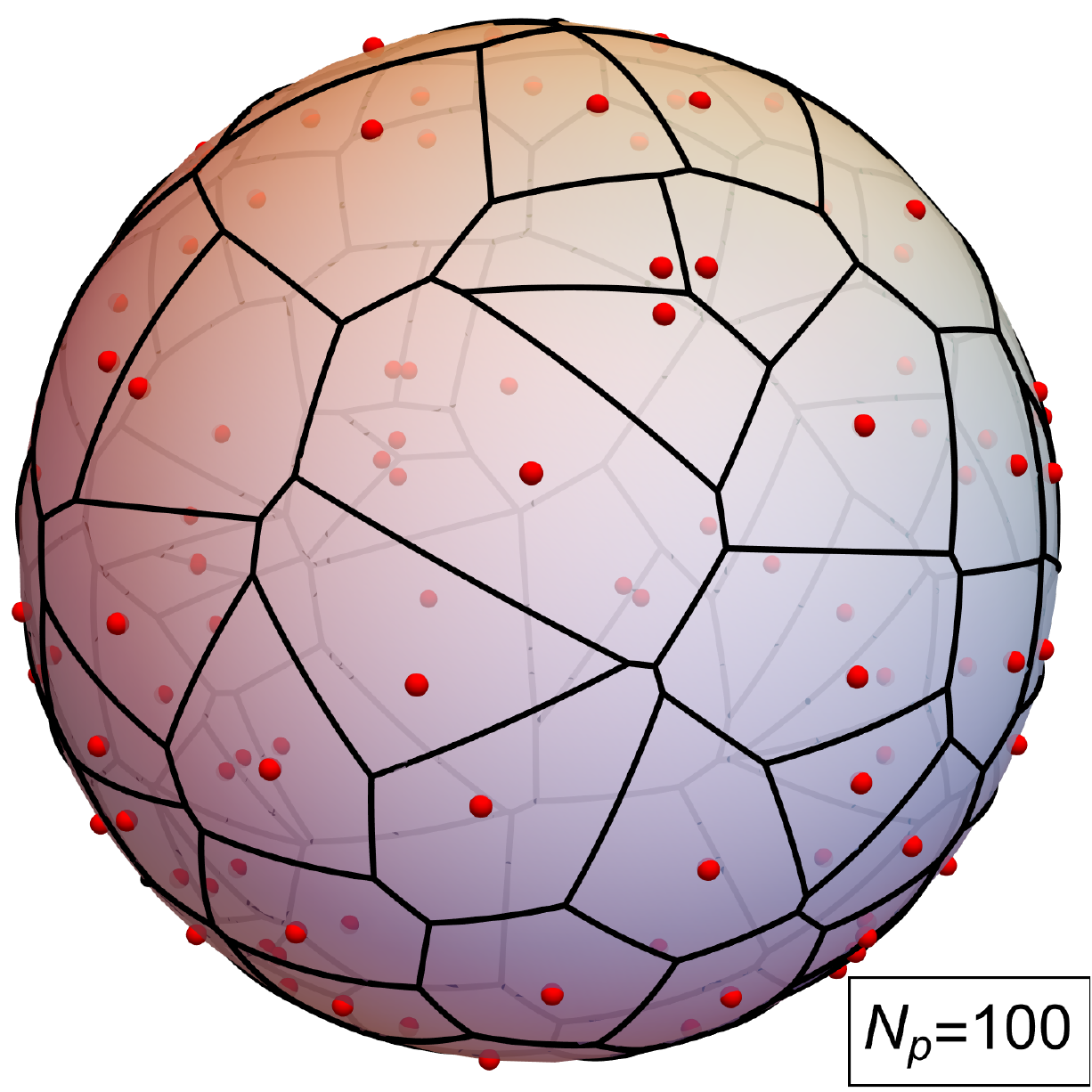}
    \includegraphics[width=0.35\columnwidth]{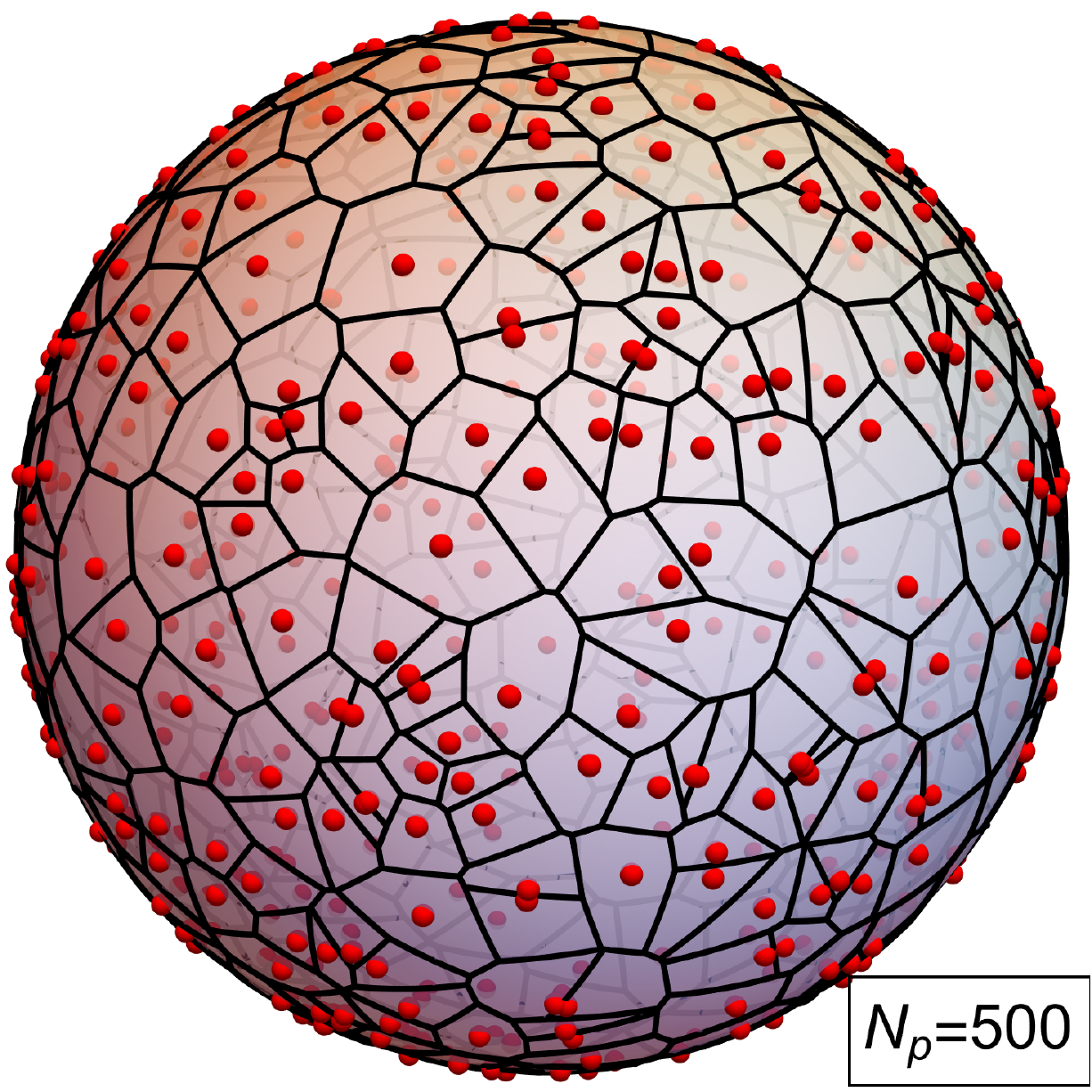}
    \includegraphics[width=0.35\columnwidth]{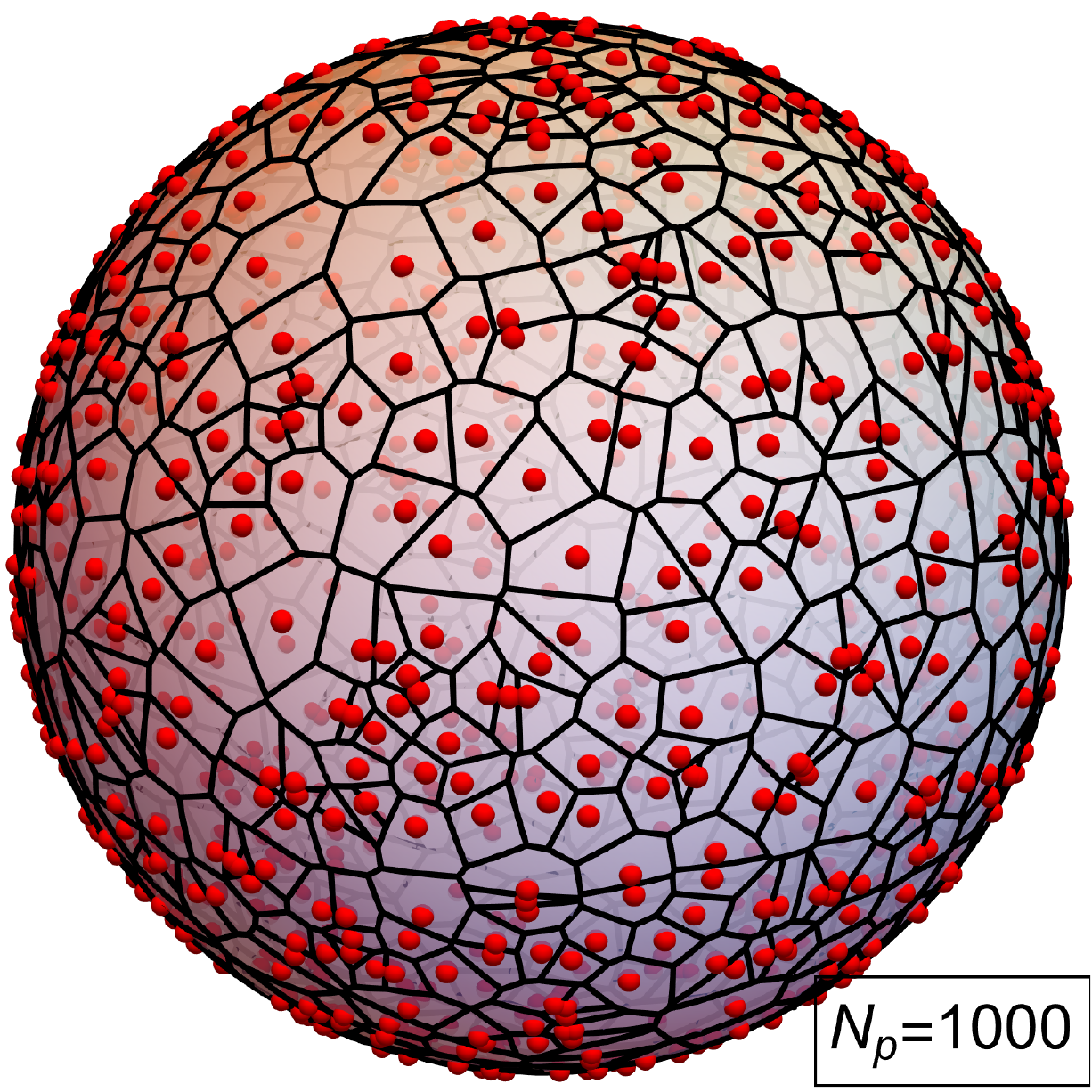}
    \includegraphics[width=0.8\columnwidth]{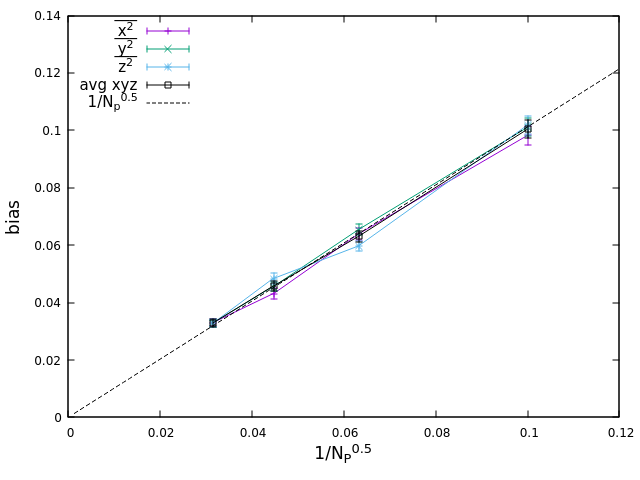}
    \caption{Three instances of the random grid generator. Voronoi meshing generated with Mathematica\texttrademark\;\;using a script available on \url{https://demonstrations.wolfram.com/VoronoiDiagramOnASphere/}. On the bottom, the anisotropy bias as a function of the discretisation.}
    \label{fig:three_balls}
\end{figure}

As one can expect, this algorithm has a very poor tiling performance. In figure \ref{fig:three_balls} we show single instances of the random uniform grid generator \eqref{eq:random_grid_generator_uniform} for points numbers $N_p=100, 500, 1000$. We superimposed the Voronoi cells of the obtained particles configuration. For all three sizes the shape of Voronoi cells fluctuates wildly: even at the larger size, we can see the presence of clusters with high concentration of points\footnote{This is a consequence of the random tesselation: the distribution of points on the sphere surface is a Poisson points statistics, and the distance between pairs of points is exponentially distributed: thus, any pair can be arbitrary close with finite probability.}. In the bottom figure, we show the anisotropy bias $\rho\,=\,\overline{x^2+y^2+z^2}/3$, where $x, y, z$ are the cartesian coordinates of a given $\Vec{r}_i$ and $\overline{\cdot}$ is an average over different grids, for growing number of points. This number should be exactly $1/\sqrt{N}_p$ for large $N_p$, since single components are normal variables. Our data agree very well with the theoretical expectation.

The reason we chose this algorithm is the following: we want to see how in the worst case scenario (the random grid) physical observables depend on the discretisation, depending on temperature and external field. Then, we can use our results as a reference for future work, aimed at developing a good tiling algorithm taylored for disordered systems problems on the $3$-dimensional sphere.

\subsection{Numerical simulations}

We performed a series of simulations in a range of temperature $T\in[0.0,T_c(0)]$ and field $H\in[0.0,1.5]$, using discretizations $N_p\,=\,100, 250, 500, 1000$. We used $N_{pop}=10^4$ population size in all simulations. For each temperature $T$, starting from a sufficiently high value of $H=H_{ini}$, we performed an annealing in the external field: the fixed point of each run at $H=H'$ was used as an initial condition for the next run at $H=H'-\Delta H$. In almost all runs we iterated \eqref{eq:BP_equations_vector_pda_discr} $t_{pda}\,=\,25$ sweeps time, in some runs we used $t_{pda}\,=\,50$. 

\begin{figure}
    \centering
    \includegraphics[width=0.8\columnwidth]{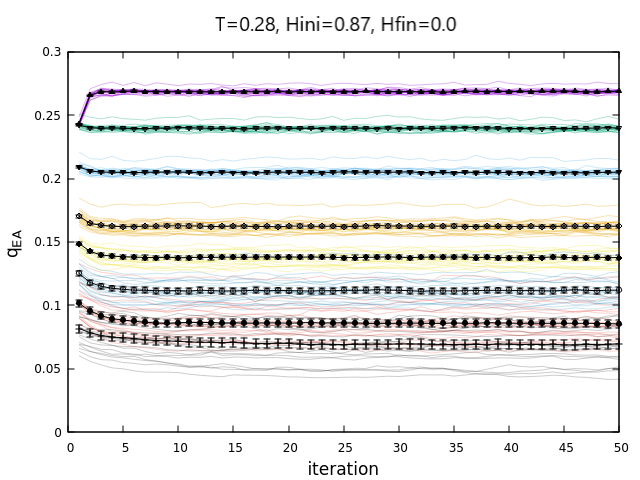}
    \caption{The self-overlap as a function of sweep iteration time. The top purple curve corresponds to $H=0.87$, the bottom to $H=0$. Data are from simulations with $N_p=500$.}
    \label{fig:overlap_sparse}
\end{figure}

In figure \ref{fig:overlap_sparse} we show the overlap as a function of $t$, for $T=0.28$, $H$ ranging from $H_{ini}=0.82$ to $H_{fin}=0$ and $N_p=500$ points. We can see that in all cases convergence is reached within $t\simeq 10$ sweeps.
We measured it at the end of each sweep using
\begin{eqnarray}
\label{eq:overlap_from_pops}
    && q(t)\,=\,\frac{3}{N_{g}}\sum_{i=1}^{N_{g}}\frac{1}{N_{pop}}\sum_{p=1}^{N_{pop}}\left(\frac{1}{N_{p}}\sum_{k=1}^{N_p}\eta_p(\Vec{r}_k^{(i)}, t)x_k^{(i)}\right)^2 \\
    && \eta_p(\Vec{r}, t)\,=\,e^{\beta \Vec{b}\cdot \Vec{r}}\prod_{k=1}^{c}\sum_{\Vec{r'}\in S(N_p)} e^{-\beta\Vec{r}\cdot \Vec{r}'}\nu^{(k)}(\Vec{r}', t)
\end{eqnarray}
where $\eta_p$ is the exact marginal and $\nu(\Vec{r}, t)$ are cavity marginals from the time $t$ (in sweep units) population $\{\nu_{t}\}$ and $N_g$ is the number of simulated grids: we used $N_g=100$ grids for $N_p=100$, $N_g=50$ for $N_p=250$ and $N_g=25$ for $N_p=500, 1000$.
Clearly, in the second formula \eqref{eq:overlap_from_pops} the external field $\Vec{b}$ is drawn from the uniform distribution on the sphere each time a marginal is updated.
The top curve corresponds to $H=0.87$, the bottom to $H=0.0$. For $T=0.28$, the fixed points found become RS-unstable at $H\simeq 0.09$ (see next section). Note that sample-to-sample fluctuations become very large approaching criticality.

\subsection{Measure of the dAT line}

A fixed point reached through algorithm is stable if, after a random small perturbation of the fixed point population
\begin{equation}
    \nu_{\epsilon}^{(p)}(\Vec{r}_i)\,=\,\nu_{*}^{(p)}(\Vec{r}_i)+\epsilon\delta^{(p)}(\Vec{r}_i)\qquad \frac{1}{N_p}\sum_{i=1}^{N_p}\delta^{(p)}(\Vec{r}_i)\,=\,0
\end{equation}
the pda algorithm with initial condition $\{\nu_{\epsilon}^{(p)}\}$ converges back to the original fixed point. We perform in parallel population dynamics for the perturbed population and the original fixed point population, and at each sweep time step we measure the discretised $L_2$ norm on the unit sphere of the difference $\nu_{\epsilon}(\Vec{r})-\nu_*(\Vec{r})$, averaged over populations
\begin{equation}
\label{eq:Delta_VS_t}
    \Delta(t)\,=\,\frac{1}{N_{pop}}\sum_{p'}\sqrt{\frac{1}{N_p}\sum_i |\nu_{\epsilon}^{(p')}(\Vec{r}_i, t)-\nu_*^{(p')}(\Vec{r}_i, t)|^2}
\end{equation}
where $\nu(\Vec{r}, t)$ are the marginals at the end of the $t$-th population sweep. When the fixed point is stable, $\Delta(t)$ is expected to decay to zero exponentially fast. If instead
\begin{equation}
    \Delta(t)\sim const\qquad t\rightarrow\infty
\end{equation}
then the pda dynamics from the perturbed population has reached a different fixed point\footnote{Notice that the Lyapunov exponent $\log\Delta/t$ cannot be strictly positive-modulo the perturbed marginal converges to a delta function, we do not think this is possible at finite $T$-because the marginals are defined on a compact set, the unit sphere.}. When for a given $H, T$ this happens for $\epsilon$ no matter how small, the RS phase is unstable. We tested the stability of our fixed points using always $\epsilon=10^{-6}$.

\begin{figure}
    \centering
    \includegraphics[width=0.45\columnwidth]{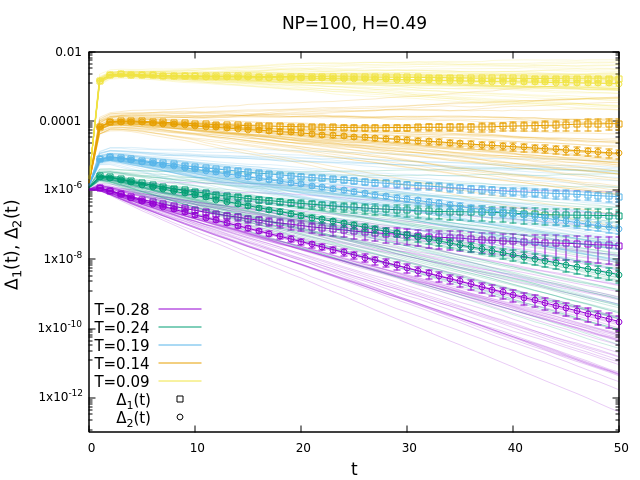}
    \includegraphics[width=0.45\columnwidth]{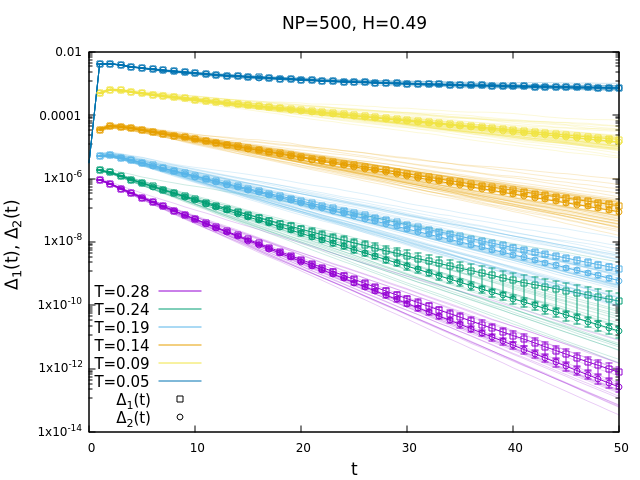}
    \includegraphics[width=0.45\columnwidth]{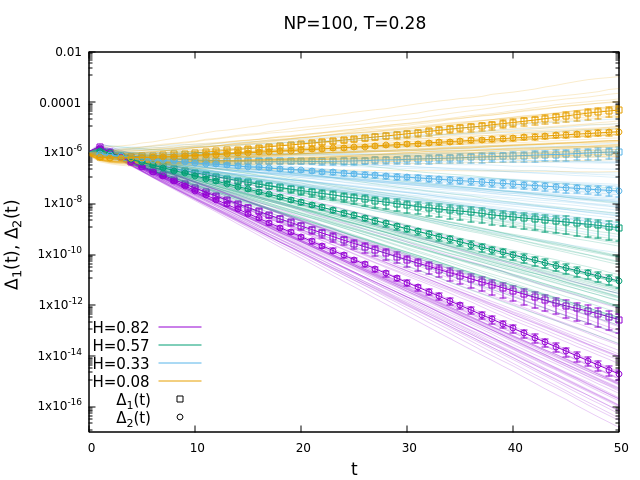}
    \includegraphics[width=0.45\columnwidth]{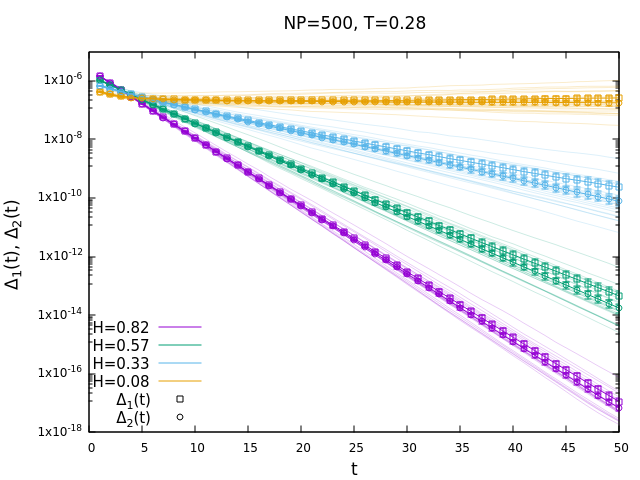}
    \includegraphics[width=0.45\columnwidth]{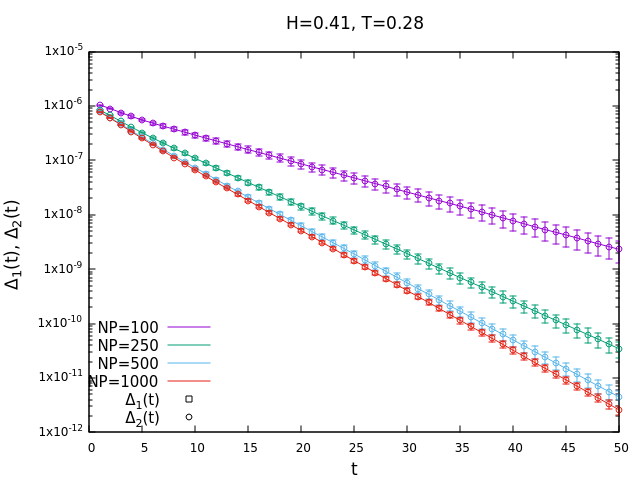}
    \includegraphics[width=0.45\columnwidth]{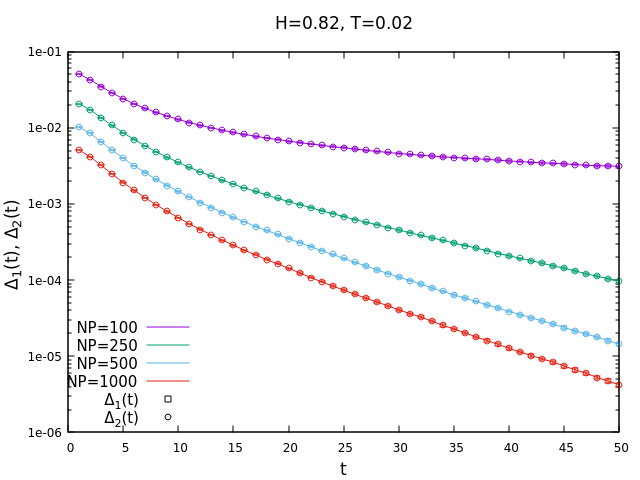}
    \caption{\textbf{Top}: the discrepancy $\Delta(t)$ at fixed temperature, left is $N_p=100$ and right is $N_p=500$. \textbf{Center}: the discrepancy $\Delta(t)$ at fixed field, left is $N_p=100$ and right is $N_p=500$. In all plots shaded curves are trajectories of $\Delta$ on different samples. \textbf{Bottom}: dependency on the number of discretiation points. Left is "high" temperature, right is "low".}
    \label{fig:delta_VS_t}
\end{figure}

In figures \ref{fig:delta_VS_t} we show averages over different grids of the discrepancy $\Delta$ at fixed $H$ (top) and at fixed $T$ (bottom), for $N_p=100, 500$. We compare the two different sample averages
\begin{eqnarray}
\label{eq:averages_delta}
    && \Delta_1(t)\,=\,\overline{\Delta}(t)\,=\,\frac{1}{N_{g}}\sum_{i=1}^{N_g}\Delta_i(t) \\
    && \Delta_2(t)\,=\,\exp(\overline{\log\Delta(t)})\,=\,\exp\left(\frac{1}{N_{g}}\sum_{i=1}^{N_g}\log\Delta_i(t)\right)
\end{eqnarray}
in order to check sample to sample fluctuations. Indeed, we observe that with $N_p=100$ the difference between $\Delta_1$ and $\Delta_2$ is significant. In some cases, like the green curve ($T=0.24$, $H=0.5$), also for $N_p=500$. Since we are interested in typical trajectories, we chose to consider only $\Delta_2(t)$. Apparently, as the discretisation is improved, sample-to-sample fluctuations seem to diminish. In the bottom pair of figures in \ref{fig:delta_VS_t} we show $\Delta_2(t)$ versus $t$ varying $N_p$. In the left picture $T=0.28, H=0.41$ the difference between the two largest $N_p$ curves is relatively small, whereas in the right picture $T=0.02, H=0.82$ finite discretisation effects are strong. This does not come as a surprise to us, since at low temperature we expect marginals to be localised around specific sectors of the spheres. 

\begin{figure}[t!]
    \centering
    \includegraphics[width=0.45\columnwidth]{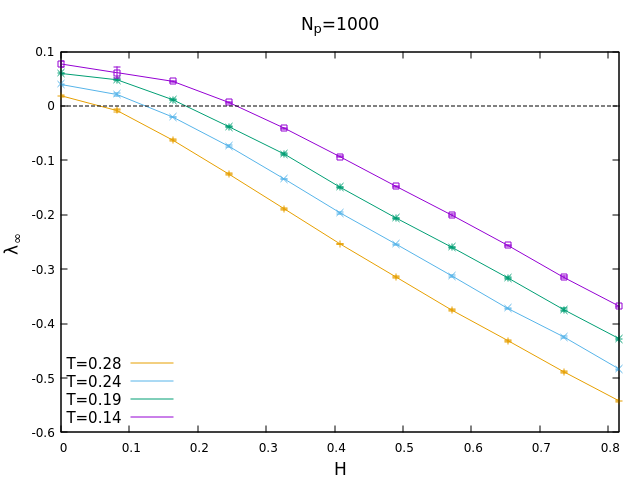}
    \includegraphics[width=0.45\columnwidth]{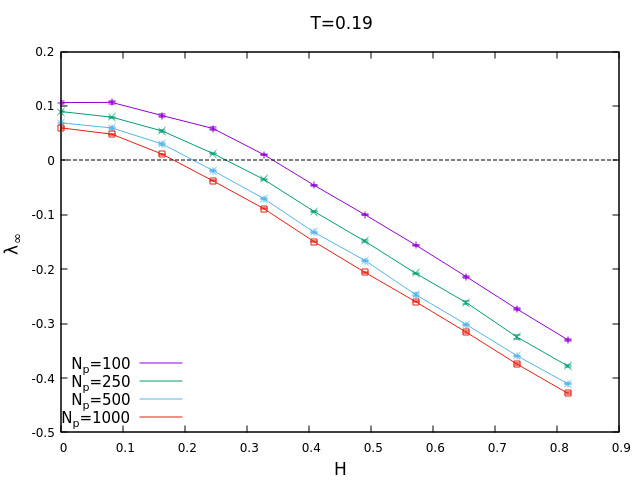}
    \caption{The exponents $\ell_{\infty}$ as functions of $H$. On the left a comparison of the curves at different temperatures, with the largest $N_p$ available. On the right, the effect of discretisation at intermediate temperature.}
    \label{fig:lyaps}
\end{figure}

In order to measure the values of $H, T$ on which the RS instability line is located, we adopt the following measure protocol:
\begin{itemize}
    \item We measure the slope of the curve $\log\Delta_2(t)$ versus $t$ in intervals of length $5$, from $t_0=5$ to $t_M=50$
    \begin{equation}
        \ell_i\equiv \frac{\log\Delta_2(5(i+1))-\log\Delta_2(5i)}{5}\qquad i=1,\dots,9 
    \end{equation}
    \item We extrapolate the asymptotic slope of the curve $\log\Delta_2(t)$ through a fit\footnote{We assume analytic corrections to the asymptotic value} of the $\{\ell_i\}$ for any $T, H, N_p$
    \begin{equation}
        \ell(t)\,=\,\ell_{\infty}-\frac{a}{t}
    \end{equation}
    \item For any $T, N_p$, we consider the curves $s_{\infty}$ versus $H$ and extrapolate from the interval where $l_{\infty}<0$ (RS stable points) the value $H_c$ such that $\ell_{\infty}=0$. We show an example of these curves in figure \ref{fig:lyaps}. In the top picture we show curves at different temperatures and $N_p=1000$, while in the bottom we show the dependency on the discretisation for the values $N_p\,=\,100, 250, 500, 1000$.
\end{itemize}

\begin{figure}[h!]
    \centering
    \includegraphics[width=0.8\columnwidth]{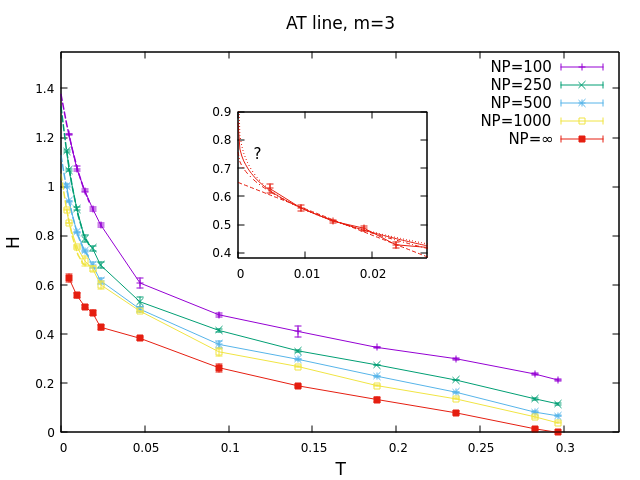}
    \caption{The instability line of the $m=3$ spin glass on the $c=3$ random regular graph. Units are for unit in norm couplings. The red line is our extrapolation of the instability line from data at finite discretisation.}
    \label{fig:dAT_line}
\end{figure}

At the end of this procedure, we have an instability line $H_c$ versus $T_c$ for each value of $N_p$ measured. 
In figure \ref{fig:dAT_line} we show the dAT lines we obtained. The red curve is our extrapolation at $N_p=\infty$, obtained through a power-law approach
\begin{equation}
    H_c(T)\,=\,H_c^{(\infty)}(T)+\frac{A}{N_p^{\gamma}}.
\end{equation}
We computed the exponent of the power law considering our data at $T=T_c(0)$ (since we know that in this case $H_c(T_c(0))=0$). We found that the exponent should be roughly $\gamma\simeq 0.72$: we assume this as the exponent of the leading corrections for all other temperatures. 

 It is worth to consider discretisation finite size effects: in the case we are analysing, the points on the sphere are morally a perfect gas. For a perfect gas living in a sphere, the typical distance $l$ between particles is given by the inverse of the number density, implying $l\sim N_p^{-1/2}$. This scaling can be considered an upper bound for the error made with discretisation, and write the true marginal corresponding to a certain $\nu$ ad $\nu^{(\infty)}\,=\,\nu+O(N_p^{-1/2})$. If in the term $(\nu_{\epsilon}-\nu_*)^2$ in \eqref{eq:Delta_VS_t} the correction does not cancel, the $O(N^{-1/2})$ propagates in \eqref{eq:averages_delta} and all the following expressions. If the $O(N_p^{-1/2})$ corrections of the marginals in \eqref{eq:Delta_VS_t} tend to cancel, then the exponent $\gamma>1/2$: it appears that this is the case for our simulations. We expect that computations performed on optimised grids of the sphere further reduce the overall error, possibly giving reliable estimates for $N_p=O(10^2)$, as it happens for instance in the clock model of the XY spin glass.

Unfortunately, we were unable to provide a reliable estimate of the zero-temperature critical field $H=H_(0)$. Indeed, as the inset in figure \ref{fig:dAT_line} shows, we do not have any argument for the scaling of the dAT line close to zero temperature. We show in the inset extrapolations using different possible exponents: the lower dashed-dotted curve is a linear extrapolation, the upper curves are extrapolation with an exponent less than unity, $H_c=H_c(0)-A T^{a}$, $0<a<1$.
In the dense case, one has $H_c\propto T\log(1/T)$: based on our numerical data concerning eigenvalues and eigenvector, we do not believe that this scaling is the correct one. However, we do not know the value of exponent $a$; based on our data on eigenvalues and eigenvectors, we will discuss in the next section, we believe that the zero-temperature critical point should be located in the interval $[0.81, 0.91]$, roughly corresponding to $0.2<a<1/3$. We believe that with a grid-optimised algorithm it will be possible to make a precise estimate of this point, possibly extending vector BP equations \eqref{eq:BP_equations_vector_pda} to zero temperature.

\section{Linear excitation modes of inherent structures}

The rest of this chapter is devoted to the study of linear excitation modes of inherent structures of the energy landscape. In Chapters 4 and 5 we learnt that the softest linear modes of the dense system are localised in the paramagnetic phase $H>H_c(0)=1/\sqrt{m(m-2)}$ and delocalised in the spin glass phase $H<H_c(0)$. In particular, we showed that they possess multi-fractal features at the critical point.

We would like to extend these studies to the finitely connected system analysed in this chapter. Systems defined on sparse graphs, even if still mean-field, are much closer to real systems. Even though their topological structure, induced by the underlying random regular graph, is infinite-dimensional, 
each spin interacts with a finite neighbourhood, so one hopes that the local properties of these systems resemble to some extent those of finite-dimensional systems. Clearly, in the latter the abundance of short loops for any size $N$ makes cavity method a terrible approximation; however, one wonders if the locally tree-like approximation is enough to capture some important physical features.

Glassy systems in finite dimensions show a feature that was not reproduced by our fully connected models: a vibrational density of states (VDoS) following a quartic law close to zero frequency (see section \ref{sec:boson_peak}, eq. \eqref{eq:non_phononic_VDoS}). In the models studied in chapter 4 and 5 the lower tail of the spectral density is $\lambda^{m-1}$, so for the VDoS one has $D(\omega)\sim\omega^{2m-1}$: the exponent depends on the dimension of the spins\footnote{In particular, one recovers the quartic law for $m=2.5$: unfortunately, it is not easy to imagine a system with spins having rational dimension.}. This is in contrast with what is observed in finite dimensions, where the quartic law seems to be almost ubiquitous among many different glassy models in different settings (see the discussion in \ref{sec:boson_peak}). In this chapter we show that the sparse system we study (eq. \eqref{eq:Hamiltonian_sparse}) has linear excitations of the energy minima whose VDoS follows the quartic law of finite dimensional systems. The quartic law was already observed in spin glasses: in \cite{baity2015soft} for the three-dimensional random field Heisenberg spin glass and in \cite{lupo2017critical} for the random field XY spin glass model. We confirm the result for the $m=2$ case and show that it extends also to the $m=3, 4$ cases. We consider this result quite interesting, not so much for the $\omega^4$ law in itself, but for the fact that it shows off in a first principle mean field theory. There are mean field theories that feature $\omega^4$ modes, even at the dense level (like the model studied in \cite{rainone2021mean}), but they are phenomenological. Our theory, instead, is a first-principle theory.

Another important aspect concerns the localisation properties of eigenvectors.
At variance with the dense case, in the sparse case the eigenvector has a "spatial" structure, even though an infinite-dimensional one; one can study the topological relation between different "soft spots" of the random graph, by which we mean regions where the normalisation is concentrated and thus the linear response of the system to perturbations is strong. In dense systems, one can at most classify sites according to their energetic and susceptibility contributions. We already discussed the relevance of localisation in sparse systems in section \ref{sec:sparse_matrices} of Chapter 3. As far as spin glasses are concerned, the study of localisation in random graphs is a very recent research topic. First steps have been moved in \cite{lupo2017critical}, here we want to pursue that research, by considering also the case $m=3$. Given our results of chapter 4 concerning the dense case, it is worth to see if something analogous happens in diluted models. In particular, what kind of changes, if any, eigenvectors undergo at the $T=0$ spin glass transition. We will show by means of numerical simulations that in the $m=2$ case there are apparently no significant changes in the properties of the modes, whereas for $m=3$ non-trivial topological long-range effects appear as $H\rightarrow H_c(0)$.  

The remainder of the chapter is organised as follows: in section \ref{sec:Hessian_sparse} we consider the Hessian matrix in the sparse case as a random matrix. In section \ref{sec:strong_field_sparse} we consider the large $H$ case, for which we able to define a random matrix ensemble for the Hessian. We also discuss the closure of the spectral gap, yielding an estimate of the location of the value $H=H_{gap}$ where the spectrum becomes gapless. In section \ref{sec:soft_modes_quartic_sparse} we discuss the spectrum in the gapless region $H<H_{gap}$, focusing on the softest modes. Finally, in section \ref{sec:eigvec_weak_deloc} we discuss the nature of eigenvectors in the region $H_c(0)\ll H<H_{gap}$ and $H\sim H_c(0)$. 

\subsection{The Hessian}
\label{sec:Hessian_sparse}

This model formally has the same Hessian as the dense case, eqs. \eqref{eq:Hessian_energy_fields}, \eqref{eq:Hessian_energy_form2}, modulo the sparsity of the interaction matrix
\begin{equation}
\label{eq:Hessian_SG_sparse}
    M_{ij}^{ab}[{\bf S}]\,=\,-(\hat{e}_i^a\cdot\hat{e}_j^b)J_{ij}\mathbf{1}(j\in\partial i)+|\Vec{\mu}_i({\bf S}_{\partial i})|\delta_{ij}\delta_{ab}
\end{equation}
where $\mathbf{1}$ is an indicator function representing the adjacency matrix. We remind that $\{\hat{e}_i^a\}$, $a=1,\dots,m-1$, are random bases satisfying the orthogonal constraints $\hat{e}_i^a\cdot \Vec{S}_i=0$.
The local fields $\Vec{\mu}_i$ are functions of the neighborhoods $\{{\bf S}_{\partial i}\}$
\begin{equation}
\label{eq:local_fields_sparse}
    \Vec{\mu}_i({\bf S}_{\partial i})\,=\,-\frac{\partial \mathcal{L}}{\partial \Vec{S}_i}\,=\,\sum_{j\in\partial i}J_{ij}\Vec{S}_j+\Vec{b}_i
\end{equation}
where as usual the random external field acts as a spatial noise term and $\mathcal{\mathcal{L}}$ is the Lagrangian \eqref{eq:Model_again} but with \eqref{eq:Hamiltonian_sparse} as Hamiltonian.

Being a mean-field theory on random graphs, one can on principle exploit the semi-analytical tools \eqref{eq:spectral_cavity_solution_given_sample} provided by cavity method to have a theoretical prediction. This task was done in \cite{lupo2017critical} on the XY model. Here, considered also the great difficulty one encounters in finding a good discretisation of the sphere, we decided to attack the problem of low energy linear excitations only numerically, in order to verify the results in \cite{lupo2017critical} and generalise them to the $m>2$ case.

The procedure we follow for the numerical simulations presented in this chapter is the same as that described in chapter 4. We find stationary points through the over-relaxation algorithm \eqref{eq:GD_update_move_OR}, we diagonalise the Hessian evaluated at the configuration found and, if the stationary point is a minimum, we measure the statistics of eigenvalues and eigenvectors. All figures concerning numerical simulations we show are for $c=3$ RRGs. Our discussion will mostly focus on the $m=3$ case, but we will discuss the $m=2$ and $m=4$ models as well. 

The sparse nature of \eqref{eq:Hessian_SG_sparse} implies that for any finite $H$, local fields are strongly correlated to couplings, which in addition are finite ($J=\pm 1$) in the thermodynamic limit. This is exact opposite of the dense case, where precisely because of the decorrelation between fields, couplings and spins we were able to classify the Hessian as a Rosenzweig-Porter matrix.
There is no obvious random matrix ensemble to classify our Hessian:
it seems more convenient to study the statistics of the off-diagonal matrix, the diagonal entries and their correlation and from these informations try to build a random matrix ensemble representing Hessians of sparse matrices.
It is worth to consider the limit of large external field $H\rightarrow\infty$. In this limit, the correlation between fields and interactions is suppressed by the random fields: in this special case, we are able to find a random matrix ensemble for our Hessian.

\subsection{The case of strong external field}
\label{sec:strong_field_sparse}

When $H\rightarrow\infty$, it is possible to expand local fields representing a stationary point of the Hamiltonian in $1/H$. We can rewrite them as follows
\begin{equation}
\label{eq:local_fields_sparse_rewritten}
    \Vec{\mu}_i\,=\,\Vec{\eta}_i+H\Vec{b}_i
\end{equation}
The first term in the r.h.s. is the field generated by the neighborhood of spin $i$, and enforces correlations between neighbors. The second, the random external field, has the opposite role of promoting decorrelations between neighbors. In the RSB phase the first term is dominant, but as $H>c$ the second starts to be dominant.  
Let us rewrite \eqref{eq:local_fields_sparse_rewritten} in a form suitable for the expansion in $1/H$. Since $\Vec{\mu}_i\,=\,\mu_i\Vec{S}_i$ in a stationary point, we can also write
\begin{equation}
    \mu_i\,=\,\Vec{\eta}_i\cdot\Vec{S}_i+H(\Vec{b}_i\cdot\Vec{S}_i)
\end{equation}
We have to expand $\Vec{S}_i$ and then the neighbors field $\Vec{\eta}_i$.
Let us begin with spins first:
\begin{equation*}
    \Vec{S}_i\,=\,\frac{\Vec{\mu}_i}{\mu_i}\,=\,\frac{\Vec{b}_i+(1/H)\Vec{\eta}_i}{\sqrt{(b_i^2+(2/H)\Vec{b}_i\cdot\Vec{\eta}_i+(1/H)^2\eta_i^2}}\,=\,\Vec{b}_i+\text{corrections}
\end{equation*}
The $O(1/H^1)$ term is
\begin{equation}
    \Vec{\Delta}_{i}^{(1)}\,=\,\sum_{j\in\partial i}J_{ij}[\Vec{b}_j-(\Vec{b}_j\cdot\Vec{b}_i)\Vec{b}_i]\,=\,\mathbb{P}_{b_i}^{(\perp)}\Vec{\eta}_i^{(0)}\qquad \Vec{\eta}_i^{(0)}\,=\,\sum_{j\in\partial i}J_{ij}\Vec{b}_j
\end{equation}
which is a vector orthogonal to $\Vec{b}_i$, as it should be for the leading perturbative correction of an unit vector. Thus, at order $1/H$
\begin{eqnarray*}
    && \Vec{b}_i\cdot\Vec{S}_i\,=\,1-O(1/H^2) \\
    && \\
    && \Vec{\eta}_i\,=\,\Vec{\eta}_i^{(0)}+\frac{1}{H}\sum_{j\in\partial i}J_{ij}\mathbb{P}_{b_j}^{(\perp)}\Vec{\eta}_{j}^{(0)}+O(1/H^2) \\
    && \\
    && \Vec{\eta}_i\cdot\Vec{S}_i\,=\,\sum_{j\in\partial i}J_{ij}\Vec{b}_i\cdot\Vec{b}_j+\frac{1}{H}\sum_{j\in\partial i}J_{ij}[\Vec{b}_j\cdot\mathbb{P}_{b_i}^{(\perp)}\Vec{\eta}_i^{(0)}+\Vec{b}_i\cdot\mathbb{P}_{b_j}^{(\perp)}\Vec{\eta}_j^{(0)}]+O(1/H^2) \\
    && \\
    && \mu_i\,=\,H+\sum_{j\in\partial i}J_{ij}\Vec{b}_i\cdot\Vec{b}_j+\frac{1}{H}\sum_{j\in\partial i}J_{ij}[\Vec{b}_j\cdot\mathbb{P}_{b_i}^{(\perp)}\Vec{\eta}_i^{(0)}+\Vec{b}_i\cdot\mathbb{P}_{b_j}^{(\perp)}\Vec{\eta}_j^{(0)}]+O(1/H^2)
\end{eqnarray*}
At leading order, the Hessian reads
\begin{eqnarray}
\label{eq:Hessian_leading_order_largeH}
   && M_{ij}^{ab}\,=\,-(\hat{e}_i^a\cdot\hat{e}_j^b)J_{ij}+(H+\mu_i^{(0)})\delta_{ij}\delta_{ab}+O(1/H) \\
   && \hat{e}_i^a\cdot\Vec{b}_i\,=\,0 \\
   && \mu_i^{(0)}\,=\,\Vec{\eta}_i^{(0)}\cdot\Vec{b}_i\,=\,\sum_{j\in\partial i}J_{ij}(\Vec{b}_j\cdot\Vec{b}_i).
\end{eqnarray}
Let us comment the terms in \eqref{eq:Hessian_leading_order_largeH}: the off-diagonal entries are made by the couplings and the scalar product $(\hat{e}_i^a\cdot\hat{e}_j^b)$ between respectively a vector of a random basis orthogonal to $\Vec{b}_i$ and one of another random basis orthogonal to $\Vec{b}_j$. The scalar product between two random uniform vector is distributed according to
\begin{equation}
\label{eq:distr_scalar_prods}
    P_u(u)\,=\,\frac{(1-u^2)^{\frac{m-3}{2}}}{\int_0^{\pi}d\theta(\sin\theta)^{m-2}}\theta(u+1)\theta(1-u)
\end{equation}
If $J\,=\,\pm 1$, this is also the distribution of the off-diagonal entries, since the sign of the interaction can be absorbed into the other factor. Notice that in the $m=2$ case $P_u(u)$ is very large close to $u\,=\,\pm 1$, whereas in the $m=3$ case it is the uniform distribution and for $m=4$ it is a semi-circular distribution; for $m$ going to infinity, \eqref{eq:distr_scalar_prods} tends to a delta-function\footnote{This is a known properties of high-dimensional spaces.}. If one considers also the diagonal term, the overall form of \eqref{eq:distr_scalar_prods} resembles that of a graph Laplacian:
\begin{equation*}
    L_{ij}\,=\,A_{ij}+\sum_{k\in\partial i}A_{ik}\delta_{ij}
\end{equation*}
where $A_{ij}$ is the adjacency matrix of a graph. In this zero-th order case, the only correlation between diagonal entries and off entries is given by the orthogonality conditions in \eqref{eq:Hessian_leading_order_largeH}. If one we ignore this week correlation, the random matrix in \eqref{eq:Hessian_leading_order_largeH} as the structure of eq. \eqref{eq:dirty_RRG}, which is quite common in the study of Anderson models. Under precise conditions on the distributions of the disorder in the adjacency matrix and in the diagonal, the spectral densities of models of this kind have Lifshitz tails \cite{bapst2011lifshitz}, of which we discussed in \ref{sec:sparse_matrices}. 

\begin{figure}
    \centering
    \includegraphics[width=0.8\columnwidth]{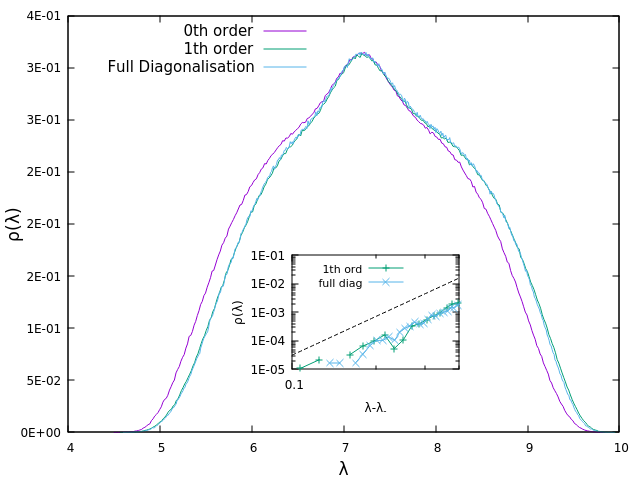}
    \caption{The spectral density for $H=12$: we compare the spectral density obtained from the full diagonalisation of $N_s=500$ Hessian matrices with the spectral density obtained by diagonalising instances of the two random matrix ensembles \eqref{eq:Hessian_leading_order_largeH}, \eqref{eq:Hessian_next_order}. We see that the ensemble at order $1/H$ reproduces very well the spectrum.}
    \label{fig:mmml}
\end{figure}

At order $1/H$, the Hessian becomes
\begin{eqnarray}
\label{eq:Hessian_next_order}
    && M_{ij}^{ab}\,=\,-(\hat{e}_i^a\cdot\hat{e}_j^b)J_{ij}+\Bigl\{H+\mu_i^{(0)}+\frac{1}{H}\sum_{j\in\partial i}J_{ij}[\Vec{b}_j\cdot\mathbb{P}_{b_i}^{(\perp)}\Vec{\eta}_i^{(0)} \\
    && +\Vec{b}_i\cdot\mathbb{P}_{b_j}^{(\perp)}\Vec{\eta}_j^{(0)}]\Bigr\}\delta_{ij}\delta_{ab} \nonumber \\
    && \nonumber \\
    && \hat{e}_i^a\cdot\left(\Vec{b}_i+\frac{1}{H}\mathbb{P}_{b_i}^{(\perp)}\Vec{\eta}_i^{(0)}\right)\,=\,0
\end{eqnarray}
The new term introduces correlations between different sites: we can see it from the second of these last equations, which feature a projection of the vector $\Vec{\eta}_i^{(0)}\,=\,\sum_{j\in\partial i}J_{ij}\Vec{b}_j$.  In figure \ref{fig:mmml} we show a plot of the spectral density for RRG with $c=3$ and $H=12$, averaged from $N_{s}\,=\,2000$ samples of systems with size $N=500$. We compare it with the spectral density obtained with the diagonalisation of as many samples of random matrices generated from \eqref{eq:Hessian_next_order}. We see that already at order $1/H$ the agreement is very good.

As $H$ is lowered, the effect of the neighbors fields $\Vec{\eta}$ becomes more and more relevant, so the perturbative expansion in $1/H$ becomes useless. To describe the Hessian in terms of random matrices following this strategy, at $H=O(c)$ we would need all terms of the expansion. Nevertheless, by discussing it we showed that the Hessian of a sparse matrix evaluated on a minimum can be seen as a strong perturbation of an initial setting where the Hessian has the structure of matrices of generalised Anderson models \eqref{eq:dirty_RRG}.

\subsubsection{The spectral gap}

The spectrum for $H\gg O(c)$ is gapped. The gap is caused by the following constraints respected by the local fields of our model
\begin{equation}
    H-c\leq \mu_i \leq H+c
\end{equation}
which can be understood by simple geometrical considerations based on \eqref{eq:local_fields_sparse}. For finite $H$, we expect physical gaps induced by Onsager reaction, so that $\mu_{gap}=H-c+O(1/H)$. This relation is true as long as $H\gg c$.

\begin{figure}
    \centering
    \includegraphics[width=0.6\columnwidth]{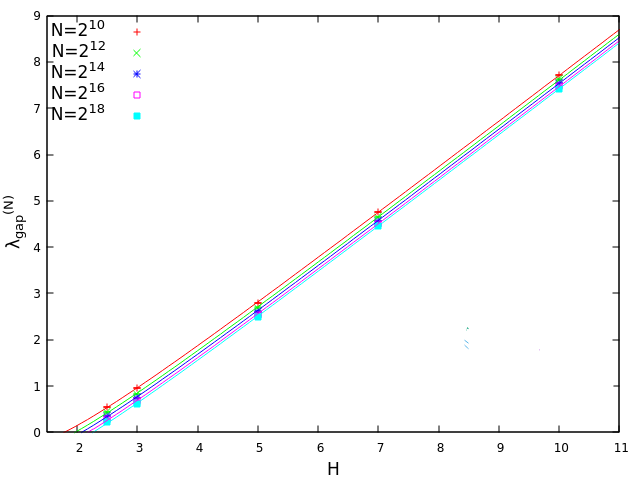}
    \caption{Extrapolation of the field at which the spectrum becomes gapless. The units of $H$ in this figure are those with couplings $|J_{ij}|=1/\sqrt{3}$.}
    \label{fig:lambdaGAP}
\end{figure}

The spectral gap vanishes at a value of external field $H=H_{gap}$ which is $O(c)$. We can make a rough estimate of this value through the following argument: consider the eigenvalue equation
\begin{equation}
    \label{eq:specEQ}
    -\sum_{j\in\partial i}J_{ij}\mathbb{P}^{(\perp)}(\Vec{S}_i)\vec{\psi}_j(\lambda)+(\mu_i-\lambda)\vec{\psi}_i=0
\end{equation}
where for an eigenvector $\boldsymbol{\psi}(\lambda)\equiv (\vec{\psi}_1,\dots,\vec{\psi}_N)$, being each $\vec{\psi}_i$ a $m$-component vector. Rearranging this expression, we can write
\begin{equation}
    \lambda\,=\,\mu_i-\sum_{j\in\partial i}J_{ij}\frac{\mathbb{P}^{(\perp)}(\Vec{S}_i)\vec{\psi}_j\cdot \vec{\psi}_i}{\psi_i^2}\,=\,\mu_i-\sum_{j\in\partial i} J_{ij}\frac{\psi_j}{\psi_i}(\hat{\psi}_i\cdot\hat{\psi}_j)
\end{equation}
which is true for any $i$ (now $i$ is again a site index) and in particular for the spin for which $\mu_i$ is minimum. We can write a lower bound for $\lambda$
\begin{eqnarray}
    && \lambda>H-c-K\,=\,\lambda_{gap}^{(0)} \\
    && K\,=\,\max \sum_{j\in\partial i} J_{ij}\frac{\psi_j}{\psi_i}(\hat{\psi}_i\cdot\hat{\psi}_j)
\end{eqnarray}
where with $\lambda_{gap}^{(0)}$ we mean an estimate of the spectral gap using the rough lower bound $\mu_{gap}\,=\,H-c$ for the local fields, i.e. ignoring $O(1/H)$ corrections. By imposing $\lambda_{gap}^{(0)}=0$, we find an estimate for the value of external field at which the spectrum becomes gapless
\begin{equation}
    H_{gap}^{(0)}\,=\,c+K.
\end{equation}
The quantity $K$ must be evaluated using constraints that come from physical considerations. In particular, if the maximum is at a site where the eigenvector is strongly localised, typically one has $\psi_j\sim \epsilon\psi_i$, for some $0<\epsilon<1$. Indeed, in sparse systems localised eigenvectors typically decay exponentially from a sharp well defined peak on a single site. One can safely assume that the normalisation is yield by the site where the eigenvector is peaked and its neighbors, so that $\epsilon\sim 1/c$. This would return
\begin{equation}
\label{eq:field_gap_order_zero}
    H_{gap}^{(0)}\,\simeq\,c+\sqrt{c}.
\end{equation}
We expect for the true gap closure field to satisfy $H_{gap}<H_{gap}^{(0)}$. In figure \ref{fig:lambdaGAP} we show an extrapolation of the external field through our data for $H=17, 12, 8.7, 5.2$. With a careful extrapolation including a $1/H$ correction to $\lambda_{gap}^{(0)}$, we find $H_{gap}=4.12(4)$. If instead we use a simple linear extrapolation of $\lambda_{gap}^{(0)}$, we find $H_{gap}=4.73(4)$, which is quite close to \eqref{eq:field_gap_order_zero}.

\subsection{The gapless region}
\label{sec:soft_modes_quartic_sparse}

It is time to study the gapless region $H<H_{gap}$. Qualitatively, the appearance of low energy excitations, i.e. $\lambda_{gap}=0$ is related to the fact that the alignment of spins is no more dominated by the external fields. In the RS phase $H_{c}<H<H_{gap}$ there is an even competition between internal and external fields. As $H$ is lowered down to the critical region, internal fields become dominant and spins get strongly correlated or anti-correlated to their neighbors, respectively for ferromagnetic and antiferromagnetic couplings.
In figures we show the probability distributions of the local overlap with the internal field $q_{\eta}=\Vec{\eta}_i\cdot \Vec{S}_i/\eta_i$ (main plot) and the local overlap with the external field $\Vec{b}_i\cdot\Vec{S}_i$ (inset). The figures shown are for $m=3$ and $N=2^20$ spins, with $H=2.60, 1.73, 1.30, 0.87, 0.52$: the first three values are in the RS gapless region, the last two, according to our rough estimate $H_c(0)\in [0.81, 0.91]$, in the spin glass phase. The distribution of the overlap with external fields do not show any particular feature: for large $H$ it tends to a delta function in $q_{\nu}$-i.e. spins strongly correlated to external fields, for small $H$ to a uniform distribution-i.e. spins uncorrelated to external fields. The distribution of the overlap with the internal field is more interesting: as $H$ is lowered towards criticality it develops a maximum at a value quite close to unity: for instance, at $H=0.87$ one has $q_{\eta}^{(*)}\approx 0.95$. As the spin glass phase is entered, the maximum becomes a sharp peak. We did not observe this feature in the distribution of the $q_{\nu}$ for large $H$. 


\begin{figure}
    \centering
    \includegraphics[width=0.8\columnwidth]{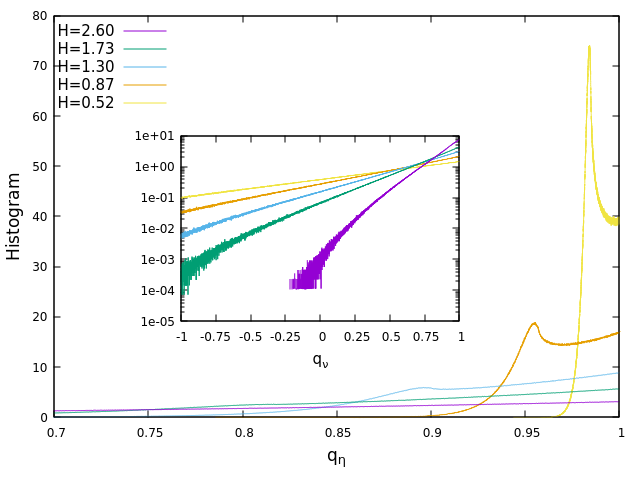}
    \caption{Empirical pdfs of the overlap with the internal field and that with the external field (inset).}
    \label{fig:alignments}
\end{figure}




\subsubsection{Lower edge: the quartic law}

We measured numerically the spectrum of energy minima for four orders of magnitudes of sizes $N=2^8,\dots,2^{20}$ and for several values of $H$. We used a statistics $N\times N_s\geq 3\times 10^7$.
For some of the lowest sizes, we measured the full spectrum, whence for $N\geq 2^{10}$ we computed only the $100$ smallest eigenvalues using the Arnolid method through the Python library \emph{Arpack}\texttrademark.

\begin{figure}
    \centering
    \includegraphics[width=0.8\columnwidth]{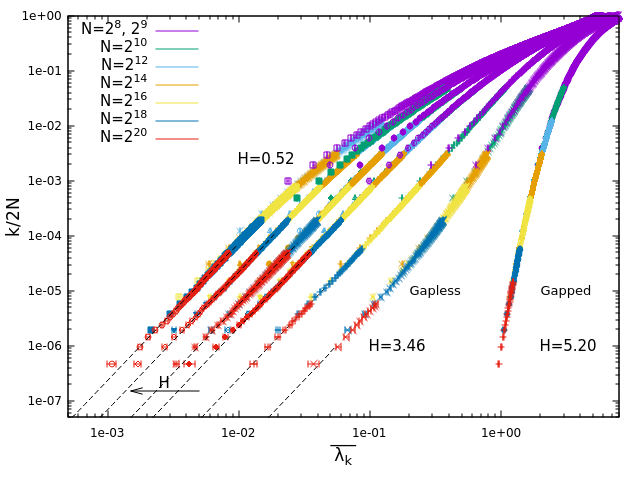}
    \caption{Measure of the empirical cumulative distribution for values of $H=5.20, 3.46, 2.60, 1.73, 1.30, 0.87, 0.52$. We used data ranging in five orders of magnitudes of sizes. Our data are fairly consistent with $a=2.5$ in the whole gapless region.}
    \label{fig:avgCumul_m=3}
\end{figure}

\begin{figure}
    \centering
    \includegraphics[width=0.75\columnwidth]{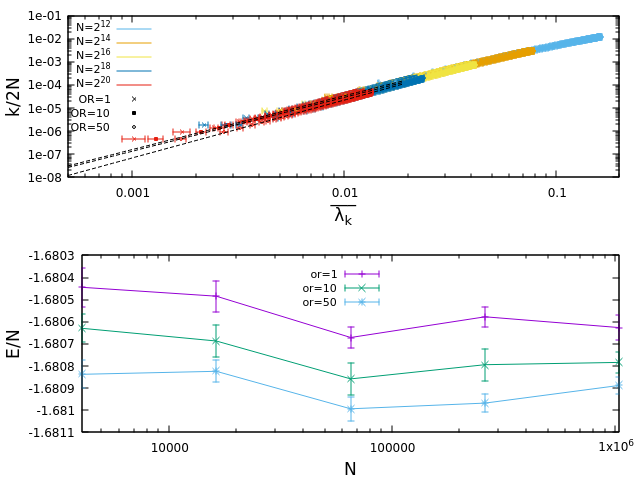}
    \caption{\textbf{Top}: the lower tail of the empirical cdf for $H=0.87$, showing data from minimisations with $O=1, 10, 50$. Even though the three curves are almost indistinguishable, the exponent of the lower tail is $2.5$ for data with $O=50$ but $2.3$ for $O=1, 10$. \textbf{Bottom}: The average energies reached for $N=2^{12},\dots, 2^{20}$ using $O=1,10,50$. As already shown in chapter 4, the greater $O$ the lower in energy.}
    \label{fig:avgCumul_robustness_depth}
\end{figure}

The tails of the spectral densities of our data seem to robustly show $\rho(\lambda)\sim \lambda^{3/2}$, which corresponds to a density of states $D(\omega)\sim \omega^4$. We remind the reader that the relation between the two exponents in $\rho(\lambda)\sim\lambda^a$ and $D(\omega)\sim\omega^b$ is
given by
\begin{equation*}
    D(\omega)=2\omega \rho(\omega^2)\Longrightarrow b=2a+1 
\end{equation*}
We show this in figure \ref{fig:avgCumul_m=3}, where we convey a measure of the empirical cumulative distribution of the eigenvalues, the relative rank $r\equiv k/N$ of each eigenvalue versus its sample average $\overline{\lambda}_k$. Indeed, $r$ clearly tends to the cumulative function for $N\rightarrow\infty$.
For a wide range of values of $H$, both in the RS gapless phase $H_c<H<H_{gap}$ and in the spin glass phase $H<H_{c}$, the lower tail of our empirical spectral cumulative is consistent with $C(\lambda)\sim \lambda^{5/2}$. 
For a given $H$ in the spin glass phase, this behavior is robust provided we typically reach minima sufficiently deep in the energy landscape. In figure \ref{fig:avgCumul_robustness_depth} we report on the top the same measure as before for the single value $H=0.87$ but comparing spectra of minima obtained with different values of the over-relaxation parameter, representing different depths in the energy landscape, as shown in the bottom picture of the same figure. 
While the lower tail of the cumulatives related to the deepest minima, obtained with $O=50$, is $\lambda^{2.5}$, for minima obtained with $O=1, 10$ the tail is better described by $\lambda^{2.3}$. We interpret this result following \cite{gurarie2003bosonic}: the pdf of the curvature on the minima of a quartic random polynomial, constrained in a double-well configuration, is $P(\lambda)\sim\lambda$ for small curvatures, corresponding to $D(\omega)\sim\omega^3$; if however one is interested in the curvature of the global minimum, then $P(\lambda)\sim\lambda^{3/2}$ and $D(\omega)\sim\omega^4$. 
If we assume the validity of this reasoning for our model, then having an exponent lower than $2.5$ for $O=1, 10$ implies that in these minimisations we did not always reach the local ground state within a basin of the energy landscape. 
The idea that in spin glass models with finite-connectivity the energy landscape is locally double-well-like is supported by \cite{baity2015soft}. Even though in that work the authors consider a three-dimensional Heisenberg model, we believe that the same scenario reoccurs in our model: indeed, if the properties of these double-well configurations are dictated by a small group of spins, like a spin and its neighborhood, then there should not be much difference with a model defined on a tree-like random graph.
Despite the statistics of the smallest eigenvalues is susceptible of the depth reached in a local valley of the energy landscape, we will show in next section that the properties of the related eigenvectors are totally robust.

\begin{figure}
    \centering
    \includegraphics[width=0.75\columnwidth]{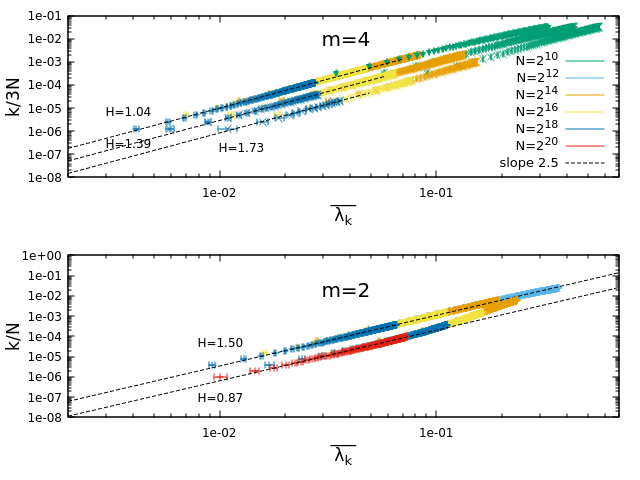}
    \caption{Measures of the cumulative distributions of the $m=2$ and $m=4$ systems. The spectra are from minima obtained with $O=50$. The data agree very well with $C(\lambda)\sim\lambda^{2.5}$.}
    \label{fig:avgCumul_robust_dims}
\end{figure}

The quartic law is robust with respect to the dimension of spins: in figures \ref{fig:avgCumul_robust_dims} we show the cumulative distributions for $m=2, 4$. 
Our data are consistent with an exponent $a=2.5$ for the cumulative distribution. In finite dimensional systems the quartic law is robust with respect to dimensionality \cite{kapteijns2018universal}. In these models on sparse graphs however only the dimension of the degrees of freedom can be varied. Thus, these analogies between finite dimensional and random graph models seem to suggest that what matters is the local behavior of a spins with its finite neighborhood, rather than the underlying geometry.


\subsection{Properties of the eigenvectors}
\label{sec:eigvec_weak_deloc}

When studying vector spin glasses in the fully connected case in chapters 4 and 5 we characterised the properties of their excitations by means of eigenvector moments \eqref{eq:eigvec_moments}. 
In sparse systems we can define more sophisticated tools to study the localisation properties of eigenvectors. While eigenvector moments give a global information on the nature of the mode, namely if it is localised or delocalised, properly defined correlation functions could shed some light on the dependency of the mode on the underlying graph. Given that eigenvector components are related to the local linear response of the system, by identifying local maxima of the eigenvector on the graph one can classify soft spots, i.e. regions strongly susceptible to external perturbations. The condensate we found in the paramagnetic phase of the fully connected model in a random graph becomes a soft spot, around which the mode is exponentially localised. Since the dense model has a delocalisation transition at the $T=0$ critical point, it is natural to see if this phenomenon is present also in the sparse model.

The section is organised as follows: in the first subsection we start with an overall view of the localisation properties of the system, discussing our measures of the IPR in all the spectral phases (gapped, RS gapless, RSB gapless) previously identified. In the second subsection, we define a strategy to unveil the presence of long-range order approaching the SG transition. In the third, we show our results and compare the $m=2, 3, 4$ systems.

\subsubsection{The lowest modes are always localised}

\begin{figure}
    \centering
    \includegraphics[width=0.75\columnwidth]{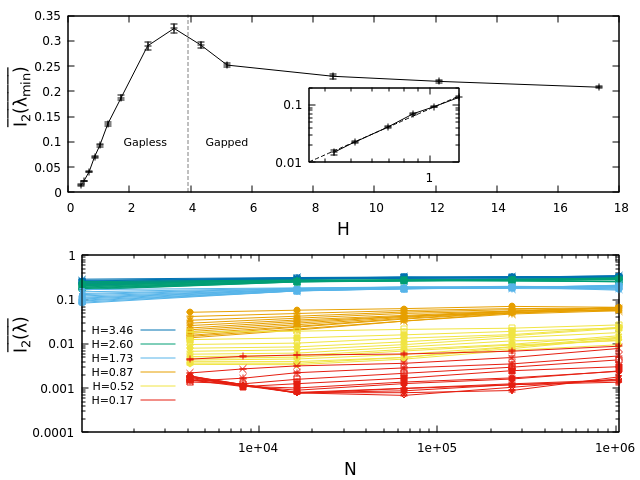}
    \caption{\textbf{Top}: The sample-averaged IPR related to the smallest eigenvalue, as a function of $H$. The localisation decreases as $H$ is lowered. \textbf{Bottom}: the sample-averaged IPRs of the smallest twelve modes versus the number of spins. Cold colors are measures in the gapless RS phase, warm colors measures in the RSB phase. The asymptotic IPRs are finite in both cases.}
    \label{fig:iprmin_VS_H}
\end{figure}

We begin our analysis of eigenvector by considering the IPR of our modes:
\begin{equation*}
    I_2(\lambda_k)\,=\,\sum_{i=1}^N |\Vec{\psi}_i(\lambda_k)|^4
\end{equation*}
In figure \ref{fig:iprmin_VS_H} (top plot) we show the sample-averaged IPR of the smallest eigenvalue as a function of the external field strength $H$, ranging from $H=17$ in the gapped phase to $H=0.17$ deep in the spin glass phase. We consider $N=2^{18}$, the largest size available for all $H$.
The IPR tends to a constant for large $H$, has a peak supposedly at $H=H_{gap}$\footnote{It is not at $H=3.46$ for sure, since for this value we observe the glassy lower tail in figure \ref{fig:avgCumul_m=3}: thus a posteriori $3.46<H<4.33$.} and decrease with decreasing field in the whole gapless region. In the inset we show a zoom of the small $H$ region, with a quadratic fit of our data.

The infinite-size IPR is finite for any $H$ we measured. 
In figure \ref{fig:iprmin_VS_H} (bottom plot) we show the IPRs of the $n=12$ smallest modes. Even for the smallest $H$ we simulated, $H=0.17$, the IPRs seem to converge to a finite value as the size increases.

These measures unambiguously show that modes whose rank is $O(1)$ with $N$ growing are always localised, deep in the spin glass phase as well. However, a finite IPR does not imply necessarily that the system is localised around a single spin. It is possible that with the transition soft modes localise around multiple soft spots, separated by distances comparable with the global scale of the random graph $L=\log N/\log(c-1)$.

\subsubsection{Probing multiple soft spots}

In order to measure the emergence of multiple soft spots we abide to the following strategy:
\begin{itemize}
    \item First, we consider the following correlation functions
    \begin{eqnarray}
    \label{eq:corr_funcs_eigvecs}
        && \mathcal{C}_2^{(k)}(d)\,=\,\sum_{i\in\partial^d 0}|\Vec{\psi}_i(\lambda_k)|^2\qquad \sum_{d=0}^{\infty}C_2^{(k)}(d)\,=\,1 
    \end{eqnarray}
    where $\partial^d 0$ is the set of nodes with distance $d$ from node $0$ (the $d$-th shell). We label the site where each eigenvector has its absolute maximum with $i=0$. The Correlation function in \eqref{eq:corr_funcs_eigvecs} tells us how the normalisation is distributed in the various shells centered on node $0$. When an eigenvector is localised in one single core, typically these function decay exponentially with distance. The emergence of a maximum of these functions at distances $d=O(\log N)$ is a signal of the presence of other relevant spots. 
    In order to distinguish between average and typical behavior, we consider both $\overline{\mathcal{C}_2(d)}$ and $\exp\overline{\log\mathcal{C}_2(d)}$.
    \item Second, for each sample and eigenvector among the first $n=100$ we consider the first $l=10$ maxima of the density profiles $\{|\Vec{\psi}_i(\lambda_k)|^2\}$. Related to these, we measure the following quantities
    \begin{itemize}
        \item The histogram of the distances $d_{1k}$ between the absolute maximum on the eigenvector and secondary maxima, focusing in particular on $d_{12}$.
        \begin{eqnarray}
            && P_{d_{1k}}(d)\,=\,\frac{1}{N_s}\overline{\sum_{k=1}^{N_s}\delta(d-d_{1k})} \\
            && \nonumber \\
            && d_{1k}\equiv d(\Psi_1^2, \Psi_k^2)\equiv \{\text{length of shortest path connecting $1$ and $k$}\} \nonumber \\
            && \nonumber \\
            && \Vec{\Psi}_k(\lambda)\in \{\Vec{\psi}_i\in \boldsymbol{\psi}(\lambda):|\Vec{\psi}_i|^2>|\Vec{\psi}_j|^2\;\;\forall j\in \partial i\} \nonumber
        \end{eqnarray}
        \item The values of the maxima: their distribution with distance and their sample averages.
        \item Considering the shortest path connecting the absolute maximum and a secondary maximum, we measure for each such pair the absolute minimum along the path and the distance between the absolute minimum and the secondary maximum. 
    \end{itemize}
    By measuring directly different soft spots we can get a connection with the underlying topology given by the random graph.
\end{itemize}

\subsubsection{Weak delocalisation at the spin glass transition}

\begin{figure}
    \centering
    \includegraphics[width=0.8\columnwidth]{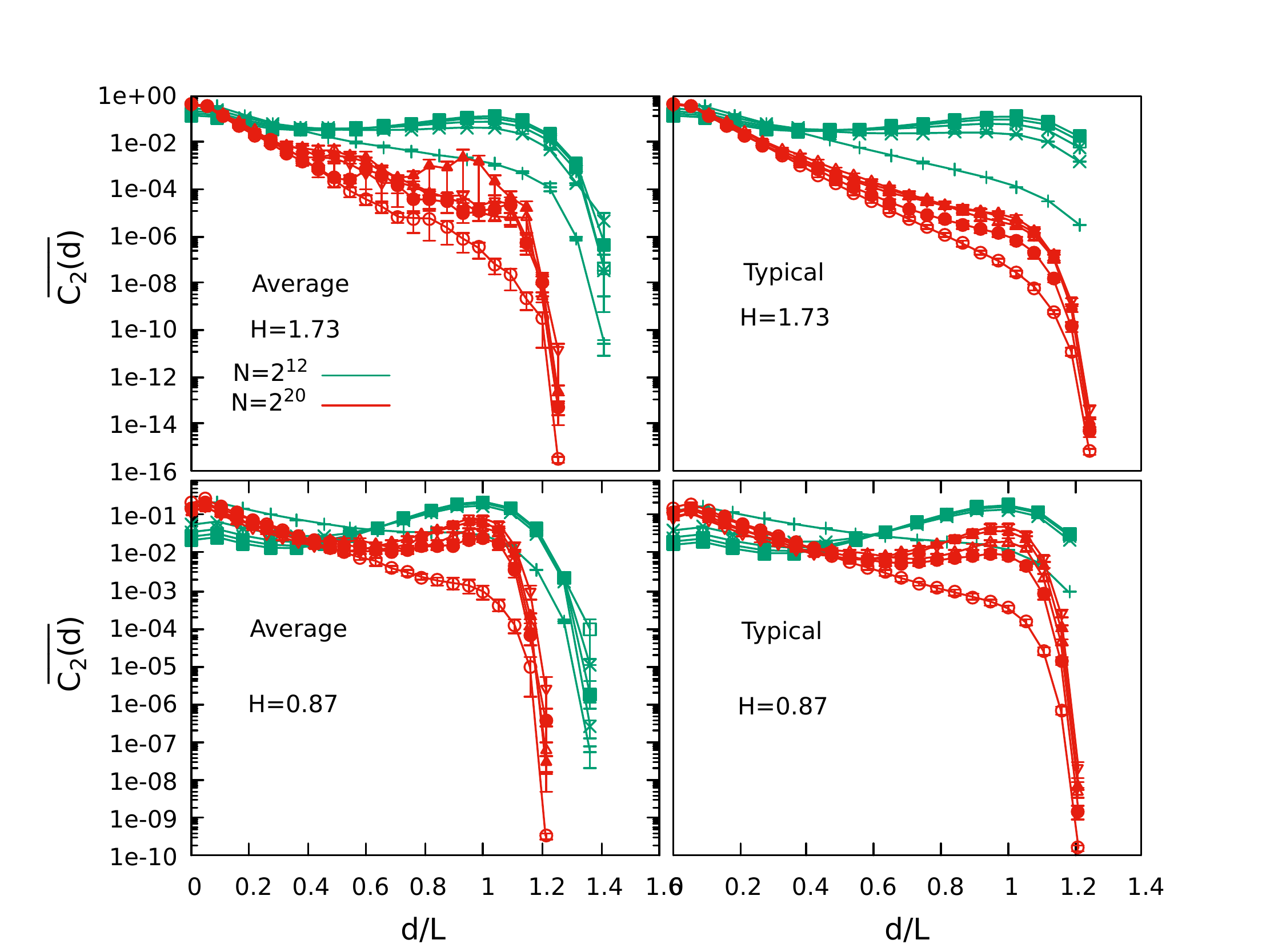}
    \caption{The correlation function defined by \eqref{eq:corr_funcs_eigvecs}. Left figures are sample-averages, right ones evaluations of typical values through the exponentiation of the averaged logarithm of the correlations. Modes with rank $1\ll k \ll N$ feature correlations with a maximum at distances $O(\log N)$ (For each group of curves, the lowest one refers to mode $n=1$, the others in ascending order to the remainders).}
    \label{fig:corrs}
\end{figure}

We start showing our measures of \eqref{eq:corr_funcs_eigvecs} through figure \ref{fig:corrs}. The top figure on the left contains curves $\overline{C_(d)}$ VS $d$ for $H=1.73$, $N=2^{12}, 2^{20}$ and the modes with rank $n=1, 20, 40, 60, 80$. The top figure on the right same plot but with $\exp\overline{\log{C}_(d)}$. In the bottom figures same for $H=0.87<H_c(0)$. In both cases a maximum at distance $\sim\log(N)/\log(2)$ appears for $N=2^{12}=4096$. Conversely, for the largest size $N=2^{20}=1048576$ one has typically a maximum for modes $1\ll k\ll N$ in the case $H=0.87$ but not in the case $H=1.73$. This hints a difference between the modes in the RS phase and those in the RSB one. However, since the behavior of the smallest mode seems to be very similar in both cases, we should study the properties of eigenvectors by rescaling the rank of our modes with the size of the system: $k_N=k/N^a$, for some $0<a<1$.

\begin{figure}
    \centering
    \includegraphics[width=0.8\columnwidth]{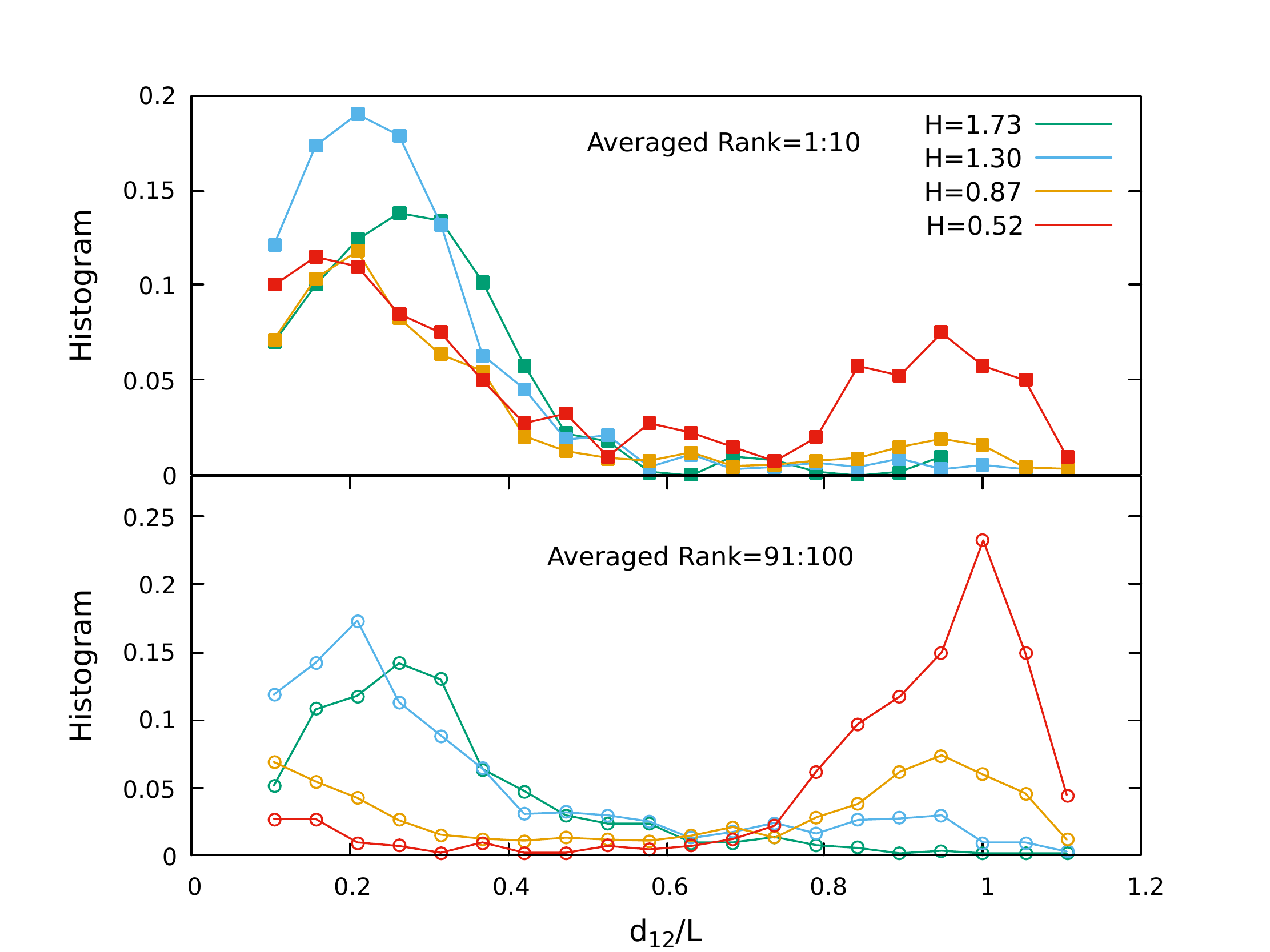}
    \caption{Histogram of the distance on the RRG between the two most relevant spots of each sample, for samples with size $N=2^{20}=1048576$. Figure on top is an average over the first ten modes, on bottom an average over the last ten within the $n=100$ modes available.}
    \label{fig:histo_d12_compare_H_Nmax}
\end{figure}

\begin{figure}
    \centering
    \includegraphics[width=0.45\columnwidth]{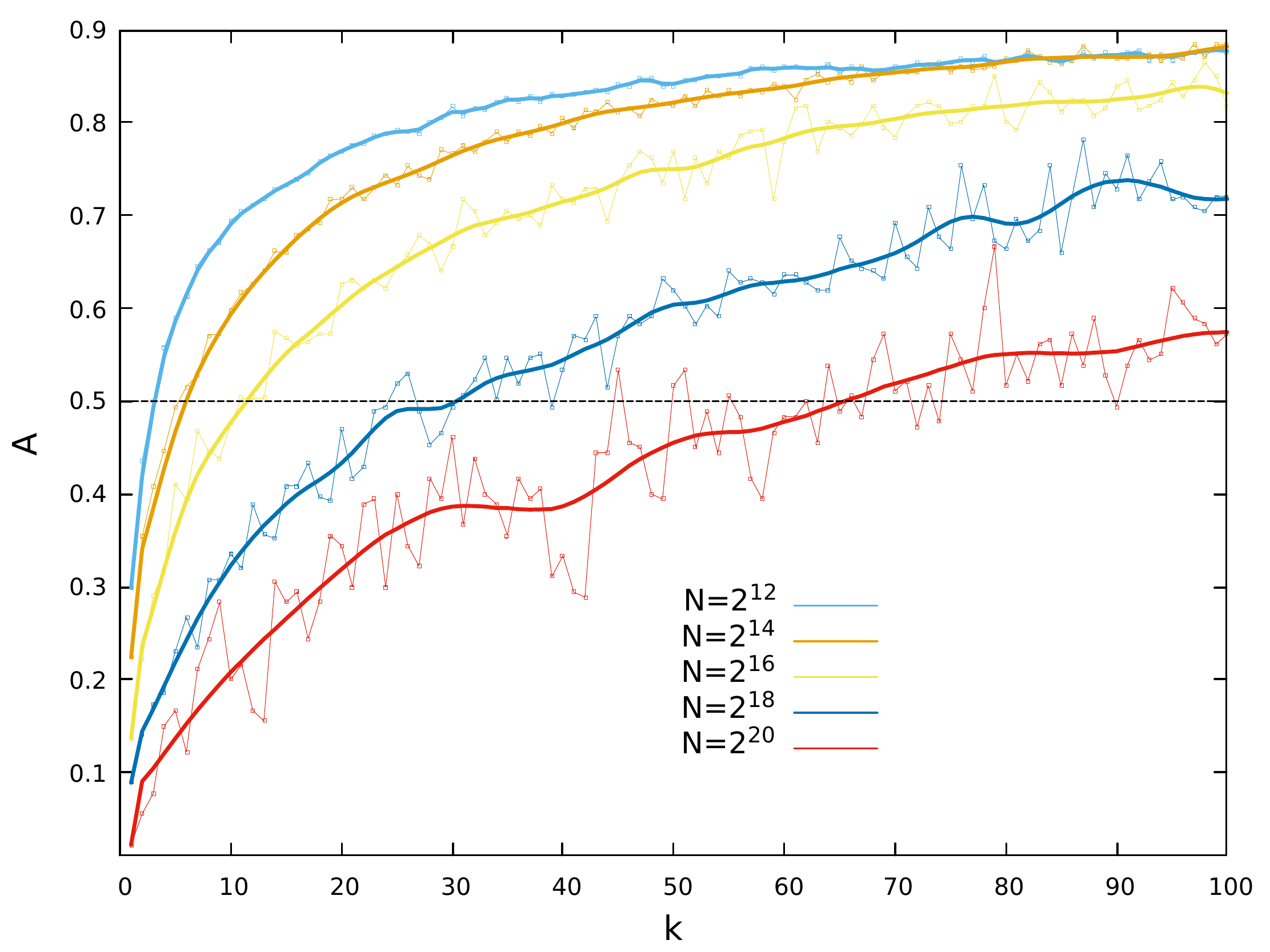}
    \includegraphics[width=0.45\columnwidth]{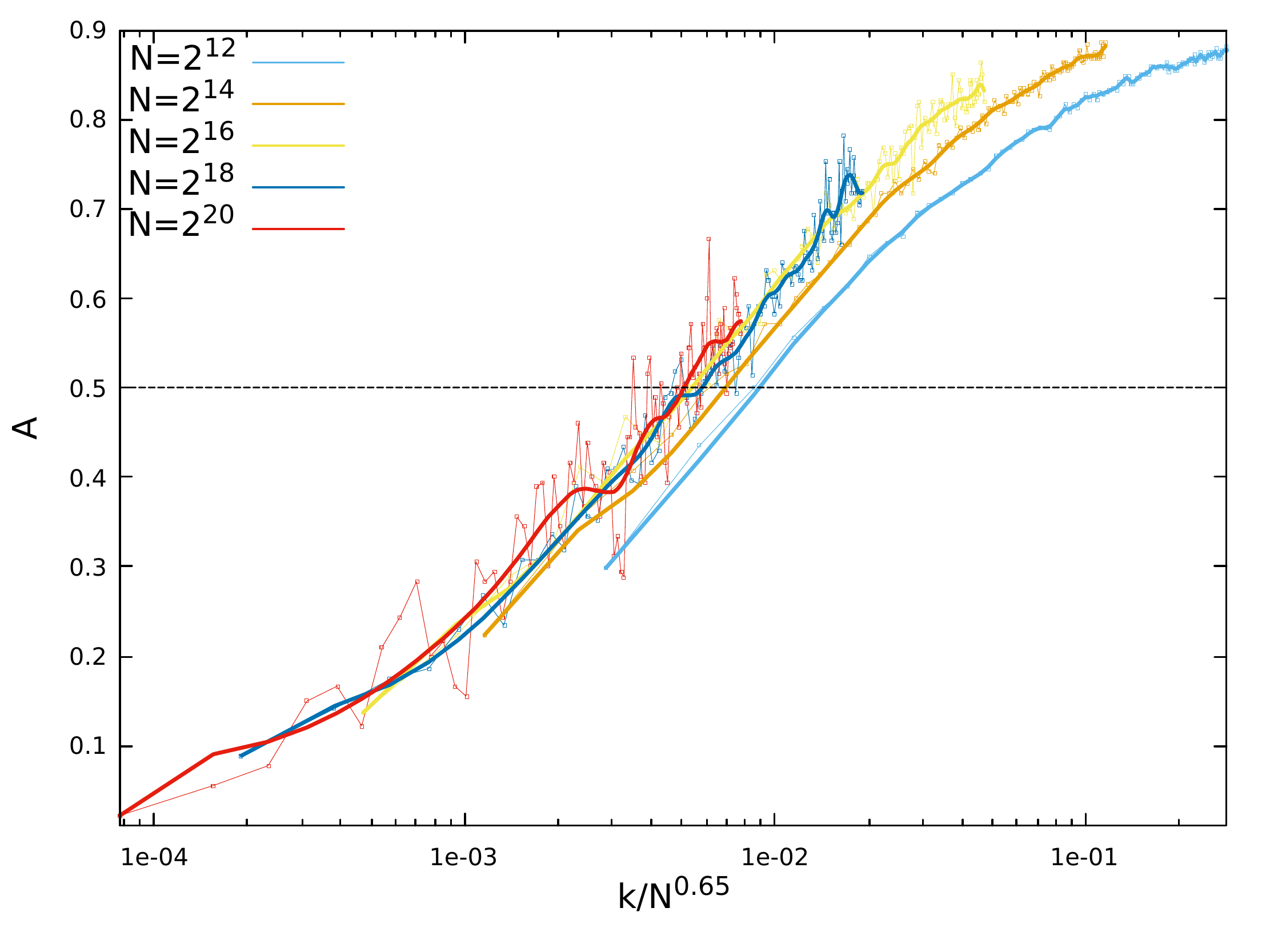}
    \caption{The probability of observing two soft spots at distance $O(\log N)$, as a function of modes rank (left) and the rescaled rank (right) for $H=0.87$. By rescaling the rank with a suitable non-trivial power of $N$, the curves show a good collapse.}
    \label{fig:Pgreater_RSB}
\end{figure}

In order to do this, we proceed with the second step of our strategy. We begin by showing in figure \ref{fig:histo_d12_compare_H_Nmax} measures of the histograms $P(d_{12})$ VS $d_{12}/d_g(N)$, where ($c=3$)
\begin{equation}
    d_g(N)\,=\,\operatorname{Floor}\left(\frac{\log(N/3)}{\log(2)}\right)+1
\end{equation}
is the distance at which $\mathcal{N}(d)$ in \eqref{eq:path_length_density} is maximised. We show histograms of the $m=3$ system for the largest size measured, $N=10^6$, and for the values $H=1.7, 1.3, 0.87, 0.5$. We can appreciate two maxima, one at distances $d=O(1)$ and the second at $d=O(\log N)$: as $H$ is lowered down to the spin glass phase, the relative importance between the two maxima change, with the maximum at large distances becoming more and more important. These two maxima identify two different kind of samples: in those with $d_{12}=O(1)$ the second dominant contribution to normalisation belong to the same regular subtree\footnote{The regular tree rooted on the absolute maximum.} of the absolute maximum, whereas when $d_{12}=O(\log N)$ the two dominant spots $\Psi_1^2$ and $\Psi_2^2$ belong to different subtrees.
We estimated that on average in a RRG the regular subtree that emanates from any node has depth $d_t=L/2\sim \frac{1}{2}\frac{\log N}{\log (c-1)}$, i.e. the half of the global scale of the graph; details in Appendix \ref{sec:typ_subtree}. 
So, we decided to split the histogram in two parts, $d_{12}<d_{t}$ and $d_{12}>d_{t}$, and measure 
\begin{equation}
\label{eq:Area_under_logN_max}
    \mathcal{A}\equiv \mathcal{P}[d_{12}\geq L/2]=\sum_{d\geq L/2} P_{d_{12}}(d)
\end{equation}
for each mode at given $N, H$. Eq. \eqref{eq:Area_under_logN_max} is just the area of the "long distance" region in figures \ref{fig:histo_d12_compare_H_Nmax}, corresponding to the probability to observe a sample with $\Psi_1^2$ and $\Psi_2^2$ at distance $O(\log N)$. We study this probability as a function of modes rank, and in particular of the rescaled rank $k_N=k/N^a$, for $0<a\leq 1$.
In figures \ref{fig:Pgreater_RSB} we show these probabilities as functions of the rank $n$ of the mode, for the value $H=0.87$ in the spin glass phase. 
The figure on the left is the probability as a function of the rank, on the right, we rescaled the rank with a suitable power of $N$. This measure shows that, for any $k$, the probability of finding a sample with the two main soft spots separated decreases as the size is increased.
On the other hand, the picture on the right tells us that by rescaling the rank with a suitable power (for our data $a=0.65$ works well), the probability of finding samples with separated $\Psi_1^2$ and $\Psi_2^2$ does not depend on $N$ for large $N$.
A similar phenomenology holds for soft spots of lower rank, i.e. for the distance between $\Psi_1^2$ and $\Psi_k^2$ with $k>2$.

\begin{figure}
    \centering
    \includegraphics[width=0.45\columnwidth]{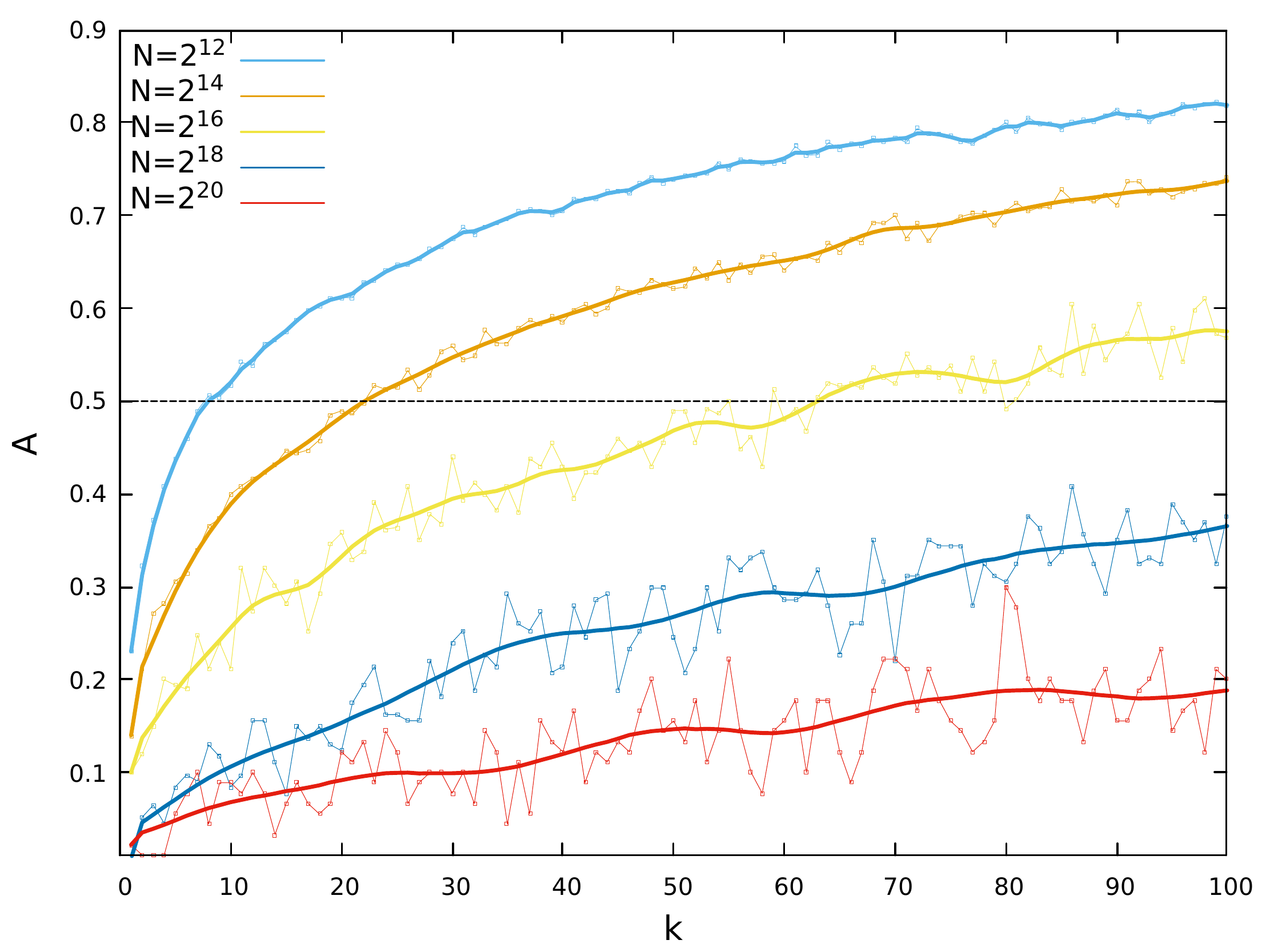}
    \includegraphics[width=0.45\columnwidth]{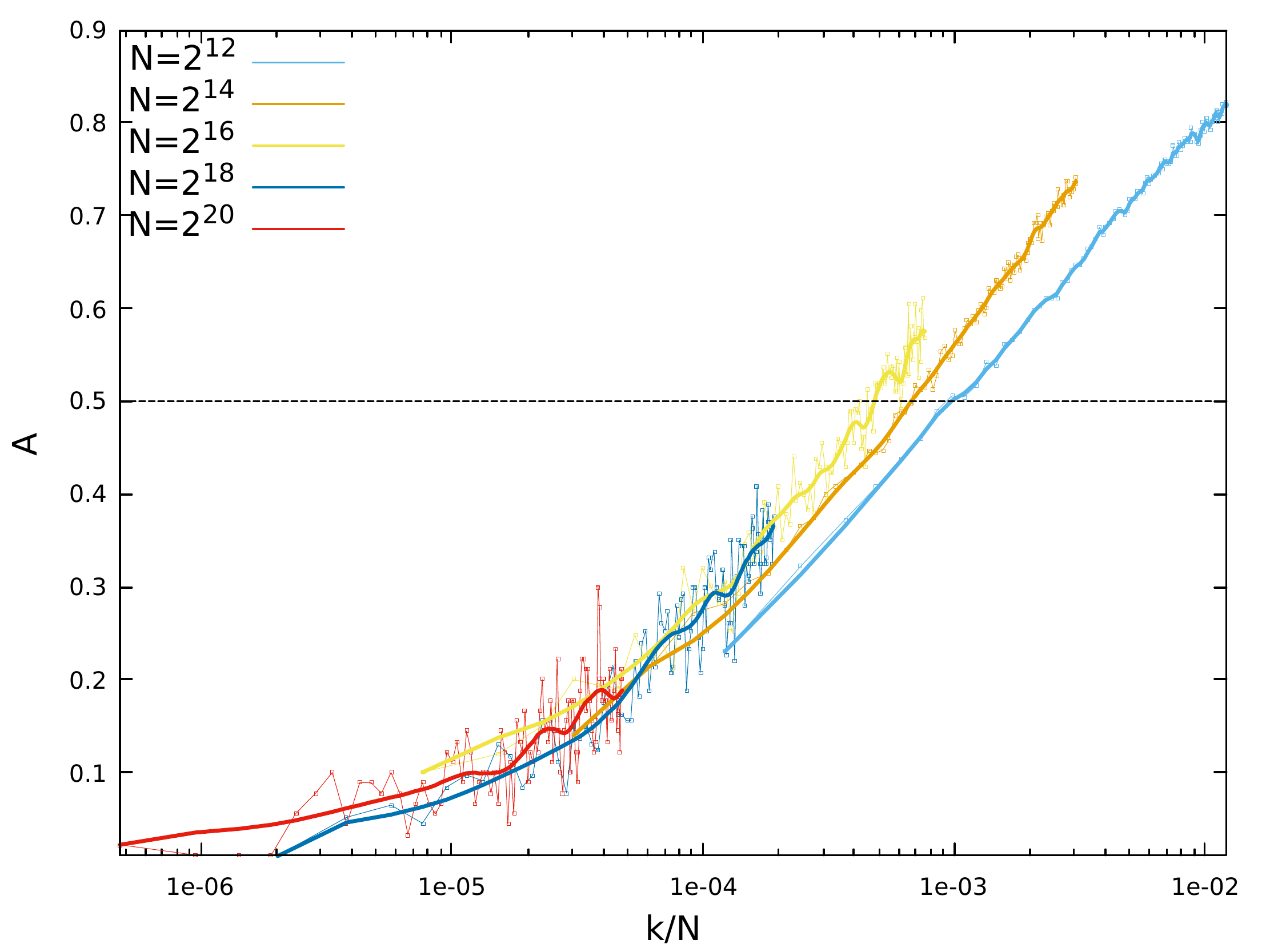}
    \caption{The probability of observing two soft spots at distance $O(\log N)$, as a function of modes rank (left) and the rescaled rank (right) for $H=1.30$. In the RS phase we can have a size independent probability only by rescaling $k$ with $N$.}
    \label{fig:Pgreater_RS}
\end{figure}

\begin{figure}
    \centering
    \includegraphics[width=0.45\columnwidth]{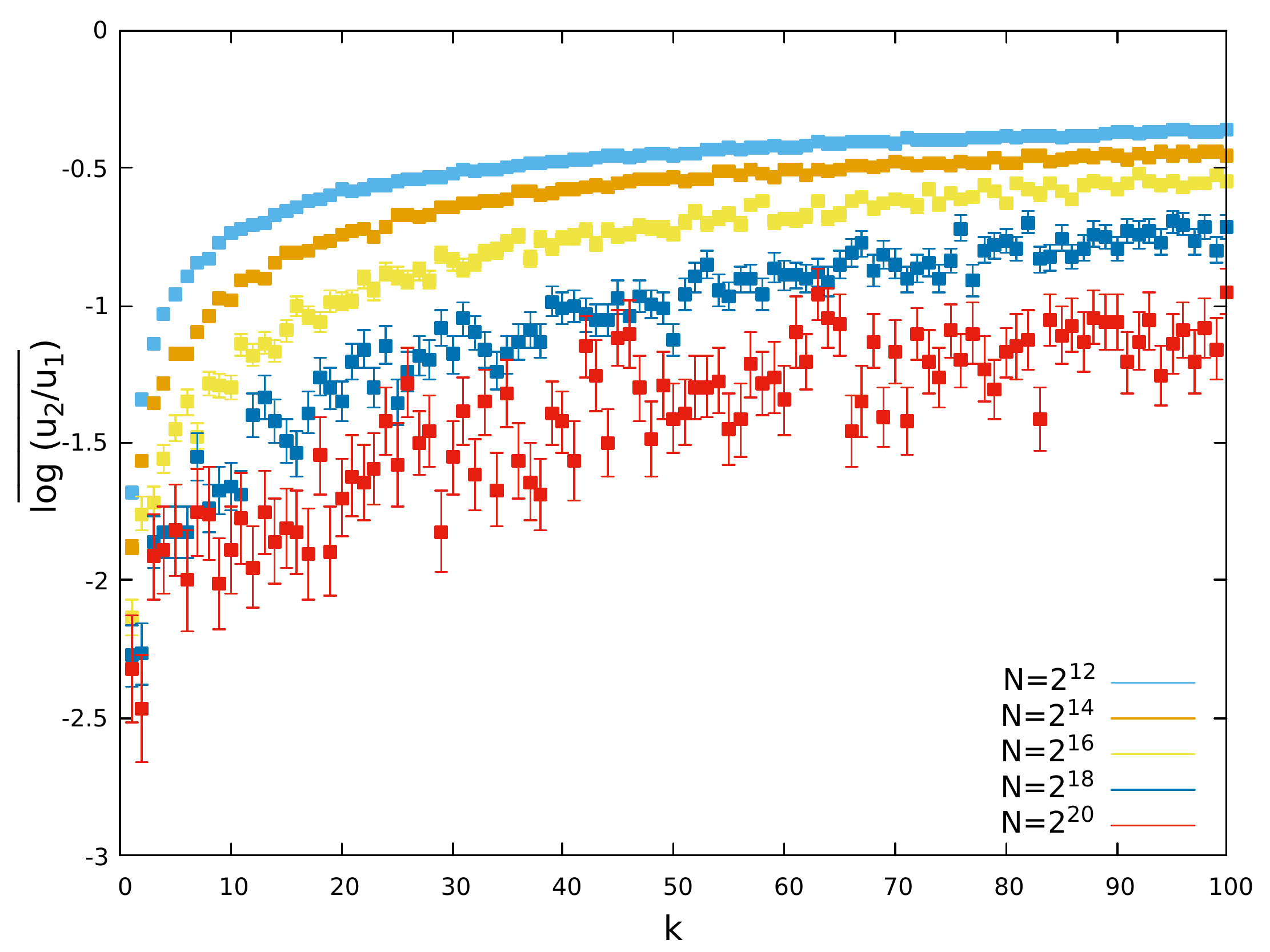}
    \includegraphics[width=0.45\columnwidth]{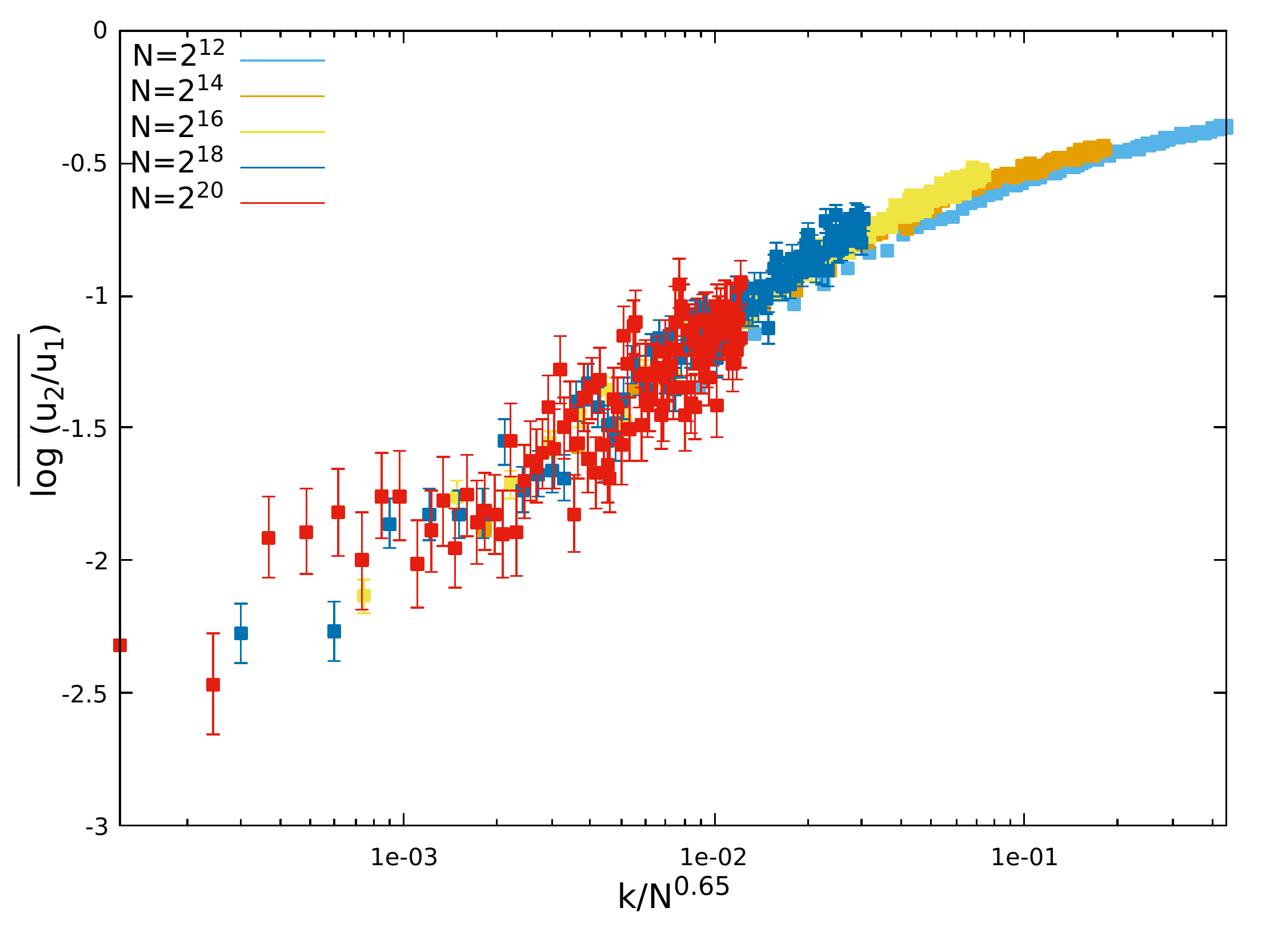}
    \includegraphics[width=0.45\columnwidth]{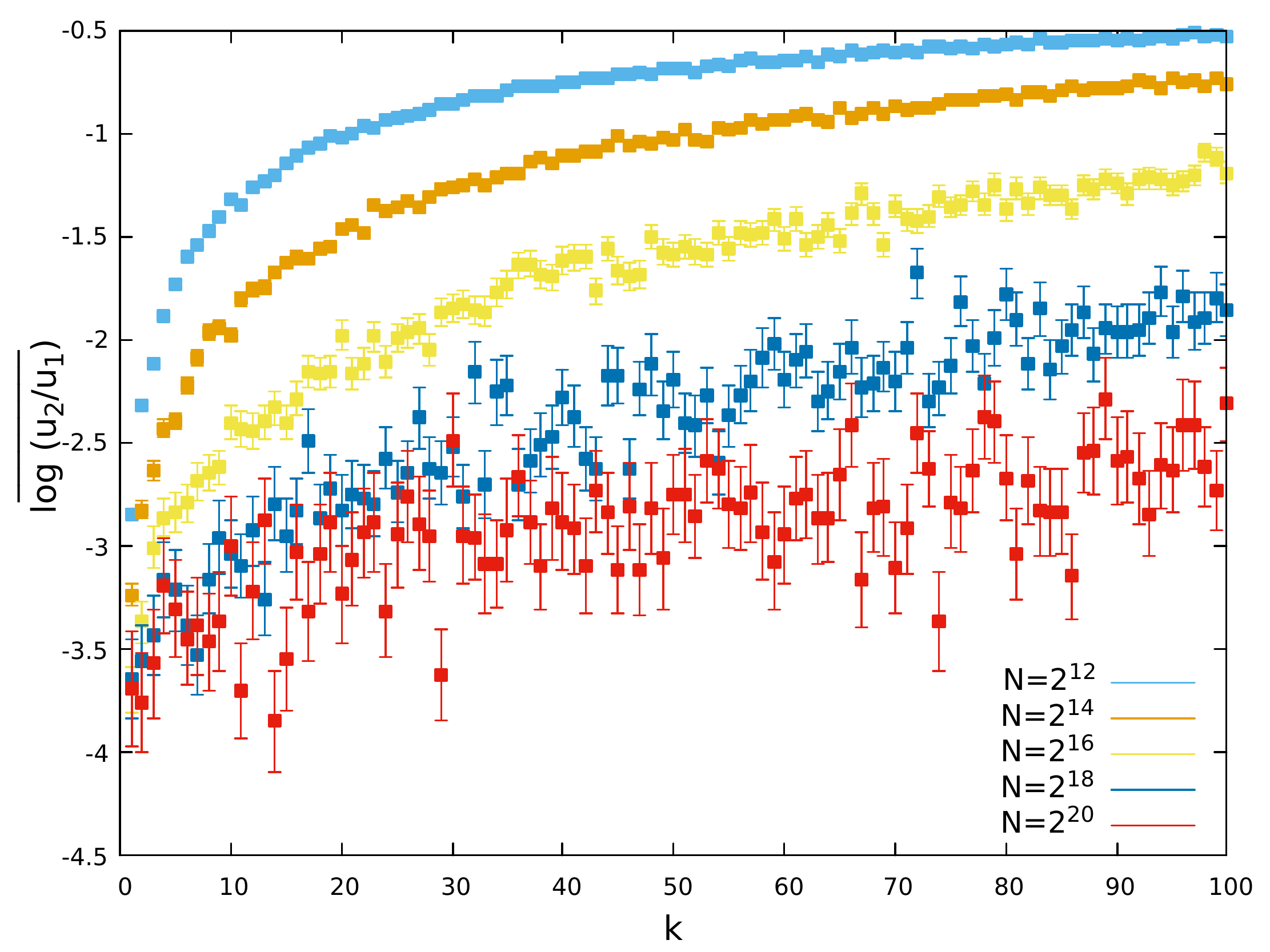}
    \includegraphics[width=0.45\columnwidth]{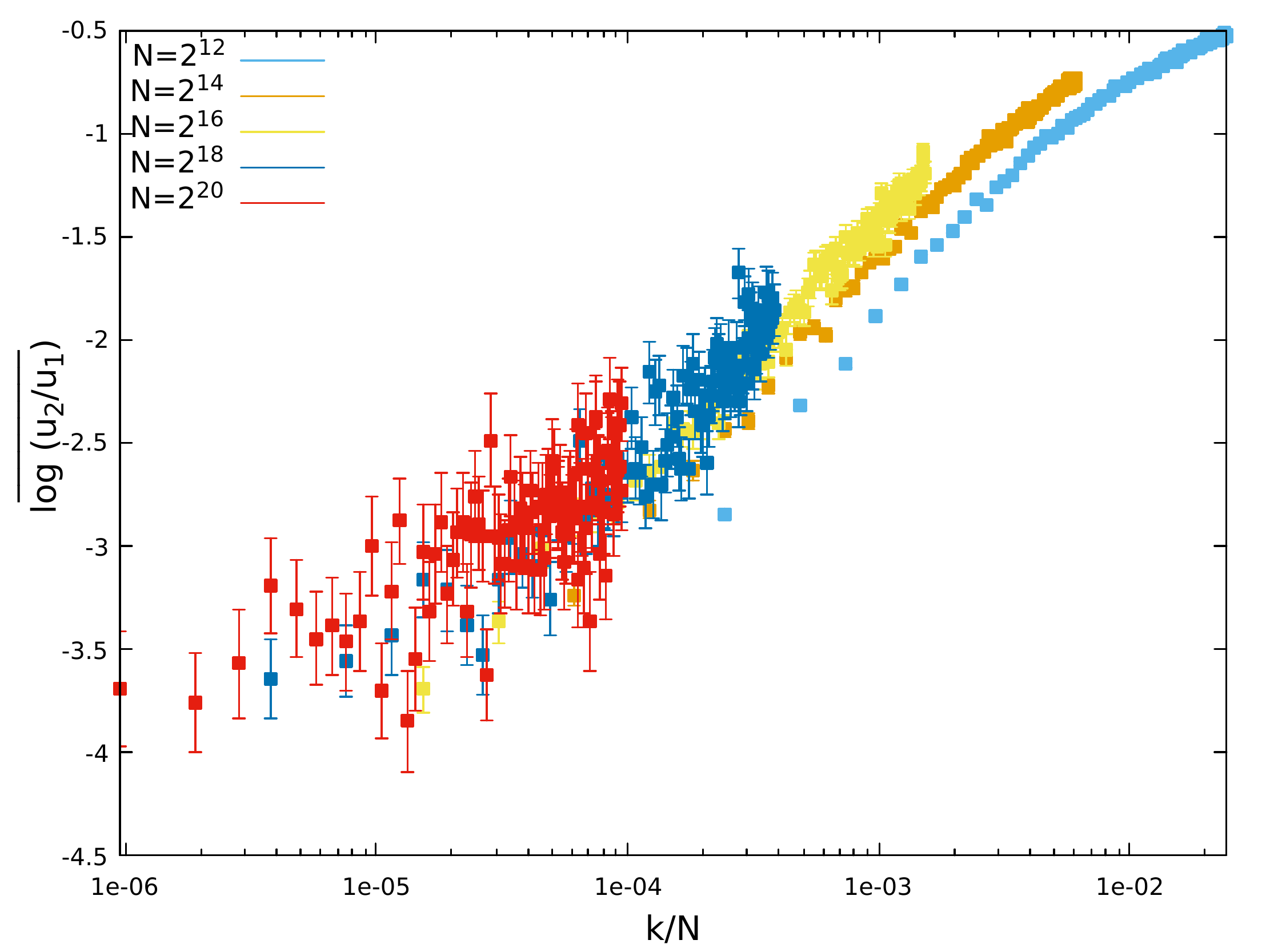}
    \caption{The sample-averaged logarithm of the ratio between the two main spots. Top-left is $H=0.87$ (putative RSB) with $k=O(1)$, top-right same but with $k=O(N^{0.65})$. Bottom-left is $H=1.30$ (RS) with $k=O(1)$, bottom-right same but $k=O(N)$.}
    \label{fig:logRatio}
\end{figure}

This evidence suggests that with the transition there is a weak delocalisation phenomenon: modes with multiple soft spots appear when $k=O(N^a)$, with $a<1$. Since each finite eigenvalue in the infinite-size limit is in one-to-one corrispondence with $k/N$, we conclude that this separation of spots holds for eigenvectors whose related eigenvalues in the thermodynamic limit are arbitrarily close to the lower edge\footnote{They are eigenvalues with rank $k=O(N^a)$, thus $CDF\approx k/N=N^{a-1}$ which vanishes for $a<1$ and $N$ going to infinity.}. Note that in this situation if one considers a fixed position in the spectrum by setting $k=O(N)$, the probability $A$ increases with size.
This is at variance with the same measures in the RS phase: in figures \ref{fig:Pgreater_RS} we show $\mathcal{A}$ but for $H=1.3$: here only for $k=O(N)$ the quantity $A$ is size-independent.
In figure \ref{fig:logRatio} we show the typical ratio of the two main soft spots, calling $u_1=\Psi_1^2$ and $u_2=\Psi_2^2$, as a function of the rank $k$. For finite ranks the second spot becomes smaller and smaller with respect to the main one, as size increases. With proper rescaling, the quantity becomes size-independent. In the RS phase, as before, the only scaling yielding collapse is $k=O(N)$.

\begin{figure}
    \centering
    \includegraphics[width=0.8\columnwidth]{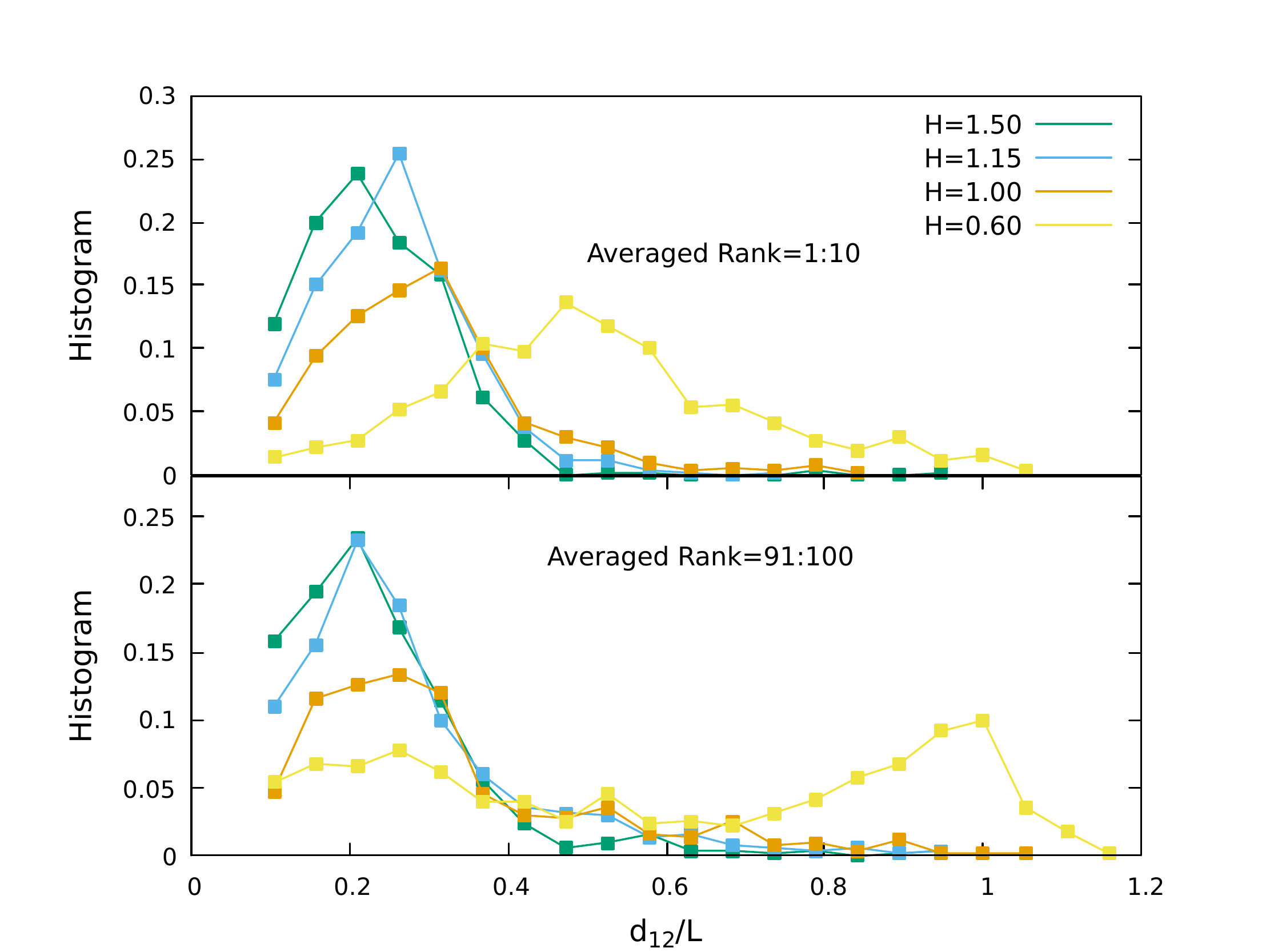}
    \caption{Histograms of the distance between the two most relevant soft spots in the XY model. The probability of observing separated couples is far lower than in the Heisenberg case.}
    \label{fig:histo_d12_compare_H_Nmax_XY}
\end{figure}

\begin{figure}
    \centering
    \includegraphics[width=0.8\columnwidth]{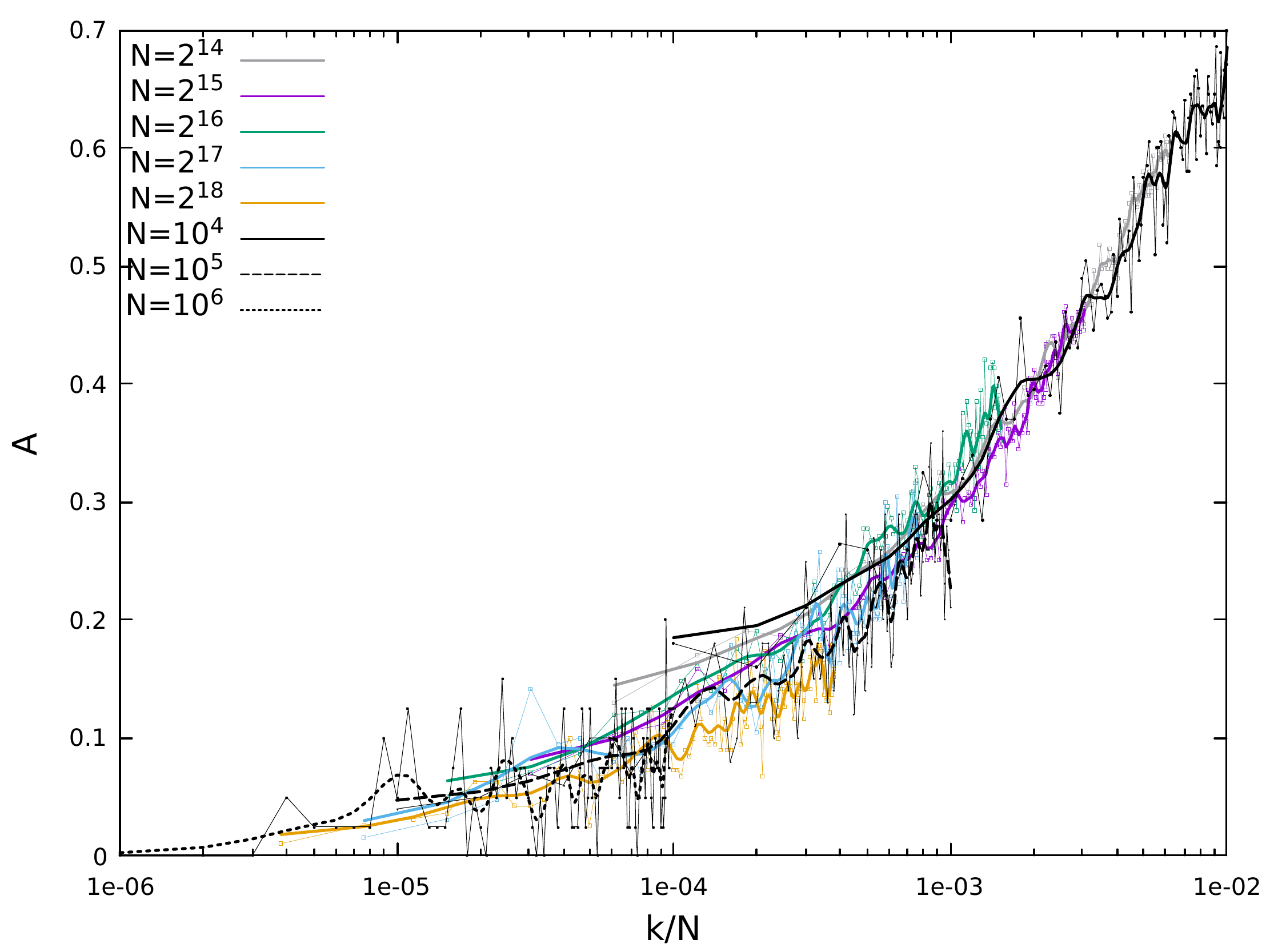}
    \caption{The probability of having $d_{12}\geq L/2$ for the XY model with $H=1.00<H_c(0)=1.15(1)$ \cite{lupo2017approximating}. For this value of $H$ down in the spin glass phase the probability is finite only when $k=O(N)$.}
    \label{fig:Pgreater_XY}
\end{figure}

In figure \ref{fig:histo_d12_compare_H_Nmax_XY}, we show the same plot as in \ref{fig:histo_d12_compare_H_Nmax} but for the XY model, considering $H=1.50$ (RS phase), $H=1.15$ (critical point), $H=1.00, 0.600$ (spin glass phase\footnote{The critical field estimated in \cite{lupo2017approximating} is $H_c(0)=1.15(1)$}). The histograms in these figure are computed from data gently offered by Dr. Lupo, from his PhD thesis work \cite{lupo2017critical}. 
Apparently, the phenomenon just described for the Heisenberg model seems to be very weak in this case: only for the quite low value $H=0.60$ we observe a bimodal pattern in the histogram. At the critical point $H_c(0)=1.15(1)$ nothing seems to happen. Finally, in figure \ref{fig:Pgreater_XY} we show the measure of $\mathcal{A}$ for the value $H=1.00$ in the spin glass phase. Data referring to power of two sizes are from us, the others again from \cite{lupo2017critical}. The probability of observing two modes is finite only for $k=O(N)$. The absence of a clear signal at criticality seems to suggest that, by the point of view of low energy excitations, the behavior of Heisenberg and XY models is differrent.

\section{Conclusions}

In this chapter we discussed a diluted version of the model studied in chapter 4. The main focus of the chapter has been the extension of our results in the dense case to the sparse model \eqref{eq:Hamiltonian_sparse}. We found a gapless phase in $H$ with spectra populated by low energy linear excitations with a density of states following a quartic law. In the RS phase, the modes are localised on a single core, whence as criticality is approached long-range order seems to emerge under the appearance of localised modes with multiple localisation cores. This effect is not as strong in the XY model, hinting possibly a different critical behavior with respect to the Heiseberg as long as the low linear excitations are concerned. We believe these results enrich the picture around the zero temperature criticality of sparse spin glass models. 

\chapter{Conclusions and Perspectives}

In this thesis we studied three different vector spin glass models. The first model is a $m$-vector SK model with a random field on a fully-connected graph, the second is a $m$-vector $p$-spin model on a fully-connected graph and finally the third is a random $m$-vector SG on a random regular graph with connectivity $c=3$.
We focused on the low and zero temperature physical behavior, by considering the problem of linear low energy excitations. Soft modes are a central topic in condensed matter: they rule the low-temperature mechanical responses and specific heat.
Structural glasses feature an additional density of quasi-localised states in their vibrational spectrum. While many numerical results have been available in the last decades, first principle theories of glassy excitations are still missing. 
Our work can be considered as an attempt to explore this problem in a first-principle theoretical framework, that of continuous spin glasses. Vector models constitute an improvement with respect to spherical models. While the energy Hessians of the latter are represented by classical random matrix ensembles, which feature only delocalised excitations as a consequence of rotational invariance in the space of matrices, vector models Hessians are represented by rotationally non-invariant random matrix ensembles, like for instance the Rosenzweig-Porter or Deformed Wigner ensemble. In this case, localisation in the softest and/or hardest eigenvectors can occur. In the dense case, we characterised the transition from RS (stable) to RSB (marginal) phases in terms of a random matrix transition, involving the softest modes of the energy Hessian. In the sparse case we were not able to characterise excitations in terms of random matrices.

Let us make a resume of the work contained in this thesis. In the first part (Background) of the manuscript we introduced the reader to the main tools used in our research. 
\begin{itemize}
    \item In chapter 1 we discussed spin glass theory in a historical perspective. We wanted to give an idea to the reader of the steps that progressively led to the RSB theory of G. Parisi, in order to make apparent its central role as a tool to understand the physics of disordered system. After that, we considered vector models. These models posses a richer critical phenomenology, featuring weak and strong RSB according to whether one considers longitudinal or transverse alignment with respect to an external magnetic field. When a random external field with zero mean is applied, these models allow to study non-Goldstone soft modes. In addition, a random external field destroys the Gabay-Tolouse line, leaving a dAT instability line with zero temperature phase transition. This transition exists only for spins with $m>2$ components in the dense case, while it is always present in sparse models.  In the last part of the chapter we introduced Belief propagation and showed its connection to cavity method. These techniques permit to find a solution of the system in sparse tree-like graphs, solution which is obtained from self-consistent analytical equations that connect the on-site spin cavity marginals: from the solution of these equations, the actual one-point and two-point spins marginals can be computed. Sparse models can be considered as an improvement of the naive mean field theory yielded by fully connected models.
    \item In chapter 2 we introduced structural glasses and the glass transition framework, through the lenses of Random First Order Theory, a mean field approach that explains glass formation by means of the emergence of a complex 1RSB landscape. After considering glass formation by a dynamical point of view, we showed how a connection with statics and in particular with the energy landscape through the study of complexity could help to characterise the different phases of glass formation.
    
    We then considered the spherical p-spin model, a toy model for structural glasses. This model features a paramagnetic (liquid) phase at high temperatures, a dynamic (supercooled liquid) phase at intermediate temperatures and a glassy phase (spin glass) at low tempeartures. The Golstein crossover to activated dynamics in finite dimensions corresponds to a dynamic transition in the mean field case. The glass transition is characterised by the entropy crisis phenomenon, which in the p-spin case corresponds to the vanishing of the equilibrium complexity: the equilibrium measures condensates in the lowest free energy states. 
    
    Finally, we considered excitation spectra: the Hessian of spherical p-spin models is shifted (non-centered) Wigner matrix. The energy landscape is divided in two disjoint intervals: the lowest one contains energy minima that are stable with a gapped spectrum, whence the one on top is populated by saddles. The common point between the two is the level of marginal energy minima: these have gapless spectra with delocalised excitations.
    \item Chapter 3 wants to show how potential energy Hessians of disordered systems can be represented in the dense mean field case by random matrix ensembles. To this purpose, we start by discussing the general problem of harmonic vibrational excitations in crystals and structural glasses. We discuss how much they differ from each other, by pointing out the existence of low frequency quasi-localised excitations in glassy systems, that qualitatively strongly differ from phonon excitations. We introduce soft potential models, phenomenological theories that explain the appearance of these modes by representing the potential energy of a glass as a superimposition of anharmonic potentials, identifying strucural local defects of the sample.

    The second part of the chapter introduces random matrix theory. After a general introduction about gaussian ensembles, we consider the problem of deformed gaussian ensembles, where a Wigner matrix is perturbed by a diagonal random matrix. The resulting ensemble, named Rosenzweig-Porter, represents the Hessians of a large class of disordered systems.
\end{itemize}

The second part of the thesis (Vector spin glasses on fully connected graphs) is divided in two chapters: chapter 4 contains our original results about the random field Heisenberg spin glass, chapter 5 involves our findings related to the p-spin vector model.

\begin{itemize}
    \item Chapter 4 begins with a general replica computation of the free energy of the model, both in the replica-symmetric case and in the spin glass phase. We discuss the low temperature behavior in both cases, hinting at possible future research developments in the spin glass regime.
    
    After that, we re-derive the TAP free energy of the model, through a Plefka-Georges-Yepidia expansion of the Gibbs potential. This is preparatory for a brief discussion about the Hessian of the TAP free energy, specialising on the low temperature limit.

    The last part of the chapter contains our study of soft modes of the inherent structures of the energy landscape, i.e. the study of the Hessians of the Hamiltonian function evaluated on a minimum configuration. The results are based on our paper \cite{franz2021delocalization}. We derive analytical equations for the resolvent function through cavity method and test its predictions through an extensive set of numerical simulations.
    
    We begin by testing the performances of the algorithm chosen for obtaining energy minima, the over-relaxated gradient descent algorithm with random initial condition. We find that whenever the landscape is non-trivial-in finite size systems this happens close to criticality in the paramagnetic phase and in the spin glass phase-the gradient descent with over-relaxation reaches lower energy minima than the simple gradient descent. The lower the random external field strength, the broader is the range of values of the over-relaxation weight for which this statement is true. The energy advantage provided by the over-relaxated descent is relevant only in the spin glass phase has the size of the system is increased. The convergence time advantage instead is relevant also in the paramagnetic phase, for values of $H$ up to $2H_c$ (the non-trivial paramagnetic phase). The convergence time of the simple gradient descent grows with size approximately as $N^{0.5}$ in the spin glass phase, while in the non-trivial paramagnetic phase it grows with $N$ through an $H$ dependent exponent. For all $H, N$, it does exist a value of $O$ such that for any stronger over-relaxation the convergence time is size-independent. This impacts the overall time complexity of the minimisation. This non-trivial dependence of the convergence time on the over-relaxation also in the paramagnetic phase suggests that also in this case the landscape is non-convex, provided that $H$ is close enough to $H_c$. In this scenario, scenarios with multiple energy levels and abundance of saddles occur with finite probability.

    Once the performances of the algorithm have been tested, we begin the numerical study of the Hessians, evaluated for each sample with given $H$ and $N$ at a fixed point configuration of the over-relaxated descent. We focused on the $m=3$ model, with several values of $H$ in the paramagnetic phase, at the critical point and few in the spin glass phase. We measured sizes ranging from $N=2^6=64$ to $N=2^{11}=2048$: these are sizes typically accessible in numerical simulations of dense systems. Diagonalisation is achieved through the Lapack library: spectra having all eigenvalues positive were identified as minima and saved. We studied the properties of eigenvalues and eigenvectors: we tested the prediction of the theory both in the bulk of the spectrum and in the lower edge. In the paramagnetic phase the theory predicts a condensation of the softest eigenvectors on single sites of the sample, related to the smallest cavity fields of the system: in a fully-connected model, these unambiguously identify the most magnetically susceptible spins. At the spin glass transition, the condensate disappears and the lower edge modes undergo a delocalisation transition. We find that our data from simulations agree well with bulk predictions but are affected by strong finite size effects at the lower edge: these effects prevent to observe the condensation in near critical ($H\sim H_c$ but larger) finite size systems. Only in the trivial paramagnetic phase condensation can be observed.

    Our results in this chapter yield a naive mean field description of glassy excitations. The existence of localisation phenomena in dense mean field models revisits common beliefs, stemming from the study of excitation of popular mean field models of disordered systems, such as the spherical p-spin and the perceptron. The zero temperature spin glass transition is characterised as a random matrix transition: this gives us a greater insight on the proeprties of the zero temperature critical point. In particular, the softest modes at criticality and in spin glass phase are multi-fractal: these suggests even in presence of delocalisation a highly non-trivial organisation of the local responses.

    \item In chapter 5 we applied the results obtained in chapter 4 to a mean field model of structural glasses, the $m$-vector $p$-spin. The findings in this chapter are contained in our work \cite{franz2022linear}. The results discussed are entirely analytical.

    We begin the study by characterising the dynamical phase of the model, through the computation of Monasson potential and the equilibrium Complexity of energy minima. The dynamical temperature is identified by the vanishing of the Replicon eigenvalue of the Hessian of the replicated action, whereas the glass trnasition temperature is identified by the vanishing of the equilibrium complexity.

    Following that, we focus on the zero temperature limit. We compute the zero temperature limit of Monasson free energy and the related complexity of the energy minima of the landscape. By studying the Hessian of the energy, we identify two bands in the energy landscape: the first is identified by $E_{gs}\geq E<E_{mg}$ and is dominated by minima whose excitation spectra are gapless and related lower edge eigenvectors condensated. We call these minima "stable" because within them no avalanche phenomena can occur. Depper minima in the landscape have increasingly localised soft excitations, in agreements with observation in numerical simulations of computer glasses.
    The second, identified by $E_{mg}\leq E<E_{th}$, is dominated by marginal minima: these feature excitation gaussian-like excitation spectra, with delocalised soft eigenvectors. In these minima the system responds finitely to infinitesimal perturbations, leading to avalanches. We did not compute the energy level $E_{th}$, whose existence we postulate by comparison with the spherical mixed p-spin.

    Besides gapless stable minima, in the stable phase there are rare ultra-stable minima, with a lower complexity with respect to typical gapless ones. These are mean field representatives of ultra-stable glasses: these are obtained by circumventing usual super-cooling vitrification through a physical vapor deposition process. Ultra-stable minima are identified through an enhanced stability condition on the local fields. The resulting excitation spectra are gapped and related excitations all delocalised, with no multi-fractality. This trivial behavior is a consequence of the enforced excess stability. Ultra-stable minima populate the whole stable energy band, being absent at the ground state level.
\end{itemize}

In this chapter we showed the generality of our results in chapter 4. We provided a non-trivial mean field framework for glassy excitations. In this framework, many properties of finite-dimensional glassy excitations are reproduced, even though in a mean field fashion. What is missing is the universal quartic law of the density of states. This property, as we learn in chapter 6, is peculiar of sparse systems. Actually, it is reproduced in a recently studied dense system, but that theory is a mean field transposition of the phenomenological soft potential model. Our theory instead is a first-principle mean field theory. 

The third part of the thesis (Vector spin glasses on sparse graphs) contains our yet unpublished results concerning the excitation properties of the random field vector spin glass on a random regular graph. The work is almost entirely numerical, considering only systems with connectivity $c=3$ and focusing mostly on $m=3$.

The chapter begins with a replica computation: we want to stress the difference between the sparse and dense model, by showing the emergence of a new order parameter. The order parameter of the diluted model is a functional that takes account of all possible spins correlations: it is a reflection of the uniqueness of each neighborhood of the underlying random graph and of the strong correlation standing between each spin and its neighborhood. The most convenient way to deal with this new object is to use the cavity method.

We consider the problem of determining the instability line of the $m=3$ system in the plane $H, T$, where $H$ is the strength of the random external field. Auxiliary to this problem, there is that of finding a good discretisation of the sphere: while the XY model can be effectively represented through the clock model, the discretisation of the sphere is a long-standing problem named "generalised Thomson problem". The points on the sphere are represented as particles, and different grids can be obtained by imposing different kinds of repulsive interactions. In general, the best discretisation for a given problem on the sphere is system dependent, so a thorough study of it taylored to the random field Heisenberg spin glass should be pursued. In this thesis we considered the case of random grid: while this choices yields poor tiling of the sphere and unfortunately enhances the discretisation effects especially at low temperature, with some efforts predictions on the instability line can be carried out. From our knowledge of the zero field critical temperature, we were able to extrapolate the instability line, though failing in computing accurately the zero temperature critical field: we instead identified a likely interval of $H$ where, comparing with our numerical data on eigenvectors, the transition occurs.

In the last part of the chapter we report our finding concerning the excitations of energy minima, focusing on the $m=3$ system. At variance with the dense case, we were not able to represent the Hessian in terms of a simple random matrix ensemble. This is possible only in the large $H$ limit, where our Hessians are similar to a generalised Anderson matrix. For sufficiently low $H$ correlations between diagonal and off-diagonal entries are too strong.

In this system, the spectrum is gapped for $H\gg c$: for $H\sim c$ and lower the spectrum becomes gapless and soft excitations appear. We identify a RS gapless phase and a RSB gapless phase. In the gapless RS phase we find density of states featuring the quartic law of finite-dimensional glassy systems. This observation stands also in the spin glass phase, provided that the minimisation goes sufficiently deep in the energy landscape: indeed, by averaging spectra related to minima with different relative depth in their respective basins, we observe a slight bias towards lower exponents. Only with sufficiently strong over-relaxation we robustly observe the quartic law.
We test the robustness of the quartic scaling also in $m=2$ and $m=4$ system, finding a very good agreement with an exponent four. This results suggests that the statistics of the softest eigenvalues does not depend on $m$, provided it is finite.

Finally, we considered the properties of eigenvectors. We find that for any $H>0$ the softest eigenvectors are localised: their related density functions decay exponentially from a single microscopic core of the sample. However, in the spin glass phase we observe in the $m=3$ system the existence of excitations with a multi-core localisation pattern: eigenvectors amplitudes peaked in different regions of the random graph, separated by a long-range distance of the order of $\log N$. We interpret this as a signal of long-range order. In the XY model ($m=2$) this signal is far weaker, suggesting that the $m=3$ and $m=2$ system may have a different qualitative behavior at the spin glass transition.

Results concerning sparse vector spin glasses are relatively few. Following the thesis work of Dr. Lupo about the XY model, this chapter has to be considered a continuation of his efforts. To our knowledge, the sparse Heisenberg model was not studied in past works as we did in this thesis: we believe then that our work can provide some insight on the nature of the zero temperature spin glass transition in sparse random graphs.


In these years, we studied most exclusively problems related to linear excitations, but some projects considering the problem of non-linear excitations were started. In particular, two interesting lines of research are the following:
\begin{itemize}
    \item In the sparse case, the existence of double-well structures can be probed by considering excitations beyond the Hessian, i. e. to expand the Hamiltonian of the system to higher order. Following \cite{lerner2021low}, a study of cubic and quartic excitation modes of the energy landscape can be pursued. This is particularly relevant for the sparse spin glass model studied in chapter 6.
    \item The problem of finding algorithms able to solve Parisi equations in the whole spin glass phase is an interesting line of research. To my knowledge, no systematic numerical studies for vector models have been carried out. Connected to this problem, it is particularly interesting the study of avalanches, i. e. abrupt reorganisations of spins orientations following a change of configuration induced by an external driving force. The study of avalanches can improve the understanding of the spin glass phase, both in the dense and in the sparse case.
\end{itemize}
In addition to these considerations, we believe that a detailed study of the best-discretisation of the sphere and of the sparse model in different networks can be an interesting continuation of our research.

\printbibliography

\appendix

\chapter{Bessel Functions}
\label{sec:Bessel}

Bessel functions are generalisations of oscillatory functions. They are ubiquitous in all transport processes on variable media, like the propagation of elastic waves on a membrane or along a string with variable thickness. They were introduced by Friedrich Bessel as canonical solutions of the following differential equation
\begin{equation*}
    x^2\frac{d^2 y}{dx^2}+x\frac{dy}{dx}+(x^2-\alpha^2)y=0.
\end{equation*}
A canonical Bessel function of order $\alpha$ (a generic complex number) is written as $J_{\alpha}(x)$. It holds the following integral representation for integer $n$
\begin{equation*}
    J_{n}(x)\,=\,\frac{1}{2\pi}\int_{-\pi}^{\pi}e^{in\tau-x\sin\tau}d\tau
\end{equation*}
This representation holds for non-integer orders if $\Re x>0$. In this thesis, modified Bessel functions were ubiquitous. They are related to canonical Bessel functions through
\begin{equation*}
    I_{\alpha}(x)\,=\,i^{-\alpha}J_{\alpha}(ix)
\end{equation*}
and are also called "hyperbolic Bessel functions". In the following, a formularium for the $I_{\alpha}(z)$:
\begin{itemize}
    \item Asymptotic expansion large $|z|$, $|\operatorname{Arg}(z)|\leq \frac{\pi}{2}$:
    \begin{equation*}
        I_{\alpha}(z)\sim \frac{e^z}{\sqrt{2\pi z}}\left(1-\frac{4\alpha^2-1}{8z}+\frac{(4\alpha^2-1)(4\alpha^2-9)}{2!(8z)^2}-\frac{(4\alpha^2-1)(4\alpha^2-9)(4\alpha^2-25)}{3!(8z)^3}+\dots\right)
    \end{equation*}
    \item Recurrence relations
    \begin{equation*}
        \frac{2\alpha}{z}I_{\alpha}(z)\,=\,I_{\alpha-1}(z)-I_{\alpha+1}(z)
    \end{equation*}
    \begin{equation*}
        \frac{d I_{\alpha}}{dz}=\frac{I_{\alpha-1}(z)+I_{\alpha+1}(z)}{2}
    \end{equation*}
    \item Phase space volume of the local field $\Vec{x}$ acting on a spin 
    \begin{equation*}
        \mathcal{K}_m(x)\equiv \int_{S_m(1)}d^m S\;e^{\Vec{x}\cdot\Vec{S}}
    \end{equation*}
    \begin{equation*}
        \mathcal{K}_m(x)\,=\,\frac{(2\pi)^{m/2}I_{m/2-1}(x)}{x^{m/2-1}}
    \end{equation*}
    \item Activation function (expresses relation between magnetisation and effective field, see TAP equations \eqref{eq:TAP_eqs}):
    \begin{equation*}
        \mathcal{g}_m(x)\equiv \frac{d\log \mathcal{K}_m}{dx}\,=\,\frac{I_{m/2}(x)}{I_{m/2-1}(x)}
    \end{equation*}
    \begin{equation*}
        \mathcal{g}_m(x)\sim
        \begin{cases}
            \frac{x}{m}\qquad x\ll 1 \\
            1-\frac{m-1}{2x}+O(x^{-2})\qquad x\gg 1, m>1 \\
            1-2e^{-2x}+O(e^{-4x})\qquad x\gg 1, m=1
        \end{cases}
    \end{equation*}
    \item Special cases
    \begin{equation*}
        \mathcal{K}_1(x)\,=\,2\cosh(x)\qquad \mathcal{K}_3(x)\,=\,4\pi\frac{\sinh(x)}{x}
    \end{equation*}
    \begin{equation*}
        \mathcal{g}_1(x)=\tanh(x)\qquad \mathcal{g}_3(x)=\coth x-\frac{1}{x}
    \end{equation*}
\end{itemize}

\chapter{Expansion of $\rho(\lambda)$ close to $\lambda=0$}
\label{sec:Expansion_rho_band_edges}

This is a reprint from \cite{franz2021delocalization}.

In the following primed quantities will indicate real parts and double primed imaginary parts of complex variables. 
In general, the solutions to resolvent equations will be complex if $\lambda$ lies in the spectrum of the Hessian. Let us define then 
$x=G_0-G(\lambda)-\lambda=x'+i x''$ and $\tP_h(h)=(1-1/m)P_h(h)$. 
Detailing the real and immaginary part of the resolvent equation, we have
\begin{eqnarray}
  \label{eq:10}
&&  G'=\int dh\; \tilde P_h(h) \frac {h+x'}{(h+x')^2+(x'')^2}\\
&&  G''=\int dh\; \tilde P_h(h) \frac {G''}{(h+x')^2+(x'')^2}
\end{eqnarray}
These equations can be easily solved numerically if we know the
distribution of the cavity field $h$, in particular this is possible 
in the paramagnetic
phase since the distribution $P_h(h)$ is known exactly. 
In this appendix we study analytically the spectral edge. We would like first to illustrate a simple mechanism 
implying the absence of spectral gap, for any choice of the 
parameters in the model, and then
 to show that in the whole paramagnetic phase 
 the spectral density presents a pseudo-gap
$\rho(\lambda)\sim x''(\lambda) \sim\lambda^{m-1}$ at small $\lambda$. 

We can prove that the spectrum is ungapped with the following argument. From the definition of $x$, we have $x=0$ for $\lambda=0$,  while $A<1$ in the whole paramagnetic phase. We should then have $x'=G_0-G'(\lambda)-\lambda \approx -(1+\chi_{SG})\lambda <0$ for small but positive $\lambda>0$. 
But in that case, admitting that  $x''=0$, the resulting
integral for $G$ in (\ref{eq:10}) would be divergent. In order to have
a convergent result for $x'<0$ one clearly needs a small imaginary part
$x''\ne 0$. 
Let us then proceed to estimate
the spectrum in the vicinity of $0$. To this aim we observe that
defining $\epsilon=1-A$, the resolvent equation can be rewritten as
\begin{eqnarray}
  \label{eq:13}
  \lambda +\epsilon x&=&-\int dh\; \tilde P_h(h) \frac {x^2}{h^2(h+x)}\\ 
&=&-\int dh\; \tilde P_h(h) \frac {x^2 h+x |x|^2}{h^2|h+x|^2}\nonumber\\
&=&- x^2 J- x |x|^2 I\nonumber
\end{eqnarray}
with 
\begin{eqnarray}
  \label{eq:14}
&&  J=\int dh\; \tilde P_h(h) \frac {1}{h|h+x|^2}\\
&&  I=\int dh\; \tilde P_h(h) \frac {1}{h^2|h+x|^2}\nonumber
\end{eqnarray}
giving 
\begin{eqnarray}
  \label{eq:15}
&&  I |x|^2=-\epsilon-2 x' J\\
&&\lambda=|x|^2 J \nonumber
\end{eqnarray}
It is clear that for $\lambda\to 0$, in order to compensate for the $\lambda$-independent term in the first of (\ref{eq:15}) at small $x'$ and $x''$,  the integrals $I$ and $J$ must be dominated by divergent contributions.
Using $x''/((h+x')^2+(x'')^2)\approx
\pi\delta(h+x')$, valid for $|x'|\gg x''$ we can estimate the leading behavior of the integrals as:
\begin{eqnarray}
  \label{eq:16}
&&  J\approx \pi \frac{ \tilde P_h(|x'|)}{|x'||x''|}\\
&&  I\approx \pi \frac{ \tilde P_h(|x'|)}{|x'|^2|x''|}\nonumber
\end{eqnarray}
so that 
\begin{eqnarray}
  \label{eq:17}
&&  \epsilon=\pi \frac{\tilde  P_h(|x'|)}{|x''|}\\
&& J=\epsilon/|x'|\nonumber\\
&& |x'|=\lambda/\epsilon \nonumber\\
&&\rho(\lambda)=\frac{m}{m-1}|x''|/\pi=P_h(\lambda/\epsilon)/\epsilon\sim
   \lambda^{m-1}/\epsilon^m \nonumber
\end{eqnarray}
This analysis is valid as long as  $|x'|\ll \epsilon$ and $x''\ll |x'|$, i.e. $\lambda/\epsilon \ll \lambda^{m-1}/\epsilon^m $ or $\lambda\ll \epsilon^{\frac{m-1}{m-2}}$. As we approach the critical point, when
$|x'|\sim \epsilon$ the singular contribution to $J$
would not be divergent any more and the analysis needs to be revised.

\chapter{Statistics of the lowest eigenvalues}
\label{sec:stat_eigs_edges}

The probability distribution of the smallest eigenvalue can be derived very simply once one remembers that in the RS phase $\rho(\lambda)\sim \frac{P_h(\lambda/\Lambda)}{\Lambda^m}$ for small $\lambda$. We have

\begin{equation*}
    \mathcal{P}(\lambda_{min}\leq \lambda)\approx 1-\left(1-\int_0^{\lambda}d\lambda'\frac{P_h(\lambda'/\Lambda)}{\Lambda}\right)^{N-1}
\end{equation*}
\begin{equation*}
    1-\left(1-\int_0^{\lambda}d\lambda'\frac{P_h(\lambda'/\Lambda)}{\Lambda}\right)^{N-1} \approx 1-\exp\left(-\frac{N\lambda^{m}}{m Z_m \Lambda^m}+O(N\lambda^{2m})\right)
\end{equation*}
where we used $P_h(x)\approx x^{m-1}/Z_m$. As declared in the main test, the smallest eigenvalue is a Weibull variable.

\chapter{Transformation of eigenvector moments after a base change}
\label{sec:change_base_eigvec}

Let us consider a random base $\mathcal{B}=\{\Vec{S},\Vec{\epsilon}_1, \dots, \Vec{\epsilon}_{m-1}\}$, where $\Vec{\epsilon}_a\cdot S=0$ for any $a=1,\dots, m-1$. A generic vector orthogonal to $\Vec{S}$ can be written in this base as $\Vec{v}=a_1\Vec{\epsilon}_1+\dots a_{m-1}\Vec{\epsilon}_{m-1}$: if its coordinates are gaussian variables with zero mean and variance $\sigma^2$, then their transformation in the canonical base $\{\Vec{e}_1,\dots,\Vec{e}_{m}\}$ of $\mathbb{R}^m$ are gaussian variables with zero mean and variance $(\sigma')^2=(1-S_{\alpha}^2)\sigma^2$: upon performing the angular average, one finds that their second and fourth moments read
\begin{equation*}
    m_2'=(1-1/m)\sigma^2\qquad m_4'=\frac{3(m^2-1)}{m(m+2)}
\end{equation*}
Generalising this formula to arbitrary gaussian moments, one can compute the geometrical factors emergening when representing non-rotational invariants eigenvector moments \eqref{eq:eigvec_moments} in one of the two bases introduced above. For instance, the IPR from \eqref{eq:eigvec_moments} in the canonical base reads
\begin{eqnarray}
  \label{eq:34}
  \IPR = \frac {3(m^2-1)}{N m^2(m+2)} \int \frac{P_h(h) dh}{|h+G_0-\lambda-G(\lambda)|^4}\;.
\end{eqnarray}
In the base $\mathcal{B}_1\times \cdots \times\mathcal{B}_N$ instead the IPR reads
\begin{eqnarray}
  \label{eq:34}
  \IPR_0 = \frac {3(m-1)}{N m^2} \int \frac{P_h(h) dh}{|h+G_0-\lambda-G(\lambda)|^4}\;
\end{eqnarray}
thus the ratio between the two quantities is $IPR/IPR_0=(m+2)/(m+1)$.

\chapter{Derivation of Monasson Potential}
\label{sec:der_mon_pot}

This is a reprint from \cite{franz2022linear}.

The computation of ${\cal G}=-\beta n f+\Sigma$ follows standard paths \cite{monasson1995structural}, for completeness we sketch it here the main steps: 
\begin{eqnarray}
Z_n=e^{N{\bf G}}=\overline{\sum_{\bm{S}_a} e^{-\beta \sum_{a=1}^n H({\bm S}_a)}\prod_{a<b}^{1,n}\delta({\bm S}_a\cdot{\bm S}_b-q N)}\nonumber
\end{eqnarray}
where $\underset{\bm S}{\sum}(\cdot)\equiv \overset{N}{\underset{k=1}{\prod}}\,\int d\Vec{S}_k\,\delta(S_k-1)(\cdot)$.
Performing the average and using $\overline{H(\bm S)H(\bm S')}= N f(q_{S,S'})$ one gets
\begin{eqnarray}
&&e^{NG}=\exp\left\{\frac{N\beta^2}{2}[ 
n f(1)+n(n-1)f(q)]
\right\}\zeta(q)\nonumber\\
&&\zeta(q)=\sum_{\bm S_a} \prod_{a<b}^{1,n}\delta({\bm S}_a\cdot{\bm S}_b-q)\nonumber
\end{eqnarray}
The quantity $\zeta$ after one Hubbard-Stratonovich transformation and the integration on spins becomes:
\begin{eqnarray}
  \label{eq:4}
  \zeta&&={\rm St_{\hat{q}}}\;\left\{\exp\left[-N\frac{ n(n-1)}{2} {\hat q} q+\frac{{\hat
  q}}{2}\sum_{a\ne b} {\bm S}^a \cdot  {\bm S}^b\right]\right\}\nonumber\\
&&={\rm St_{\hat{q}}}\left\{\exp\left[-N\frac{ n(n-1)}{2} {\hat q} q -N \frac{n}{2} {\hat q}\right]
 \left[\int D_{\hat q}\Vec{h}\,Y(h)^n\right]^N\right\}.\nonumber\\
&&  Y(h)=(2\,\pi)^{m/2}\frac{I_{\frac{m-2}{2}}(h)}{h^{\frac{m-2}{2}}}\nonumber \\
&& \int D_{\hat{q}}\Vec{h}\,(\cdot)\equiv\int \frac{d\Vec{h}}{(2\pi\hat{q})^{m/2}}\,e^{-\frac{h^2}{2\hat{q}}}\,(\cdot)\nonumber
\end{eqnarray}
Putting everything together and using the saddle point equation
$\hat{q}=\beta^2 f'(q)$ we get 
\begin{align}
  \label{eq:43}
{\cal G}(n,T)=&\frac{n \beta^2}{2}\left[  f(1) + (n - 1) (f(q) -  q f'(q)) -  f'(q) \right] \nonumber\\
 &+\log\left[\frac{\int_0^\infty dh\; h^{m-1} e^{-\frac{h^2}{2f'(q)}}Y(\beta h)^n} {\int_0^\infty dh\; h^{m-1} e^{-\frac{h^2}{2f'(q)}}}
 \right]\;,
\\
 &Y(u)= (2\,\pi)^{m/2}\frac{I_{\frac{m-2}{2}}(u)}{u^{\frac{m-2}{2}}}\;.\nonumber
\end{align}
The physical overlap is found by extremizing $\mathcal{G}$ with respect to $q$ and is given by 
\begin{eqnarray}
  \label{eq:45}
  q=\frac{\int_0^\infty dh\; h^{m-1} \exp\left[-\frac{h^2}{2f'(q)}\right] Y(\beta
  h)^{n-2}Y'(\beta h)^2}{\int_0^\infty dh\; h^{m-1} \exp\left[-\frac{h^2}{2f'(q)}\right] Y(\beta h)^n}.\quad
\end{eqnarray}
When $T>T_d$, there is only the $q=0$ solution, the system is in a paramagnetic phase with a unique equilibrium state and
\begin{equation}
    \beta g_{para}\,=\,\frac{\beta^2 f(1)}{2}+\log S_m.
\end{equation}
In the range $T_K<T<T_d$, \eqref{eq:45} has a non-trivial solution, corresponding to a non-zero Configurational Entropy: configurations inside the same state have a non-zero overlap, whereas two configurations belonging to two different states have zero overlap. The stability of the non-trivial $q$ is determined by the positiveness of the Replicon Eigenvalue of the Replica Free-Energy Hessian:
\begin{eqnarray}
 \Lambda&&=1-\beta^2f''(q)\Bigl\langle\Bigl\{\frac{m}{(\beta h)^2}\left[\frac{Y'(\beta h)}{Y(\beta h)}\right]^2+ \\
&& \Bigl\{\frac{Y''(\beta h)}{Y(\beta h)}-\left[\frac{Y'(\beta h)}{Y(\beta h)}\right]^2 -\frac{Y'(\beta h)}{(\beta h)\,Y(\beta h)}\Bigr\}^2+\nonumber \\
&& \frac{2 Y'(\beta h)}{(\beta h)\,Y(\beta h)}\times\Bigl\{\frac{Y''(\beta h)}{Y(\beta h)}-\left[\frac{Y'(\beta h)}{Y(\beta h)}\right]^2-\frac{Y'(\beta h)}{(\beta h)\,Y(\beta h)}\Bigr\}\Bigr\rangle \nonumber \\
&& = 1-\beta^2f''(q)\left[\Bigl\langle\Bigl(\frac{d}{d \beta h}\frac{Y'(\beta h)}{Y(\beta h)}\Bigr)^2\Bigr\rangle+(m-1)\Bigl\langle\Bigl(\frac{Y'(\beta h)}{h Y(\beta h)}\Bigr)^2\Bigr\rangle\right]
\end{eqnarray}
The internal free-energies of TAP states and their Complexity are obtained by eqs.\eqref{eq:45} and they read
\begin{eqnarray}
\label{eq:gS}
& g=-\frac{\beta}{2}[f(1)+(2 n-1) f(q)-(2 n-1)q f'(q) \\
& -f'(q)]-\frac{1}{\beta}\langle\ln Y(\beta h)\rangle_{n}\nonumber \\
& \Sigma=-\frac{n^2\beta^2}{2}[f(q)-q f'(q)]+\ln\zeta-n\langle\ln Y(\beta h)\rangle_{n}
\end{eqnarray}
where $\zeta$ is defined in \eqref{eq:43} and $\langle\cdot\rangle_n$ is an average with respect to
\begin{eqnarray}
  P(h)=\frac{ h^{m-1} \exp\left[-\frac{h^2}{2f'(q)}\right] Y[\beta
  h]^n}{\int_0^\infty dh\; h^{m-1} \exp\left[-\frac{h^2}{2f'(q)}\right] Y[\beta
  h]^n}. \nonumber
\end{eqnarray}
Setting $q$ equal to the correct physical value, one can explore different families of metastable states by varying $n$ at fixed $T$ in the range $[T_K, T_d]$, whereas the equilibrium values in the same interval are computed by setting $n=1$. The equilibrium Replicon vanishes at $T_{d}$ as $(T_d-T)^{1/2}$: at higher temperatures, the thermodynamic equilibrium is completely determined by the paramagnetic state $\boldsymbol{m}=0$. The equilibrium Complexity vanishes at $T_K$ as $T-T_K$: for lesser temperature, the Equilibrium Complexity remains zero, meaning that the Gibbs measure is concentrated on the lowest free-energy states.

The $T=0$ limit is performed sending $T$ and $n$ to zero with $y=n/T$ fixed: the result obtained for the Monasson free energy and the Replicon are retrieved by considering the asymptotic expansions of $Y(x)$, $Y'(x)/Y(x)$ and $Y''(x)/Y(x)$:

\begin{eqnarray}
    Y(x) &&\,\overset{x\rightarrow\infty}{\sim}\frac{(2\pi)^{m/2}\,e^x}{x^{m/2-1}}\,\left[\sqrt{\frac{1}{2\pi x}}+O\left(\frac{1}{x}\right)^{3/2}\right] \\
    \frac{Y'(x)}{Y(x)} &&\,\overset{x\rightarrow\infty}{\sim} 1-\frac{m-1}{2\,x}+O\left(\frac{1}{x}\right)^{2} \\
    \frac{Y''(x)}{Y(x)} &&\,\overset{x\rightarrow\infty}{\sim} 1-\frac{m-1}{x}+O\left(\frac{1}{x}\right)^{2}.
\end{eqnarray}

\chapter{Complexity of Ultra-stable minima}
\label{sec:compl_us_minima_app}

This is reprinted from \cite{franz2022linear}.

In this Appendix we show in greater detail all the computations concerning the Complexity of the Ultra-Stable Minima of the energy.
First of all, we set $u-\chi_{h_0}=\Delta$, and rewrite the Monasson free energy with the cavity gap as
\begin{eqnarray}
\label{eq:G0gapAPP}
& \mathcal{G}_0(y; h_0)= \frac{y^2}{2}[f(1)-f'(1)]-\frac{f''(1)}{2 f'(1)}\Delta^2 \\
& -y f''(1) \Delta-\frac{f''(1)}{2}\Delta(\Delta+2\chi_{h_0})+\ln \zeta(y; h_0) \nonumber \\
& \nonumber \\
& \zeta_{\Delta} = \frac{\int_{h_0}^{\infty}dh\,h^{m-1}e^{-\frac{h^2}{2 f'(1)}+\frac{h}{f'(1)}[f''(1)\Delta+yf'(1)]}}{{\int_0^{\infty}\,dh\,h^{m-1}e^{-\frac{h^2}{2f'(1)}}}} \nonumber\\
& \nonumber \\
& P_{\Delta}(h)\,=\,\frac{\theta(h-h_0)}{Z(y; \Delta)}h^{m-1}e^{-\frac{h^2}{2 f'(1)}+\left[y f'(1)+f''(1)\Delta\right]\frac{h}{f'(1)}}
  \nonumber
\end{eqnarray}
By combining the equations in the main text defining $\chi_{h_0}$ and $u$ and approximating the sums with integrals, we find that $\Delta$ satisfies the self-consistent equation
\begin{equation}
    \Delta\,=\,\frac{h_0^{m-1}\,e^{-\frac{h_0^2}{2 f'(1)}+[y+f''(1)\Delta]h_0}}{\int_{h_0}^{\infty}dh\,h^{m-1}\,e^{-\frac{h^2}{2 f'(1)}+[y+f''(1)\Delta]h}}\equiv P_{\Delta}(h_0)
\end{equation}
In particular, for small $h_0$ one has ($Z_0(y)=1/p_0$ defined in eq.\eqref{eq:p0})
\begin{equation}
\label{eq:Deltah0small}
    \Delta\,=\,\frac{h_0^{m-1}}{Z_0(y)}(1+h_0 y)+O(h_0^{m+1}).
\end{equation}
The expression of $\Sigma(y; h_0)$ is obtained by applying the definition $\Sigma(y; h_0)\,=\,y E(y; h_0)+\mathcal{G}_0(y; h_0)$, and the full expression is
\begin{eqnarray}
    & \Sigma(y; h_0)\,=\,\Sigma(y; 0)-\left[\frac{f''(1)^2}{f'(1)}+f''(1)\right] \\
    & \times\left[\frac{1}{2}+\frac{y f'(1) (\langle h \rangle_{\Delta}-h_0)}{f'(1)+(\langle h \rangle_{\Delta}-h_0) f''(1)\Delta}\right]\Delta^2+\nonumber \\
    & \nonumber \\
    & -y[\chi_{h_0}+yf''(1)]\frac{f'(1)\,(\langle h \rangle_{\Delta}-h_0)}{f'(1)+(\langle h \rangle_{\Delta}-h_0)f''(1)\Delta}\Delta-\chi\Delta\nonumber \\
    & \nonumber \\
    & +\frac{y f'(1) [f''(1)(m-1)-\chi_{h_0}\langle h\rangle_{\Delta}]}{f'(1)+f''(1)(\langle h \rangle_{\Delta}-h_0)\Delta}\Delta\nonumber \\
    & \nonumber \\
    & -\frac{y f'(1) [\langle h \rangle_{\Delta}-\langle h \rangle_0]-\langle h \rangle_{0}(\langle h \rangle_{\Delta}-h_0)\Delta}{f'(1)+(\langle h \rangle_{\Delta}-h_0)\Delta}+\ln[\zeta_{\Delta}(y)/\zeta_0(y)]\nonumber
\end{eqnarray}
where $\langle \cdot \rangle_{\Delta}$ is a mean according to $P_{\Delta}$ in \eqref{eq:G0gapAPP}. This nasty expression can be simplified a lot by expanding for low cavity gap: by substituting \eqref{eq:Deltah0small} one gets
\begin{equation}
\label{eq:Sigmah0small}
     \Sigma=\,\Sigma_0-\left[\frac{1+y\,\langle h\rangle_0}{m\,Z_0}\right]h_0^m+O(h_0^{m+1}).
 \end{equation}
For $h_0=O(1)$, $\Sigma$ becomes proportional to $h_0^{(max)}(y)-h_0$, thus vanishing at a certain maximal cavity gap. This last quantity is $O(1)$ far from $y_{gs}$; as this point is approached, the maximal cavity gap is expected to vanish, since  ultra-stable minima cannot be lower in energy than the ground state level. Taking $\Sigma=0$ in \eqref{eq:SigmaGapSmallh0}, we can consider $\Sigma_0$ small and expand it linearly in $y_{gs}-y$, getting
\begin{eqnarray}
    &\left[\frac{1+y\,\langle h\rangle_0}{m\,Z_0}\right](h_0^{max})^m\simeq \frac{d\Sigma_0}{dy}(y_{gs})(y_{gs}-y)\nonumber\\
    & h_0^{(max)}\simeq A\,(y_{gs}-y)^{1/m} \\
    & A\,=\,\left[(m\,Z_0)\frac{\Sigma_0'(y_)}{1+y\langle h \rangle_0}\right]^{1/m}\Bigl |_{y=y_{gs}}
\end{eqnarray}
that is, a singularity approaching $y_{gs}$.

\chapter{Response function of Ultra-Stable minima}
\label{sec:resp_US_comp}

This is reprinted from \cite{franz2022linear}.

This appendix is devoted to the computation of the linear response function of the system when perturbed in a ultra-stable configuration at zero temperature: we show that the linear response function in this case is given by the order parameter $u$, which satisfies
\begin{equation*}
    u\,=\,\chi_{h_0}+P_{\Delta}(h_0)
\end{equation*}
Suppose to perturb the system with an external field $\Vec{\epsilon}_i$ on each site: the static linear response function is given by
\begin{eqnarray}
    & \mathcal{R}\,=\,\frac{1}{N}\sum_{i,\alpha}R_{ii}^{\alpha\alpha} \\
    & R_{i j}^{\alpha \beta}\,=\,\frac{\partial \overline{\langle S_i^{\alpha}  \rangle}}{\partial \epsilon_j^{\beta}}\Bigl|_{\epsilon=0}
\end{eqnarray}
where off-diagonal terms of the response matrix are neglected since their disorder average is zero. Here $\langle\cdot\rangle$ is an average according to Kac-Rice-Moore measure:
\begin{eqnarray}
\label{eq:KRM}
    P_{KRM}\,\propto\,e^{-y{\cal H}}
  \prod_{i,\alpha}\delta\left(
{\cal H}_i^{\alpha'}-\mu_i S_i^\alpha
\right)\left| \det\left( H''-\text{diag}(\mu)\right)\right|
  \nonumber
\end{eqnarray}
Then, one has for the response
\begin{eqnarray}
\label{eq:R2}
    & R_{ii}^{\alpha\alpha}\,=\,\langle (S_i^{\alpha})^2 \rangle-\langle S_i^{\alpha} \rangle^2+i\langle S_i^{\alpha} \hat{S}_i^{\alpha} \rangle
    \\ 
    & \rightarrow\mathcal{R}\,=\,\frac{1}{N}\overset{N}{\underset{k=1}{\sum}}\overline{\langle \Vec{S}_k\cdot i\Vec{\hat{S}}_k \rangle}\nonumber
\end{eqnarray}
where $\hat{S}_i^{\alpha}$ are Lagrange multipliers that ensures the $\boldsymbol{S}$ configuration is one of minimum of $\mathcal{H}$ (they are obtained from the Fourier Representation of the delta function in \eqref{eq:KRM}). After performing similar passages to those used to derive the zero-temperature Monasson potential of ultra-stable minima, one finds for the relevant part of the integrals involved in the second eq. of \eqref{eq:R2}
\begin{eqnarray*}
   & \prod_l\int d\Vec{\mu}_l\int d\Vec{\hat{S}}_l\;(\Vec{S}_k\cdot i\Vec{\hat{S}}_k)e^{-\frac{f'(1)}{2}\hat{S}_l^2-i[\mu_l-u-y f'(1)](\Vec{S}_k\cdot i\Vec{\hat{S}}_k)}\propto \\
   & \nonumber \\
   & \propto\,-\frac{\int d\boldsymbol{\mu}\frac{\partial\, e^{-\sum_l\frac{c_l^2}{2 f'(1)}}}{\partial c_k}}{\int d\boldsymbol{\mu}\,e^{-\sum_l\frac{c_l^2}{2 f'(1)}}}\Bigl|_{c_l\equiv\mu_l-u-y f'(1)}\,=\,\frac{\overline{\mu}-u-y f'(1)}{f'(1)}
\end{eqnarray*}
The remainder of the integrals and factors cancel out with the normalization, and in the end we get
\begin{equation}
    \mathcal{R}\,=\,\frac{1}{N}\sum_{k,\alpha}\,\overline{R_{kk}^{\alpha\alpha}}\,=\,\frac{1}{f'(1)}[\overline{\mu}-u-y f'(1)]\equiv u.
\end{equation}
To conclude this Appendix, we show that $u$ is always smaller than the susceptibility $\chi$ of the typical minimum configurations. From the definition of $\chi_{h_0}$ (eq.\eqref{eq:chih0})
\begin{eqnarray*}
    & \chi_{h_0}\,=\,\chi-\frac{(m-1)\int_0^{h_0}dh\,h^{m-2}e^{-\frac{h^2}{2 f'(1)}+[y f'(1)+f''(1)\Delta]\frac{h}{f'(1)}}}{\int_{h_0}^{\infty}dh\,h^{m-1}e^{-\frac{h^2}{2 f'(1)}+[y f'(1)+f''(1)\Delta]\frac{h}{f'(1)}}} \\
    & \nonumber \\
    & \equiv \chi-Q(h_0)<\chi
\end{eqnarray*}
one finds
\begin{equation*}
    u\,=\,\chi-[Q(h_0)-P_{\Delta}(h_0)].
\end{equation*}
We notice that $Q(h_0)\,=\,(m-1)\int_0^{h_0}\,dh\,\tilde{g}(h)$ and $P_{\Delta}(h_0)\,=\,h_0 \tilde{g}(h_0)$, and thus we must determine if $(m-1)\int_0^{h_0}\,dh \tilde{g}(h)-h_0 \tilde{g}(h_0)>0$; this inequality is indeed always verified for $m>2$, since in this circumstance $Q$ is a convex function: we conclude that $u<\chi$. In particular, for small $h_0$ it holds
\begin{equation}
    u\,=\,\chi-\frac{1}{Z_0}[(m-1)(m-2)-1]h_0^{m-1}+O(h_0^m).
\end{equation}

\chapter{Spectrum of Ultra-Stable minima}
\label{sec:spec_US}

This is reprinted from \cite{franz2022linear}.

When a cavity gap $h_0$ is present, one has a spectral gap $\lambda_0>0$ if the quantity $\Re x(\lambda)\,=\,f''(1)[\chi_{h_0}-G_R(\lambda)]-\lambda$ satisfies $|\Re x(\lambda_0)|<h_0$: in these circumstances, the spectral gap is determined by solving
\begin{eqnarray}
\label{eq:gapSpeceqs}
& 1\,=(m-1)f''(1)\,\int_{h_0}^{\infty}dh\,\frac{P_{h_0}(h)}{[h+\Re x(\lambda_0)]^2} \\
& \lambda_0\,=\,(m-1)(\Re x(\lambda_0))^2\,\int_{h_0}^{\infty}dh\,\frac{P_{h_0}(h)}{h\,[h+\Re x(\lambda_0)]^2}.\nonumber
\end{eqnarray}
We shall now consider the small $h_0$ limit of these last equations and the two cases $y>y_{mg}$ and $y=y_{mg}$. Let's begin with $y>y_{mg}$: the first integral in \ref{eq:gapSpeceqs} is dominated by the values of $h$ close to the cavity gap $h_0$; here $P_{h_0}(h_0)\sim h_0^{m-1}$, thus integrating in a small region $[h_0,c\,h_0]$ we get ($x_0\equiv x(\lambda_0)$)
\begin{eqnarray}
\label{eq:x_0scal}
& 1\sim \frac{(1-1/c)(-x_0)^{(m-1)}}{Z_{h_0}(h_0+x_0)} \nonumber \\
& x_0\,\sim\,-h_0+\frac{(1-1/c)}{Z_{h_0}}|x_0|^{m-1}
\end{eqnarray}
which ensures us that $|x_0|<h_0$. Then, rearranging the second of \ref{eq:gapSpeceqs}
\begin{equation*}
    \lambda_0\,=\,f''(1)\chi_{h_0}-x_0-f''(1)(m-1)\left\langle\frac{1}{h+x_0}\right\rangle_{h_0}
\end{equation*}
expanding $1/(h+x_0)$ in $x_0/h$ and simplifying:
\begin{equation*}
    \lambda_0\,=\,\Lambda|x_0|-f''(1)(m-1)x_0^2\left\langle\frac{1}{h^3}\right\rangle_{h_0}+O(x_0^3)
\end{equation*}
and plugging into this last equation eq. \eqref{eq:x_0scal}, it is found at leading order in $h_0$ 
\begin{equation}
    \lambda_0\,=\,\Lambda\,h_0+O(h_0^2)
\end{equation}
We consider now the case $y=y_{mg}$, i.e. $\Lambda=0$. Here one finds from the first of \eqref{eq:gapSpeceqs}
\begin{eqnarray}
\label{eq:gapSpecCrit}
    && \left\langle\frac{1}{h^2}\right\rangle_0\,=\,\left\langle\frac{1}{(h+x_0)^2}\right\rangle_{h_0}
\end{eqnarray}
which after a few manipulation yields
\begin{eqnarray}
\\
    && |x_0|\,=\begin{cases}
    \frac{h_0^{m-2}}{2\,(m-2)\,Z_0\,\langle 1/h^3 \rangle_0}+O(h_0^{m-1}),\,\,m>3 \\
    \\
    \frac{h_0}{2\,|\ln h_0|}+O(h_0^{2}),\,\,m=3
    \end{cases}
\end{eqnarray}
From the second of \eqref{eq:gapSpeceqs} then expanding $\Lambda_{h_0}$, setting $\Lambda=0$ and keeping terms up to order $x_0^2$, we find
\begin{equation}
    \lambda_0\,=\begin{cases}
    \left[\frac{f''(1)(m-1)}{4\, (m-2)^2\,Z_0^2\,\langle 1/h^3 \rangle_0}\right] h_0^{2(m-2)}+O(h_0^{2(m-1)}),\,\,\,m>3 \\
    \\
    \left[\frac{f''(1)}{2\, \,Z_0}\right] \frac{h_0^{2}}{|\ln h_0|}+O(h_0^{4}),\,\,\, m=3.
    \end{cases}
\end{equation}

We shall now consider the scaling of the spectral density and of the IPR close to $\lambda_0$. The equations for $I$ and $J$ found for instance in \ref{sec:Expansion_rho_band_edges} are still valid if one replaces the ungapped $P_0(h)$ with the gapped one $P_{h_0}(h)$:
\begin{eqnarray*}
    && \lambda+\Lambda_{h_0}x\,=\,-x^2\,J_{h_0}-x|x|^2\,I_{h_0} \\
    && J_{h_0}(\lambda)\,=\,f''(1) (m-1) \left\langle\frac{1}{h |h+x|^2}\right\rangle_{h_0} \\
    && I_{h_0}(\lambda)\,=\,f''(1) (m-1) \left\langle\frac{1}{h^2 |h+x|^2}\right\rangle_{h_0}
\end{eqnarray*}
Differently from the gapless case, here the integrals $J_{h_0}$ and $I_{h_0}$ are always finite in the limit $\lambda\rightarrow\lambda_0$, for any $y_{mg}\leq y\leq y_{gs}$: at $\lambda=\lambda_0$, it follows directly from $h_0+x_0>0$. For $\lambda>\lambda_0$, one finds $h+x\simeq h+x_0-\frac{m-3}{m-2}(\lambda-\lambda_0)+\Im x$, since $d\Re x(\lambda_0)/d\,\lambda=-\frac{m-3}{m-2}$; so the integrals are well defined if and only $h_0+x_0+\Im x>\frac{m-3}{m-2}(\lambda-\lambda_0)$, so necessarily $\Im x\gg O(\lambda-\lambda_0)$. In fact, one finds that the spectral density has a square root behavior close to the spectral edge:

\begin{eqnarray}
    & \rho\simeq\sqrt{\frac{(1-C_{h_0})(\lambda-\lambda_0)}{J_{h_0}}} \\
    & C_{h_0}\,=\,|x_0|\left(\frac{m-3}{m-2}\right)[2\,J_0+|x_0|\,(dJ(x_0)/dx)] \\
    & J_{h_0}\,=\,f''(1)(m-1)\left\langle\frac{1}{h(h+x_0)^2}\right\rangle_{h_0}
\end{eqnarray}

As a consequence, the related lower edge eigenvectors of ultra-stable minima are found to be fully delocalised. Indeed, the IPR close to the spectral edge for $y>y_{mg}$ behaves as
\begin{eqnarray}
    && N\,IPR(\lambda)\propto \int_{h_0}^{\infty}\,\frac{dh\,P_{h_0}(h)}{|h+x|^4}\,=\,\int_{h_0}^{\infty}\,\frac{dh\,P_{h_0}(h)}{(h+x_0)^4} \nonumber \\ 
    && +O(\lambda-\lambda_0) \approx \frac{|x_0|^{m-1}}{3\,Z_0\,(h_0+x_0)^3}\sim h_0^{-2\,(m-1)}
\end{eqnarray}
At the critical point we find by similar manipulations
\begin{eqnarray}
    N\,IPR(\lambda_0)\sim \begin{cases}
    1/h_0,\,\,\,m=3 \\
    \ln h_0,\,\,\, m=4 \\
    const,\,\,\, m\geq 5.
    \end{cases}
\end{eqnarray}When a cavity gap $h_0$ is present, one has a spectral gap $\lambda_0>0$ if the quantity $\Re x(\lambda)\,=\,f''(1)[\chi_{h_0}-G_R(\lambda)]-\lambda$ satisfies $|\Re x(\lambda_0)|<h_0$: in these circumstances, the spectral gap is determined by solving
\begin{eqnarray}
\label{eq:gapSpeceqs}
& 1\,=(m-1)f''(1)\,\int_{h_0}^{\infty}dh\,\frac{P_{h_0}(h)}{[h+\Re x(\lambda_0)]^2} \\
& \lambda_0\,=\,(m-1)(\Re x(\lambda_0))^2\,\int_{h_0}^{\infty}dh\,\frac{P_{h_0}(h)}{h\,[h+\Re x(\lambda_0)]^2}.\nonumber
\end{eqnarray}
We shall now consider the small $h_0$ limit of these last equations and the two cases $y>y_{mg}$ and $y=y_{mg}$. Let's begin with $y>y_{mg}$: the first integral in \ref{eq:gapSpeceqs} is dominated by the values of $h$ close to the cavity gap $h_0$; here $P_{h_0}(h_0)\sim h_0^{m-1}$, thus integrating in a small region $[h_0,c\,h_0]$ we get ($x_0\equiv x(\lambda_0)$)
\begin{eqnarray}
\label{eq:x_0scal}
& 1\sim \frac{(1-1/c)(-x_0)^{(m-1)}}{Z_{h_0}(h_0+x_0)} \nonumber \\
& x_0\,\sim\,-h_0+\frac{(1-1/c)}{Z_{h_0}}|x_0|^{m-1}
\end{eqnarray}
which ensures us that $|x_0|<h_0$. Then, rearranging the second of \ref{eq:gapSpeceqs}
\begin{equation*}
    \lambda_0\,=\,f''(1)\chi_{h_0}-x_0-f''(1)(m-1)\left\langle\frac{1}{h+x_0}\right\rangle_{h_0}
\end{equation*}
expanding $1/(h+x_0)$ in $x_0/h$ and simplifying:
\begin{equation*}
    \lambda_0\,=\,\Lambda|x_0|-f''(1)(m-1)x_0^2\left\langle\frac{1}{h^3}\right\rangle_{h_0}+O(x_0^3)
\end{equation*}
and plugging into this last equation eq. \eqref{eq:x_0scal}, it is found at leading order in $h_0$ 
\begin{equation}
    \lambda_0\,=\,\Lambda\,h_0+O(h_0^2)
\end{equation}
We consider now the case $y=y_{mg}$, i.e. $\Lambda=0$. Here one finds from the first of \eqref{eq:gapSpeceqs}
\begin{eqnarray}
\label{eq:gapSpecCrit}
    && \left\langle\frac{1}{h^2}\right\rangle_0\,=\,\left\langle\frac{1}{(h+x_0)^2}\right\rangle_{h_0}
\end{eqnarray}
which after a few manipulation yields
\begin{eqnarray}
\\
    && |x_0|\,=\begin{cases}
    \frac{h_0^{m-2}}{2\,(m-2)\,Z_0\,\langle 1/h^3 \rangle_0}+O(h_0^{m-1}),\,\,m>3 \\
    \\
    \frac{h_0}{2\,|\ln h_0|}+O(h_0^{2}),\,\,m=3
    \end{cases}
\end{eqnarray}
From the second of \eqref{eq:gapSpeceqs} then expanding $\Lambda_{h_0}$, setting $\Lambda=0$ and keeping terms up to order $x_0^2$, we find
\begin{equation}
    \lambda_0\,=\begin{cases}
    \left[\frac{f''(1)(m-1)}{4\, (m-2)^2\,Z_0^2\,\langle 1/h^3 \rangle_0}\right] h_0^{2(m-2)}+O(h_0^{2(m-1)}),\,\,\,m>3 \\
    \\
    \left[\frac{f''(1)}{2\, \,Z_0}\right] \frac{h_0^{2}}{|\ln h_0|}+O(h_0^{4}),\,\,\, m=3.
    \end{cases}
\end{equation}

We shall now consider the scaling of the spectral density and of the IPR close to $\lambda_0$. Equations \eqref{eq:lambdaIJ} are still valid if one replaces the ungapped $P_0(h)$ with the gapped one $P_{h_0}(h)$:
\begin{eqnarray*}
    && \lambda+\Lambda_{h_0}x\,=\,-x^2\,J_{h_0}-x|x|^2\,I_{h_0} \\
    && J_{h_0}(\lambda)\,=\,f''(1) (m-1) \left\langle\frac{1}{h |h+x|^2}\right\rangle_{h_0} \\
    && I_{h_0}(\lambda)\,=\,f''(1) (m-1) \left\langle\frac{1}{h^2 |h+x|^2}\right\rangle_{h_0}
\end{eqnarray*}
Differently from the gapless case, here the integrals $J_{h_0}$ and $I_{h_0}$ are always finite in the limit $\lambda\rightarrow\lambda_0$, for any $y_{mg}\leq y\leq y_{gs}$: at $\lambda=\lambda_0$, it follows directly from $h_0+x_0>0$. For $\lambda>\lambda_0$, one finds $h+x\simeq h+x_0-\frac{m-3}{m-2}(\lambda-\lambda_0)+\Im x$, since $d\Re x(\lambda_0)/d\,\lambda=-\frac{m-3}{m-2}$; so the integrals are well defined if and only $h_0+x_0+\Im x>\frac{m-3}{m-2}(\lambda-\lambda_0)$, so necessarily $\Im x\gg O(\lambda-\lambda_0)$. In fact, one finds that the spectral density has a square root behavior close to the spectral edge:

\begin{eqnarray}
    & \rho\simeq\sqrt{\frac{(1-C_{h_0})(\lambda-\lambda_0)}{J_{h_0}}} \\
    & C_{h_0}\,=\,|x_0|\left(\frac{m-3}{m-2}\right)[2\,J_0+|x_0|\,(dJ(x_0)/dx)] \\
    & J_{h_0}\,=\,f''(1)(m-1)\left\langle\frac{1}{h(h+x_0)^2}\right\rangle_{h_0}
\end{eqnarray}

As a consequence, the related lower edge eigenvectors of ultra-stable minima are found to be fully delocalised. Indeed, the IPR close to the spectral edge for $y>y_{mg}$ behaves as
\begin{eqnarray}
    && N\,IPR(\lambda)\propto \int_{h_0}^{\infty}\,\frac{dh\,P_{h_0}(h)}{|h+x|^4}\,=\,\int_{h_0}^{\infty}\,\frac{dh\,P_{h_0}(h)}{(h+x_0)^4} \nonumber \\ 
    && +O(\lambda-\lambda_0) \approx \frac{|x_0|^{m-1}}{3\,Z_0\,(h_0+x_0)^3}\sim h_0^{-2\,(m-1)}
\end{eqnarray}
At the critical point we find by similar manipulations
\begin{eqnarray}
    N\,IPR(\lambda_0)\sim \begin{cases}
    1/h_0,\,\,\,m=3 \\
    \ln h_0,\,\,\, m=4 \\
    const,\,\,\, m\geq 5.
    \end{cases}
\end{eqnarray}

\chapter{Typical subtrees in random regular graphs}
\label{sec:typ_subtree}

\begin{figure}
    \centering
    \includegraphics[width=0.9\textwidth]{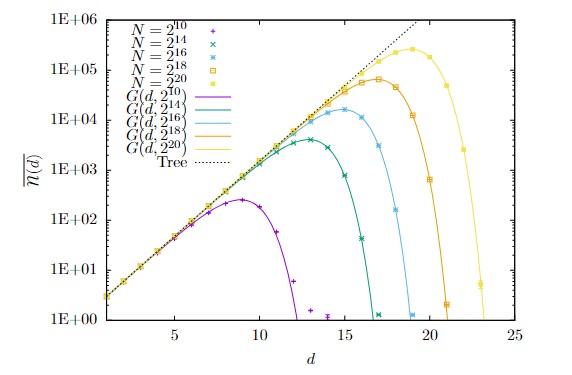}
    \caption{Number of neighbors at distance $d$ in a Random Regular Graph with connectivity $c=3$. Numerical measures are compared with the analytical prediction.}
    \label{fig:NsitesDist}
\end{figure}

In this appendix we estimate the distance from a reference vertex at which the first loop containing the same vertex appears. In order to do this, we consider the exact number of neighbors at distance $d$ on a random regular graph and the same quantity in a regular tree: these are given for $d>0$ by the following formulae \cite{tishby2022mean}
\begin{equation}
\label{eq:growth_RRG}
    \mathcal{N}(d)\,=\,Ne^{\eta}\left\{e^{-\eta(c-1)^{d-1}}-e^{-\eta(c-1)^{d}}\right\}
\end{equation}
\begin{equation}
\label{eq:growth_regular_tree}
    \mathcal{N}_{tree}(d)\,=\,c(c-1)^{d-1}.
\end{equation}
where $\eta=\frac{c}{N(c-2)}$. In figure \ref{fig:NsitesDist} we compare numerical measures of the number of neighbors at distance $d$ in a RRG with formula \eqref{eq:growth_RRG}.
The distance of first loop $d_*$ is estimated through the equation
\begin{equation}
\label{eq:condition_first_loop}
    \mathcal{N}_{tree}(d_*)-\mathcal{N}(d_*)=2.
\end{equation}
Indeed, imagine to generate a graph by starting from a generic root: after forming the first generation, one can progressively assign links in sequence, leaf by leaf, forming new generations. By definition, a loop is created whenever two leaves belonging to the same generation are connected with each other: this implies two less leaves for that shell, compared with what would be expected in a regular tree. 
We can get an explicit expression for $d_*$ by expanding Eq. \eqref{eq:condition_first_loop} for $\eta (c-1)^d$ small, corresponding to
\begin{equation}
\label{eq:L_def}
    d \ll \frac{\log\left[\frac{(c-2)N}{c}\right]}{\log(c-1)}\equiv L
\end{equation} 
where this last quantity defines the global scale of the graph. Retaining terms up to order $1/N$, Eq. \eqref{eq:condition_first_loop} can be rewritten as a second degree algebraic equation
\begin{equation*}
    \mathcal{N}_{tree}(d_*)^2-2\mathcal{N}_{tree}(d_*)-\frac{4(c-2)}{c}N=0
\end{equation*}
and we extract $d_*$ for large $N$ from its solution, finding
\begin{equation}
\label{eq:d_star}
    d_*\,=\,\frac{1}{2}\frac{\log\left[\frac{(c-2)N}{c}\right]}{\log(c-1)}+\left[1-\frac{\log(c/2)}{\log(c-1)}\right]=\frac{L}{2}+const
\end{equation}
that is, first loops typically form at a scale which is half the global scale of the system. Notice that the reasoning behind this derivation can be easily generalised, yielding the typical distance at which $O(N^a)$ loops are created, for some $0<a<1$. One can easily verify that
\begin{equation}
    d_*^{(a)}\,=\,\left(\frac{1+a}{2}\right)L+const.
\end{equation}

\end{document}